\documentclass[a4paper,10pt]{article}
\pdfoutput=1 

\usepackage{jheppub} 
\usepackage{graphicx,floatflt,amssymb,rotate,epsfig}
\usepackage[utf8]{inputenc}
\usepackage[inline]{enumitem}
\usepackage{mathrsfs}
\usepackage{amssymb}
\usepackage{amsmath}
\usepackage{color}
\usepackage{graphicx}
\usepackage{subfig}
\usepackage{multirow}
\usepackage{mathtools}
\usepackage{float}
\usepackage{pstricks}
\usepackage[toc,page]{appendix}
\usepackage{pifont}
\usepackage{braket}
\usepackage{bbold}
\usepackage{hyperref}
\usepackage{newfloat}
\usepackage{pdflscape}
\DeclareCaptionType{equ}[][]

\usepackage{relsize}
\def\BaBar{{\mbox{\slshape B\kern-0.1em{\smaller A}\kern-0.1em B\kern-0.1em{\smaller A\kern-0.2em R}}}}

\newcommand{\ba}{\begin{array}}
	\newcommand{\ea}{\end{array}}
\def\beq{\begin{equation}}
\def\eeq{\end{equation}}
\def\bea{\begin{eqnarray}}
\def\eea{\end{eqnarray}}
\def\nn{\nonumber}

\def\roughly#1{\mathrel{\raise.3ex\hbox
		{$#1$\kern-.75em\lower1ex\hbox{$\sim$}}}}
\def\lsim{\roughly<}
\def\gsim{\roughly>}
\def\sla#1{\raise.15ex\hbox{$/$}\kern-.57em #1}

\def\bd{B_d^0}

\def\order{\lower 1.8ex \hbox{\LARGE\~{}}}
\def\btopilnu{B \to \pi\ell\nu}

\def\btopi0lnu{B^+ \to \pi^0 \ell\nu}

\def\btorho0nu{B^+ \to \rho^0 \ell\nu}
\def\btorholnu{B \to \rho \ell\nu}

\newcommand*{\rom}[1]{\expandafter\@slowromancap\romannumeral #1@}

\def\bd0tau{B\to D \tau\nu_{\tau}}

\def\be {\begin{equation}}
\def\ee {\end{equation}}

\usepackage[normalem]{ulem}

\definecolor{darkgreen}{cmyk}{1,0,1,0.4}

\definecolor{pink}{cmyk}{0.4,1,0.3,0}

\def\com2#1{\textcolor{red}{\it{#1}}}

\title{\boldmath Study of the $b \to d \ell\ell$ transitions in the Standard Model and test of New Physics sensitivities}

\author[a]{Aritra Biswas,}
\emailAdd{iluvnpur@gmail.com}
\author[a]{Soumitra Nandi,}
\emailAdd{soumitra.nandi@iitg.ac.in}
\author[b]{Sunando Kumar Patra,}
\emailAdd{sunando.patra@gmail.com}
\author[a]{and Ipsita Ray}
\emailAdd{ipsitaray02@gmail.com}

\affiliation[a]{Department of Physics, Indian Institute of Technology Guwahati, Assam 781039, India}
\affiliation[b]{Department of Physics, Bangabasi Evening College}

\abstract{ After incorporating all the available experimental data and the most up-to-date Lattice and light cone sum rule (LCSR) inputs on the form factors, we analyze the exclusive $b\to u\ell\nu_{\ell}$ and $b\to d\ell \ell$ decays simultaneously. We have extracted the shapes of all the associated form factors using which we have provided predictions in the standard model for the branching ratios, direct CP asymmetries and isospin asymmetry for $B\to\pi\ell \ell$ and various angular observables for $B\to\rho\ell \ell$ transitions. Also, we have tested the sensitivities of these observables towards physics beyond the standard model (BSM).For the $B\to\rho\ell \ell$ decays, we have defined tagged and untagged observables and predicted them in the SM and BSM. In the context of BSM, we have found some compelling information. The respective predictions in a few benchmark scenarios are given, which can be tested in the experiments at the LHCb and the Belle.}

\arxivnumber{}   

\begin{document}
	\maketitle
	\section{Introduction}
	
	The neutral and charged current semileptonic $B$ meson decays have gained much attention over the last decades from both theoretical and experimental fronts. Deviations of a few $\sigma$ in some observables in exclusive $b\to c \ell(\tau)\nu$ and $b\to s\ell\ell$ (with $\ell= e$ or $\mu$) decay modes from their corresponding Standard Model (SM) predictions have been observed \cite{hflavWeb,pdgWeb}. These two sectors are under constant scrutiny from experimental and theoretical studies, and a lot of progress has been made on both sides. Although originally formed as observables sensitive to lepton flavor universality (LFU) violating new physics (NP), the deviations are not significant to claim a contribution from a dynamics beyond the SM. To conclude further, more precise predictions and measurements on these and other relevant observables are required. In this respect, the $b\to sll$ decays secure an important position where plenty of observables have been measured by different experimental collaborations~\cite{Aaij:2020nrf,Aaij:2016flj,Sirunyan:2018jll,Aaij:2015esa,Abdesselam:2019wac,Aaij:2019wad,ATLAS:2018gqc}, thus providing a scope for a thorough understanding on how to handle the SM uncertainties and as a result define ``cleaner" observables sensitive to NP.

	Similar to the $b\to s\ell\ell$ transitions, the $b\to d\ell\ell$ transitions are also flavor-changing neutral current (FCNC) in nature, and are loop-suppressed in the SM. However, unlike the $b\to s\ell\ell$ modes, we have less amount of information available on the relevant exclusive decay modes from experiment\footnote{This might be due to the low branching ratio of the $b\to d\ell\ell$ modes.} as well as from the theory, like the non-availability of the lattice inputs on $B\to \rho$ form factors. Note that $b\to d\ell\ell$ decays are sensitive to the CKM element $V_{td}$ hence to the CP-violating phase $\beta$, whereas the $b\to s\ell\ell$ decays are sensitive to $V_{ts}$ with the corresponding CP-phase $\beta_s \approx 0$. Therefore, we expect a relatively large CP violation in the $b\to d\ell\ell$ decays within the SM, which provides further motivation towards a systematic, robust and precise analysis of these modes within the framework of the SM before using them as potential building blocks to constrain the parameters associated with an NP scenario after the arrival of sufficient data. In addition, in the SM, the decay rates are suppressed in $b\to d\ell\ell$ as compared to the $b\to s \ell\ell$ channels, which is an effect of CKM suppression in $b\to d \ell\ell$ decays. For example, the decay rate $\Gamma(B\to \pi\mu\mu)$ is suppressed by at least two orders of magnitude compared to $\Gamma(B\to K \mu\mu)$. Therefore, it would be easy to discriminate the NP effect in $b\to d\ell \ell$ transitions compared to that in $b\to s \ell\ell$ decays.

	The exclusive decays corresponding to $b\to d\ell\ell$ transition which we will consider in this paper are $B \to\pi \ell\ell$ and $B \to \rho\ell\ell$. So far, we do not have data available on $B\to \rho\ell\ell$ decays. However, there are some data available on the branching fraction (BR) and CP-asymmetry in $B^{\pm} \to\pi^{\pm}\mu\mu$ decays reported by LHCb~\cite{LHCb:2012de,LHCb:2015hsa}. In theory, the main aspects challenging a precise estimate for these modes and related observables lie in handling the form factors and the long-distance effects. The non-perturbative methods like lattice and light-cone sum rules (LCSR) are useful in constraining the form factors' high and low $q^2$ (dilepton mass-invariant) behaviours, respectively. An exclusive neutral current $B\to\pi\ell\ell$ transition is characterized by three form factors: $f_{+,0,T}(q^2)$. For these form factors, the lattice inputs are available for zero, and non-zero recoils \cite{FermilabLattice:2015cdh,Lattice:2015tia,Colquhoun:2022atw,Flynn:2015mha}. In refs. \cite{Ball:2004ye, Duplancic:2008ix}, the form factors are computed for the first time to next-to-leading order (NLO) in twist 3 using pion distribution amplitude.
	In refs.~\cite{Wang:2015vgv,Khodjamirian:2017fxg,Lu:2018cfc}, following the LCSR approach, the estimates on $f_{+,0}(q^2=0)$ are obtained, whereas $f_{+,0,T}(q^2)$ are estimated in refs. \cite{Gubernari:2018wyi,Leljak:2021vte} for a few values of $q^2$ (including $q^2=0$). Note that LCSR estimates are more reliable near the maximum recoil. The estimates in refs.\cite{Wang:2015vgv,Lu:2018cfc} and \cite{Gubernari:2018wyi} are obtained using the B-meson light-cone distribution amplitudes (LCDA), whereas refs.~\cite{Khodjamirian:2017fxg,Leljak:2021vte} uses the $\pi$-meson LCDA. Ref. \cite{Khodjamirian:2017fxg} uses expressions up to twist-four accuracy at leading order (LO) in $\alpha_s$ and expressions
		up to twist-three accuracy at NLO in $\alpha_s$. Ref. \cite{Wang:2015vgv} includes NLO correction and ref. \cite{Lu:2018cfc} includes higher-twist corrections up to twist-6 at LO in $\alpha_s$. In ref.\cite{Gubernari:2018wyi}, the authors provided expressions for the two-particle and three -particle contributions up to twist four. Also, experimental data on the differential rates are available for these modes  \cite{Ha:2010rf,Lees:2012vv,delAmoSanchez:2010af,Sibidanov:2013rkk}. For a recent update on the extraction of $|V_{ub}|$ from $B \to \pi \ell\nu_{\ell}$ modes the readers may look at the following refs.~\cite{hflavWeb,Biswas:2021qyq,Martinelli:2022tte}. All three form factors mentioned above are correlated in the analysis of Fermilab-MILC \cite{Lattice:2015tia}. Also, the estimates done in LCSR analyses have correlations between the three form factors. From the above discussion, it is clear that a combined study of the $B \to\pi \ell\ell$ and $B \to \pi \ell\nu_{\ell}$ decay modes using all the available inputs are relevant and necessary, which so far is missing in the literature.

	The situation is a bit grim for the form factors associated with the $B\to\rho$ transitions. Both the $B \to \rho\ell\nu_{\ell}$ and $B\to\rho \ell\ell$ decays are sensitive to the four form factors: $V(q^2)$, $A_0(q^2)$, $A_1(q^2)$ and $A_2(q^2)$. In addition, the amplitudes in $B\to\rho \ell\ell$ decays are dependent on a few additional form factors: $T_1(q^2)$, $T_2(q^2)$, and $T_3(q^2)$. No lattice estimates exist for any of these form factors. On top of that, no experimental data has been reported on $B\to\rho \ell\ell$ decays to date, though a few data points exist for the differential rates in $B \to \rho\ell\nu_{\ell}$. All these form factors are estimated in the LCSR approach, which are available for a few values of $q^2$, including the value at $q^2=0$ \cite{Straub:2015ica,Gubernari:2018wyi}, and all these form factors are fully correlated. Also, see ref.~\cite{Gao:2019lta} for a recent update on the estimate of $B\to \rho$ form factors at $q^2=0$ in the LCSR with B-meson LCDA amplitudes with higher twists-six accuracy. A simultaneous study of the available inputs in $B \to \rho\ell\nu_{\ell}$ and $B\to\rho \ell\ell$ decays might be useful.   
	
	In light of the above-mentioned arguments, we look into the exclusive $B\to\pi(\rho) \ell\nu_{\ell}$ and $B\to\pi(\rho) \ell\ell$ decay modes in the scope of this article. We first carry out a combined fit incorporating all available lattice, LCSR and experimental inputs for the decay modes mentioned above. Though there is no direct correlation between the form factors in $B\to \rho$ and $B\to \pi$ transitions, both the decay rates $\Gamma(B\to \pi\ell\nu_{\ell})$ and $\Gamma(B\to \rho\ell\nu_{\ell})$ are proportional to $|V_{ub}|^2$ which is the only common link between the $B\to \pi$ and $B\to \rho$ decay channels we are considering here. Armed with an estimate of $|V_{ub}|$ and the corresponding form factors from the fit, we provide the SM estimates of several observables subject to the $B\to\pi(\rho) \ell\ell$ decays. These include the optimized ($P_i$) and asymmetric ($A_i$) observables, the forward-backward symmetry ($A_{FB}$), the longitudinal polarization ($F_L$), the direct CP asymmetries corresponding to the neutral and charged B-decays ($A_{CP}^{0,+}$), the CP-averaged Isospin asymmetry ($A_I$) and the BR's \cite{Bobeth:2007dw,Altmannshofer:2008dz,Descotes-Genon:2015hea,Descotes-Genon:2012isb}. In addition, we have given predictions for the observables associated with the tagged analysis of the $B^+$, $B^-$, $B^0$ and $\bar{B^0}$ decay modes as mentioned above. Ref.~\cite{Kindra:2018ayz} provides predictions for a few observables related to the $B \to \rho \ell\ell$ decays using the fit results for the form factors obtained in \cite{Straub:2015ica} which uses only LCSR inputs. We have also tested the NP sensitivities of all these observables in a few well-motivated benchmark NP scenarios. 
	
	The long-distance effects, which are the other dominant sources of uncertainty in $b\to dll$ transitions arise primarily due to the presence of $c\bar{c}$ and $u\bar{u}$ resonant states within the allowed range of the dilepton invariant-mass. The commonly available description of such effects are heavily model dependent. The most reliable way to tackle such uncertain contributions is hence to restrict the dilepton-invariant mass range so that the resonant regions are excluded. Therefore, we restrict our analysis within the range $0.1 \lsim q^2 \lsim 6$ (GeV$^2$) and all the predictions are given in bins of size $\sim$ 2 GeV$^2$.

	The paper is organized as follows. In section~\ref{theory}, we briefly discuss the differential decay distribution for the charged and neutral current $B\to\pi,\rho$ decays. Section~\ref{FF} is dedicated to a detailed discussion of the form factors that we use in our analysis. Section~\ref{results} provides an in-depth account of our analysis and the results that we present. Finally, in section \ref{summary} we have summarised our main results. 
	
	\section{Theoretical Background}\label{theory}
	\subsection{Tree level processes}
	\subsubsection{$\btopilnu$}
	
	As mentioned earlier, the differential decay rate w.r.t. $q^2$ for $\bar{B}\to\pi \ell \nu_{\ell}$ decays are function of the form factors $f_{+,0}(q^2)$. In particular, for $\bar{B^0}\to\pi^+$ semileptonic transitions, we have \cite{Sakaki:2013bfa} \footnote{The corresponding charged $B$ will decay to a neutral pion and hence will be scaled by a factor of $1/2$ at the decay width level since $\pi^0=\frac{u\bar{u}-d\bar{d}}{\sqrt{2}}$}
	\begin{align}
	\frac{d\Gamma}{dq^2}\left(\bar{B^0}\to\pi^+l^-\bar{\nu}_l\right)
	=  &\frac{G_F^2|V_{ub}|^2}{24\pi^3m_{B^0}^2q^4}\left(q^2-m_l^2\right)^2\left|p_{\pi}(m_{B^0},m_{\pi^+},q^2)\right|\times \nonumber\\ 
	&\left[\left(1+\frac{m_l^2}{2q^2}\right)m_{B^0}^2 \left|p_{\pi}(m_{B^0},m_{\pi^+},q^2)\right|^2\left|f_+\left(q^2\right)\right|^2 + 
	\frac{3m_l^2}{8q^2}\left(m_{B^0}^{2}-m_{\pi^+}^{2}\right)^2\left|f_0\left(q^2\right)\right|^2\right].
	\end{align}
	where $\left|p_\pi(m_B,m_\pi,q^2)\right|=\sqrt{\lambda(m_B,m_\pi,q^2)}/2 m_B$ with $\lambda(m_B,m_\pi,q^2)=((m_B-m_\pi)^2 - q^2)((m_B + m_\pi)^2 - q^2)$.
	
	Therefore, to extract $|V_{ub}|$, we need information on the form-factors at different values of $q^2$. The shape of the form factors will be obtained using lattice-QCD and LCSR. At present the lattice estimates are available on $f_{+/0}(q^2)$ at zero and non-zero recoils \cite{Flynn:2015mha,Lattice:2015tia,Colquhoun:2022atw}. There is also a recent update on the values of these form-factors from LCSR at $q^2=0$ and at values other than $q^2=0$ \cite{Gubernari:2018wyi,Leljak:2021vte}.
	
	\subsubsection{$\btorholnu$}
	
	The differential decay width for a pseudoscalar to vector semileptonic decay is a function of the form factors $A_{0,1,2}(q^2)$ and $V (q^2)$ \cite{Sakaki:2013bfa}
	
\begin{equation}
	\begin{split}
	\frac{d\Gamma}{dq^2}\left(\bar{B^0}\to\rho^+l^-\bar{\nu}_l\right) = {G_F^2 |V_{ub}|^2 \over 192\pi^3 m_B^3} q^2 \sqrt{\lambda_\rho(q^2)} \left( 1 - {m_l^2 \over q^2} \right)^2 \times\biggl\{ \biggr.  \left[ \left( 1 + {m_l^2 \over2q^2} \right) \left( H_{V,+}^2 + H_{V,-}^2 + H_{V,0}^2 \right) + {3 \over 2}{m_l^2 \over q^2} \, H_{V,t}^2 \right]  \biggl.\biggr\} \,,
	\end{split}
	\end{equation}
	
	where the hadronic amplitudes are given as: 
	
		\begin{subequations}
		\begin{align}
		&H_{V,\pm}(q^2)  =  (m_B+m_{\rho}) A_1(q^2) \mp { \sqrt{\lambda_{\rho}(q^2)} \over m_B+m_{\rho} } V(q^2) \,, \\
		& \nonumber \\
		&H_{V,0}(q^2) = { m_B+m_{\rho} \over 2m_{\rho} \sqrt{q^2} } \left[-(m_B^2-{m_{\rho}}^2-q^2) A_1(q^2) \right.\left. + { \lambda_{\rho}(q^2) \over (m_B+m_{\rho})^2 } A_2(q^2) \right] \,, \\
		& \nonumber \\
		&H_{V,t}(q^2)  = -\sqrt{ \lambda_{\rho}(q^2) \over q^2 } A_0(q^2)  
		\end{align}
	\end{subequations}.
	
	The currently available state of the art LCSR inputs for $B\to\rho$ decays are taken from the refs.~\cite{Straub:2015ica,Gubernari:2018wyi} 
	
	\subsection{Loop level processes} \label{subsec:loop}
	
	The effective Hamiltonian for semileptonic $b\to d\ell \ell$ transitions at the scale $\mu\sim m_b$ after integrating out the heavy degrees of freedom within the Standard Model is given by~\cite{Altmannshofer:2008dz} : 
	
	\begin{equation} \label{eq:Heff}
	{\cal H}_{eff} = - \frac{4\,G_F}{\sqrt{2}}\left(
	\lambda_t {\cal H}_{eff}^{(t)} + \lambda_u {\cal
		H}_{eff}^{(u)}\right)
	\end{equation}
	where $G_F$ is the Fermi constant and $\lambda_i$ = $V_{ib} V_{id}^*$ are the CKM factors. The combination $\lambda_c$ = $V_{cb} V_{cd}^*$ is eliminated by using the unitarity relation $\lambda_u + \lambda_c + \lambda_t$ = 0. The effective hamiltonians ${\cal H}_{eff}^{(t)}$ and ${\cal H}_{eff}^{(u)}$ can be written as:
	\begin{eqnarray} \label{eq:operator}
	{\cal H}_{eff}^{(t)} 
	& = & 
	C_1 \mathcal O_1^c + C_2 \mathcal O_2^c + \sum_{i=3}^{10} C_i 
	\mathcal O_i \,
	\\
	{\cal H}_{eff}^{(u)} 
	& = & 
	C_1 (\mathcal O_1^c-\mathcal O_1^u)  + C_2(\mathcal O_2^c-\mathcal
	O_2^u)\,.
	\end{eqnarray}
	The detail of the operator basis can be seen from~\cite{Altmannshofer:2008dz} which includes the current-current ($i=1,2$), QCD penguin ($i=3,..,6$), electromagnetic dipole ($\mathcal O_7$), chromomagnetic dipole ($\mathcal O_8$), and the semileptonic operators $\mathcal O_9$ and $\mathcal O_{10}$, respectively. The corresponding Wilson coefficients ($C_i's$) are calculated at the scale $\mu$ = $m_W$ upto next-to-next-to leading order (NNLO) and expressed as a perturbative expansion in the strong coupling constant $\alpha_s(\mu_W)$ \cite{Asatrian:2001de,Bobeth:1999mk,Asatrian:2003vq}. They are then evolved down to scale $\mu$ = $m_b$ using renormalization group equations which require a calculation of anomalous dimension matrices $\gamma(\alpha_s)$ upto three-loop accuracy. 
	After incorporating the QCD corrections, the coefficients $C_9$ and $C_7$ will modify and appear as combinations of the other $C_i's$. Therefore, it is convenient to define the effective Wilson coefficients: ${C}_{9}^{ eff}$ and ${C}_{7}^{ eff}$. The four-quark current-current operators $\mathcal O^q_1$, $\mathcal O^q_2$ (with $q=u$ and $c$) and quark-penguin operators $\mathcal O_{3..6}$ will contribute to ${C}_{9}^{ eff}$ which also includes the leading order (LO) charm-loop effects. Note that ${ C}_{9}^{ eff}$ is in general complex. A part of the contributions from the operators $\mathcal O_{3..6}$ also appear in ${ C}_{7}^{eff}$. The perturbatively calculable non-factorizable corrections to the operators $\mathcal O_1$ and $\mathcal O_2$ at order $\alpha_s$ are taken from \cite{Asatrian:2001de,Asatryan:2001zw} and added as a correction to ${ C}_{9}^{ eff}$. 
	
	As we have discussed in the introduction, the amplitude for the decay $B\to P \ell\ell$ (P=pseudoscalar meson) is sensitive to the QCD form factors $f_{+,0,T}(q^2)$ while the form factors relevant for $B\to V \ell\ell$ (V=vector meson) decays are $A_{0,1,2}(q^2)$, $V(q^2)$, and $T_{1,2,3}(q^2)$ respectively \cite{Altmannshofer:2008dz}. In addition to the QCD form factors, the above decay amplitudes are sensitive to some non-factorizable corrections. In particular, those corrections will be essential for a precision study. These corrections are related to the four-quark operators, $\mathcal O_7$, and $\mathcal O_8$, respectively.  In particular, these effects include the contributions from the ``hard" corrections to the weak vertex, hard-spectator corrections and the contributions from the weak annihilation (WA) diagrams. We do not have such corrections for the semileptonic operators. In the QCD-factorization (QCDF) approach, one can systematically calculate these various effects. In this paper, we have included the non-factorizable corrections calculated in the refs.~\cite{Beneke:2000wa,Beneke:2001at,Beneke:2004dp}. In the QCDF, the QCD corrections to the decays mentioned above are calculated in leading power in the inverse heavy quark mass and to the next-to-leading order (NLO) in $\alpha_s$. The most general factorization formula for a heavy to light transition amplitude at leading order in $1/m_b$ reads schematically:
	\begin{equation}
	\mathcal T_i  = C_i  \xi_M  + \phi_B \otimes T_i\otimes \phi_M,
	\end{equation}
	where $\xi_M$ is the ``soft" form factor and $T_i$ is the hard-scattering kernel convoluted with heavy and light meson light-cone wave functions $\phi_B$ and $\phi_M$, respectively. The vertex corrections to the four-quark operators contribute to $C_i$, which could have factorizable and non-factorizable parts: 
	\begin{equation}
	C_i = C_i^{(0)} + \frac{\alpha_s}{4 \pi}\big( C_i^{(f)} + C_i^{(nf)} \big) .
	\end{equation}  
	Similarly, the contribution from the WA diagram will appear in $T_i$ at the leading order in the expansion of the strong coupling constant. The WA contributions are obtained by computing the hadronic matrix elements of the four quark current-current and quark-penguin operators $\mathcal{O}_{1,..,6}$. As we will discuss later, the WA contributions to the amplitudes contain the imaginary part, which is the essential source of the strong phase required for the CP-violation or the isospin asymmetries. At order $\alpha_s$, there will be factorizable and non-factorizable contributions in $T_i$. These corrections are obtained by computing matrix elements of the four-quark operators ($\mathcal{O}_{1,..,6}$) and the chromomagnetic operator $\mathcal{O}_{8}$. As can be noted from \cite{Beneke:2001at}, the QCDF calculation for the matrix element of $\mathcal{O}_{8}$ at the subleading order in the $\Lambda/M_B$ expansion suffers from the logarithmic endpoint singularity. In the QCDF, these divergences are usually parameterized by some model-dependent parameters; for example, see the treatment in \cite{Kagan:2001zk,Feldmann:2002iw}. This approach will give us a conservative estimate of the theoretical uncertainty related to the absolute value and phase of this non-factorizable soft contribution. In particular, the associated strong phase governs the size of CP asymmetries. In practice, there will be large theoretical uncertainties in predicting associated CP observables. However, in the case of $b\to d$ transition, the dominant contribution to the CP or isospin asymmetric observables will come from the strong phases associated with the leading order WA contributions. The soft contribution due to $\mathcal{O}_8$ is a subleading effect, hence, will be power suppressed compared to the WA contributions \cite{Khodjamirian:2012rm}. Because of the above reason, in this analysis, we have not included this power suppressed subleading contribution from $\mathcal{O}_8$.  
	
	Here we would like to mention that in ref. \cite{Lyon:2013gba} the contributions from the WA diagrams have been computed in LCSR considering a complete operator basis with 2-particle twist-3 light meson distribution amplitude (DA). The resulting WA contributions have strong phases important for studying CP violation. In LCSR, the hard and soft-gluon contributions can be separated systematically. Also, in ref.~ \cite{Dimou:2012un}, the contribution from $\mathcal O_8$ has been evaluated in LCSR, which gives rise to long-distance contributions responsible for strong phases. At the moment, the estimates have large errors. However, as we have mentioned above, this is a subleading effect as compared to the contributions from the WAs. To check the impact of this contribution on the CP asymmetries, we will include this effect in our analysis in future work.
	
Here, we will analyse the $B\to \pi\ell\ell$ and $B\to \rho\ell\ell$ decay channels, the detailed calculations of the factorizable and non-factorizable corrections can be seen from refs.~\cite{Beneke:2000wa,Beneke:2001at,Beneke:2004dp}. The naively factorizable amplitudes will be defined by the full QCD form factors and to this amplitudes we will add only the non-factorizable corrections. As can be seen from \cite{Beneke:2000wa}, in the QCDF framework, the factorizable corrections arise when the full QCD form factors are expressed in terms of $\xi's$. Therefore, in our set-up we don't need to separately add them.  
	
	\subsubsection{$B\to\pi\ell \ell$}
	
	Our treatment of the differential decay rate in $B\to\pi\ell \ell$ is closely related to that in $B\to K\ell\ell$ decay \cite{Bobeth:2007dw}. The amplitude reads:
	\begin{align}
	\label{eq:Bpill}
	\mathcal{M} (B \to \pi \ell^+ \ell^-) 
	& =  \frac{G_F}{\sqrt 2} \frac{\alpha_{\rm em}}{\pi} 
	\Biggl[ \left(\bar \ell \gamma_{\mu} \ell \right) (p^{\mu} + p'^{\mu}) \Bigg((\lambda_t C_{9,\pi}^{(t)} + \lambda_u C_{9,\pi}^{(u)}) f^+_{B\pi}(q^2)
	\nonumber\\
	& + \frac{2 m_b}{m_B + m_\pi} \lambda_t f_{B\pi}^T (q^2) C_7^{\rm eff} \Bigg)
	+  \left(\bar \ell \gamma_{\mu} \gamma_5 \ell \right) (p^{\mu} + p'^{\mu}) \lambda_t f^+_{B\pi}(q^2) 
	C_{10} \Biggr]\,, 
	\end{align}
	where the coefficients $C_{9,\pi}^{(t)}$ and $C_{9,\pi}^{(u)}$ include the non-factorizable corrections and they are defined as   
	\begin{align} \label{eq:C9}
	& C_{9,\pi}^{(t)} (q^2) = C_9 + \frac{2 m_b}{M_B} \frac{\mathcal T_{\pi}^{(t)} (q^2)}{\xi_{\pi} (q^2)} \,, \\
	& \nonumber \\
	& C_{9,\pi}^{(u)} (q^2) = \frac{2 m_b}{M_B} \frac{\mathcal T_{\pi}^{(u)} (q^2)}{\xi_{\pi} (q^2)}.
	\end{align}
	Note that in QCDF, the factorization scheme is such that the relation $\xi_{\pi} \equiv f_+$ holds exactly to all order in perturbation theory. The most general form of the amplitude $\mathcal T_{\pi}$ is given by \cite{Beneke:2001at}
	\begin{equation} 
	\mathcal T_{\pi}^{i}
	= \xi_{\pi} \Big[ C_{\pi}^{(0,i)} + \frac{\alpha_s C_F}{4 \pi} \big( C_{\pi}^{(f,i)}   + C_{\pi}^{(nf,i)} \Big)\Big] 
	+\frac{\pi^2 f_B f_{\pi}}{N_c M_B}  \sum_{\pm}
	\int_0^\infty \frac{d\omega}{\omega}\Phi_{B,\pm}(\omega)\int_0^{1}du\, \phi_{\pi}(u)
	T_{\pi,\pm}^{(i)}(u,\omega)
	\label{eq:taupi}
	\end{equation}
	with $i= u$ or $t$ quark. The hard kernal $T_{\pi,\pm}^{(i)}(u,\omega)$ can be expressed as 
	\begin{equation}
	T_{\pi,\pm}^{(i)}(u,\omega) = T_{\pi,\pm}^{(0,i)}(u,\omega) + \frac{\alpha_s C_F}{4 \pi} \big( T_{\pi,\pm}^{(f,i)}(u,\omega) +  T_{\pi,\pm}^{(nf,i)}(u,\omega) \Big).
	\label{eq:Tpi}
	\end{equation}
	Here, $T_{\pi,\pm}^{(0)}, T_{\pi,\pm}^{(f)}$, and $ T_{\pi,\pm}^{(nf)}$ are the WA (non-factorizable correction at order $\alpha_s^0$), factorizable, and the non-factorizable corrections due to hard spectator scattering.   
	In eq.~\ref{eq:taupi}, the leading order contributions to the $\mathcal T_{\pi}^{(i)}$ are defined as
	\begin{equation}
	C_{\pi}^{(0,t)} =  \frac{M_B}{2 m_b} Y(q^2),\ \ \ \ 
	C_{\pi}^{(0,u)} = \frac{M_B}{2 m_b} Y^u(q^2),
	\end{equation} 
	where the contributions from the $\mathcal{O}_{1,...,6}$ matrix elements are incorporated in $Y(q^2)$ which are commonly included in $C_9^{eff}$. Since we will be using the form factors defined in full QCD, we need to add only the non-factorizable corrections in eq.~\ref{eq:Bpill} which are defined in the above equations. The detailed mathematical expressions for $C_{\pi}^{(nf,i)}$, $T_{\pi,\pm}^{(0,i)}(u,\omega)$ (WA contribution), and  $T_{\pi,\pm}^{(nf,i)}(u,\omega)$ are obtained from refs.~\cite{Beneke:2001at,Beneke:2004dp}\footnote{For the $B\to P$ transitions the expressions for the non-factorizable corrections are extracted from the results of \cite{Beneke:2001at,Beneke:2004dp} using the following replacement: $C_{\pi}^{nf}=  -C_{||}^{nf}$, $T_{\pi,\pm}^{(0)}(u,\omega) = - T_{||,\pm}^{(0)}(u,\omega) $, and $T_{\pi,\pm}^{(nf)}(u,\omega) = - T_{||,\pm}^{(nf)}(u,\omega) $ }.
	
	Information from long-distance physics is encoded in the light-cone
	distribution amplitudes, $\Phi_{B,\pm}$ for the B meson and $\phi_{\pi}$ for the final state mesons. For B meson, the light-cone distribution amplitudes are written as~\cite{Grozin:1996pq,Beneke:2001at},
	\begin{align}
	\Phi_{B,+}(\omega)
	= \frac{\omega}{\omega_0^2}e^{-\omega/\omega_0}, \quad
	\Phi_{B,-}(\omega)
	= \frac{1}{\omega_0}e^{-\omega/\omega_0}.
	\label{eq:BLDA}
	\end{align}
	These enter the decay amplitude through the moments $\lambda_{B,+}^{-1}(q^2)$ and $\lambda_{B,-}^{-1}(q^2)$ \cite{Grozin:1996pq,Beneke:2001at} 
	\begin{equation} \label{eq:lam+}
	\lambda_{B,+}^{-1}
	= \int_0^\infty d\omega \frac{\Phi_{B,+}(\omega)}{\omega} =\omega_0^{-1}
	\end{equation}
	and
	\begin{equation} \label{eq:lam-}
	\lambda_{B,-}^{-1}(q^2)
	= \int_0^\infty d\omega \frac{\Phi_{B,-}(\omega)}{\omega-q^2/M_B-i\epsilon}
	=\frac{e^{-q^2/(M_B\omega_0)}}{\omega_0}[-{\rm Ei}\left(q^2/M_B\omega_0\right) +i\pi ].
	\end{equation}
	Here Ei$(z)$ is the exponential integral function. $\lambda_{B,-}^{-1}$ appears via the weak annihilation term, and the imaginary part in Eq.~(\ref{eq:lam-}) acts as an important source of strong phase necessary for CP violation. In the leading twist, for a light meson ($M$) the most general expression of the light-cone distribution amplitude is given by \cite{Ball:2007rt,Ball:1998sk},
	\begin{equation}
	\phi_{M}(u)
	=6u(1-u)\Big[ 1 +a_{1}^M\, C_1^{(3/2)}(2u-1) 
	+a_{2}^M\, C_2^{(3/2)}(2u-1)\Big],
	\label{eq:lightWF}
	\end{equation}
	where $C_n^{(3/2)}(x)$ are Gegenbauer polynomials and $a_i^{M}$ are the respective coefficients.
	
	\subsubsection{$B\to\rho\ell \ell$}\label{subsubsec:Btorhointro}
	
	The theory description of $B\to \rho \ell\ell$ is quite similar to that given for $B\to K^*\ell\ell$ decay. The  decay distribution for a four body decay $\bar{B} \to \bar{V}(\to \bar{M}_1 \bar{M}_2) l^+ l^-$ can be completely described in terms
	of four kinematic variables: the lepton invariant mass squared ($q^2$) and the three angles \cite{Kruger:1999xa,Altmannshofer:2008dz}.
	\begin{equation}\label{eq:d4Gamma}
	\frac{d^4\Gamma}{dq^2 dcos\theta_l dcos\theta_{\rho} d\phi} =
	\frac{9}{32\pi} J(q^2, \theta_l, \theta_{\rho}, \phi)
	\end{equation}
	where, 
	\begin{align} \label{eq:angulardist}
	J(q^2, \theta_l, \theta_{\rho}, \phi)& = 
	J_1^s sin^2\theta_{\rho} + J_1^c cos^2\theta_{\rho}
	+ (J_2^s sin^2\theta_{\rho} + J_2^c cos^2\theta_{\rho}) cos2\theta_l
	\nonumber \\       
	& + J_3 sin^2\theta_{\rho} sin^2\theta_l cos2\phi 
	+ J_4 sin2\theta_{\rho} sin2\theta_l cos\phi 
	\nonumber \\       
	& + J_5 sin2\theta_{\rho} sin\theta_l cos\phi
	\nonumber \\      
	& + (J_6^s sin^2\theta_{\rho} + J_6^c cos^2\theta_{\rho}) cos\theta_l 
	+ J_7 sin2\theta_{\rho} sin\theta_l sin\phi
	\nonumber \\ 
	& + J_8 sin2\theta_{\rho} sin2\theta_l sin\phi
	+ J_9 sin^2\theta_{\rho} sin^2\theta_l sin2\phi.
	\end{align} 
	
	Like in the case of $B\to K^*(\to K\pi )\ell^+\ell^-$, one can express the $B \to \rho(\to \pi\pi) \ell^+\ell^-$ decay amplitudes in terms of seven helicity/transversity amplitudes like $A_{\bot L,R}$, $A_{\| L,R}$, $A_{0L,R}$ and $A_t$, which are then used to form angular coefficients (J's) relevant in defining the CP-symmetric and asymmetric observables measured by the different experimental collaborations. The detailed expressions of the transversity amplitudes and the corresponding angular coefficients are obtained from the  ref.~\cite{Altmannshofer:2008dz}. These amplitudes are expressed as functions of the Wilson coefficients and the QCD form factors. We need to add the corresponding non-factorizable corrections to these amplitudes, which are as given below
	\begin{eqnarray}\label{eq:corrQCDF}
	\Delta A_{\perp,L,R}^{QCDF} &=& \sqrt{2} N \frac{2 m_b}{q^2} (m_B^2-q^2) \big(\mathcal T_{\perp}^{(t), WA+nf} + \hat{\lambda}_u \mathcal T_{\perp}^{(u), WA + nf)}\big), \nn \\
	\Delta A_{||,L,R}^{QCDF} &=& - \Delta A_{\perp,L,R}^{QCDF}, \nn \\
	\Delta A_{0,L,R}^{QCDF} &=&  \frac{N (m_B^2-q^2)^2 }{m_{\rho} m_B^2 \sqrt{q^2} } m_b \big(\mathcal T_{||}^{(t),WA+nf} + \hat{\lambda}_u \mathcal T_{||}^{(u),WA + nf)}\big). 
	\end{eqnarray}
	Here, $\hat{\lambda}_u = \lambda_u/\lambda_t$ and following the QCDF the transversity amplitudes $\mathcal T_a $ can be written as 
	\begin{equation} 
	\mathcal T_a^{(i)}
	= \xi_a \Big{(}C_a^{(0,i)} + \frac{\alpha_s C_F}{4 \pi} C_a^{(1,i)}\Big{)} 
	+\frac{\pi^2 f_B f_{\rho,a}}{N_c M_B} \Xi_a \sum_{\pm}
	\int_0^\infty \frac{d\omega}{\omega}\Phi_{B,\pm}(\omega)\int_0^{1}du\, \phi_{\rho,a}(u)
	T^{(i)}_{a,\pm}(u,\omega),
	\label{eq:mathcalTv}
	\end{equation} 
	where $C_F$ = 4/3, $N_c$ = 3 and $f_B$ refers to the B meson decay constant. $``a"$ stands for the polarisation of the vector meson in the final state ($a\to ||$ or $\perp$), $\Xi_{\perp}$ = 1, $\Xi_{||}$ = $m_\rho / E$, $E = (M_B^2 - q^2) / (2 M_B)$ refers to the energy of the final state meson and $\xi_a$ refer to the form factors in the heavy quark and high energy limit, and $i = u, t$. For the different polarizations, after estimating the WA, factorizable and non-factorizable contributions, the hard kernel $T_{a,\pm}(u,\omega)$ can be expanded as given in Eq.~\ref{eq:Tpi}. As we have discussed earlier, here we need to consider the contributions from the WA diagram at leading order in $\alpha_s$, and the non-factorizable corrections (order $\alpha_s$) from eq.~\ref{eq:mathcalTv}, the detailed expressions of which are obtained from refs.~\cite{Beneke:2001at,Beneke:2004dp}. Note that $T_{||,-}^0$ will be the only non-zero contribution from the leading order WA diagram. The detailed expressions for the light-cone distribution functions $\phi_B$, and $\phi_{\rho,a}$ can be obtained following eq.~\ref{eq:BLDA}, and \ref{eq:lightWF} for different polarizations, respectively.

	The angular decay rate distribution for the CP-conjugated process $B \to V(\to M_1 M_2) l^+ l^-$ is given by 
	\begin{equation}\label{eq:d4Gammaconjugate}
	\frac{d^4\bar{\Gamma}}{dq^2 dcos\theta_l dcos\theta_{\rho} d\phi} =
	\frac{9}{32\pi} \tilde{J}(q^2, \theta_l, \theta_{\rho}, \phi)
	\end{equation}
	which has obtained from eq.~\ref{eq:d4Gamma} with the replacement $J_i \to \tilde{J}_i \equiv \xi_i \bar{J}_i$ where, $\xi_i$ = 1 for i $\in$ {1,2,3,4,7} and -1 for i $\in$ {5,6,8,9} \cite{Altmannshofer:2008dz}. Here, the angular coefficients $\tilde{J_i}$ is formed from the helicity amplitude: $\tilde{A}_H \equiv A_H({\bar B} \to f)$, on the other hand the $\bar{J_i}$ is formed from $\bar{A}_H \equiv A_H(\bar{B}\to \bar{f})$. From the decay rates of a $B$ and $\bar B$ to a CP-eigenstate or to a CP-conjugated states, one can define the following: 
	\begin{equation}\label{eq:rateplus}
	\frac{d(\Gamma  + \bar{\Gamma}) }{dq^2 dcos\theta_l dcos\theta_{\rho} d\phi} = \sum_i [J_i + \tilde{J_i}] f_i(\theta_l,\theta_{\rho}\phi ) =  \sum_i [J_i + \xi_i \bar{J_i}] f_i(\theta_l,\theta_{\rho}\phi ),
	\end{equation}
	\begin{equation}\label{eq:rateminus}
	\frac{d(\Gamma  - \bar{\Gamma}) }{dq^2 dcos\theta_l dcos\theta_{\rho} d\phi} = \sum_i [J_i - \tilde{J_i}] f_i(\theta_l,\theta_{\rho}\phi ) = \sum_i [J_i - \xi_i \bar{J_i}] f_i(\theta_l,\theta_{\rho}\phi ).
	\end{equation}
	In the experiment like LHCb or Belle, it will be possible to extract the associated CP-averaged ($S_i$) and CP-asymmetric ($A_i$) observables which are given by 
	\begin{equation}
	S_i = \frac{J_i + \bar{J}_i }{d\Gamma/{dq^2}}, \ \ \ \ A_i = \frac{J_i - \bar{J}_i }{d\Gamma/{dq^2}},
	\end{equation}
	respectively. Here, $d\Gamma/{dq^2}$ is the combined differential decay rate of the $B^0$ and ${\bar B^0}$ decays. The CP-averaged observables are also related to the longitudinal polarization fraction $F_L$ of the $\rho$-meson which is defined as $F_L = S_1^c = -S_2^c$. Also, the forward-backward asymmetry $A_{FB}$ of the dilepton system is defined as $A_{FB} = 3 S_6/4$ \cite{Aaij:2015oid,Aaij:2020nrf}. Also, using the observables like $S_i$s, $A_{FB}$ and $F_L$ one can define several other optimised observables which are relatively clean (free from form factor) and defined as $P_i$ and $P_i^{\prime}$, for details see \cite{Matias:2012xw,Descotes-Genon:2012isb,Aaij:2015oid}. In this analysis, we have given predictions for all these observables along with some other relevant observables. Note that the effect of mixing will become relevant in the case of $B$ decays to CP-eigenstates. Hence the time-dependent decay amplitudes need to be considered, and we will discuss the details in the follow-up sections.  
	
	Note that for the geometries of the decay, like the choices of the different angles in the decay planes, we have followed the convention used in most of the theory papers \cite{Altmannshofer:2008dz,Bobeth:2008ij,Descotes-Genon:2013vna,Descotes-Genon:2015uva}. For example, the choices of the angles for $B({\bar B}) \to K^{\ast} (\bar K^{\ast})\ell^+\ell^-$ in the theory convention differs from the LHCb choice \cite{LHCb:2013zuf}, which has been clearly addressed in \cite{Gratrex:2015hna}. As mentioned above in the theory convention, $\tilde{J}_{5,6,8,9}$ will change sign between CP conjugated modes, while in the LHCb convention $\tilde{J}_{7,8,9}\to - \bar{J}_{7,8,9}$. For $\bar{B}\to \bar{K}^{\ast}\ell\ell$ decays, the two conventions are related using 
	\begin{equation}
	\theta_K^{LHCb} =  \theta_K^{Theory}, \ \ \theta_{\ell}^{LHCb} =  \pi - \theta_{\ell}^{Theory},\ \ \phi^{LHCb} = - \phi^{Theory},
	\end{equation}
	which results in the sign difference: 
	\begin{equation}\label{eq:signdiff1}
	J_{4,6,7,9}|_{LHCb} = - J_{4,6,7,9}|_{Theory}.
	\end{equation}
	The rest of the $J_i$s will be same in both the conventions. As discussed in \cite{Gratrex:2015hna}, for the $B\to {K}^{\ast}\ell\ell$ the difference in the conventions will lead to the sign differences only in the following $J_i$s: 
	\begin{equation}\label{ea:signdiff2}
	J_{4,5,9}|_{LHCb} = - J_{4,5,9}|_{Theory}.
	\end{equation}	
	These differences in the sign of the angular coefficients in both the conventions will lead to the following relations between the CP-averaged and CP-asymmetric observables: 
	\begin{align}
	(A,S)_{4,6,7,9}|_{LHCb} &= - (A,S)_{4,6,7,9}|_{Theory},\nn \\ 
	(A,S)_{1,2,3,5,8}|_{LHCb} &=  (A,S)_{1,2,3,5,8}|_{Theory}. 
	\end{align}  
For the self-tagging modes, following the theory convention one could extract [$S_{1,2,3,4,7}$, $A_{5,6,8,9}$] and [$A_{1,2,3,4,7}$, $S_{5,6,8,9}$] from the CP-averaged and CP-asymmetric decay rates defined in eq.~\ref{eq:rateplus} and eq.\ref{eq:rateminus}, respectively. Hence, from the untagged decay rates, one could extract only $S_{1,2,3,4,7}$ and $A_{5,6,8,9}$, which is the case for LHCb as well as was observed in the case of $B_s\to \phi(\to K^+K^-)\ell^+\ell^-$ decays.  
	
	\begin{table}[t] 
		\centering
		\renewcommand*{\arraystretch}{1.3}
		\begin{tabular}{c|ccccc}
			$q^2$ (GeV$^2$)  &  $19$  &  $20.5$  & $22.6$ &  $25.1$   \\
			\hline
			$f_+\left(q^2\right)$  &  $1.176\pm 0.054$  &  $1.494\pm 0.052$  &  $2.241\pm 0.064$  &  $4.455\pm 0.153$  \\
			$f_0\left(q^2\right)$  &  $0.458\pm 0.025$  &  $-$  &  $0.651\pm 0.020$  &  $0.864\pm 0.023$  \\
			$f_T\left(q^2\right)$ &  $1.133\pm 0.064$ &  $1.443\pm 0.058$ &  $2.169\pm 0.076$ &  $4.313\pm 0.176$ \\
		\end{tabular}
		\caption{\small Synthetic datapoints for $B\to \pi$ transition form factors from Lattice MILC \cite{FermilabLattice:2015cdh,Lattice:2015tia} and corresponding correlations given in eqn.~\ref{eq:MILCcorr}.}
		\label{tab:MILCsynth}
	\end{table}
	
	\begin{table}[t]
		\centering
		\renewcommand*{\arraystretch}{1.3}
		\begin{tabular}{c|ccccc}
			$q^2$ (GeV$^2$)  &  $-10$  &  $-5$ &  $0$  &  $5$  &  $10$\\
			\hline
			$f_+\left(q^2\right)$  &  $0.170\pm 0.022$  &  $0.224\pm 0.022$  &  $0.297\pm 0.030$  &  $0.404\pm 0.044$  &  $0.574\pm 0.062$  \\
			$f_0\left(q^2\right)$  &  $0.211\pm 0.029$  &  $0.251\pm 0.024$  &  $-$ & $0.356\pm 0.040$ &  $0.441\pm 0.052$  \\
			$f_T\left(q^2\right)$ &  $0.170\pm 0.021$ &  $0.222\pm 0.020$ &  $0.293\pm 0.028$ & $0.396\pm 0.039$ &  $0.560\pm 0.053$ \\
			\hline
		\end{tabular}
		\caption{\small LCSR inputs at various values of $q^2$ and the respective correlations are taken from ref.~\cite{Leljak:2021vte}.}
		\label{tab:MELICdata}
	\end{table}
	
	\begin{table}[t] 
		\centering
		\renewcommand*{\arraystretch}{1.3}
		\begin{tabular}{c|ccccc}
			$q^2$ (GeV$^2$)  &  $0$  &  $5$  & $10$   \\
			\hline
			$A_0\left(q^2\right)$  &  $-$  &  $0.473\pm 0.048$  &  $0.672\pm 0.059$  \\
			$A_1\left(q^2\right)$  &  $0.262\pm 0.026$  &  $0.292\pm 0.029$  &  $0.333\pm 0.034$  \\
			$A_2\left(q^2\right)$ &  $0.229\pm 0.036$ &  $0.291\pm 0.050$ &  $0.383\pm 0.080$  \\
			$V\left(q^2\right)$ &  $0.327\pm 0.031$ &  $0.438\pm 0.034$ &  $0.629\pm 0.042$  \\
			$T_1\left(q^2\right)$ &  $0.272\pm 0.026$ &  $0.365\pm 0.028$ &  $0.525\pm 0.033$  \\
			$T_2\left(q^2\right)$ & $-$ & $0.301\pm 0.028$ & $0.343\pm 0.033$  \\
			$T_3\left(q^2\right)$ &  $0.181\pm 0.029$ &  $0.229\pm 0.041$ &  $0.301\pm 0.069$  \\
		\end{tabular}
		\caption{\small Synthetic datapoints for $B\to \rho$ transition form factors obtained in LCSR from the ref.~ \cite{Straub:2015ica}, the corresponding correlations are given in eqn.~\ref{eq:Bharuchacorr}.}
		\label{tab:Bharuchasynth}
	\end{table}
	
\begin{table}[t]
		\begin{center} 
			\begin{tabular}{| l l| l l |} 
				\hline  
				\hline 
				\rule[-2mm]{0mm}{7mm}
				$\mu_b$                           &  4.8 GeV & 
				$f_B$                           & 190.0 $\pm$ 1.3 MeV \cite{Aoki:2021kgd}\\ 
				$\alpha_s (\mu_b)$            & 0.214 & 
				$\lambda_{B,+}^{-1}$            & $(3 \pm 1)$~GeV$^{-1}$ \cite{Beneke:2001at} \\
				$\alpha_{\rm em}$               & $ 1/137$ &
				$G_F$             & 1.166 $\times$ $10^{-5}$ ~GeV$^{-2}$ \\
				$m_{c,pole}$  & 1.4 GeV   &  $m_{b,pole}$ &  4.8 GeV  \\
				$f_{\pi}$                      & 130.2 $\pm$ 0.8 MeV \cite{Aoki:2021kgd} &
				$a_2^{\pi}$  & $0.116^{+19}_{-20}$ \cite{RQCD:2019osh} \\ 
				$f_{\rho}^{\perp}$   & 160 $\pm$ 7 MeV \cite{Straub:2015ica} &
				$a_{2,\rho}^{\perp}$    & 0.14 $\pm$ 0.06 \cite{Straub:2015ica} \\ 
				$f_{\rho}^{||}$   & 213 $\pm$ 5 MeV \cite{Straub:2015ica} &
				$a_{2,\rho}^{||}$    & 0.17 $\pm$ 0.07 \cite{Straub:2015ica}  \\
				
				\hline
			\end{tabular} 
			\caption{Inputs used in the analysis.}
			\label{tab:input}
		\end{center} 
	\end{table} 
	
	\section{Inputs}~\label{FF}

	To get the shape of the decay rate distribution, one needs to know the shape of the corresponding form-factors in the whole $q^2$ region. Therefore, it is crucial to have a parametrization of the form-factors that satisfies real analyticity in the complex $q^2$ plane. We have followed the Bharucha-Straub-Zwicky (BSZ)~\cite{Straub:2015ica} parametrization where the conformal map from $q^2$ to z is given by:
	\begin{equation}
	z(q^2) = \frac{\sqrt{t_+-q^2}-\sqrt{t_+-t_0}}{\sqrt{t_+-q^2}+\sqrt{t_+-t_0}}\,,  
	\end{equation}
	where $t_\pm \equiv (m_B\pm m_P)^2$ and $t_0\equiv t_+(1-\sqrt{1-t_-/t_+})$. $t_0$ is a free parameter that governs the size of $z$ in the semileptonic phase space. Under this parametrization, a generic form-factor for pseudoscalar-to-pseudoscalar/vector transition reads: 
	
	\begin{equation}\label{eq:bszexp}
	f_i(q^2) = \frac{1}{1 - q^2/m_{R,i}^2} \sum_{k=0}^N a_k^i \, [z(q^2)-z(0)]^k\,,
	\end{equation}
	where $m_{R,i}$ denotes the mass of sub-threshold resonances compatible with the quantum numbers of the respective form factors and $a_k^i$s are the coefficients of expansion. This parametrization has the advantage that the value of the form factor at $q^2 = 0$ is among the fit parameters which is evident from eq.~\ref{eq:bszexp}: $f_i(q^2=0)= a_0^i$. Thus, the kinematical constraints relating the form factors at $q^2$ = 0 result in simpler relations among the form factor parameters. The details are provided in ref.~\cite{Straub:2015ica}.
	
	The lattice collaborations RBC/UKQCD \cite{Flynn:2015mha} and JLQCD \cite{Colquhoun:2022atw} provide synthetic data points for $f_{+,0}(q^2)$ with full covariance matrices (both systematic and statistical) at three $q^2$ points which we have directly used in our analysis. On the other hand, the Fermilab-MILC collaboration \cite{FermilabLattice:2015cdh,Lattice:2015tia} provides the fit-results for the coefficients of the $z$-expansions of the respective form factors. They have followed Bourrely-Caprini-Lellouch (BCL) \cite{BCL} parametrization for the $z$-expansion.
	We have used them to create synthetic datapoints, which are given in table \ref{tab:MILCsynth}, at exactly the same $q^2$ values as RBC/UKQCD, with an extra point for $f_+$  and $f_T$ at $q^2 = 20.5$ GeV$^2$, thus utilizing the full information from the lattice fit. In addition to the lattice inputs, we have used the inputs on the form factors obtained by LCSR approaches \cite{Wang:2015vgv,Lu:2018cfc,Gubernari:2018wyi,Leljak:2021vte}. The analysis in \cite{Leljak:2021vte} uses the two-particle twist-two pion light-cone distribution amplitude (LCDA), and the results are more precise than those obtained in \cite{Gubernari:2018wyi} which is an LO calculation with the ill-known B-meson LCDA.  Table \ref{tab:MELICdata} presents the LCSR data points for the form factors $f_+(q^2)$ and $f_T(q^2)$ at $q^2=-10,-5,0,5,10$ GeV$^2$, respectively, which are obtained from \cite{Leljak:2021vte}. We have also utilised the respective correlations in our fits. Note that ref.~\cite{Gubernari:2018wyi} provides the values and covariance matrices for the form factors $f_+(q^2)$ and $f_T(q^2)$ at $q^2=-15,-10,-5,0,5$ and for the form-factor $f_0(q^2)$ at $q^2=-15, -10,-5, 5$ GeV$^2$ which play a sub-dominant role in our fits, since the estimated errors in \cite{Gubernari:2018wyi} are larger as compared to the one obtained in ref.~\cite{Leljak:2021vte}. Due to a similar reason, the inputs from the refs.~\cite{Wang:2015vgv,Lu:2018cfc} do not have impact on our results. However, we have included all these inputs in our fits.

	For $B\to\rho$ transitions, the LCSR data points along with the respective covariance matrix are given in ref.~\cite{Gubernari:2018wyi} for $q^2=-15, -10,-5,0, 5$ GeV$^2$, which we directly use in our fits. In ref.~\cite{Straub:2015ica}, the fit-results for the coefficients of the $z$-expansion are given. Using their fit results, we generate correlated synthetic data-points for the form factors $A_1$, $A_2$, V, $T_1$ and $T_3$ at $q^2$ = 0,5,10 GeV$^2$ and for $A_0$ and $T_2$ at $q^2$ = 5,10 GeV$^2$ \footnote{At $q^2$ = 0, $A_0$ is related to $A_1$ and $A_2$ and $T_2$ is related to $T_1$, thus they aren't independent. So, in order to keep the covariance matrix positive semi-definite, we don't include the datapoints for $A_0$ and $T_2$ at $q^2$ = 0.}, respectively, which are presented in table \ref{tab:Bharuchasynth}. The LCSR results in ref.~\cite{Straub:2015ica} have been derived up to twist-3 ${\mathcal O}(\alpha_s )$ using the $\rho$ meson LCDA, and the extracted values are relatively more precise than the ones obtained in \cite{Gubernari:2018wyi}. In ref. \cite{Gubernari:2018wyi}, for the computation of the $B\to\rho$ form factors the narrow-width approximation of the $\rho$ meson has been assumed, and uses the B-meson LCDA. Note that the $\rho$-meson is an unstable particle and decays strongly to pairs of pseudoscalar mesons. In ref. \cite{Straub:2015ica}, the authors have justified the use of $\rho$ meson DA which is characterized by the longitudinal and transverse component of the decay constant $f_{\rho}^{\parallel}$ and $f_{\rho}^{\perp}$, respectively. The inputs on $f_{\rho}^{\parallel}$ are obtained from the measurements of the decay widths: $\Gamma(e^+e^- \to \rho^0(\to \pi\pi))$ and $\Gamma(\tau^{+} \to \rho^{+}(\pi\pi)\nu)$ respectively \cite{ckmfitter2}.
	In these experiments, the $\rho$-meson is detected as a Breit-Wigner peak in the invariant mass distribution of produced pions ($\rho\to \pi\pi$), while the transverse component $f^{\perp}_{\rho}$ is obtained from the lattice estimates of the ratio $f_{\rho}^{\parallel}/f^{\perp}_{\rho}$. In principle, this decay should be analysed via the $B \to \pi \pi$ type form factor which is done using a two-pion distribution amplitude within LCSR. The authors of ref \cite{Straub:2015ica} have justified that except the behaviour around $m_{\pi \pi}^2$ locally, this contribution can be effectively absorbed into $f_{\rho}^{||}$ on integration over the $\rho$ mass window in the experimental analysis. As argued in \cite{Straub:2015ica}, as long as the treatment of the $\rho (\to \pi\pi)$ meson is the same as is used in the experimental extractions of $f_{\rho}^{\parallel}$, there is no systematic effect. The LCSR should not suffer from sizeable additional uncertainties. For completeness, we have also included the inputs from the ref.~\cite{Gao:2019lta} predicted in the LCSR approach. However, since the predictions have large errors compared to that in the  ref.~\cite{Straub:2015ica}, they have a negligible impact on our results.   
	
	Other inputs relevant for our analysis are given in table~\ref{tab:input}. The values of the WCs are taken at the scale $\mu_b$ = 4.8 GeV \cite{Altmannshofer:2008dz}. We have presented the results truncating the BSZ expansion (eq.~ \ref{eq:bszexp}) at $N = 3$ for $B\to\pi$ transitions and at $N = 2$ for $B\to\rho$ transitions. We have checked that if we go to the next order $N = 3$ for $B\to\rho$ transition, the newly added higher-order coefficients of the expansion remain mostly unconstrained, and they have a negligible impact on the precision extraction of $|V_{ub}|$ and other related observables which we will discuss in the next section.

	\begin{table}[t]
		\centering
		\begin{tabular}{|cc|cc|}
			\hline
			\multicolumn{2}{|c|}{$B\to \pi$ form factors} & \multicolumn{2}{c|}{$B\to \rho$ form factors}   \\
			\hline
			Parameters & Fitted Values & Parameters & Fitted Values \\
			\hline
			$\text{$a_0^{f_+}$ }$  &  $\text{0.260$\pm $0.008}$  & $\text{$a_1^{A_0}$}$  &  $\text{-0.879$\pm $0.153}$ \\
			\hline
			$\text{$a_1^{f_+}$}$  &  $\text{-0.639$\pm $0.065}$  & $\text{$a_2^{A_0}$}$  &  $\text{1.074$\pm $0.951}$ \\
			\hline
			$\text{$a_2^{f_+}$}$  &  $\text{-0.067$\pm $0.210}$ & $\text{$a_0^{A_1}$}$  &  $\text{0.242$\pm $0.013}$ \\
			\hline
			$\text{$a_3^{f_+}$}$  &  $\text{0.485$\pm $0.160}$ & $\text{$a_1^{A_1}$}$  &  $\text{0.468$\pm $0.086}$ \\
			\hline
			$\text{$a_1^{f_0}$}$  &  $\text{0.301$\pm $0.063}$  &  $\text{$a_2^{A_1}$}$  &  $\text{0.307$\pm $0.281}$ \\
			\hline
			$\text{$a_2^{f_0}$}$  &  $\text{0.350$\pm $0.181}$  & $\text{$a_0^{V}$}$  &  $\text{0.309$\pm $0.017}$  \\
			\hline
			$\text{$a_3^{f_0}$}$  &  $\text{0.354$\pm $0.168}$  & $\text{$a_1^{V}$}$  &  $\text{-0.742$\pm $0.115}$ \\
			\hline
			$\text{$a_0^{f_T}$}$  &  $\text{0.252$\pm $0.011}$  &  $\text{$a_2^{V}$}$  &  $\text{1.216$\pm $0.827}$ \\
			\hline
			$\text{$a_1^{f_T}$}$  &  $\text{-0.701$\pm $0.101}$  & $\text{$a_0^{A_2}$}$  &  $\text{0.220$\pm $0.017}$  \\
			\hline
			$\text{$a_2^{f_T}$}$  &  $\text{-0.455$\pm $0.396}$  & $\text{$a_1^{A_2}$}$  &  $\text{-0.397$\pm $0.096}$  \\
			\hline
			$\text{$a_3^{f_T}$}$  &  $\text{-0.015$\pm $0.365}$  &  $\text{$a_2^{A_2}$}$  &  $\text{0.405$\pm $1.002}$  \\
			\hline
			$\text{$|V_{ub}| \times 10^3$}$  &  $\text{3.60$\pm $0.10}$  & $\text{$a_0^{T_1}$}$  &  $\text{0.262$\pm $0.015}$   \\
			\hline
			$\text{$V_{tb} V_{td} cos \beta $}$  &  $\text{0.0079$\pm $0.0001}$ & $\text{$a_1^{T_1}$}$  &  $\text{-0.643$\pm $0.091}$ \\
			\hline
			$\text{$V_{tb} V_{td} sin \beta $}$  &  $\text{0.00324$\pm $0.00008}$  & $\text{$a_2^{T_1}$}$  &  $\text{0.909$\pm $0.645}$  \\
			\hline
			$\text{$V_{ud} cos \gamma $}$  &  $\text{0.404$\pm $0.019}$  & $\text{$a_1^{T_2}$}$  &  $\text{0.556$\pm $0.084}$  \\
			\hline
			$\text{$V_{ud} sin \gamma $}$  &  $\text{0.887$\pm $0.009}$  & $\text{$a_2^{T_2}$}$  &  $\text{0.714$\pm $0.336}$ \\
			\hline
			$\text{$f_B f_{\pi}$}$  &  $\text{0.0247$\pm $0.00023}$  &  $\text{$a_0^{T_3}$}$  &  $\text{0.190$\pm $0.014}$   \\
			\hline
			$\text{$\lambda_{B,+}^{-1}$}$  &  $\text{2.988$\pm $0.999}$  &  $\text{$a_1^{T_3}$}$  &  $\text{-0.374$\pm $0.081}$  \\
			\hline 
			& & $\text{$a_2^{T_3}$}$  &  $\text{1.267$\pm $0.798}$     \\
			\hline
		\end{tabular}
		\caption{Fit results using all the available inputs.} 
		\label{tab:LCSRlatfitres}
	\end{table}

	\section{Analysis and results}~\label{results}
	
	The BaBar and Belle collaborations have performed measurements on the partial branching fractions in bins of dilepton invariant mass squared ($q^2$) for the exclusive $b\to u\ell\nu$ transitions \cite{delAmoSanchez:2010af,Ha:2010rf,Lees:2012vv,Sibidanov:2013rkk} with $\pi$, $\rho$, $\omega$ and $\eta$ mesons in the final state. On the $b \to d \ell\ell$ decays, the LHCb collaboration has observed the $B^+\to\pi^+\mu^+ \mu^-$ decay for the first time at 5.2 $\sigma$ \cite{LHCb:2012de} and in \cite{LHCb:2015hsa}, it has provided measurements on the partial branching fractions in bins of $q^2$, for which we restrict ourselves to the low $q^2$ bins (upto 8 $GeV^2$). This is done to avoid the uncertainties resulting from nonperturbative corrections due to the near-threshold $c\bar{c}$ intermediate states.
	
	The theoretical expressions for the differential decay widths for $\btopilnu$ and $\btorholnu$ contain $|V_{ub}|$ as the overall normalizing constant. As one can see from eq.~\ref{eq:Heff}, the loop-induced processes $B\to\pi\ell \ell$ and $B\to\rho\ell \ell$ are sensitive to the products of CKM matrix elements $\lambda_u =V_{ub} V_{ud}^*$ and $\lambda_t =V_{tb} V_{td}^*$ respectively. Thus, to make the $|V_{ub}|$ dependence explicit, we define the product of the elements as
	\begin{equation}
	V_{ub} V_{ud}^* = |V_{ub}| |V_{ud}|e^{-i \gamma}, \ \ \ \ 
	V_{tb} V_{td}^* = |V_{tb}| |V_{td}|e^{i \beta}, 
	\end{equation}
	where angles are defined as $\gamma$ = $\text{arg}(- V_{ud} V_{ub}^*/V_{cd} V_{cb}^*)$ and $\beta$ = $\text{arg}(- V_{cd} V_{cb}^*/V_{td} V_{tb}^*)$.
	
	We perform a statistical frequentist analysis combining the available experimental inputs on $\btopilnu$, $\btorholnu$ and $B\to\pi\ell \ell$ modes \cite{Ha:2010rf,delAmoSanchez:2010af,Lees:2012vv,Sibidanov:2013rkk,LHCb:2015hsa}, and the available theory inputs from lattice and LCSR on the respective form factors in these decays and $B\to \rho\ell\ell$.  We have extracted the coefficients of the BSZ expansion defined in eq.~\ref{eq:bszexp} along with the magnitude of $|V_{ub}|$.  Here, we take the parameters $|V_{ud}| \text{cos} (\gamma)$, $|V_{ud}| \text{sin} (\gamma)$, $|V_{tb}| |V_{td}| \text{cos} (\beta)$, $|V_{tb}| |V_{td}| \text{sin} (\beta)$, the product $f_B f_{\pi}$ ($f_B$ and $f_{\pi}$ are decay constants of the $B$ and $\pi$ mesons) and the inverse moment of $B$ meson ($\lambda_{B,+}^{-1}$) as nuisance parameters in the fits. The inputs on the CKM parameters are taken from \cite{ckmfitter1}. As we have mentioned in our introduction, in this fit, we have considered all possible correlations between the form factor obtained in lattice and LCSR analyses. The fit result is given in table \ref{tab:LCSRlatfitres} and we will provide a separate file for the respective correlations. Utilizing the result from the fit, we provide predictions for the branching ratios, direct CP asymmetries and isospin asymmetry for $B\to\pi\ell \ell$ and various angular observables for $B\to\rho\ell \ell$ transitions, as discussed below.
	
	\subsection{$B\to\pi\ell \ell$}
	
	\begin{table} [t]
		\centering
		\begin{tabular}{c|cccc}
			\hline
			& \multicolumn{4}{c}{Predictions in a few $q^2$-bins ( \text{GeV}$^2$)} \\
			Observables 	& \cline{1-4} \\
			& $\text{[0.1-1]}$  &  $\text{[1-2]}$  &  $\text{[2-4]}$  &  $\text{[4-6]}$  \\
			\hline 		 
		  	$B(B^0\to\pi^0\ell \ell) \times 10^8$ & $\text{0.0499(28)}$  &  $\text{0.0503(28)}$  &  $\text{0.0985(54)}$  &  $\text{0.0976(51)}$\\
		  	$B(\bar{B}^0\to\pi^0\ell \ell) \times 10^8$ & $\text{0.0283(21)}$  &  $\text{0.0330(23)}$  &  $\text{0.0674(44)}$  &  $\text{0.0696(43)}$    
		     \\
			$B(B^-\to\pi^-\ell \ell) \times 10^8$ & $\text{0.0444(28)}$  &  $\text{0.0538(43)}$  &  $\text{0.1308(92)}$  &  $\text{0.1465(92)}$     \\
			$B(B^+\to\pi^+\ell \ell) \times 10^8$ & $\text{0.1950(70)}$  &  $\text{0.1551(65)}$  &  $\text{0.251(12)}$  &  $\text{0.223(11)}$     \\
			$\langle A_{CP}^0 \rangle$ & $\text{-0.276(15)}$  &  $\text{-0.2074(69)}$  &  $\text{-0.1873(56)}$  &  $\text{-0.1681(55)}$     \\
			$\langle A_{CP}^+ \rangle$ & $\text{-0.629(12)}$  &  $\text{-0.486(16)}$  &  $\text{-0.316(11)}$  &  $\text{-0.2077(72)}$     \\
			$ \langle A_I \rangle$ & $\text{-0.479(11)}$  &  $\text{-0.3992(50)}$  &  $\text{-0.3625(23)}$  &  $\text{-0.3445(11)}$     \\
			$ \langle \bar{R}_{\pi}^0 \rangle$ & $\text{0.99195(13)}$  &  $\text{1.00007(13)}$  &  $\text{1.00039(13)}$  &  $\text{1.00050(12)}$     \\
			$ \langle R_{\pi}^- \rangle$ & $\text{0.99043(34)}$  &  $\text{1.00022(19)}$  &  $\text{1.00044(14)}$  &  $\text{1.00051(13)}$     \\
			\hline
		\end{tabular}
		\caption{\small Predictions of observables for $B\to\pi\ell \ell$ decays in the SM, obtained using the fit results given in table \ref{tab:LCSRlatfitres}.} 
		\label{tab:BPipred}
	\end{table}

	\begin{figure*}[t]
		\small
		\centering
		\subfloat[]{\includegraphics[width=0.35\textwidth]{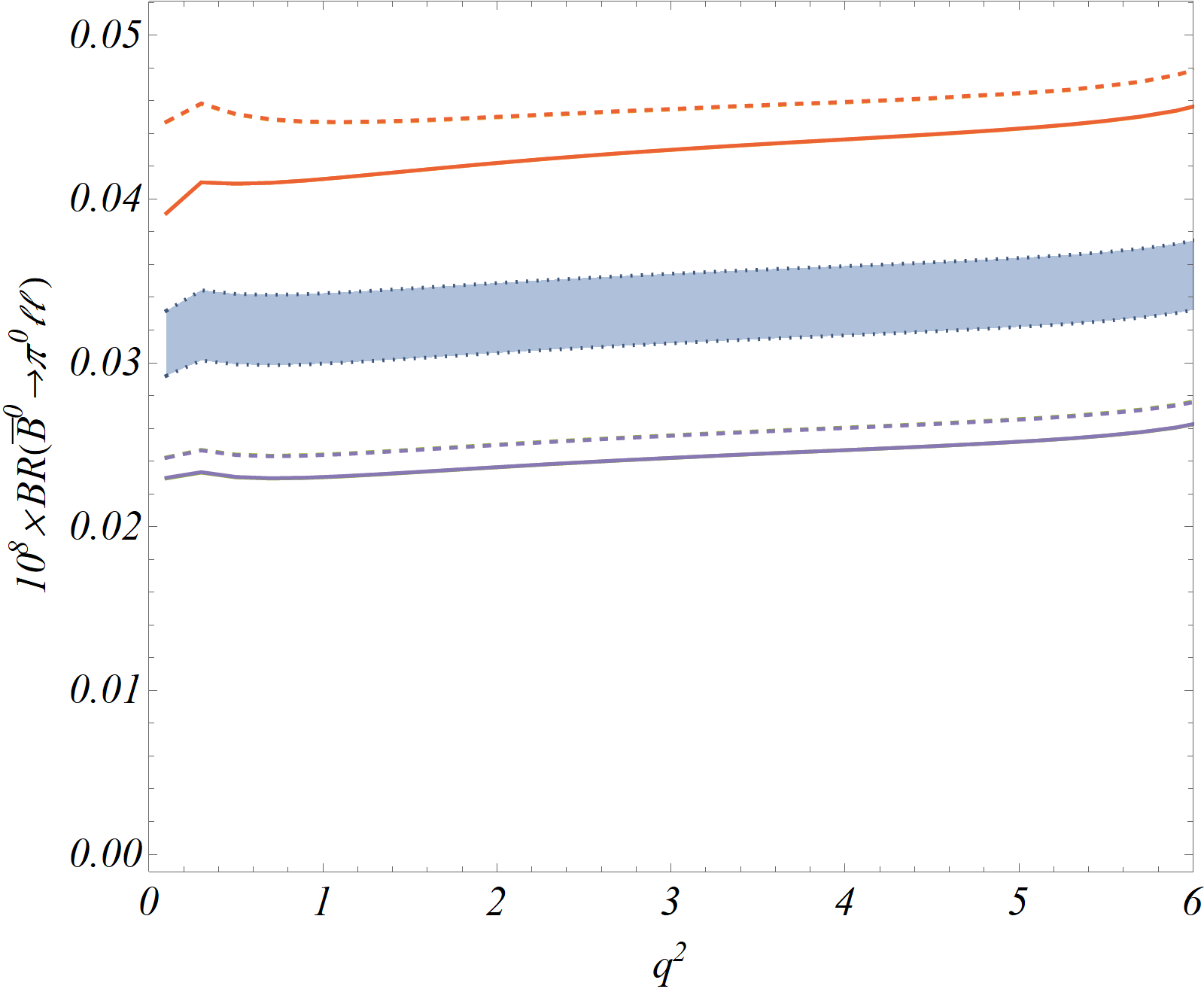}\label{fig:BR0}}
		\subfloat[]{\includegraphics[width=0.35\textwidth]{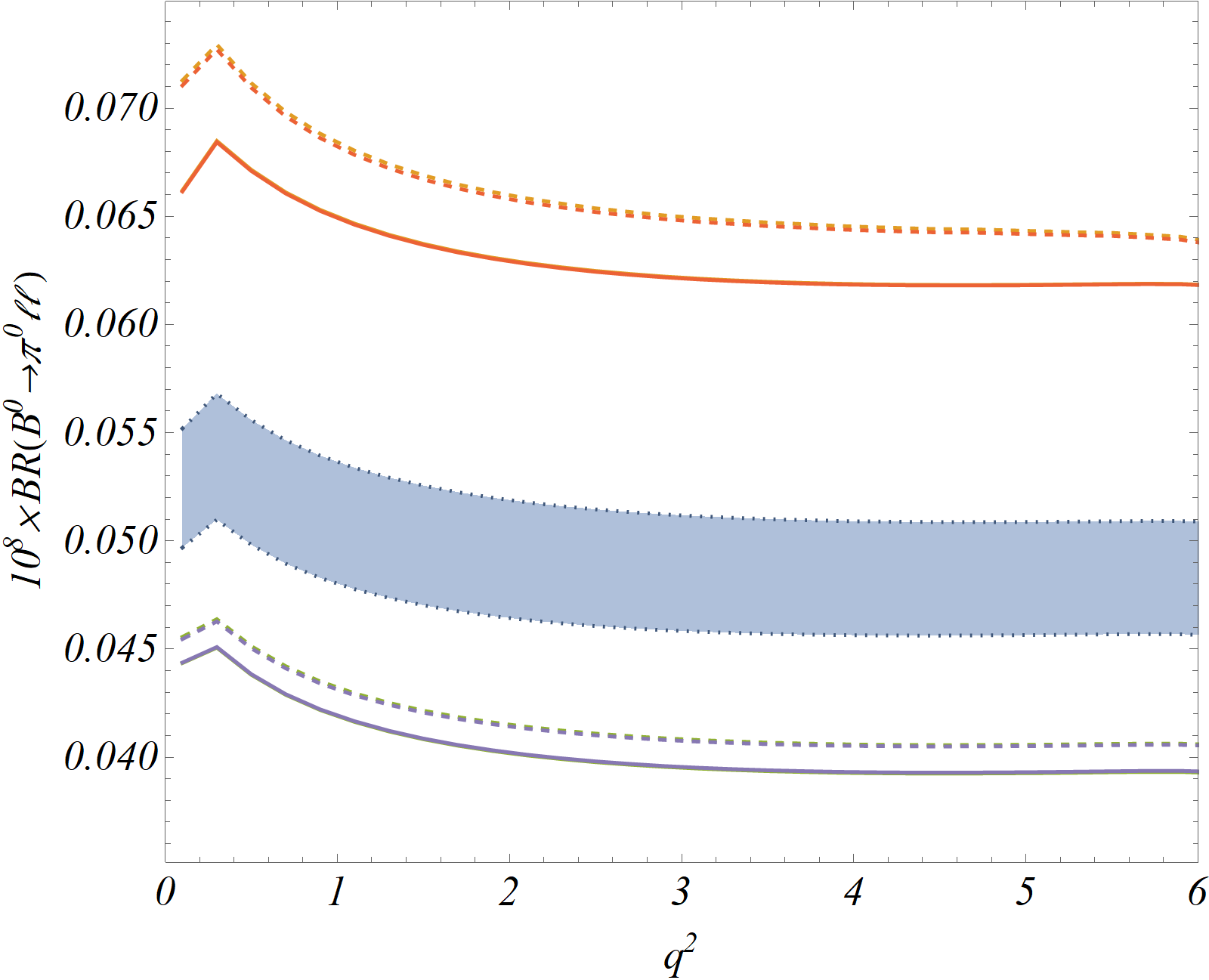}\label{fig:Brc0}}
		\subfloat[]{\includegraphics[width=0.35\textwidth]{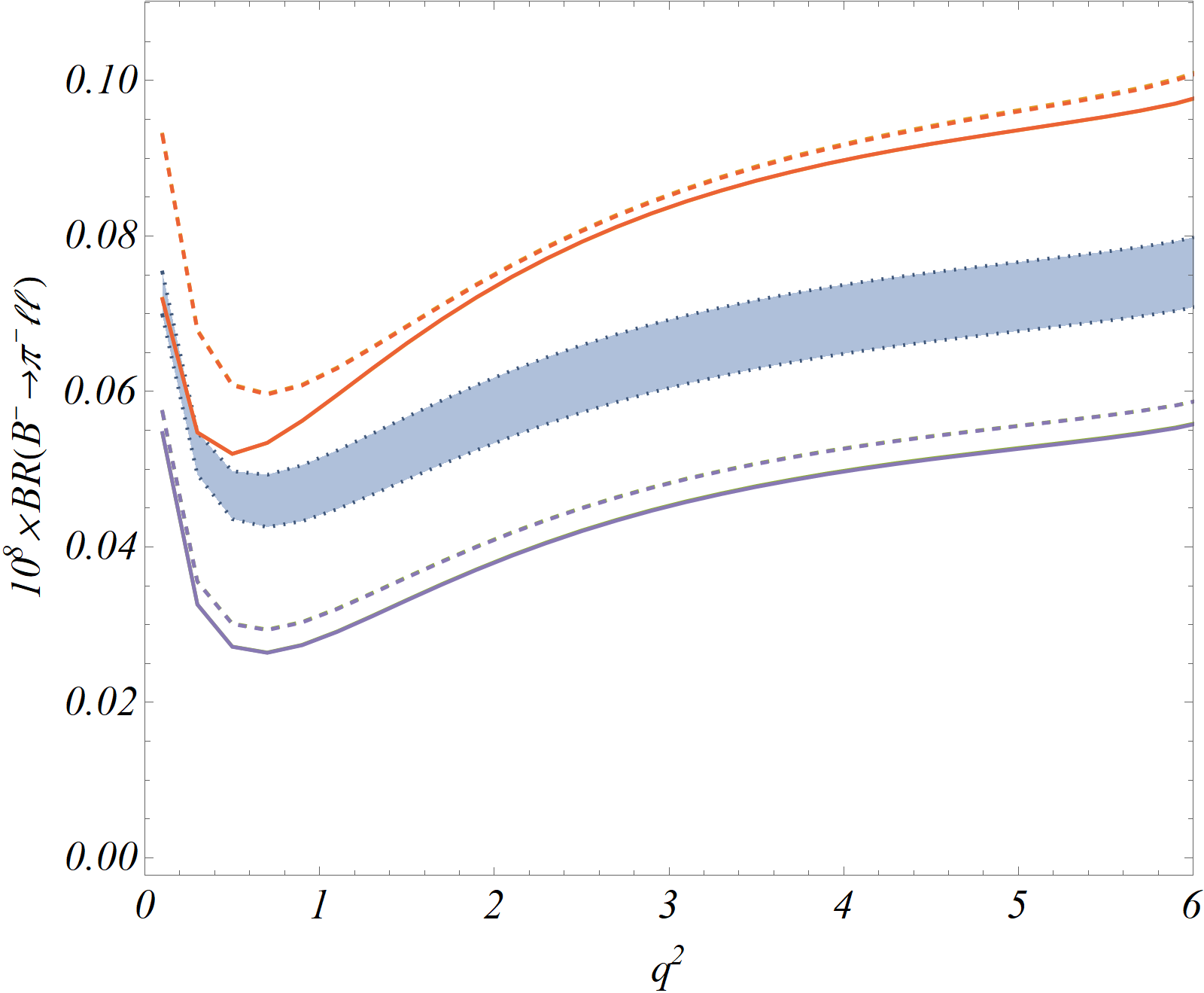}\label{fig:BRP}}\\
		\subfloat[]{\includegraphics[width=0.35\textwidth]{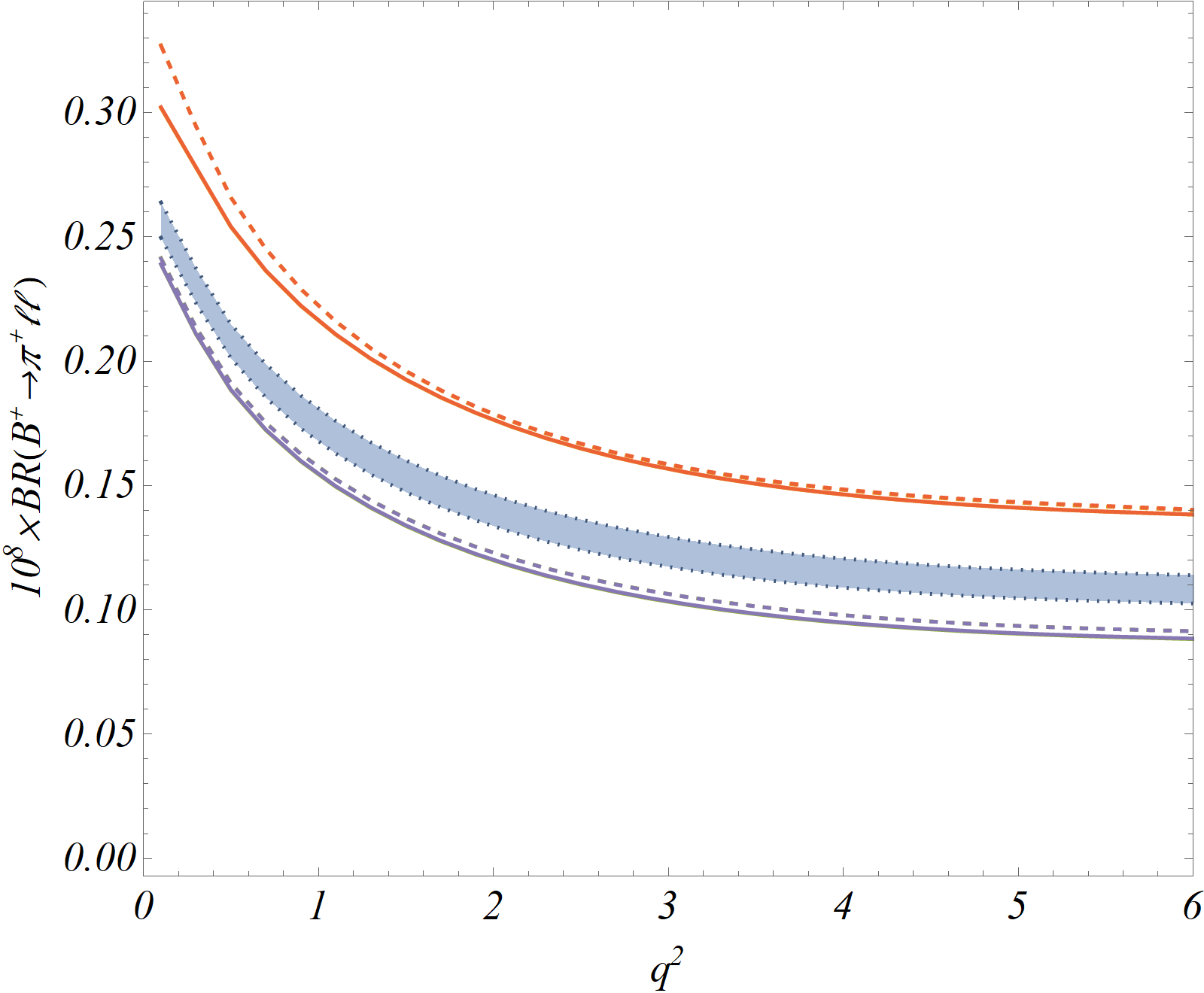}\label{fig:BRcP}}
		\subfloat[]{\includegraphics[width=0.35\textwidth]{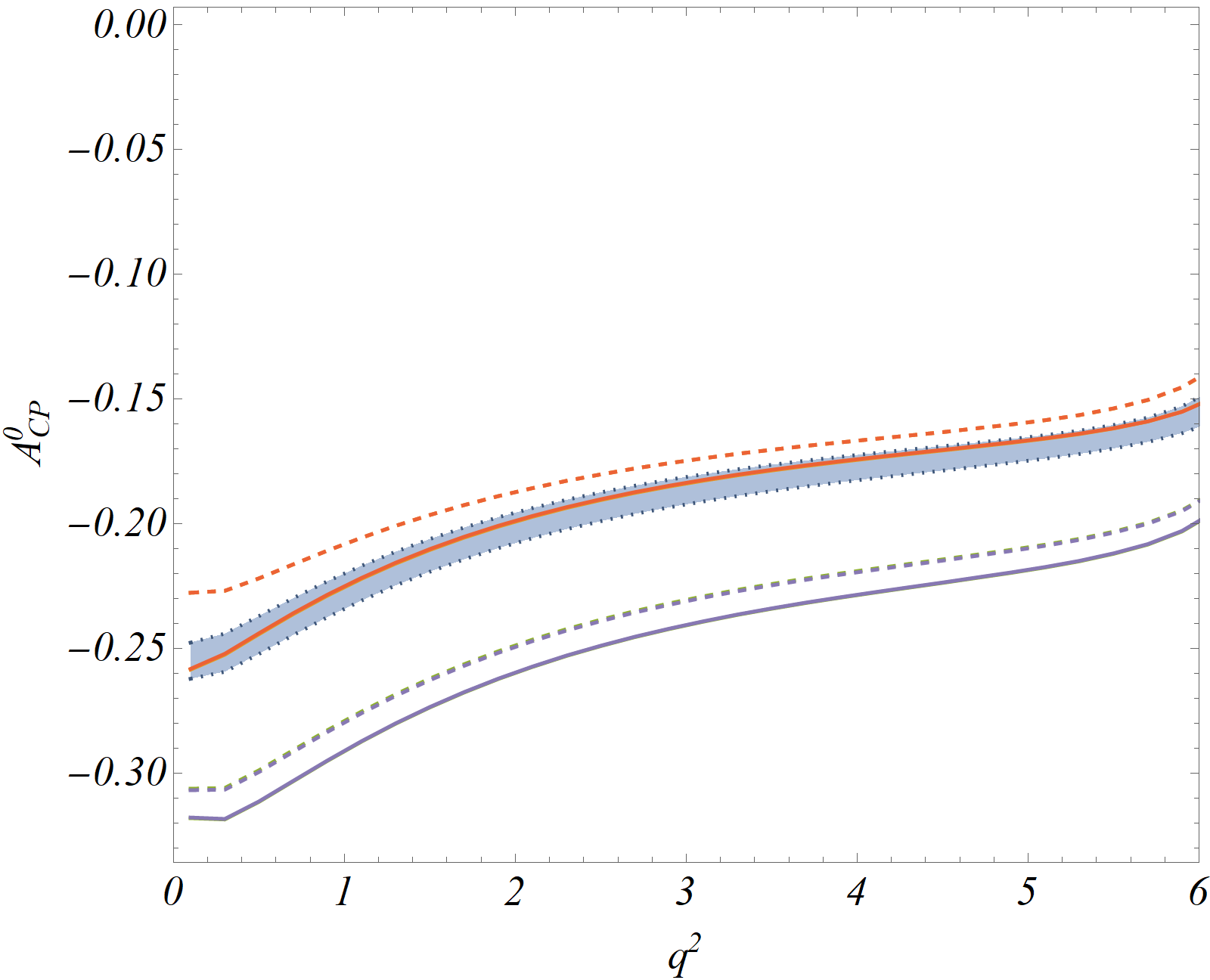}\label{fig:ACP0}}
		\subfloat[]{\includegraphics[width=0.35\textwidth]{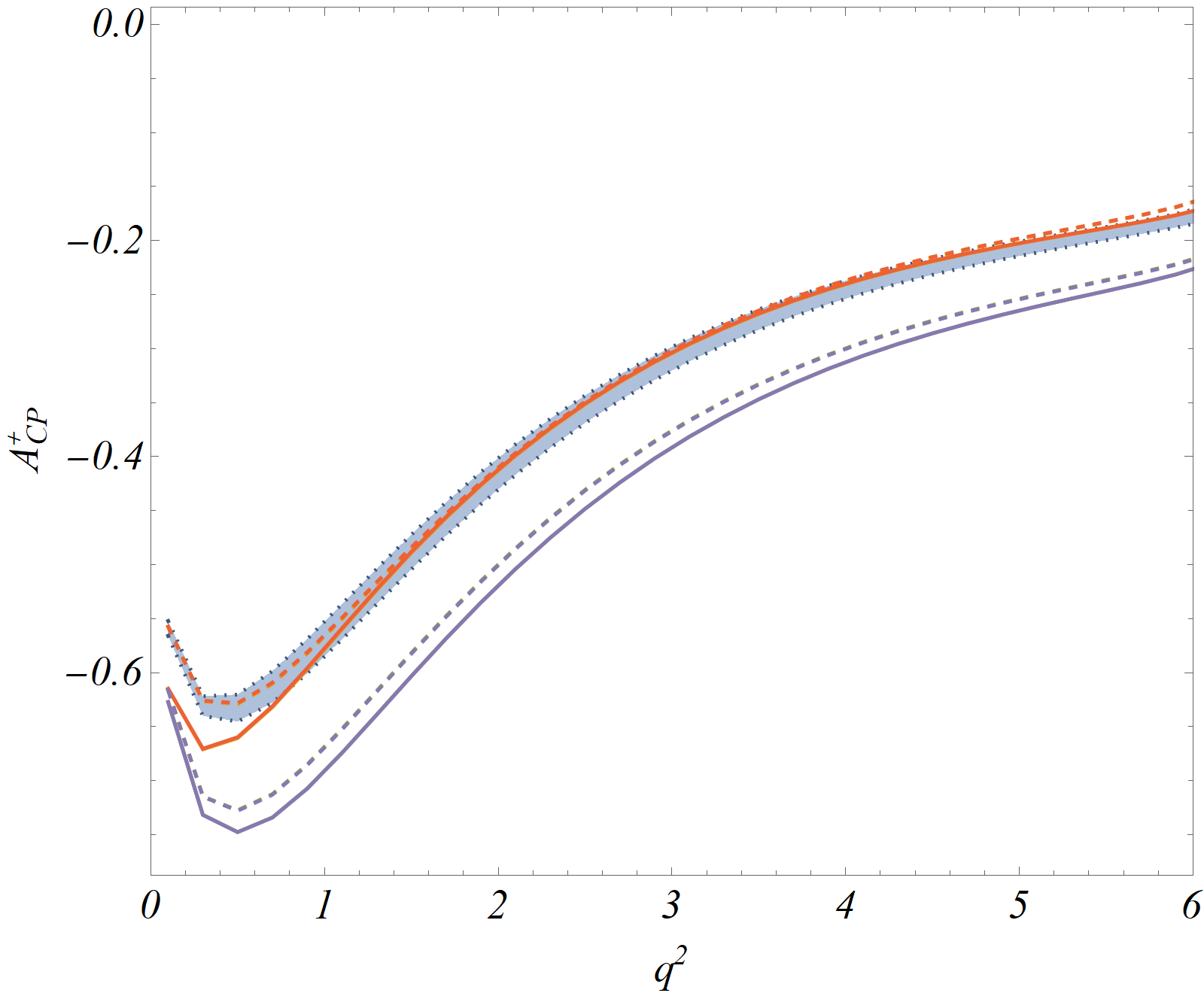}\label{fig:ACPP}}\\
		\subfloat[]{\includegraphics[width=0.35\textwidth]{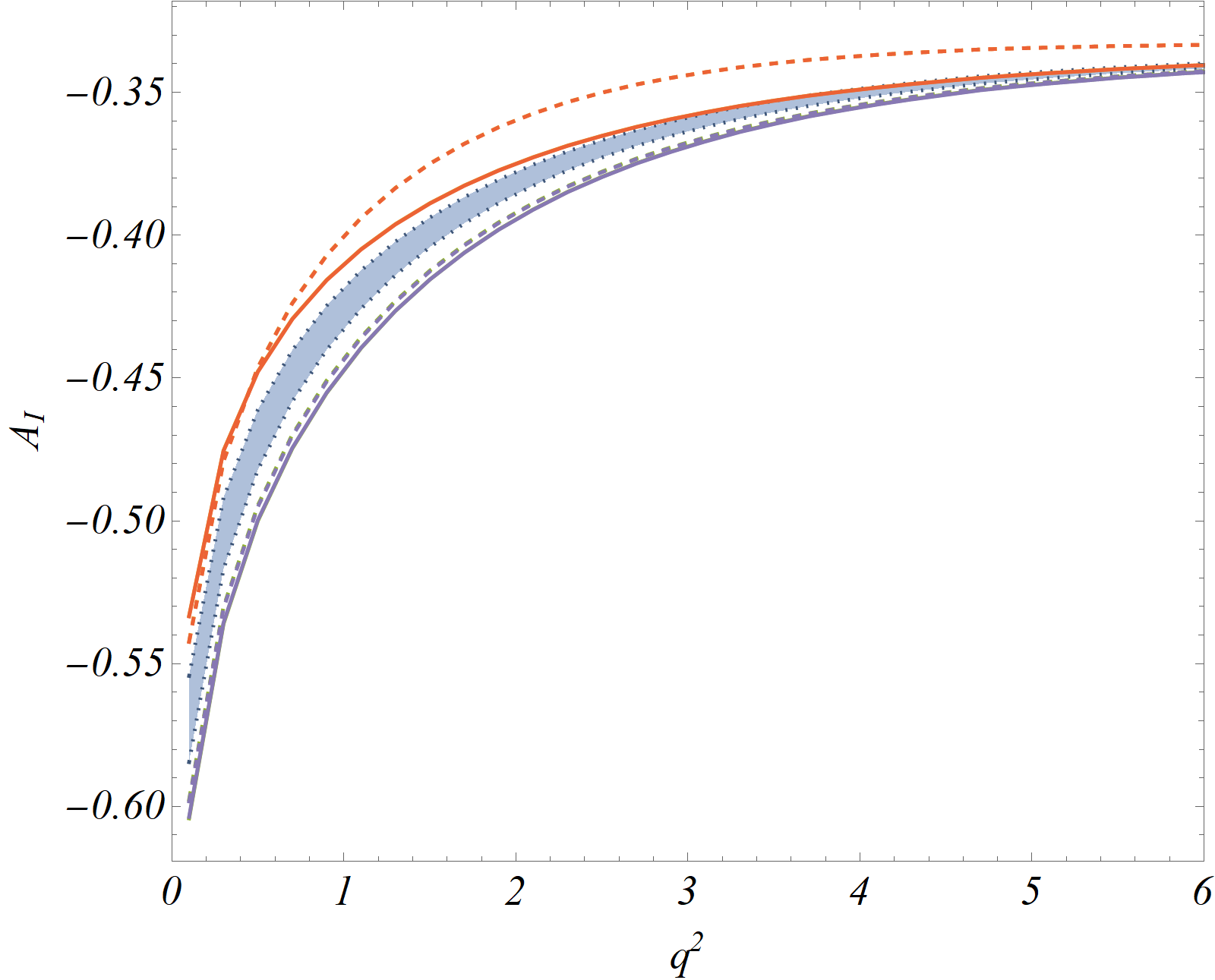}\label{fig:AI}}
		\subfloat[]{\includegraphics[width=0.35\textwidth]{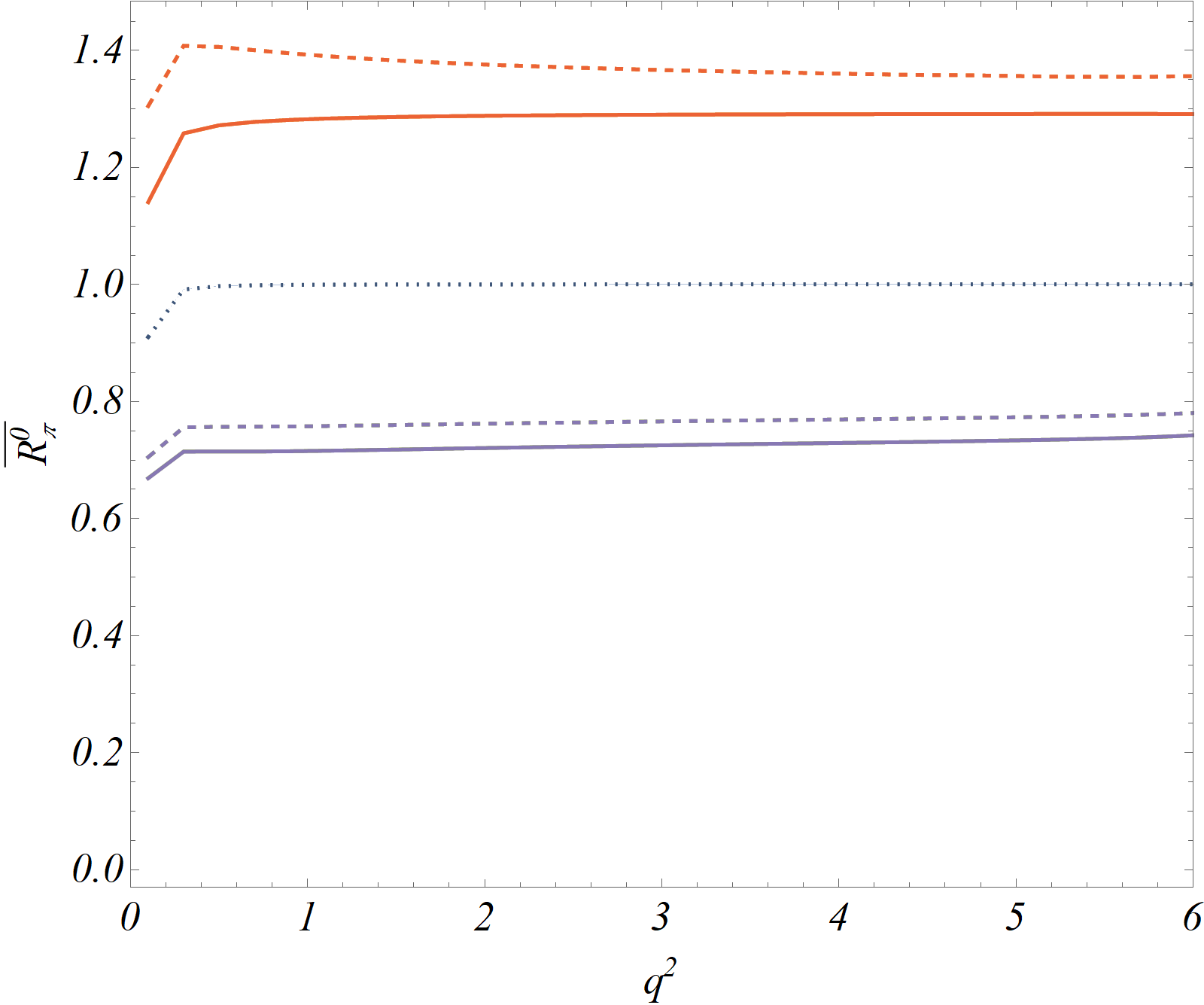}\label{fig:RPi0}}
		\subfloat[]{\includegraphics[width=0.35\textwidth]{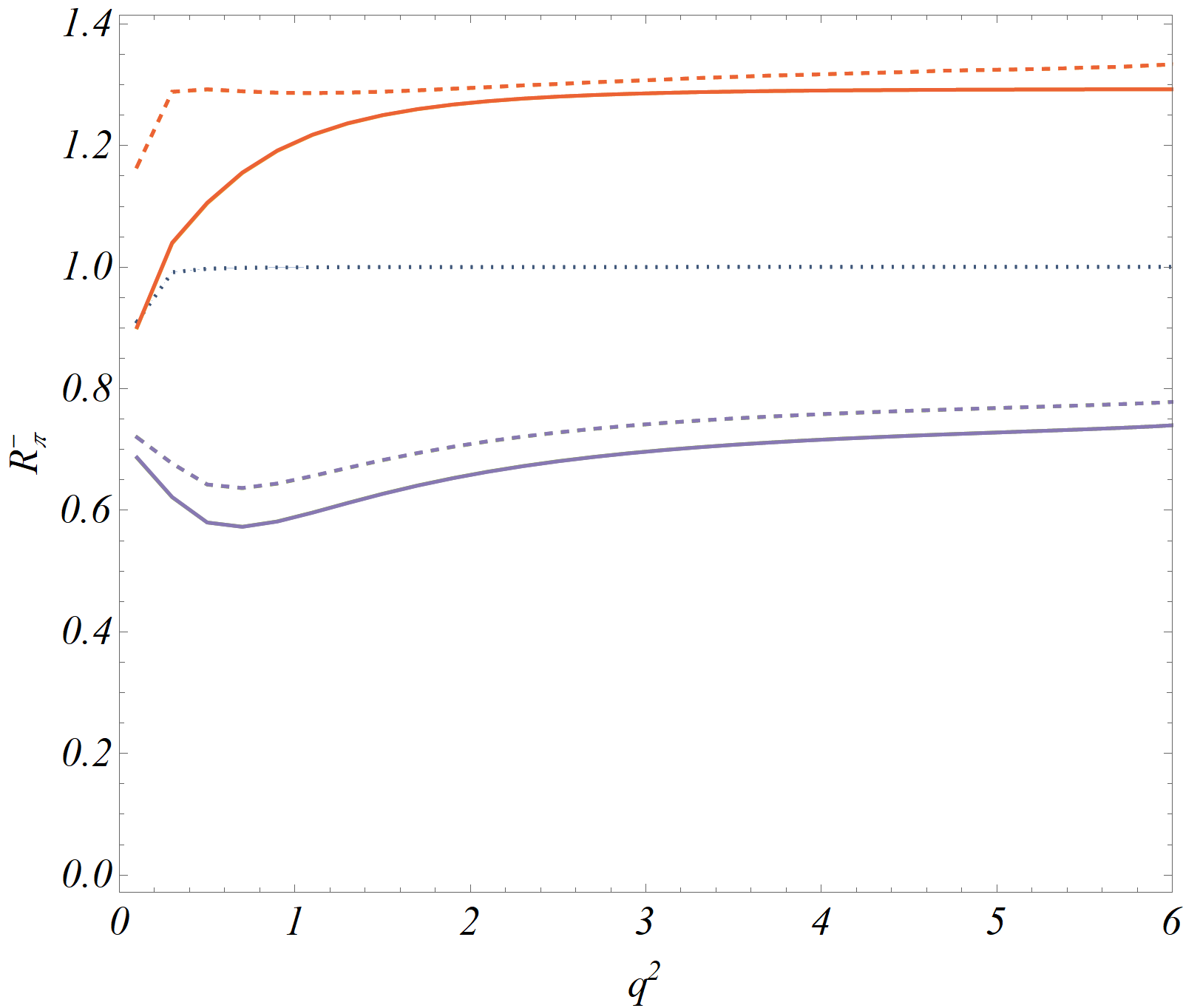}\label{fig:RPiP}}\\
		\subfloat[Legend]{\includegraphics[width=0.35\textwidth]{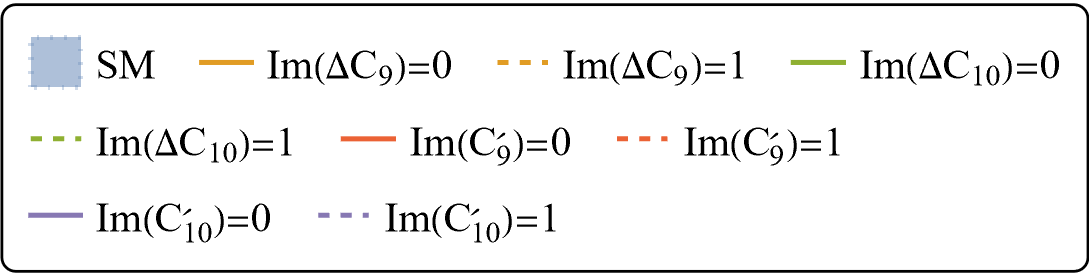}\label{fig:legpi}}
		\caption{The $q^2$ dependence for the $B\to\pi ll$ observables in the SM and the four different NP scenarios with NP Wilson coefficients $\Delta C_{9,10}$ and $C_{9,10}^\prime$. For the NP scenarios, the plots show the corresponding  dependence of the observables on the imaginary parts of the respective NP's for the two different benchmark values 1 and 0 of the imaginary parts of the WC's while the real part is fixed at 1. For detail, please see the legend.}
		\label{fig:BPiobs}
	\end{figure*}

	We study the modes $B^{\pm}\to\pi^{\pm}\ell \ell$ and $\bar{B^0}(B^0)\to\pi^0\ell \ell$ for which the $q^2$-dependent direct CP asymmetries are defined as :
	
	\begin{align}
	&A_{CP}^+ (q^2) = \frac{d\mathcal B(B^-\to \pi^-\ell\ell)/dq^2
		-d\mathcal B(B^+\to \pi^+\ell\ell)/dq^2}{d\mathcal B(B^-\to \pi^-\ell\ell)/dq^2
		+d\mathcal B(B^+\to \pi^+\ell\ell)/dq^2}\,, \\
	& \nonumber \\
	&A_{CP}^0 (q^2) = \frac{d\mathcal B(\bar B^0\to \pi^0\ell\ell)/dq^2
		-d\mathcal B(B^0\to  \pi^0\ell\ell)/dq^2}
	{d\mathcal B(\bar B^0 \to \pi^0\ell\ell)/dq^2
		+d\mathcal B(B^0\to \pi^0\ell\ell)/dq^2}
	\end{align}
	
	The definition of the $q^2$-dependent CP-averaged isospin asymmetry corresponds to the LHCb definition \cite{LHCb:2014cxe} :
	\begin{equation}
	A_I = \frac{d\mathcal B(B^0\to \pi^0\ell\ell)/dq^2 - (\tau_0/\tau_+)d\mathcal B(B^+\to \pi^+\ell\ell)/dq^2}{d\mathcal B(B^0\to \pi^0\ell\ell)/dq^2 + (\tau_0/\tau_+)d\mathcal B(B^+\to \pi^+\ell\ell)/dq^2},
	\end{equation} 
	where $\tau_0$ and $\tau_+$ are $B^0$ and $B^+$ meson lifetimes. In addition, we have incorporated the interesting observables: 
	\begin{equation}
	R_\pi^{\pm(0)} = \frac{\Gamma(B^{\pm (0)}\to \pi^{\pm(0)}\mu\mu )}{\Gamma(B^{\pm (0)}\to \pi^{\pm(0)} e e)},
	\end{equation}
	which are lepton flavour universality (LFU) conserving in the SM and are potentially sensitive to new degrees of freedom. 
	
In refs.~\cite{Ali:2013zfa,Hou:2014dza,Hambrock:2015wka}, the $q^2$ distributions of the branching fractions, CP-asymmetries and the isospin asymmetries were studied in SM. Those analyses were based on the available experimental data on $B\to \pi\ell\nu_{\ell}$ decays and with minimal theory information, and the results have large errors. 
In table \ref{tab:BPipred}, we have provided predictions for the various observables in the four $q^2$ bins. These predictions are obtained using the fit results given in table \ref{tab:LCSRlatfitres}. The uncertainty corresponds to the uncertainty in determining the form factors, decay constants and the CKM elements. These are the most precise results obtained so far. The $q^2$-distributions of the relevant branching fractions and the CP, isospin-asymmetries defined above are given in figs. \ref{fig:BPiobs}. In figs. \ref{fig:BR0}, \ref{fig:Brc0}, \ref{fig:BRP}, and \ref{fig:BRcP},  we have shown the variation of differential branching fraction with $q^2$ for both $B^{\pm}\to\pi^{\pm}\ell \ell$ and $\bar{B}(B)\to\pi^0\ell \ell$ modes, comparing which we notice that the decay spectra for $\bar{B}(B)\to\pi^0\ell \ell$ is nearly flat, whereas for $B^{\pm}\to\pi^{\pm}\ell \ell$ shows visible $q^2$ dependence. Similarly, there are differences between the charged and neutral $B$ decays in the $q^2$ variations of the CP and isospin-asymmetries. This is due to the differences in the $q^2$ behaviour of the weak annihilation contributions in the charged and neutral $B$ decays at the leading order. Note that $C_9^t$ defined in eq.~\ref{eq:C9} has dominant contribution in $\bar{B}(B)\to\pi^0\ell \ell$ decays and the contribution from $C_9^u$ is negligible. However, for $B^{\pm}\to\pi^{\pm}\ell \ell$ decays, the contribution from $C_9^u$ becomes equally important at the low $q^2$ regions. As can be seen from ref. \cite{Beneke:2004dp}, at the leading order, the contribution in $\mathcal T_{\pi}^{(t)} (q^2)$ and $\mathcal T_{\pi}^{(u)} (q^2)$ from the weak annihilation diagrams is proportional to $\lambda_{B,-}^{-1}(q^2) {\hat T}_{\pi,-}^{(0,t)} $ and $\lambda_{B,-}^{-1}(q^2) {\hat T}_{\pi,-}^{(0,u)} $, respectively. The imaginary part of  $\lambda_{B,-}^{-1}(q^2)$ (eq.~\ref{eq:lam-}) is highly $q^2$ dependent and has large values at low $q^2$ compared to that at high $q^2$ regions. On the other hand, numerically ${\hat T}_{\pi,-}^{(0,t)} <  {\hat T}_{\pi,-}^{(0,u)}$ since ${\hat T}_{\pi,-}^{(0,u)}$ is proportional to the WC $C_2 \sim 1$ whereas ${\hat T}_{\pi,-}^{(0,t)}$ is proportional to a linear combination of $C_3 \sim -0.005$ and $C_4 \sim -0.08$. Due to the reasoning as mentioned above, $\mathcal T_{\pi}^{(u)} (q^2)$ is more sensitive to the variation of $q^2$ as compared to $\mathcal T_{\pi}^{(t)} (q^2)$.

	\begin{table}[t]
		\centering
		\begin{tabular}{c|c|c}
			\hline
			$\left.\text{Bin[}\text{GeV}^2\right]$ & \multicolumn{2}{|c}{$\frac{dB}{d q^2}( B^{\pm} \to \pi^{\pm} \mu^+ \mu^-)$~ ($10^{-9}$ \text{GeV}$^{-2}$)} \\
			
			& \cline{1-2}
			
			& Exp. data  & SM prediction \\
			\hline
			0.1 - 2 & 1.89 $\pm$ 0.44 & 1.18 $\pm$ 0.04  \\
			2 - 4 & 0.62 $\pm$ 0.36 & 0.96 $\pm$ 0.06\\
			4 - 6 & 0.85 $\pm$ 0.30 &  0.93 $\pm$ 0.05 \\
			\hline
		\end{tabular}
		\caption{\small Comparison of the SM predictions for the CP averaged differential branching fractions with the experimental measurements from LHCb \cite{LHCb:2015hsa}.}
		\label{tab:brbpicomexpsm}
	\end{table}

Also, it should be noted that $\lambda_{B,-}^{-1}$ acts as a source of strong phase, in particular, at the low $q^2$ regions. Thus, the large imaginary contribution of $C_{9,P}^{(u)}$ at low $q^2$ for $B^{\pm}$ decays results in the large CP asymmetry at low $q^2$. The CP asymmetry for $B^0$ decays is small, as weak annihilation is mostly mediated by loop-suppressed QCD penguins. Isospin asymmetry is also generated by hard-spectator interactions and is more pronounced at low $q^2$, as shown in fig. \ref{fig:AI}.

Using our fit results we have also predicted the differential branching fraction $\frac{dB}{d q^2}( B^{\pm} \to \pi^{\pm} \mu^+ \mu^-)$ in three $q^2$ bins which can be compared with the respective measured values \cite{LHCb:2015hsa}. In table \ref{tab:brbpicomexpsm}, in the third column, we have shown the respective predictions in the SM, which should be compared with the data in the second column. Note that the SM predictions are consistent with the respective measured values in the rest of the two bins apart from the low bin. However, it is important to mention that the measured values have large errors, while with the currently available inputs, the predictions in the SM can be made with an accuracy $\lsim$ 5\%. To conclude it further we need to wait for precise data.

Apart from the precise predictions in the SM, we have also tested the NP sensitivities of the observables as mentioned above. In this analysis, we have considered NP effects in the following four operators:  
	\begin{align}
	\mathcal{O}_{9} &= \frac{e^2}{16 \pi^2} (\bar{d} \gamma_{\mu} P_L b)(\bar{\mu} \gamma^\mu \mu), ~~~~~~~~~~~~~~ \mathcal{O}_{9}^\prime = \frac{e^2}{16\pi^2} (\bar{d} \gamma_{\mu} P_R b)(\bar{\mu} \gamma^\mu \mu), \nn \\
	\mathcal{O}_{10} &=\frac{e^2}{16 \pi^2} (\bar{d}  \gamma_{\mu} P_L b)(  \bar{\mu} \gamma^\mu \gamma_5 \mu), ~~~~~~~~~~~~~ \mathcal{O}_{10}^\prime =\frac{e^2}{16 \pi^2} (\bar{d}  \gamma_{\mu} P_R b)(  \bar{\mu} \gamma^\mu \gamma_5 \mu).
	\label{eq:effopr}
	\end{align}
	The relevant new WCs are the following: $\Delta C_9$, $C_9^{\prime}$, $\Delta C_{10}$, $C_{10}^{\prime}$. In the NP scenarios, the $\mathcal{H}_{eff}$ defined in eq.~\ref{eq:Heff} will be modified. We have introduced the new operators via a modification in $\mathcal{H}^{(t)}_{eff}$: 
	\begin{equation}\label{eq:hefftm}
	{\cal H}_{eff}^{(t)} =  
	C_1 \mathcal O_1^c + C_2 \mathcal O_2^c + \sum_{i=3}^{10} C_i \mathcal O_i  + \Delta C_9 \mathcal O_9 + \Delta C_{10} \mathcal O_{10} + C_9^{\prime} \mathcal O_9^{\prime} +C_{10}^{\prime} \mathcal O_{10}^{\prime}   \,
	\end{equation} 
	The new WCs should be scaled accordingly. To keep our discussion limited we have not considered the NP scenarios with scalar or pseudoscalar types of operators. We have shown the $q^2$ variations of all the observables mentioned above in the four specific NP scenarios in fig.~\ref{fig:BPiobs}. Here, we will consider an observable sensitive to a particular NP scenario if we find a significant deviation in its prediction w.r.t the corresponding SM prediction. To generate the $q^2$-distributions, for all the NP cases, we have set $Re(\Delta C_i) = Re(C_i^{\prime})=1$. The imaginary component of the new WCs have been set to 0 or 1, in order to help us distinguish the sensitivities of the observables towards the new phase. The predictions of the associated observables in different NP scenarios in four $q^2$ bins are presented in tables~\ref{tab:appBtopiasy} and \ref{tab:appBtopibrR}, where we have presented our results for a few combinations of representative values of real and imaginary parts of the new WCs. Following are a few observations from fig.~\ref{fig:BPiobs}:
	\begin{itemize}
		\item None of the observables we discuss above for $B\to \pi\ell\ell$ decays will be able to distinguish the effects of $\mathcal{O}_{9}$ ($\mathcal{O}_{10}$) from that of $\mathcal{O}_{9}^{\prime}$ ($\mathcal{O}_{10}^{\prime}$), respectively. The $q^2$-distributions are the same for $\mathcal{O}_9$ ($\mathcal{O}_{10}$) and $\mathcal{O}_{9}^{\prime}$ ($\mathcal{O}_{10}^{\prime}$).
		
		\item The branching fractions $B(B \to\pi \ell \ell)$ are sensitive to all four NP scenarios. Interestingly, for the scenarios with $\mathcal{O}_{10}$ or $\mathcal{O}_{10}^{\prime}$, the branching fractions reduce than the corresponding SM predictions. While for the scenario with $\mathcal{O}_9$ or $\mathcal{O}_9^{\prime}$, we note an increase in the values of branching fractions from that of the SM.  
		
		\item As expected, it is hard to distinguish the impact of the imaginary WCs in the branching fractions.
		
		\item The CP-asymmetries are sensitive only to $\Delta C_{10}$ and $C_{10}^{\prime}$. The predicted values of both the CP-asymmetric observables in the specified NP scenarios will be less than that of the SM for these two WC's. Also, we note a negligible impact of these observables on the imaginary component of these WCs.  
		
		\item Contrary to the CP-asymmetries, the isospin asymmetry $A_I$ is sensitive to $\Delta C_9$ and $C_{9}^{\prime}$ in the region $2 < q^2 < 6$ GeV$^2$ and the predicted value will be higher than that of the SM. In the low $q^2$ regions it will be difficult to see the impact of the NP over that of SM. This observable is sensitive to the imaginary component of the respective WCs. For negative values of the respective WC, the predicted value will be consistent with the SM, and it will be hard to distinguish the effect. For positive values, the predictions will be even higher than the predictions obtained with the imaginary component as zero. The details can be seen from the table \ref{tab:appBtopiasy} in the appendix where we have given predictions for a couple of more benchmarks.

	\end{itemize}

	Apart from the limitations discussed, the above-itemized information is beneficial in distinguishing the impact of different NP scenarios we are considering here. Once the measured values are available, any deviation from the respective SM predictions will hint toward a particular type of NP scenario. For example, the NP scenario with the operator $\mathbf{O}_9$ or $\mathbf{O}_9^{\prime}$ will show deviations in $A_I$ and the respective branching fractions but not in $A_{CP}$s. On the contrary, if we see deviations in $A_{CP}$s but not in $A_I$, that will be an indication of the scenario with the operator $\mathbf{O}_{10}$ or $\mathbf{O}_{10}^{\prime}$. The predictions of these observables for a few benchmark points of the new WCs are given in tables \ref{tab:appBtopiasy} and \ref{tab:appBtopibrR} respectively, which could be tested in future experimental results. Also, these results clearly indicate the pattern of NP effects in those observables.


	\begin{table}[t]
		\small
		\centering
		\renewcommand*{\arraystretch}{1.1}
		\begin{tabular}{*{5}{c}}
			\hline
			$\left.\text{Bin[}\text{GeV}^2\right]$  &  $\text{[0.1-1]}$  &  $\text{[1-2]}$  &  $\text{[2-4]}$  &  $\text{[4-6]}$  \\
			\hline
			$\langle P_1^- \rangle$ & $\text{0.0042(33)}$  &  $\text{0.0030(47)}$  &  $\text{-0.048(21)}$  &  $\text{-0.086(34)}$   \\
			\hline
			$\langle P_2^- \rangle$ & $\text{-0.1262(50)}$  &  $\text{-0.422(13)}$  &  $\text{-0.177(28)}$  &  $\text{0.241(22)}$    \\
			\hline
			$\langle P_3^- \rangle$ & $\text{0.00040(25)}$  &  $\text{0.0024(15)}$  &  $\text{0.0048(28)}$  &  $\text{0.0052(28)}$     \\
			\hline
			$ \langle {P_4^{'}}^- \rangle$ & $\text{-0.117(29)}$  &  $\text{-0.159(19)}$  &  $\text{-0.366(15)}$  &  $\text{-0.474(12)}$     \\
			\hline
			$ \langle {P_5^{'}}^- \rangle$ & $\text{0.9343(64)}$  &  $\text{0.568(32)}$  &  $\text{-0.302(41)}$  &  $\text{-0.744(25)}$     \\
			\hline
			$ \langle {P_6^{'}}^- \rangle$ & $\text{-0.179(22)}$  &  $\text{-0.078(40)}$  &  $\text{0.012(46)}$  &  $\text{-0.048(48)}$     \\
			\hline
			$ \langle {P_8^{'}}^- \rangle$ & $\text{0.095(16)}$  &  $\text{-0.1176(72)}$  &  $\text{-0.078(14)}$  &  $\text{-0.047(18)}$     \\
			\hline
			$\langle BR^{-} \rangle \times 10^9$ & $\text{2.32(21)}$  &  $\text{1.004(89)}$  &  $\text{2.08(19)}$  &  $\text{2.48(21)}$     \\
			\hline
			$\langle A_{FB}^- \rangle$ & $\text{-0.1076(41)}$  &  $\text{-0.206(18)}$  &  $\text{-0.057(11)}$  &  $\text{0.105(13)}$     \\
			\hline
			$\langle F_{L}^- \rangle$ & $\text{0.203(19)}$  &  $\text{0.631(32)}$  &  $\text{0.765(21)}$  &  $\text{0.700(24)}$     \\
			\hline
			$\langle R_{\rho}^- \rangle$ & $\text{0.98067(29)}$  &  $\text{0.99566(39)}$  &  $\text{0.99610(38)}$  &  $\text{0.99683(27)}$    \\
			\hline
		\end{tabular}
		\caption{\small SM Predictions of observables for $B^-\to \rho^-\ell \ell$.}
		\label{tab:BMpred}
	\end{table}
	
	\begin{table}[t]
		\small
		\centering
		\renewcommand*{\arraystretch}{1.1}
		\begin{tabular}{*{5}{c}}
			\hline
			$\left.\text{Bin[}\text{GeV}^2\right]$  &  $\text{[0.1-1]}$  &  $\text{[1-2]}$  &  $\text{[2-4]}$  &  $\text{[4-6]}$  \\
			\hline
			$\langle {P_1}^+ \rangle$ & $\text{0.0038(32)}$  &  $\text{0.0034(52)}$  &  $\text{-0.048(20)}$  &  $\text{-0.087(34)}$   \\
			\hline
			$\langle {P_2}^+ \rangle$ & $\text{-0.1219(36)}$  &  $\text{-0.411(12)}$  &  $\text{-0.176(37)}$  &  $\text{0.291(20)}$     \\
			\hline
			$\langle {P_3}^+ \rangle$ & $\text{0.00057(41)}$  &  $\text{0.0031(21)}$  &  $\text{0.0059(36)}$  &  $\text{0.0055(30)}$     \\
			\hline
			$\langle {P_4^{\prime}}^+ \rangle$ & $\text{0.347(12)}$  &  $\text{0.182(15)}$  &  $\text{-0.224(19)}$  &  $\text{-0.448(11)}$     \\
			\hline
			$\langle {P_5^{\prime}}^+ \rangle$ &  $\text{0.350(15)}$  &  $\text{0.062(25)}$  &  $\text{-0.495(32)}$  &  $\text{-0.809(21)}$     \\
			\hline
			$\langle {P_6^{\prime}}^+ \rangle$ & $\text{-0.065(28)}$  &  $\text{-0.168(38)}$  &  $\text{-0.271(38)}$  &  $\text{-0.244(37)}$     \\
			\hline
			$\langle {P_8^{\prime}}^+ \rangle$ & $\text{-0.153(27)}$  &  $\text{-0.059(31)}$  &  $\text{-0.067(26)}$  &  $\text{-0.082(19)}$     \\
			\hline
			$\langle {BR}^+ \rangle \times 10^9$ & $\text{4.11(25)}$  &  $\text{2.06(14)}$  &  $\text{2.94(23)}$  &  $\text{2.78(23)}$     \\
			\hline
			$\langle {A}_{FB}^+ \rangle$ & $\text{-0.0634(40)}$  &  $\text{-0.107(11)}$  &  $\text{-0.0412(95)}$  &  $\text{0.119(14)}$     \\
			\hline
			$\langle {F}_{L}^+ \rangle$ & $\text{0.448(25)}$  &  $\text{0.782(18)}$  &  $\text{0.823(15)}$  &  $\text{0.716(23)}$     \\
			\hline
			$\langle {R_{\rho}}^+ \rangle$ & $\text{0.98448(36)}$  &  $\text{0.99768(20)}$  &  $\text{0.99721(27)}$  &  $\text{0.99718(24)}$     \\
			\hline
		\end{tabular}
		\caption{\small SM Predictions of observables for $B^+\to \rho^+\ell \ell$.}
		\label{tab:BPpred}
	\end{table}
	
	\begin{table}[t] 
		\small
		\centering
		\begin{tabular}{*{5}{c}}
			\hline
			$\left.\text{Bin[}\text{GeV}^2\right]$  &  $\text{[0.1-1]}$  &  $\text{[1-2]}$  &  $\text{[2-4]}$  &  $\text{[4-6]}$  \\
			\hline
			$\langle A_{CP} \rangle$ & $\text{-0.280(25)}$  &  $\text{-0.345(18)}$  &  $\text{-0.1728(92)}$  &  $\text{-0.0584(42)}$    \\
			\hline
			$\langle A_3 \rangle$ & $\text{      -0.0000018(262)}$  &  $\text{-0.000039(84)}$  &  $\text{0.000022(127)}$  &  $\text{0.00044(29)}$     \\
			\hline
			$\langle A_4 \rangle$ & $\text{-0.1017(62)}$  &  $\text{-0.0684(60)}$  &  $\text{-0.0144(25)}$  &  $\text{0.0043(21)}$     \\
			\hline
			$\langle A_5 \rangle$ &  $\text{0.0259(43)}$  &  $\text{0.0687(84)}$  &  $\text{0.0534(80)}$  &  $\text{0.0315(67)}$    \\
			\hline
			$\langle A_{6s} \rangle$ & $\text{0.0024(20)}$  &  $\text{0.0056(55)}$  &  $\text{0.00057(795)}$  &  $\text{-0.018(11)}$     \\
			\hline
			$\langle A_7 \rangle$ & $\text{-0.0056(40)}$  &  $\text{0.0297(35)}$  &  $\text{0.0589(47)}$  &  $\text{0.0470(40)}$     \\
			\hline
			$\langle A_8 \rangle$ & $\text{0.0502(73)}$  &  $\text{-0.0031(69)}$  &  $\text{0.0011(33)}$  &  $\text{0.0093(11)}$     \\
			\hline
			$\langle A_9 \rangle$ & $\text{0.000043(38)}$  &  $\text{0.000096(79)}$  &  $\text{0.000109(76)}$  &  $\text{0.000085(52)}$     \\
			\hline
			$\langle P_1 \rangle$ & $\text{0.0027(21)}$  &  $\text{0.0027(41)}$  &  $\text{-0.043(19)}$  &  $\text{-0.084(33)}$     \\
			\hline
			$\langle P_2 \rangle$ & $\text{-0.0826(26)}$  &  $\text{-0.3472(100)}$  &  $\text{-0.159(23)}$  &  $\text{0.257(13)}$     \\
			\hline
			$\langle P_3 \rangle$ &  $\text{0.00032(22)}$  &  $\text{0.0023(15)}$  &  $\text{0.0048(28)}$  &  $\text{0.0051(28)}$     \\
			\hline
			$\langle P_4^{'} \rangle$ &  $\text{0.1529(97)}$  &  $\text{0.048(11)}$  &  $\text{-0.271(15)}$  &  $\text{-0.452(11)}$     \\
			\hline
			$\langle P_5^{'} \rangle$ & $\text{0.422(11)}$  &  $\text{0.225(19)}$  &  $\text{-0.386(29)}$  &  $\text{-0.764(21)}$     \\
			\hline
			$\langle P_6^{'} \rangle$ & $\text{-0.080(21)}$  &  $\text{-0.120(34)}$  &  $\text{-0.137(38)}$  &  $\text{-0.148(42)}$     \\
			\hline
			$\langle P_8^{'} \rangle$ &  $\text{-0.056(12)}$  &  $\text{-0.072(19)}$  &  $\text{-0.068(19)}$  &  $\text{-0.064(18)}$     \\
			\hline
			$\langle BR \rangle \times 10^9$ & $\text{3.21(22)}$  &  $\text{1.53(11)}$  &  $\text{2.51(21)}$  &  $\text{2.63(22)}$     \\
			\hline
			$\langle F_L \rangle$ & $\text{0.360(25)}$  &  $\text{0.733(22)}$  &  $\text{0.799(17)}$  &  $\text{0.708(23)}$     \\
			\hline
			$\langle A_{FB} \rangle$ & $\text{-0.0793(40)}$  &  $\text{-0.139(13)}$  &  $\text{-0.0479(86)}$  &  $\text{0.113(11)}$     \\
			\hline
			$\langle R_{\rho} \rangle$ & $\text{0.98311(36)}$  &  $\text{0.99702(25)}$  &  $\text{0.99675(32)}$  &  $\text{0.99701(25)}$     \\
			\hline
		\end{tabular}
		\caption{\small The SM predictions of the CP-averaged and CP-asymmetric observables in $B^{\pm}\to \rho^{\pm}\ell \ell$ decays as measurable both at the LHCb and Belle. These observables are obtained using eq.~\ref{eq:d4Gammacharged}.} 
		\label{tab:Bchavpred}
	\end{table}

	\subsection{$B\to\rho\ell \ell$}
	
	In section \ref{subsubsec:Btorhointro}, we have pointed out the angular observables relevant for the $B \to \rho\ell \ell$ decays. The LHCb collaboration have already measured the CP-averaged and CP-asymmetric observables related to $B(B_s)\to K^*(\phi)\ell\ell$. decays. In this work, we will give predictions of similar observables associated with the angular anaylsis of $B^{\pm} \to \rho^{\pm}(\to \pi^{\pm} \pi^0)\ell\ell$ and $B \to \rho^0(\to \pi^+\pi^-)\ell\ell$ decays. Note that in this analysis we have considered the $\pi\pi$ system in the final state in a $P$-wave configuration.    
	
	For $B^\pm\to\rho^\pm\ell \ell$ decays, $\rho^+$ or $\rho^-$ will be reconstructed via the decays $\rho^+ \to \pi^+\pi^0$ and $\rho^- \to \pi^-\pi^0$, respectively. Note that in both the decays, the hadrons in the final state are not CP-eigenstates; they are the CP-conjugate states. The angular observables can be extracted directly from the angular distribution given in eq.~\ref{eq:d4Gamma} and eq.~\ref{eq:d4Gammaconjugate} for the $B^-$ and $B^+$ decays, respectively, or by defining the following rates  
	\begin{align}\label{eq:d4Gammacharged}
	\frac{d^4(\Gamma + \bar{\Gamma})}{dq^2 dcos\theta_l dcos\theta_{\rho} d\phi} &=
	\sum_i [J_i + \bar{J_i}] f_i(\theta_l,\theta_{\rho}\phi ), \nn \\
	\frac{d^4(\Gamma - \bar{\Gamma})}{dq^2 dcos\theta_l dcos\theta_{\rho} d\phi} &=
	\sum_i [J_i - \bar{J_i}] f_i(\theta_l,\theta_{\rho}\phi ). 
	\end{align}

	The angular coefficients in the first equation of \ref{eq:d4Gammacharged} will give us the CP-averaged observables, while that from the second one will give the CP-asymmetric observables. The definitions of these observables are taken from the ref.~\cite{Aaij:2015oid} and the references therein. Note that tagging of $B^+$ or $B^-$ is possible at both the LHCb and Belle. Hence, it is possible to directly probe the angular coefficients corresponding to the $B^+$ or $B^-$ decays, respectively for both the collaborations. The observables for $B^{-}\to \rho^{-}\ell^+\ell^-$ corresponding to tagged events are as given below.  
	\begin{itemize}
		\item The tagged decay rate distribution and the corresponding branching fraction
		\begin{equation} \label{eq:gamBP}
		\left<\frac{d\Gamma}{dq^2}\right>^{\text{Tag}}=\frac{1}{4}(3J_1^c+6J_1^s-J_2^c-2J_2^s), ~~~~~
		\left<BR\right>^{\text{Tag}}  = \tau_{B^+} \left<\frac{d\Gamma}{dq^2}\right>^{\text{Tag}}
		\end{equation}
		
		\item Observables defined from the coefficients of the angular distribution of $B^{-}\to \rho^{-}\mu^+\mu^-$ decays: 
		\begin{align} \label{eq:taggedobsCP}
		\left<P_1\right>^{\text{Tag}} &= \frac{J_3}{2 J_{2s}}, ~~~~~~~~~~~~~~~ \left<P_2\right>^{\text{Tag}} = -\frac{J_{6s}}{8 J_{2s}},~~~~~~~~~~~
		\left<P_3\right>^{\text{Tag}} = \frac{J_9}{4 J_{2s}}, \nonumber \\ 
		\left<P_4^{'}\right>^{\text{Tag}} &= -\frac{J_4}{2 \sqrt{-J_2^c J_2^s}},~~~~
		\left<P_5^{'}\right>^{\text{Tag}} = \frac{J_5}{2 \sqrt{-J_2^c J_2^s}}, ~~~~ \left<P_6^{'}\right>^{\text{Tag}} = -\frac{J_7}{2 \sqrt{-J_2^c J_2^s}},\nonumber \\ 
		\left<P_8^{'}\right>^{\text{Tag}} &= \frac{J_8}{2 \sqrt{-J_2^c J_2^s}}, ~~~~
		\left<A_{FB}\right>^{\text{Tag}} = -\frac{3}{4}\frac{J_6^s}{\left<\frac{d\Gamma}{dq^2}\right>^{\text{Tag}}},~~~~
		\left<F_L\right>^{\text{Tag}} = - \frac{J_2^c}{\left<\frac{d\Gamma}{dq^2}\right>^{\text{Tag}}} 
		\end{align}
	\end{itemize}
	
	The observables for the CP conjugate mode $B^+\to \rho^+\ell^+\ell^-$ are obtained by replacing $J_i$ $\to$ $\bar{J_i}$ in the above equations. In the following, we will define the observables obtained after combining the decay rates and the angular coefficients of $B^{\pm}\to \rho^{\pm}\ell^+\ell^-$ decays. 
	
	\begin{table}[t]
		\centering
		\begin{tabular}{|c|c|c|c|}
			\hline
			Observables & 	NP sensitivities & Observables & 	NP sensitivities \\
			\hline
			$A_3$ &  $Im(C_9^{\prime})$ & $P_1$ & $Re(C_9^{\prime})$, $Re(C_{10}^{\prime})$  \\
			\hline
			$A_4$ & $Im(\Delta C_9) $, $Im(C_9^{\prime})$  & $P_2$&   $Re(\Delta C_9) $  \\ 
			\hline
			$A_5$ & $Im(\Delta C_{10})$, $Re(C_{10}^{\prime})$, $Im(C_{10}^{\prime})$  & $P_3$ & $Re(C_9^{\prime})$    \\
			\hline
			$A_6$ &  $Im(\Delta C_{10})$ &  $P_4^{\prime}$ & $Re(\Delta C_{10})$, $Re(C_{10}^{\prime})$ \\
			\hline
			$A_7$ & $Im(\Delta C_{10})$, $Im(C_{10}^{\prime})$ & $P_5^{\prime}$ &  $Re(\Delta C_{9})$, $Re(C_{10}^{\prime})$ \\
			\hline
			$A_8$ & $Im(\Delta C_9) $,  $Im(C_{10}^{\prime})$, $Im(C_9^{\prime})$ & $P_6^{\prime}$ & Hard to distinguish from SM  \\
			\hline
			$A_9$ & $Im(C_{9}^{\prime})$, $Im(C_{10}^{\prime})$ & $P_8^{\prime}$ & Hard to distinguish from SM \\
			\hline 
			$A_{CP}$ & $Re(C_9^{\prime})$, $Re(C_{10}^{\prime})$, $Re(\Delta C_9) $, $Re(\Delta C_{10}) $ & $P_2^{+}$ & $Re(\Delta C_9)$\\  
			& (limited $q^2$ regions) & $ P_2^{-}$ & $Re(\Delta C_9)$, $Im(\Delta C_{10})$  \\
			\hline 	    
			$A_{FB}$ & $Re(\Delta C_{9})$ & $P_3^{+},P_3^{-}$ & $Im(C_9^{\prime})$, $Re(C_9^{\prime})$, $Im(C_{10}^{\prime})$ \\
			\hline 
			$F_L$ & $Re(\Delta C_{9})$ (limited $q^2$ regions ), &  ${P_6^{\prime}}^{+}$, ${P_6^{\prime}}^{-}$ & $Im(C_{10}^{\prime})$, $Im(\Delta C_{10})$  \\
			 & hard to probe, need more precision & & \\
			\hline 
			$R(\rho)$ & All the four scenarios,& ${P_8^{\prime}}^{+}$, ${P_8^{\prime}}^{-}$ & $Im(C_9^{\prime})$, $Im(C_{10}^{\prime})$, $Im(\Delta C_9) $    \\
			&  real and imaginary components & &  \\ 
			\hline
		\end{tabular}
		\caption{The observables along with the respective new physics scenarios which affect them the most. In the NP scenarios, the $q^2$ sensitivity of these observables can be visualized from the corresponding plots provided in the text. Numerical estimates for the same calculated at a few benchmark values for the NP WC's can also be read off from the relevant tables.}
		\label{tab:NPsensrho}
	\end{table}
	
	\begin{itemize}
		\item The untagged decay rate and the corresponding branching fraction:
		\begin{equation}\label{eq:gamBPav}
		\left<\frac{d\Gamma+d \bar{\Gamma}}{dq^2}\right> = \frac{1}{4} (3 J_1^c + 6 J_1^s - J_2^c -2 J_2^s)   + \frac{1}{4} (3 \bar{J}_1^c + 6 \bar{J}_1^s - \bar{J}_2^c -2 \bar{J}_2^s),~~~\langle BR \rangle = \frac{\tau_{B^+}}{2} \left<\frac{d\Gamma+d \bar{\Gamma}}{dq^2}\right>.
		\end{equation}
		
		\item The CP-asymmetric observables which are obtained from the angular distributions $\frac{d^4(\Gamma - \bar{\Gamma})}{dq^2 d\vec{\Omega}}$: 
		\begin{align}\label{eq:untaggedobsACP}
		\langle A_{CP} \rangle &= \frac{\frac{1}{4} (3 J_1^c + 6 J_1^s - J_2^c -2 J_2^s)   - \frac{1}{4} (3 \bar{J}_1^c + 6 \bar{J}_1^s - \bar{J}_2^c -2 \bar{J}_2^s) }{\left<\frac{d\Gamma+d \bar{\Gamma}}{dq^2}\right>} \nonumber \\
		\langle A_3 \rangle =& \frac{J_3-\bar{J}_3}{\left<\frac{d\Gamma+d \bar{\Gamma}}{dq^2}\right>}, ~~~~~~~~~\langle A_4 \rangle = -\frac{J_4-\bar{J}_4}{\left<\frac{d\Gamma+d \bar{\Gamma}}{dq^2}\right>}, ~~~~~\langle A_5 \rangle = \frac{J_5-\bar{J}_5}{\left<\frac{d\Gamma+d \bar{\Gamma}}{dq^2}\right>}, \nn \\
		\langle A_6^s \rangle =& -\frac{J_6^s-\bar{J}_{6s}}{\left<\frac{d\Gamma+d \bar{\Gamma}}{dq^2}\right>}, ~~~~~
		\langle A_7 \rangle = -\frac{J_7-\bar{J}_7}{\left<\frac{d\Gamma+d \bar{\Gamma}}{dq^2}\right>}, ~~~~~ \langle A_8 \rangle = \frac{J_8-\bar{J}_8}{\left<\frac{d\Gamma+d \bar{\Gamma}}{dq^2}\right>}, \nn \\ \langle A_9 \rangle =& -\frac{J_9-\bar{J}_9}{\left<\frac{d\Gamma+d \bar{\Gamma}}{dq^2}\right>}
		\end{align}
		
		\item The CP-averaged observables which are obtained from the angular distributions $\frac{d^4(\Gamma + \bar{\Gamma})}{dq^2 d\vec{\Omega}}$: 
		\begin{align} \label{eq:untaggedobsCPA}
		\langle P_1 \rangle &= \frac{2(J_3+ \bar{J}_3)}{\left<\frac{d\Gamma+d \bar{\Gamma}}{dq^2}\right> + J_2^c+ \bar{J}_2^c}, ~~~
		\langle P_2 \rangle = - \frac{1}{2} \frac{(J_6^s+\bar{J}_6^s)}{\left<\frac{d\Gamma+d \bar{\Gamma}}{dq^2}\right> + J_2^c+ \bar{J}_2^c}, ~~~\langle P_3 \rangle =  \frac{(J_9+\bar{J}_9)}{\left<\frac{d\Gamma+d \bar{\Gamma}}{dq^2}\right> + J_2^c+ \bar{J}_2^c}, \nn \\
		\langle P_4^{'} \rangle =& - \frac{(J_4+\bar{J}_4)}{\sqrt{-(J_2^c+ \bar{J}_2^c) (\left<\frac{d\Gamma+d \bar{\Gamma}}{dq^2}\right> + J_2^c+ \bar{J}_2^c})}, ~~~~~
		\langle P_5^{'} \rangle =  \frac{(J_5+\bar{J}_5)}{\sqrt{-(J_2^c+ \bar{J}_2^c) (\left<\frac{d\Gamma+d \bar{\Gamma}}{dq^2}\right> + J_2^c+ \bar{J}_2^c})}, \nn \\
		\langle P_6^{'} \rangle =& - \frac{(J_7+\bar{J}_7)}{\sqrt{-(J_2^c+ \bar{J}_2^c) (\left<\frac{d\Gamma+d \bar{\Gamma}}{dq^2}\right> + J_2^c+ \bar{J}_2^c})}, ~~~~~ \langle P_8^{'} \rangle =  \frac{(J_8+\bar{J}_8)}{\sqrt{-(J_2^c+ \bar{J}_2^c) (\left<\frac{d\Gamma+d \bar{\Gamma}}{dq^2}\right> + J_2^c+ \bar{J}_2^c})},\nn\\ &~~~~~~~~~~~~~~~~~~~~~~~\langle A_{FB} \rangle = -\frac{3}{4}\frac{J_6^s+\bar{J}_6^s}{\left<\frac{d\Gamma+d \bar{\Gamma}}{dq^2}\right>}, ~~~~~ \langle F_L \rangle = -\frac{J_2^c+ \bar{J}_2^c}{\left<\frac{d\Gamma+d \bar{\Gamma}}{dq^2}\right>}
		\end{align}
	\end{itemize}
	All the observables listed in the above equations could also be defined for the neutral $B^0(\bar{B^0}) \to \rho \ell\ell$ decays. For the details about the observables the reader may look in the refs.~\cite{Aaij:2015oid,Aaij:2015esa,Aaij:2020nrf,Descotes-Genon:2012isb}.

	\begin{figure*}[htbp]
		\small
		\centering
		\subfloat[]{\includegraphics[width=0.25\textwidth]{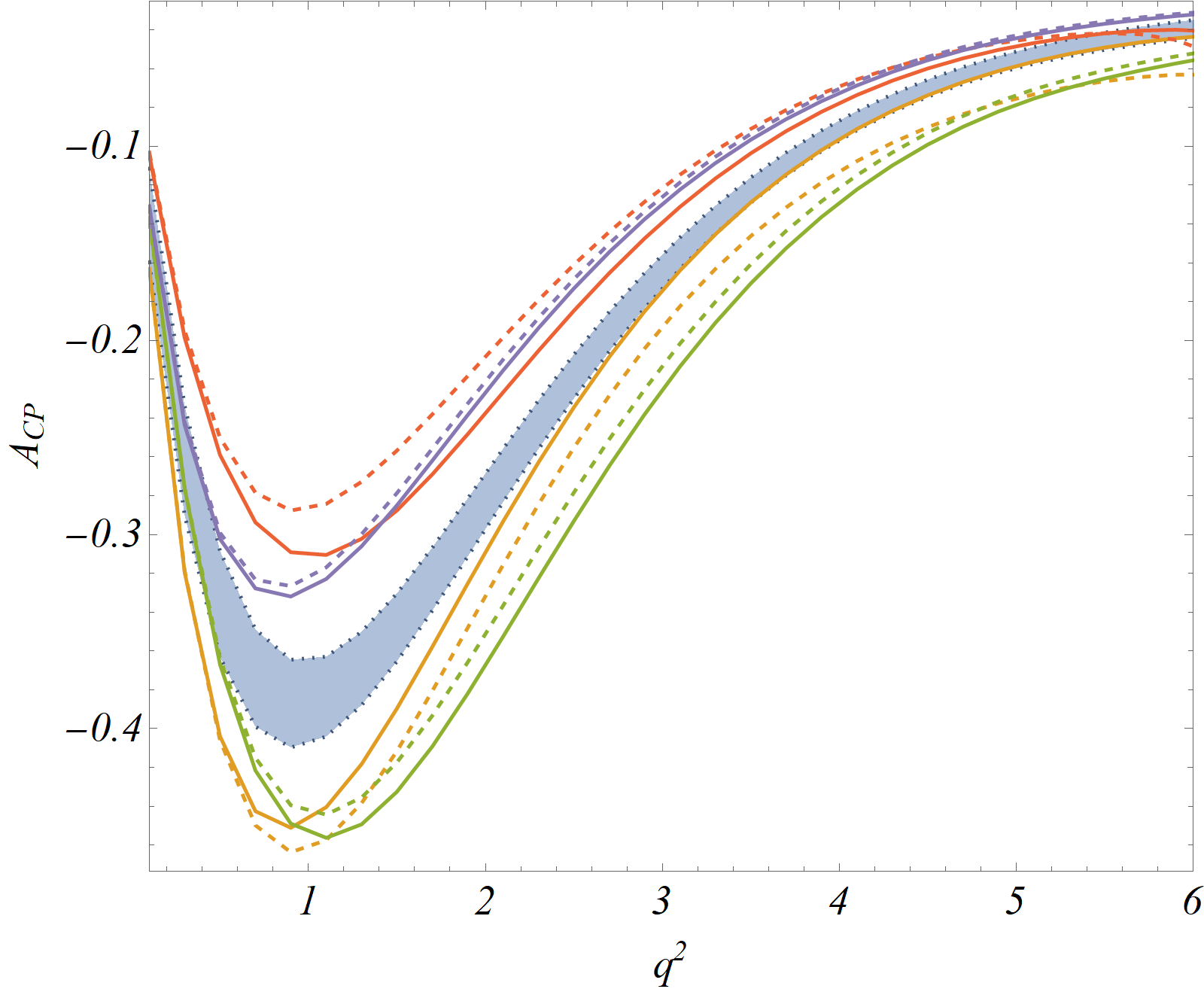}\label{fig:ACPrhoch}}~~~
		\subfloat[]{\includegraphics[width=0.25\textwidth]{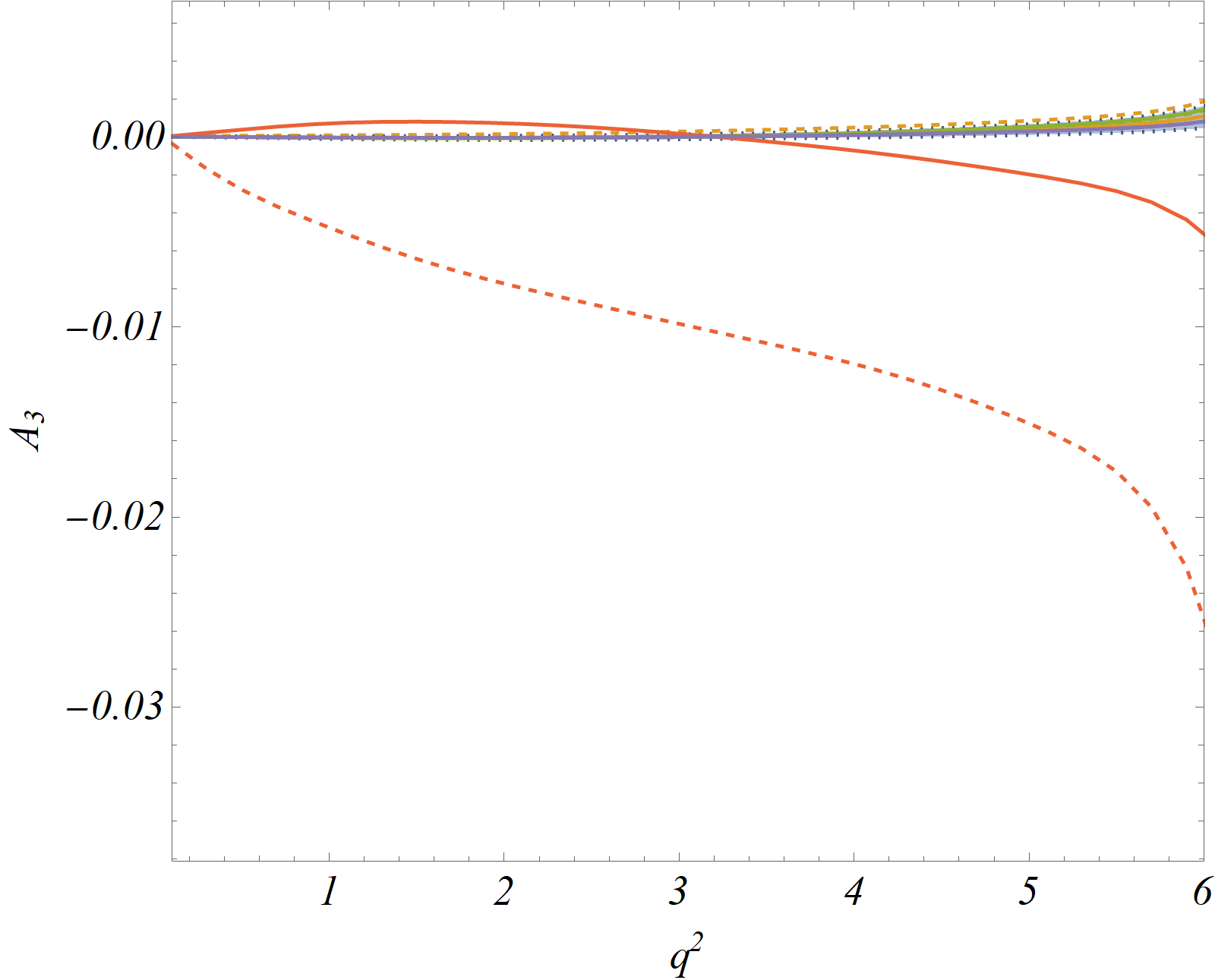}\label{fig:A3P}}~~~
		\subfloat[]{\includegraphics[width=0.25\textwidth]{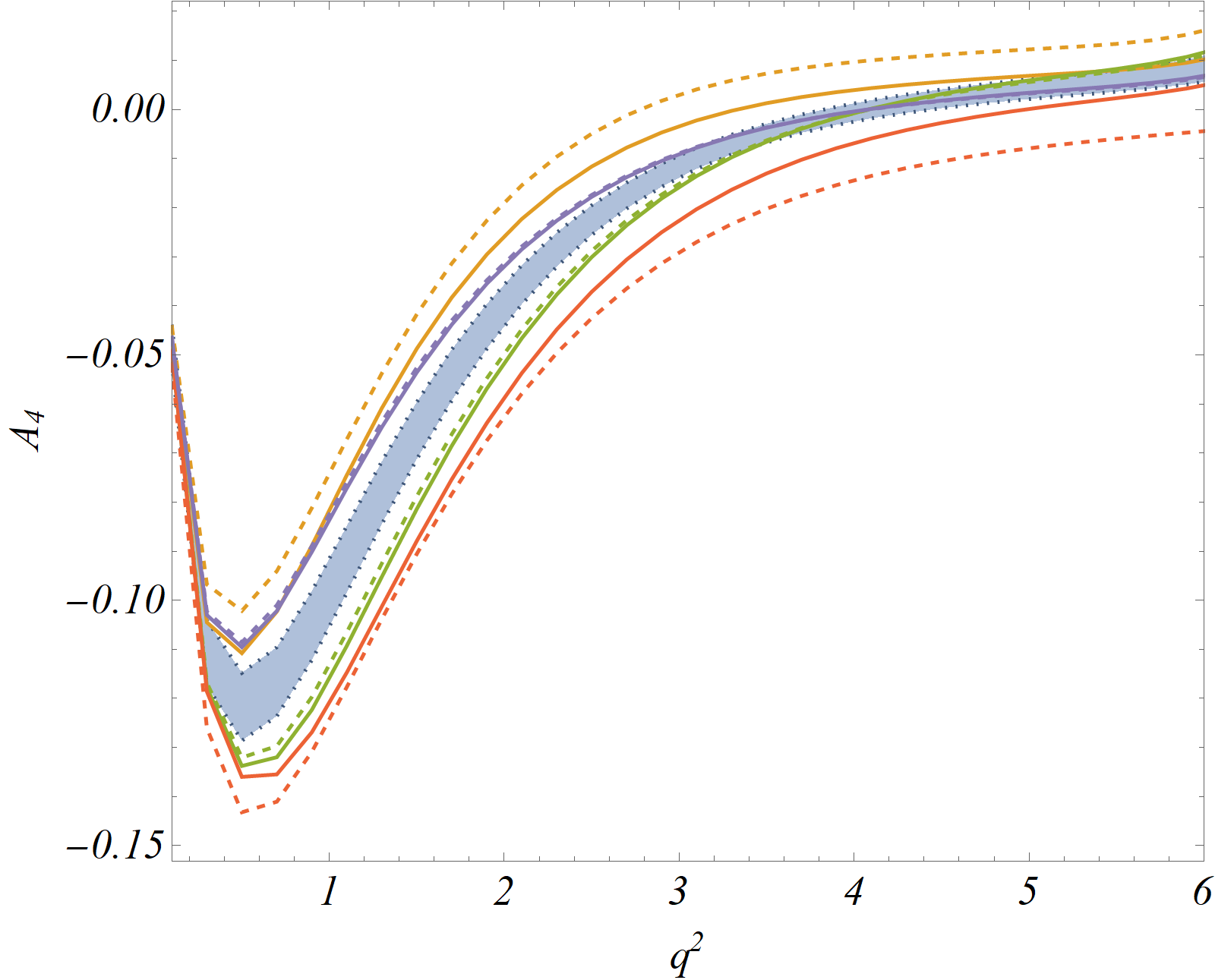}\label{fig:A4P}}~~~
		\subfloat[]{\includegraphics[width=0.25\textwidth]{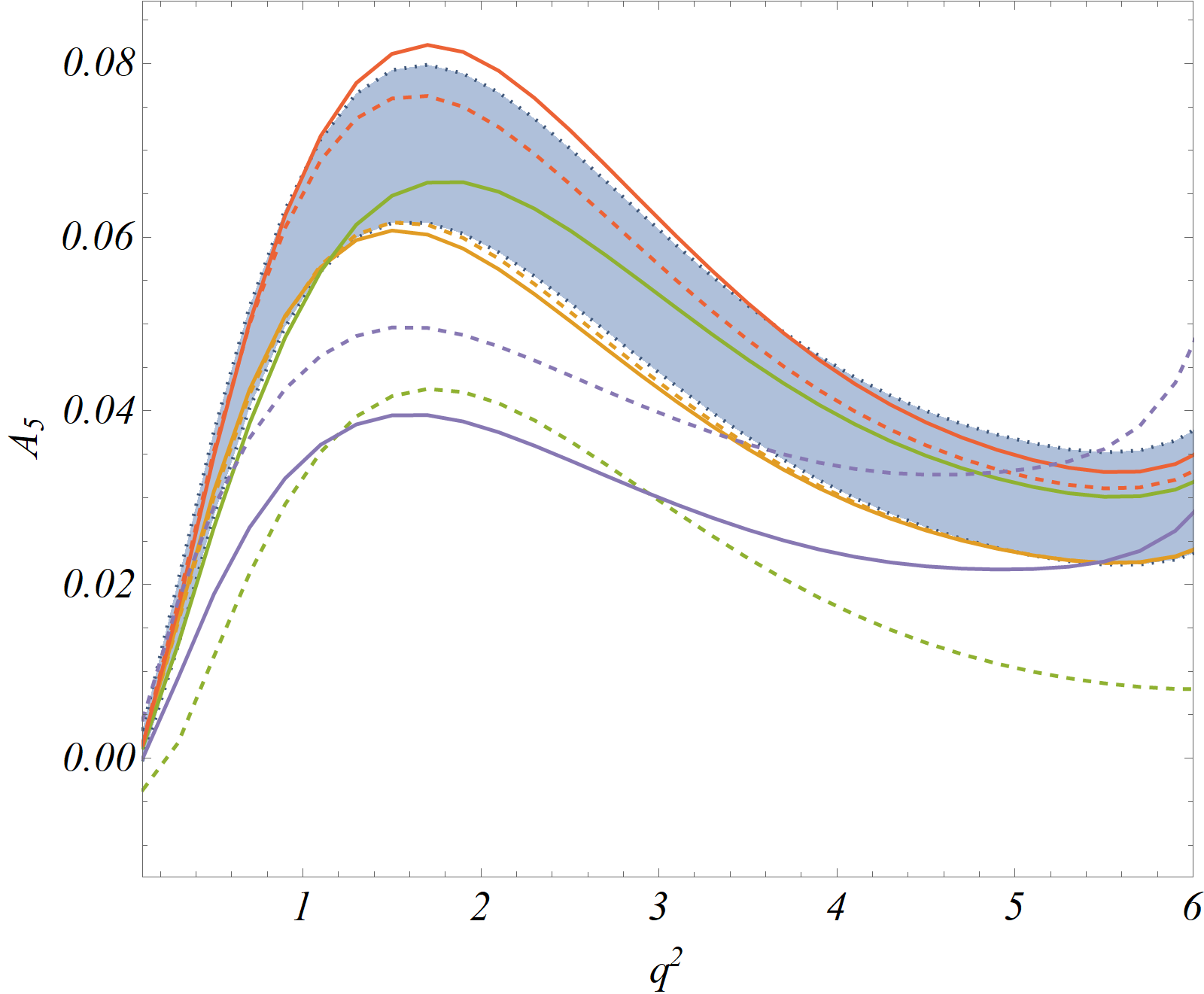}\label{fig:A5P}}\\
		\subfloat[]{\includegraphics[width=0.25\textwidth]{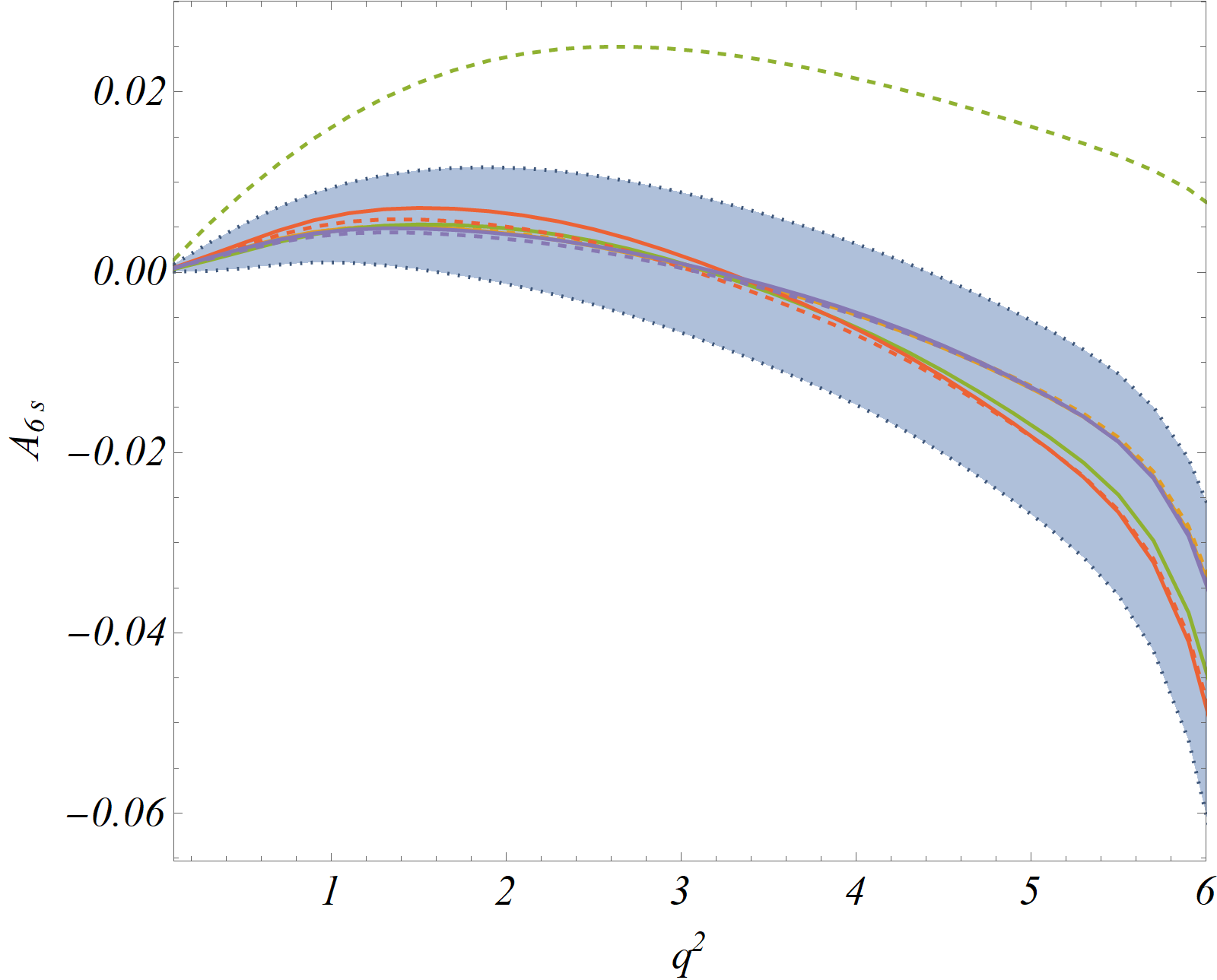}\label{fig:A6sP}}~~~
		\subfloat[]{\includegraphics[width=0.25\textwidth]{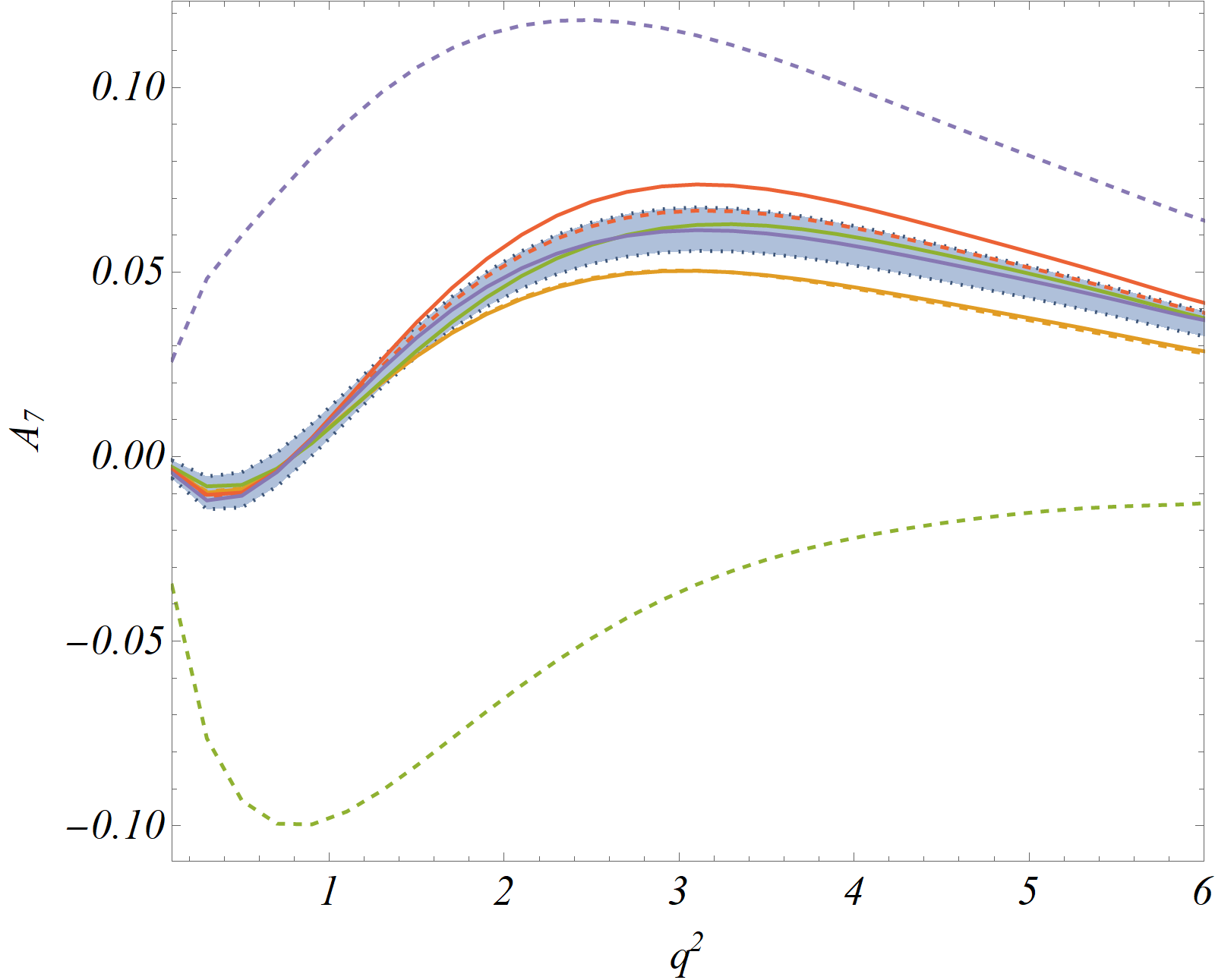}\label{fig:A7P}}~~~
		\subfloat[]{\includegraphics[width=0.25\textwidth]{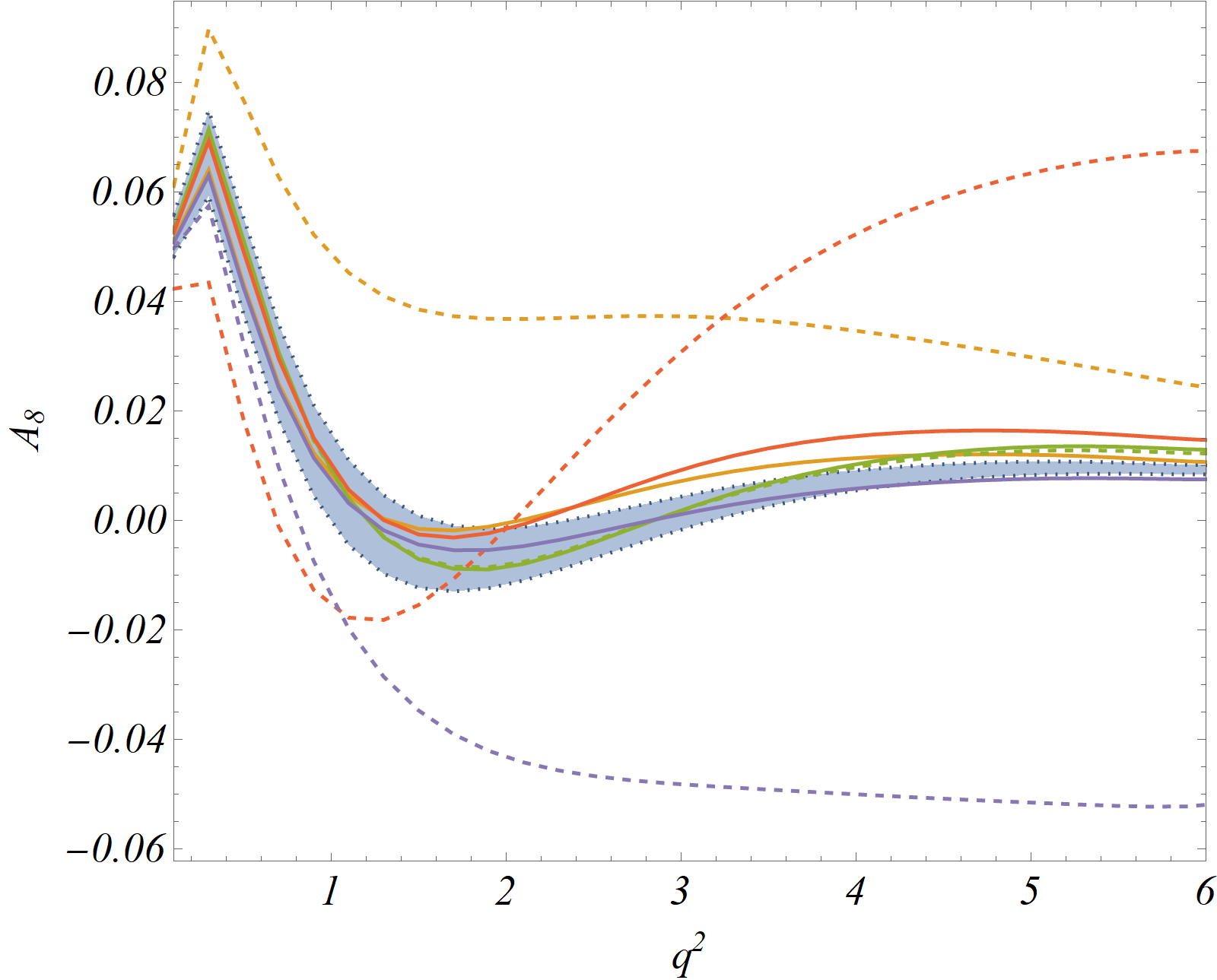}\label{fig:A8P}}~~~
		\subfloat[]{\includegraphics[width=0.25\textwidth]{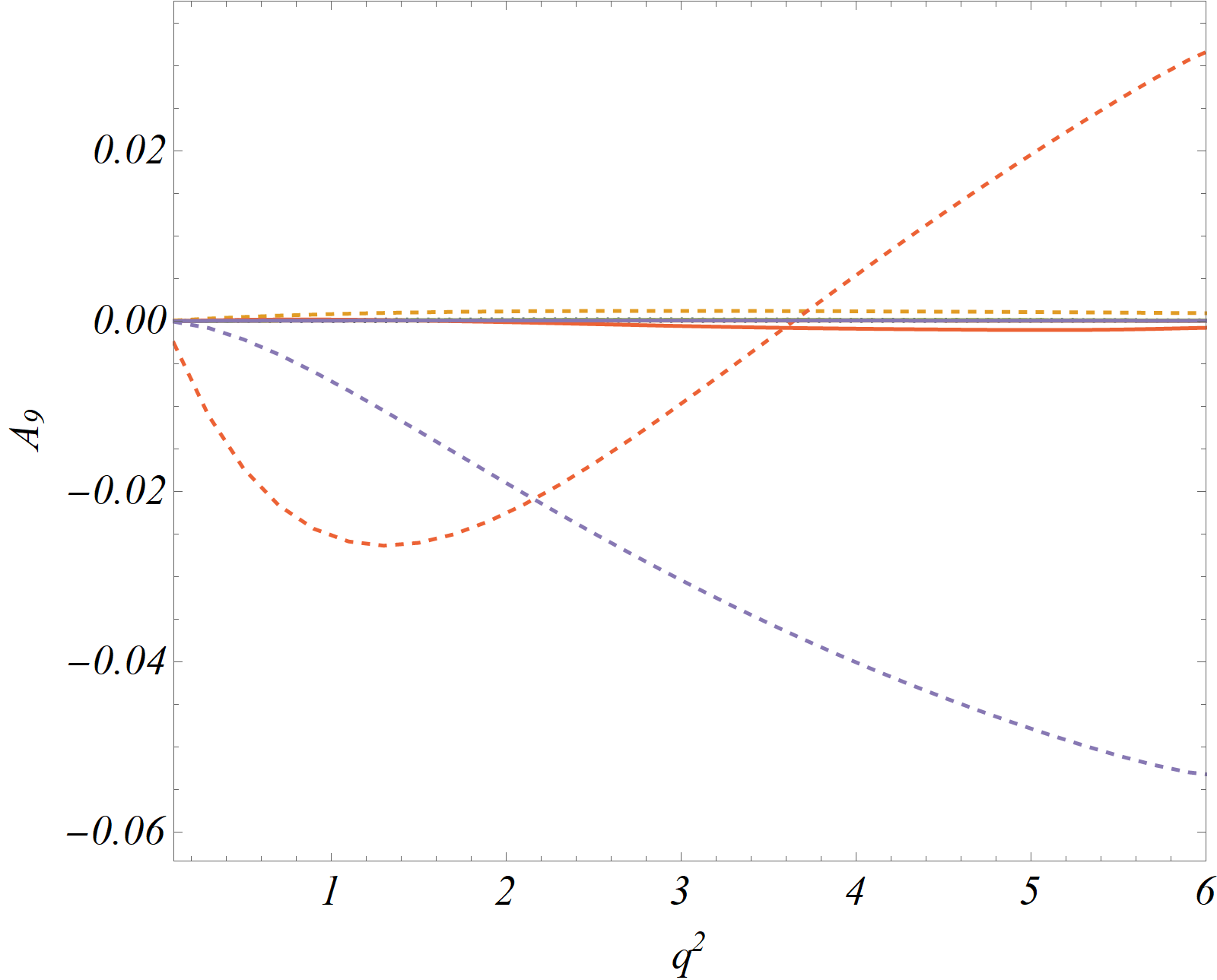}\label{fig:A9P}}\\
		\subfloat[]{\includegraphics[width=0.25\textwidth]{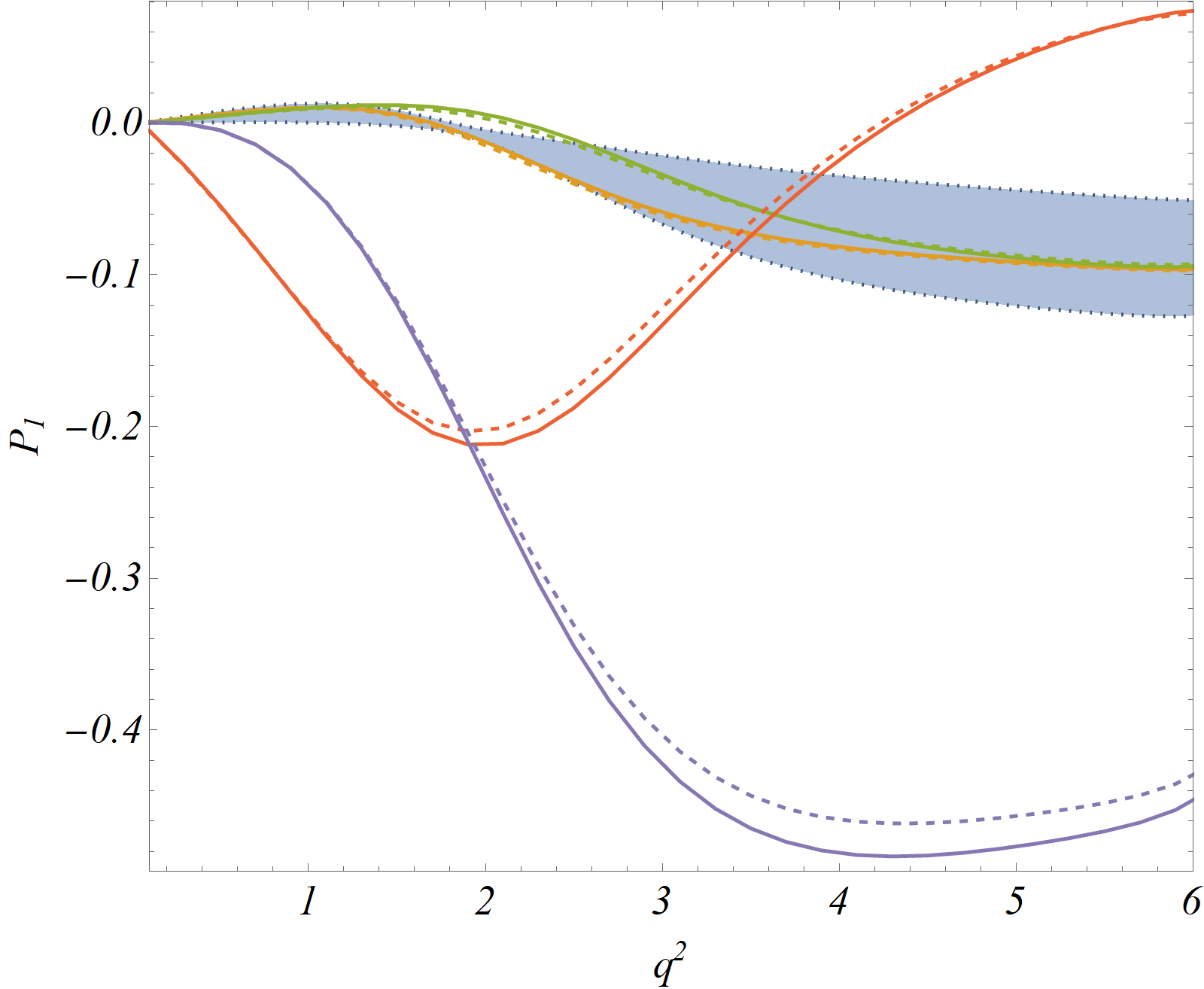}\label{fig:P1P}}~~~
		\subfloat[]{\includegraphics[width=0.25\textwidth]{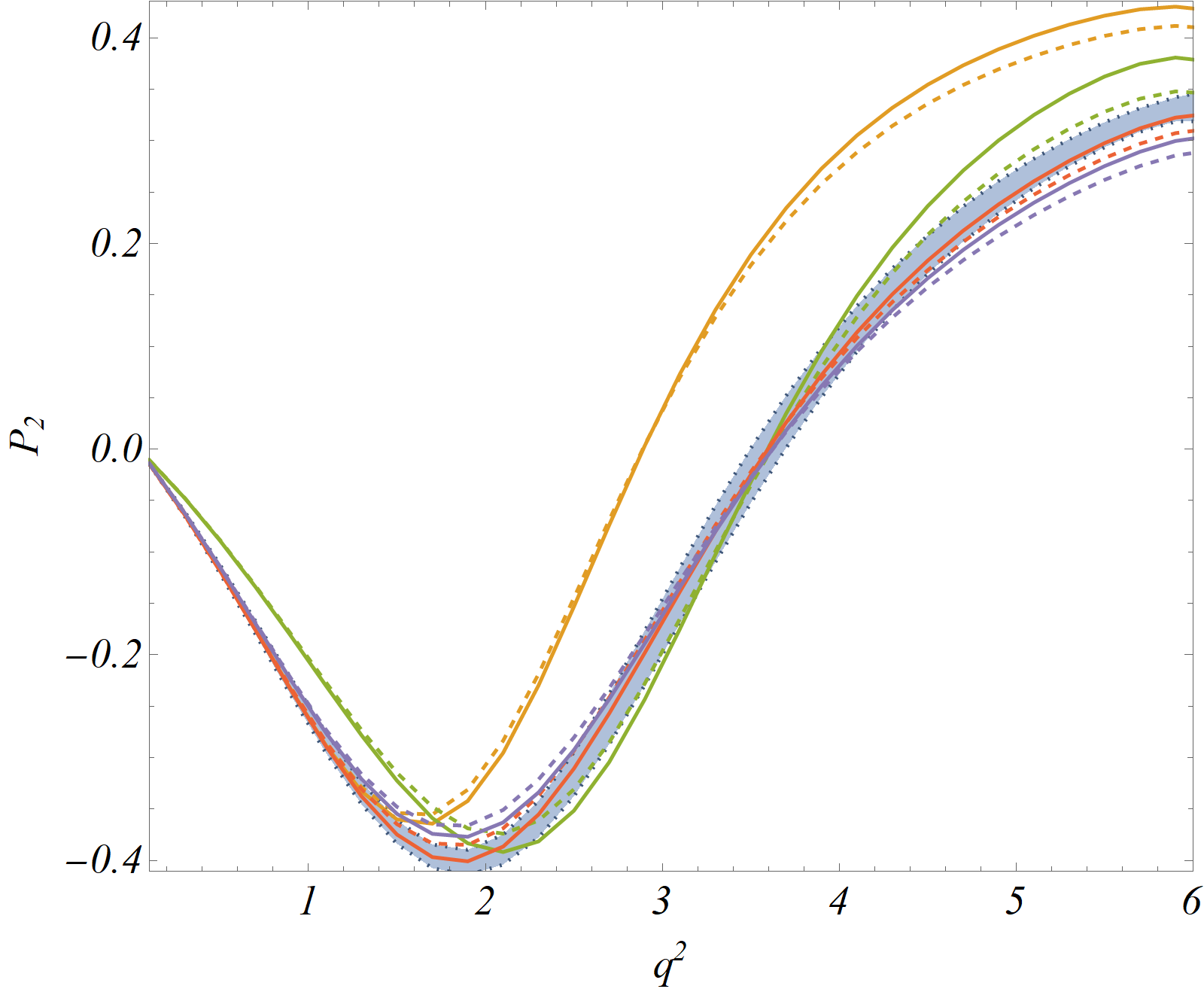}\label{fig:P2P}}~~~
		\subfloat[]{\includegraphics[width=0.25\textwidth]{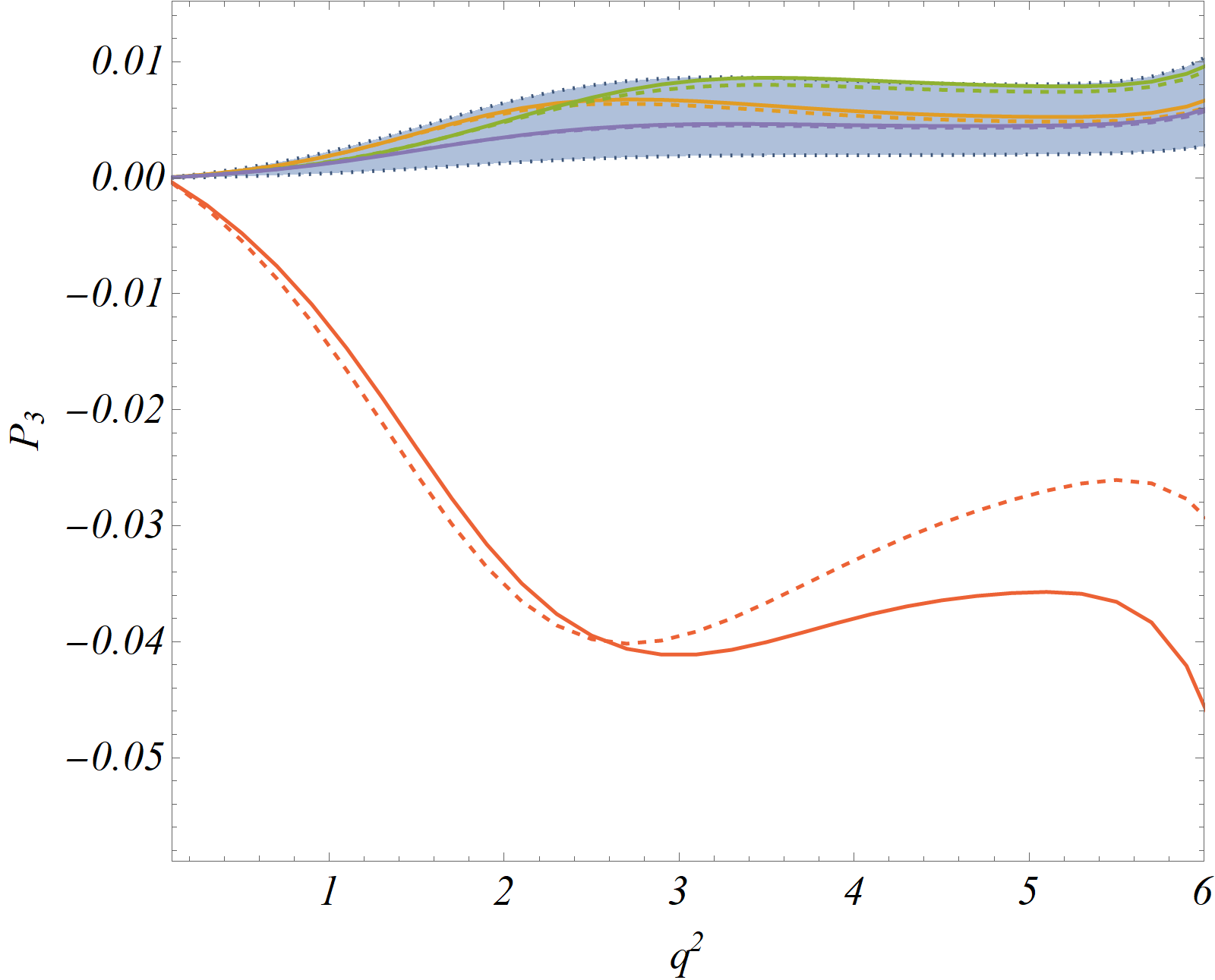}\label{fig:P3P}}~~~
		\subfloat[]{\includegraphics[width=0.25\textwidth]{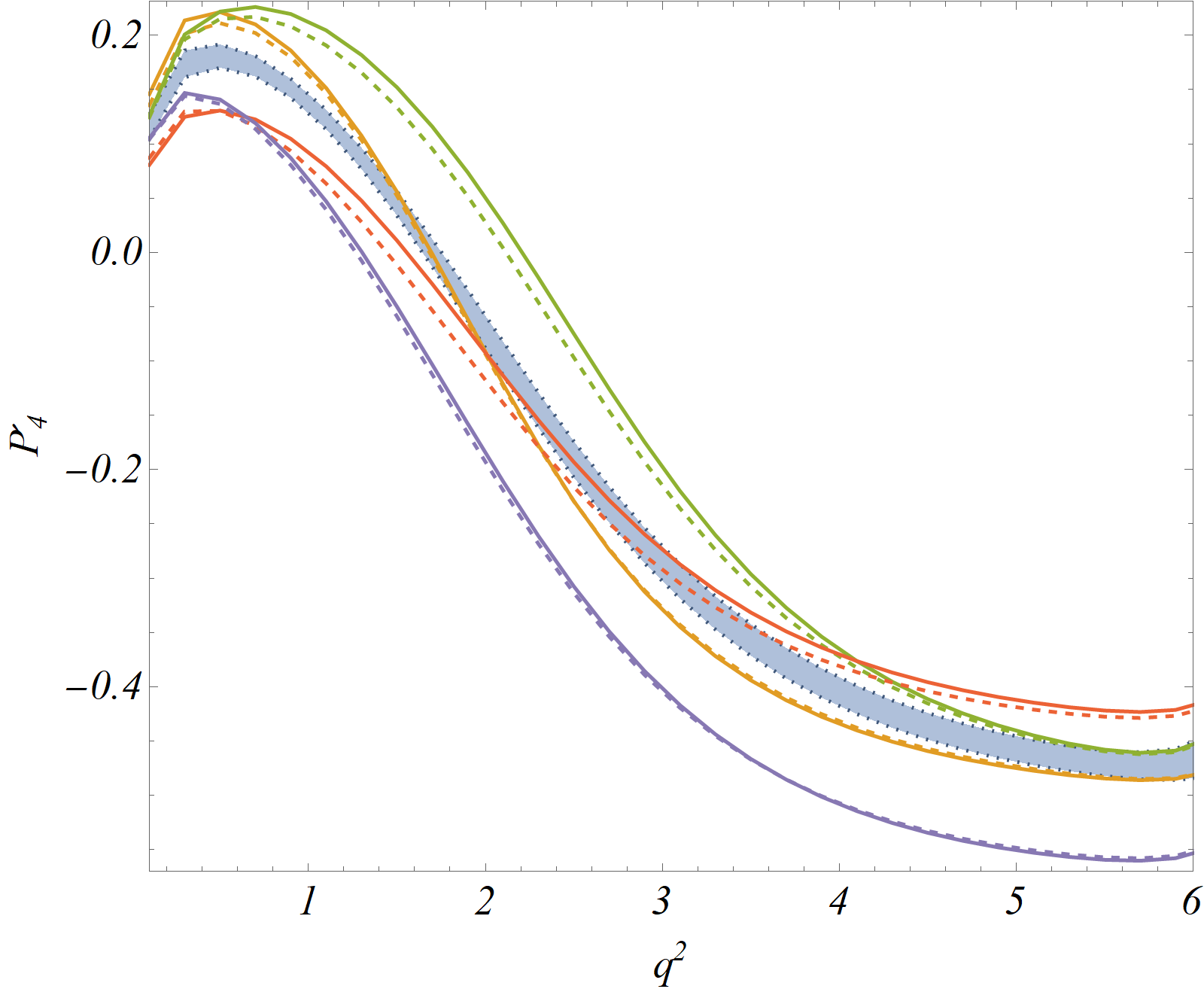}\label{fig:P4prP}}\\
		\subfloat[]{\includegraphics[width=0.25\textwidth]{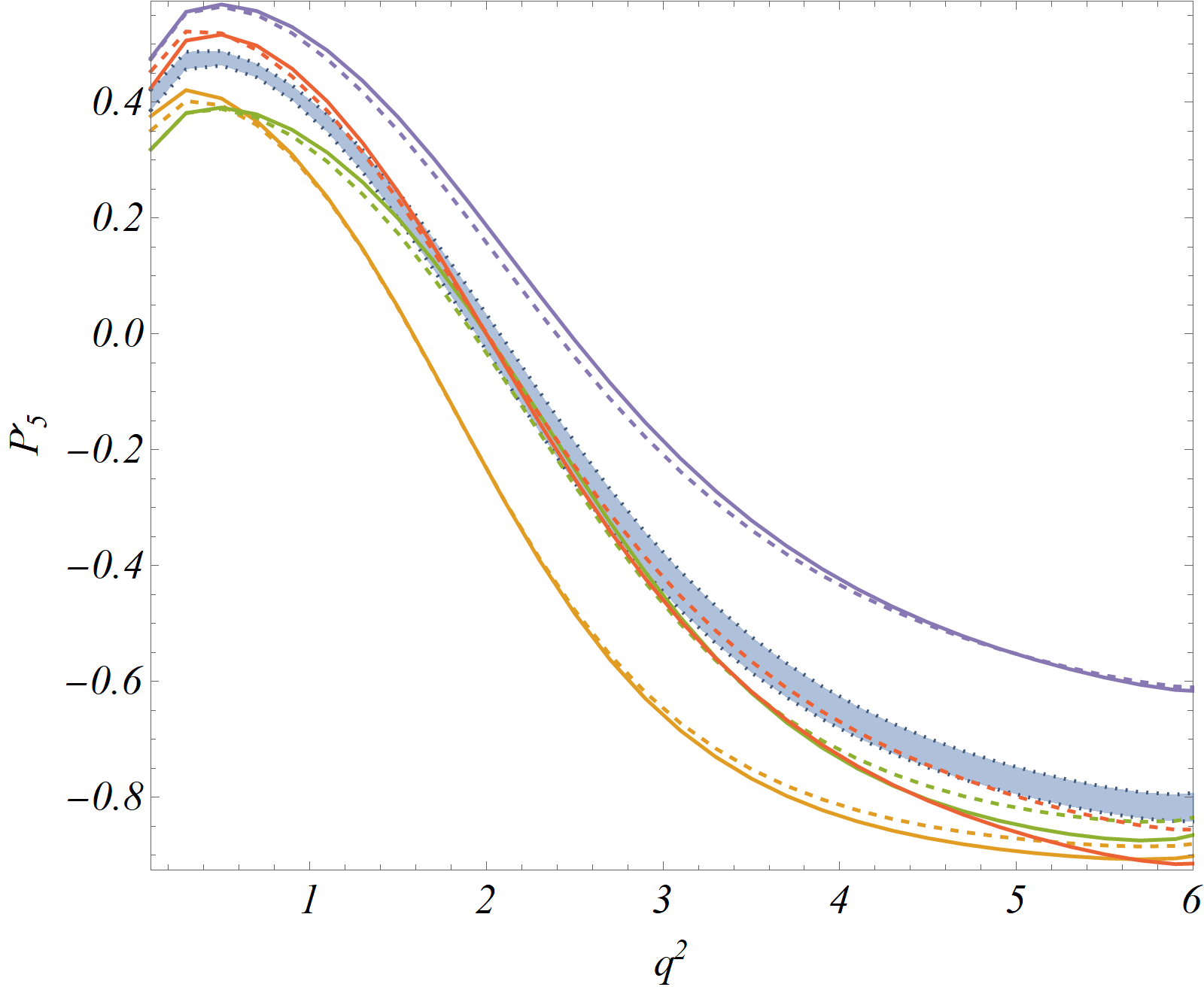}\label{fig:P5prP}}~~~
		\subfloat[]{\includegraphics[width=0.25\textwidth]{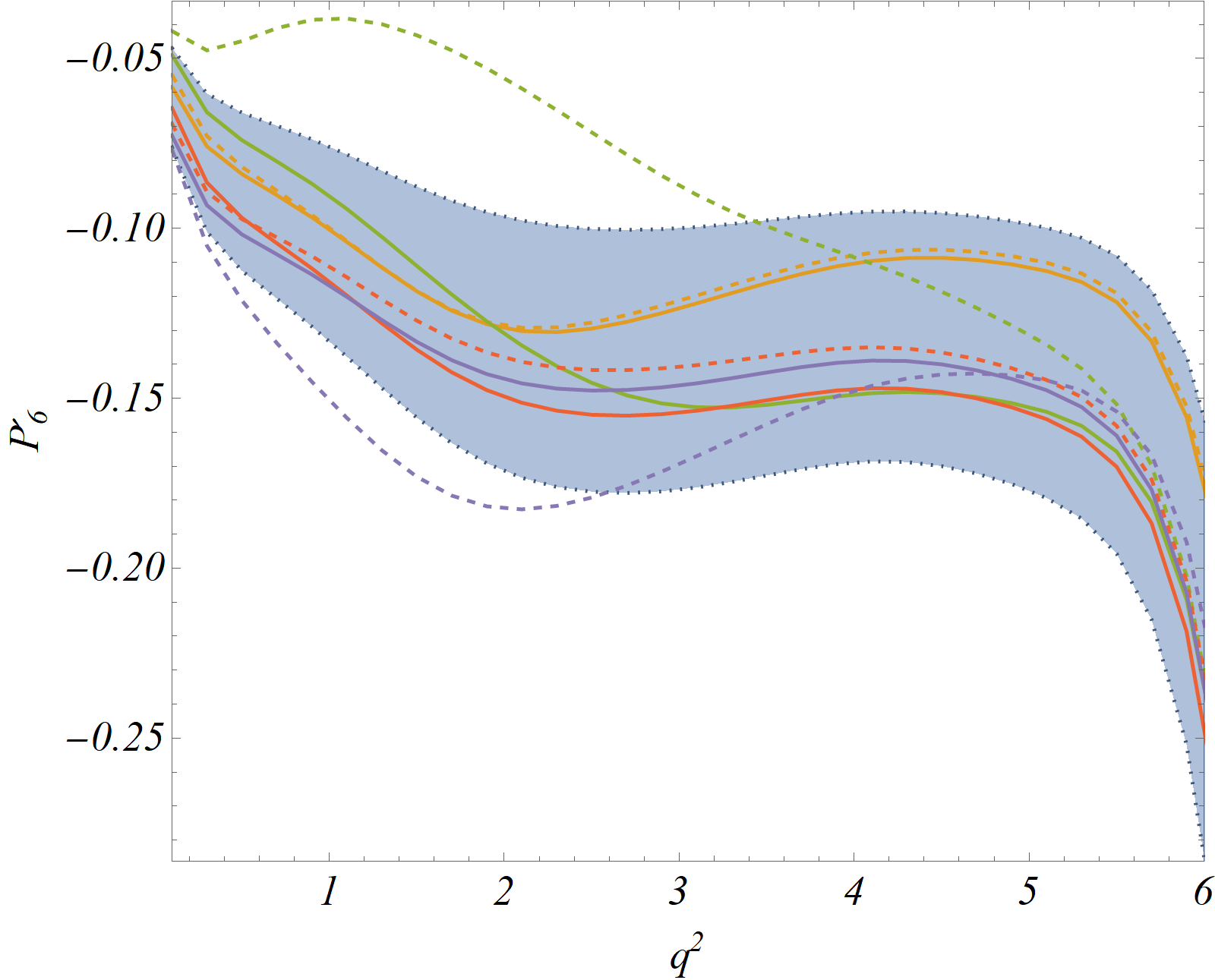}\label{fig:P6prP}}~~~
		\subfloat[]{\includegraphics[width=0.25\textwidth]{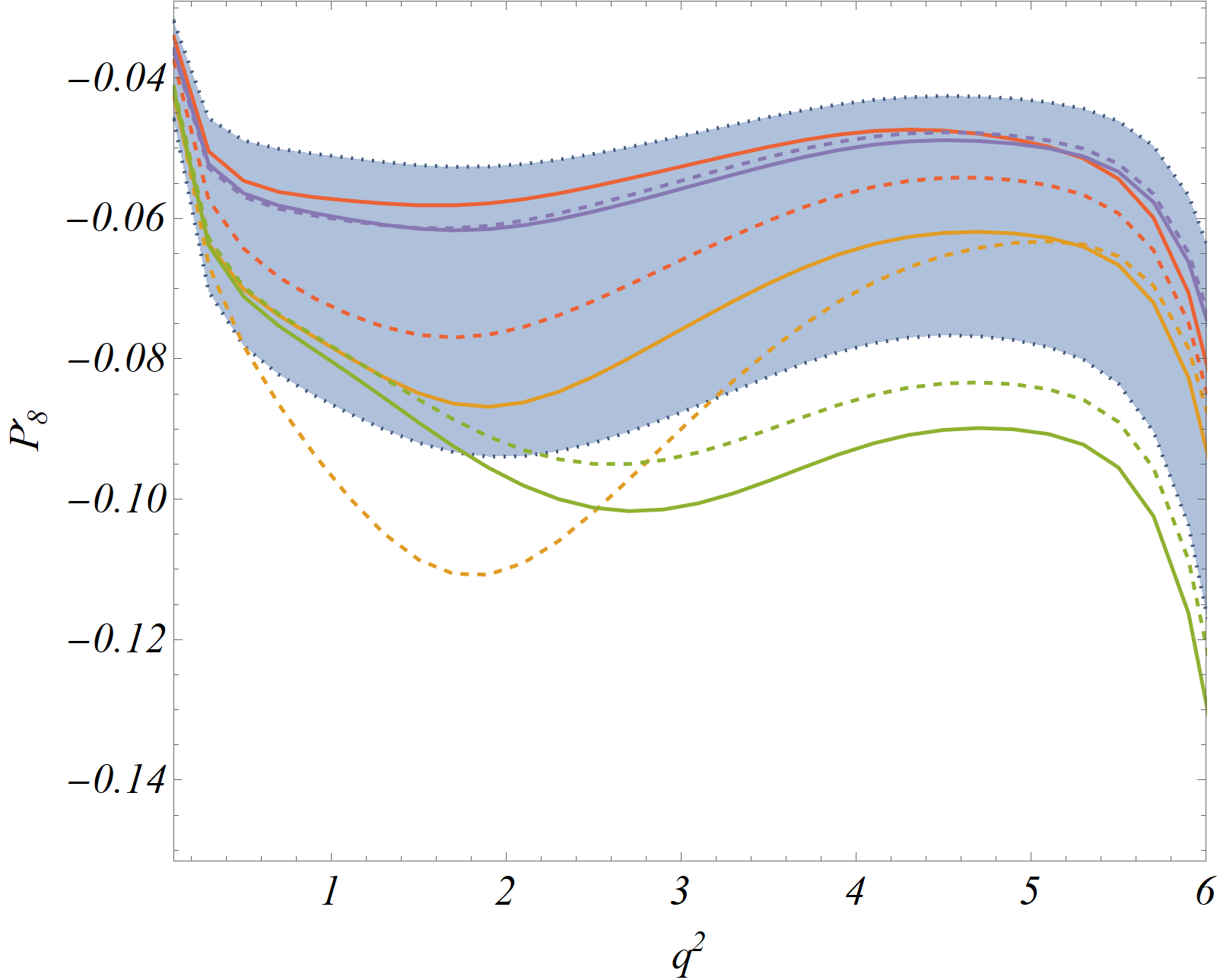}\label{fig:P8prP}}~~~
		\subfloat[]{\includegraphics[width=0.25\textwidth]{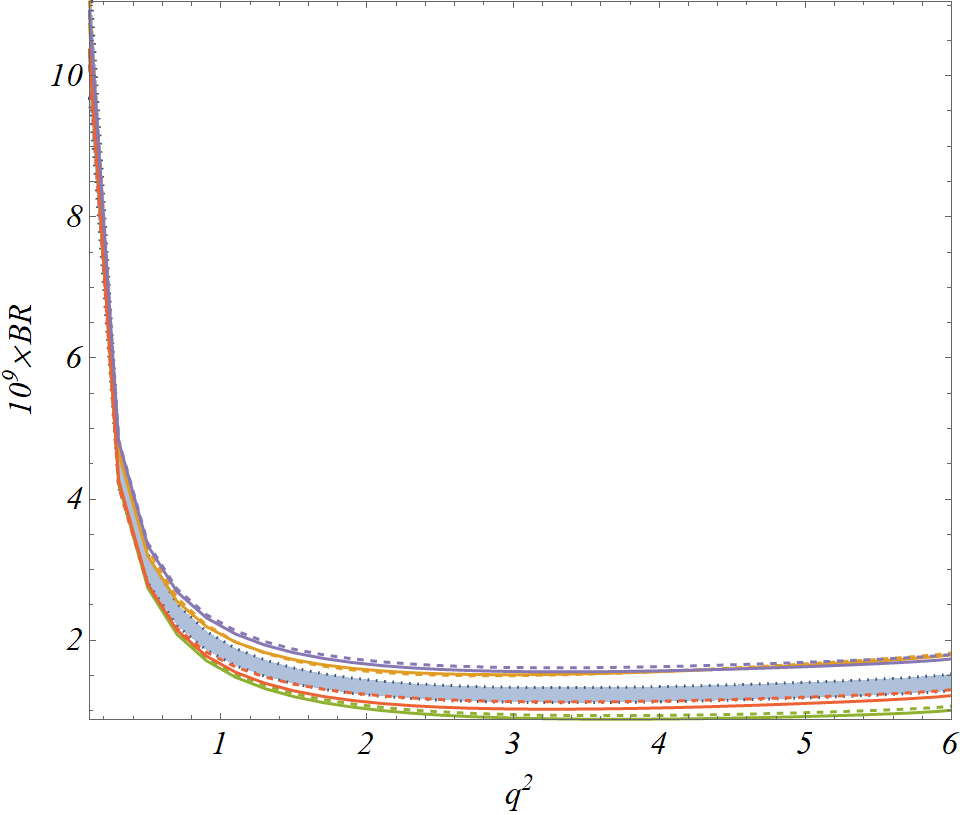}\label{fig:BRfracP}}\\
		\subfloat[]{\includegraphics[width=0.25\textwidth]{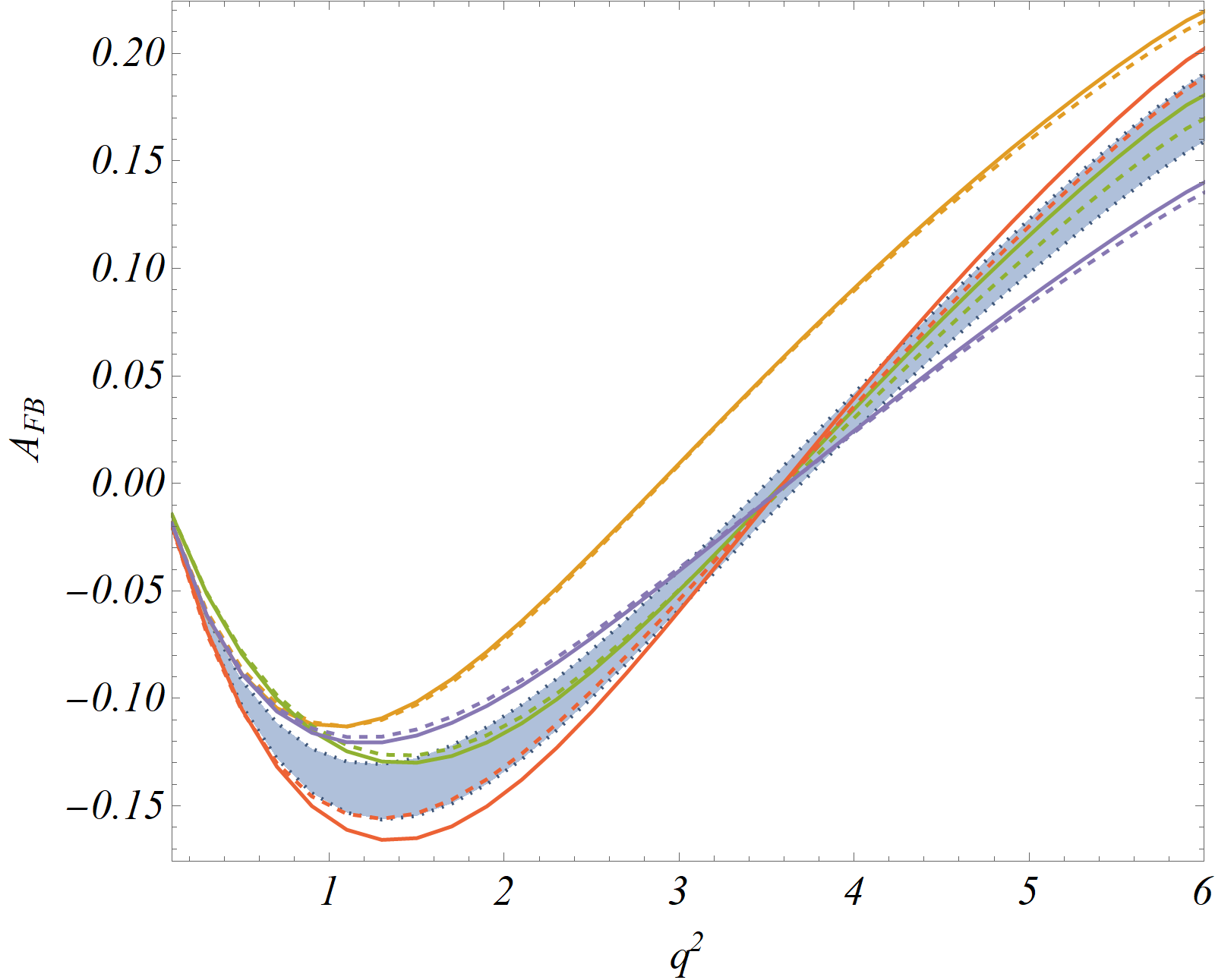}\label{fig:AFBP}}~~~
		\subfloat[]{\includegraphics[width=0.25\textwidth]{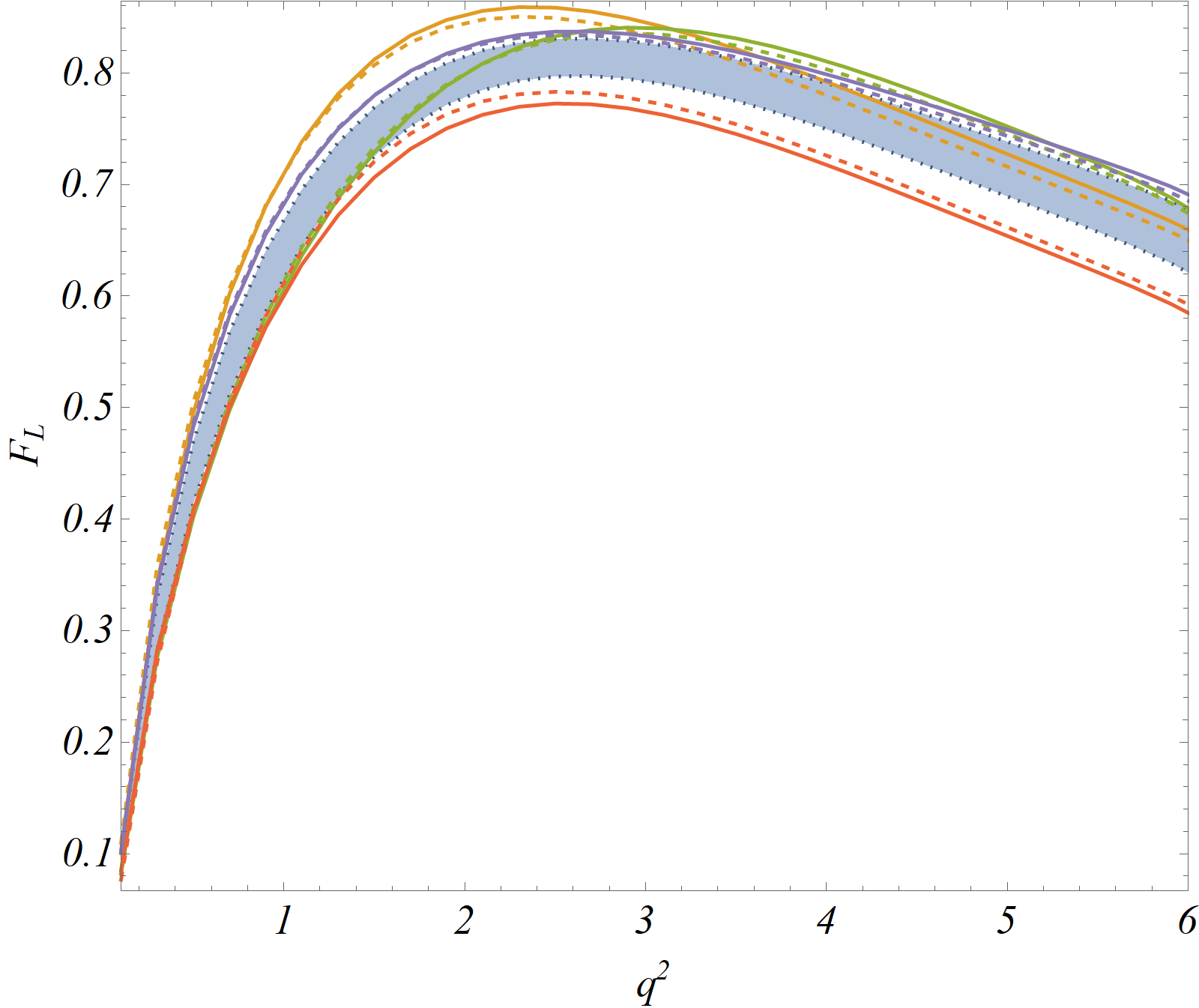}\label{fig:FLP}}~~~
		\subfloat[]{\includegraphics[width=0.25\textwidth]{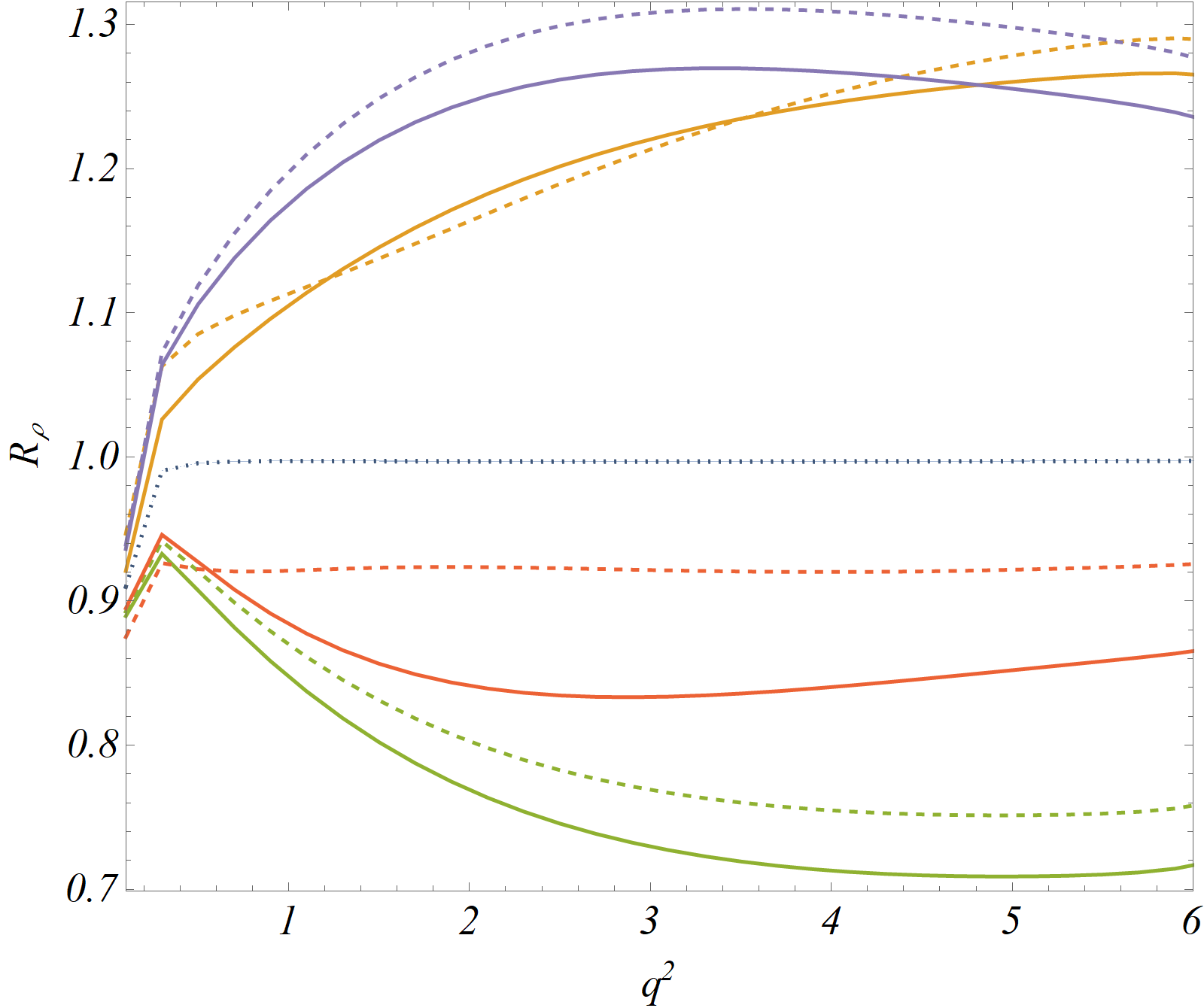}\label{fig:RrhoP}}~~~
		\subfloat[Legend]{\includegraphics[width=0.10\textwidth]{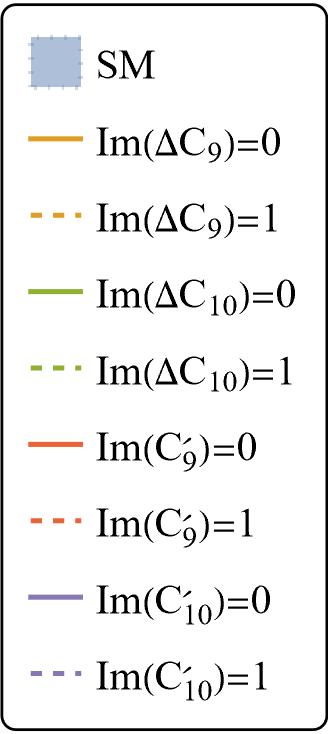}\label{fig:legrho}}
		\caption{The $q^2$ dependence of the observables in $B^{\pm}\to\rho^{\pm} ll$ decays, which are measurable at both the LHCb and Belle. The variations are shown in the SM and in the four different NP scenarios with NP Wilson coefficients $\Delta C_{9,10}$ and $C_{9,10}^\prime$. For the benchmarks, please see the legend.}
		\label{fig:BRhoobsuntagged}
	\end{figure*}
	
	\begin{figure*}[htbp]
		\small
		\centering
		\subfloat[]{\includegraphics[width=0.23\textwidth]{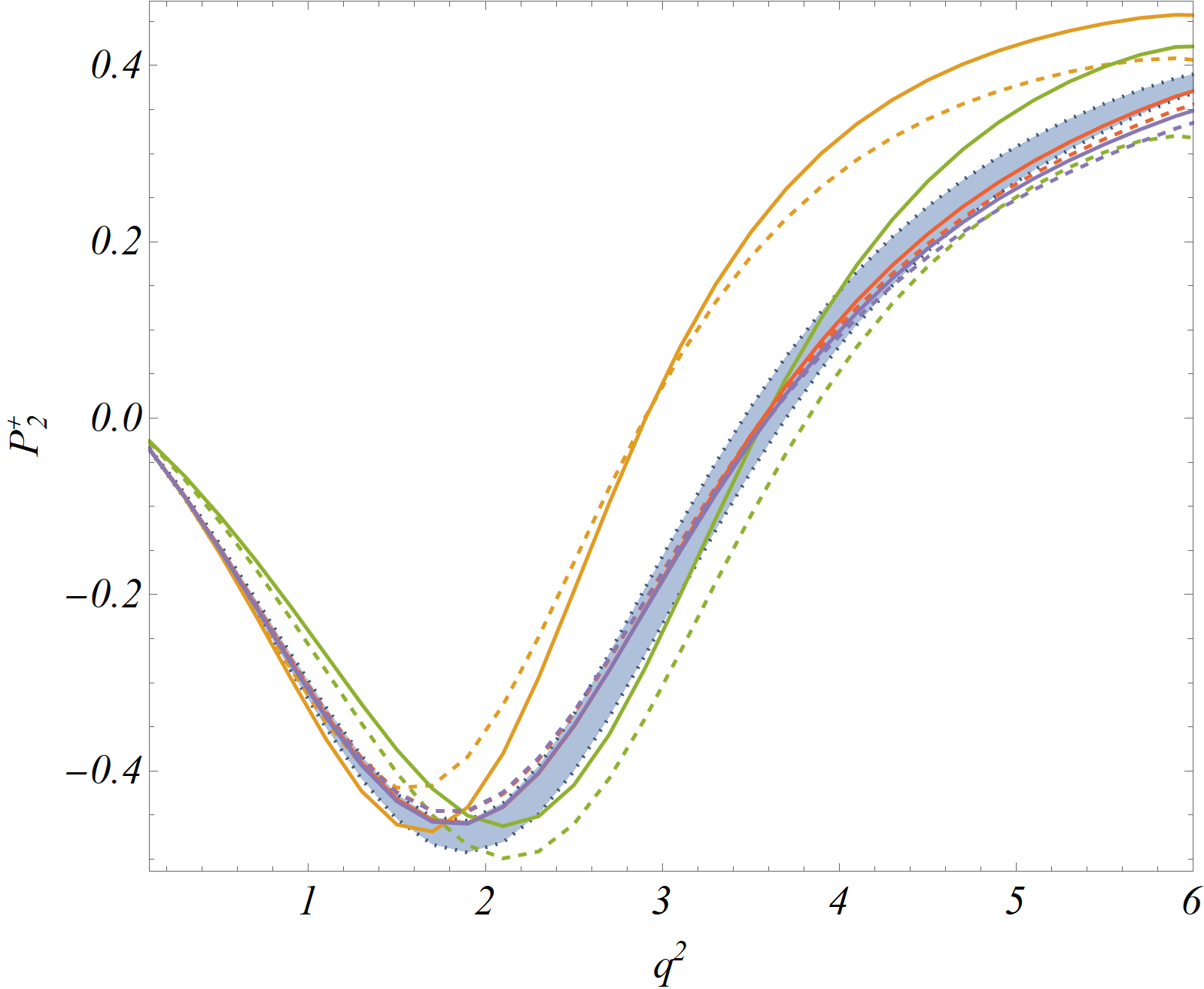}\label{fig:P2plus}}~~~~
		\subfloat[]{\includegraphics[width=0.23\textwidth]{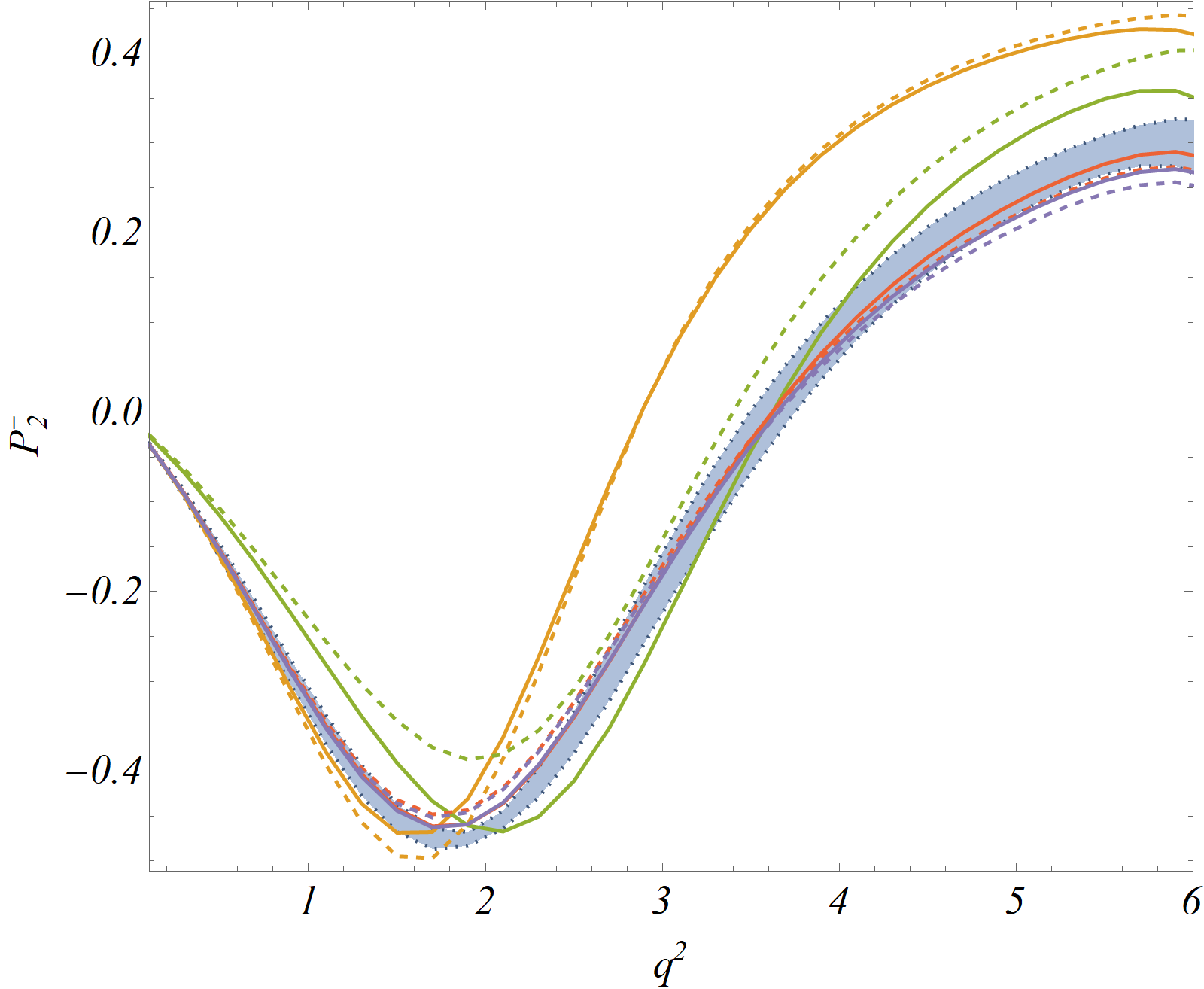}\label{fig:P2minus}}~~~~
		\subfloat[]{\includegraphics[width=0.23\textwidth]{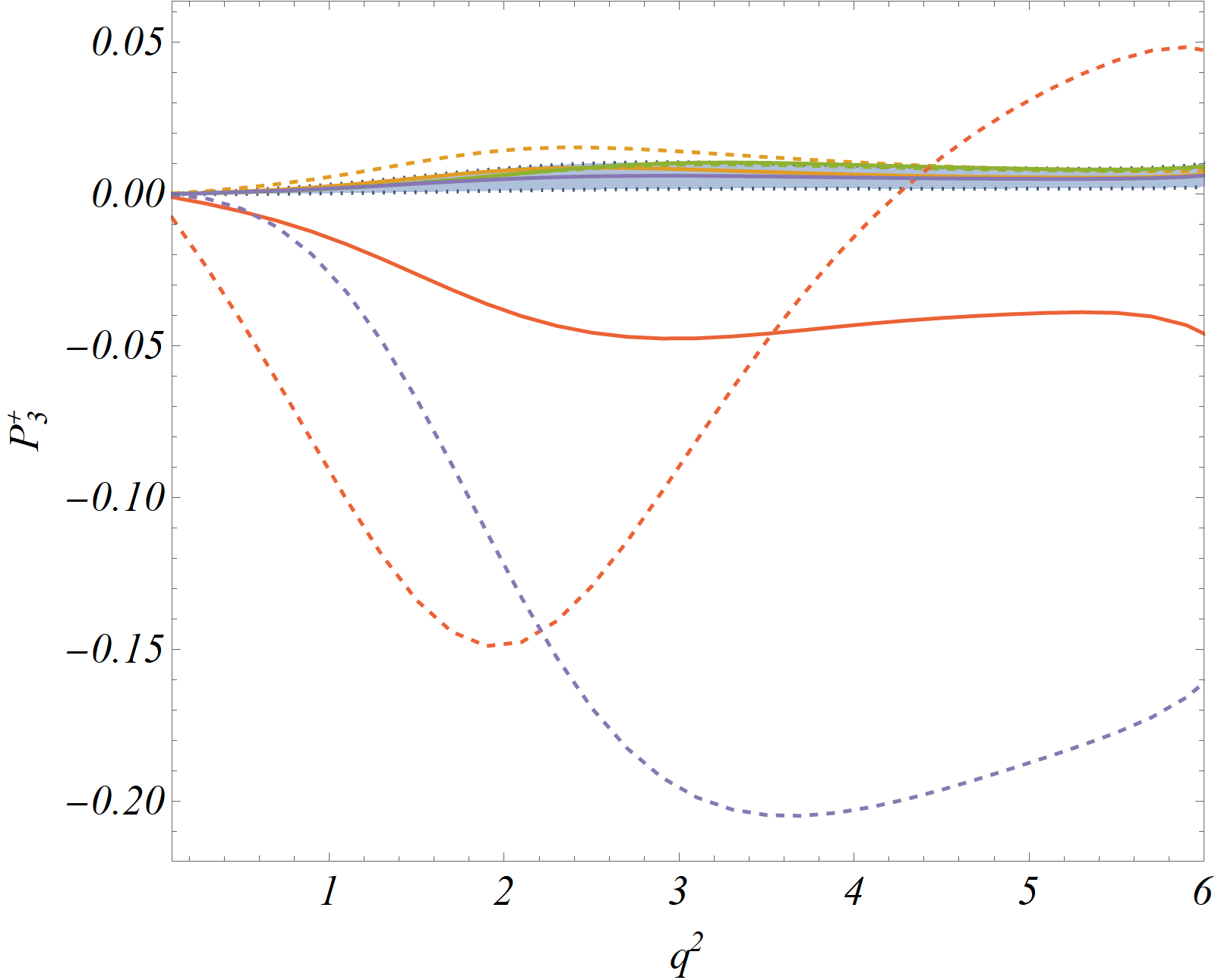}\label{fig:P3pl}}~~~
		\subfloat[]{\includegraphics[width=0.23\textwidth]{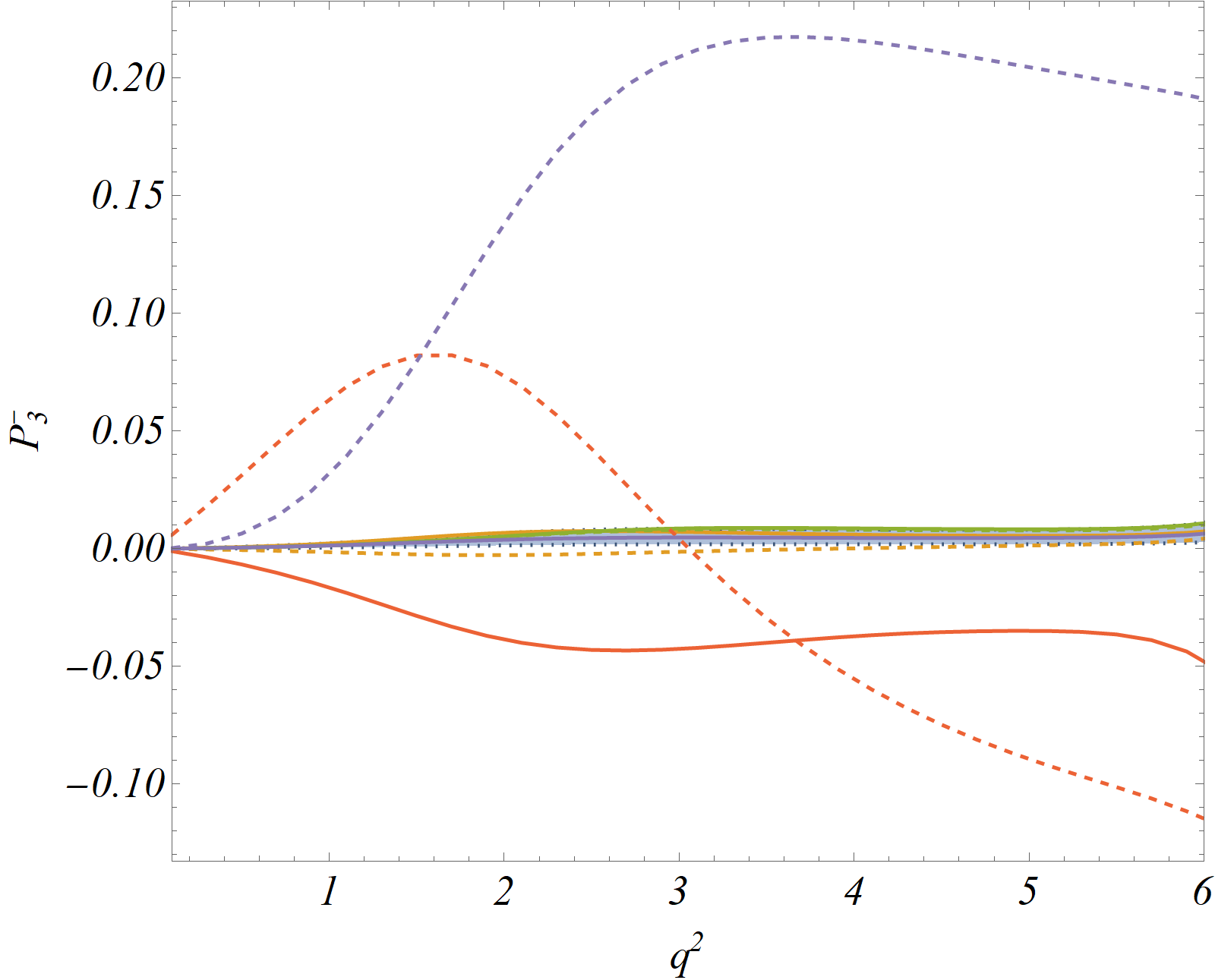}\label{fig:P3mn}}\\
		\subfloat[]{\includegraphics[width=0.23\textwidth]{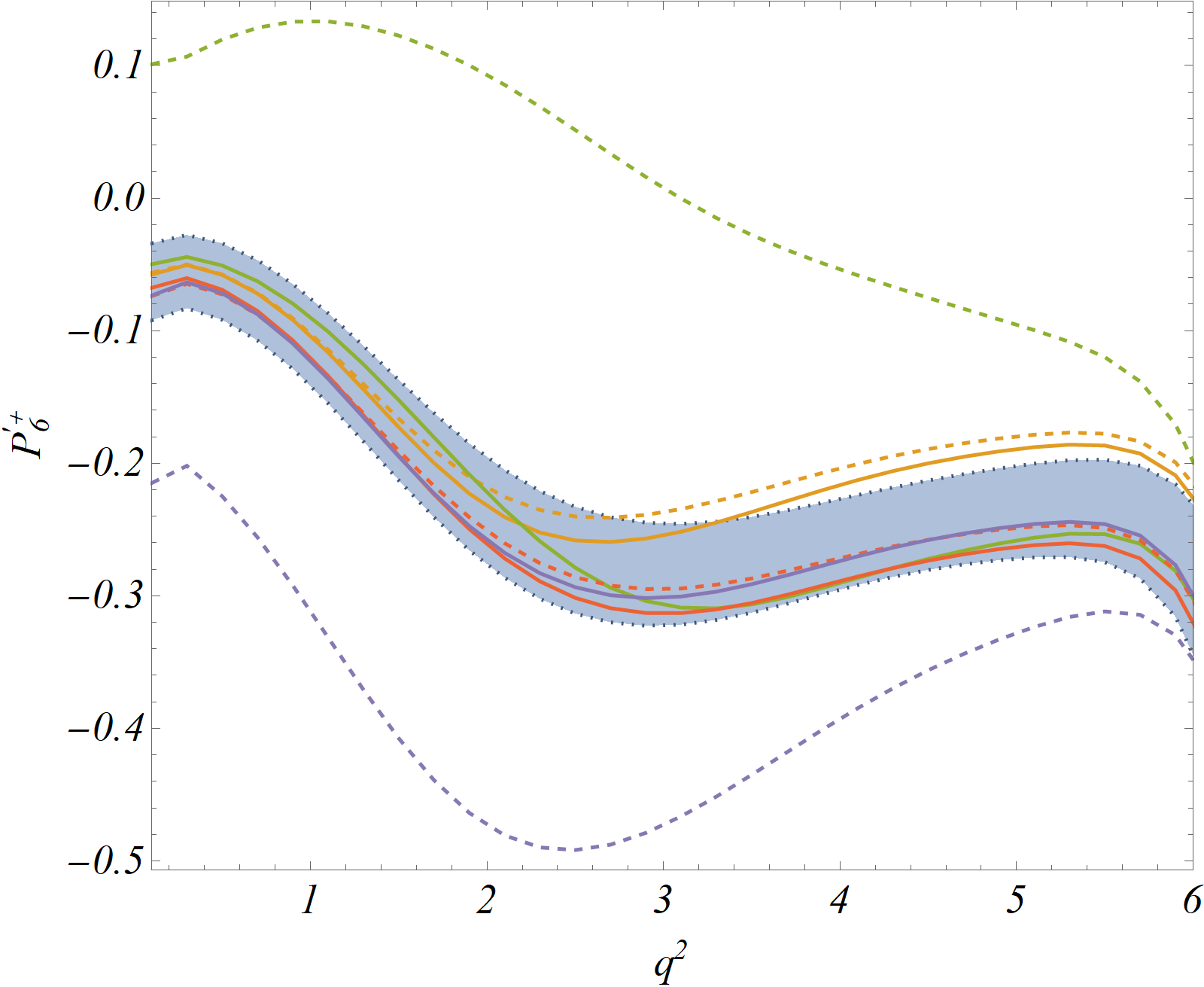}\label{fig:P6pl}}~~~
		\subfloat[]{\includegraphics[width=0.23\textwidth]{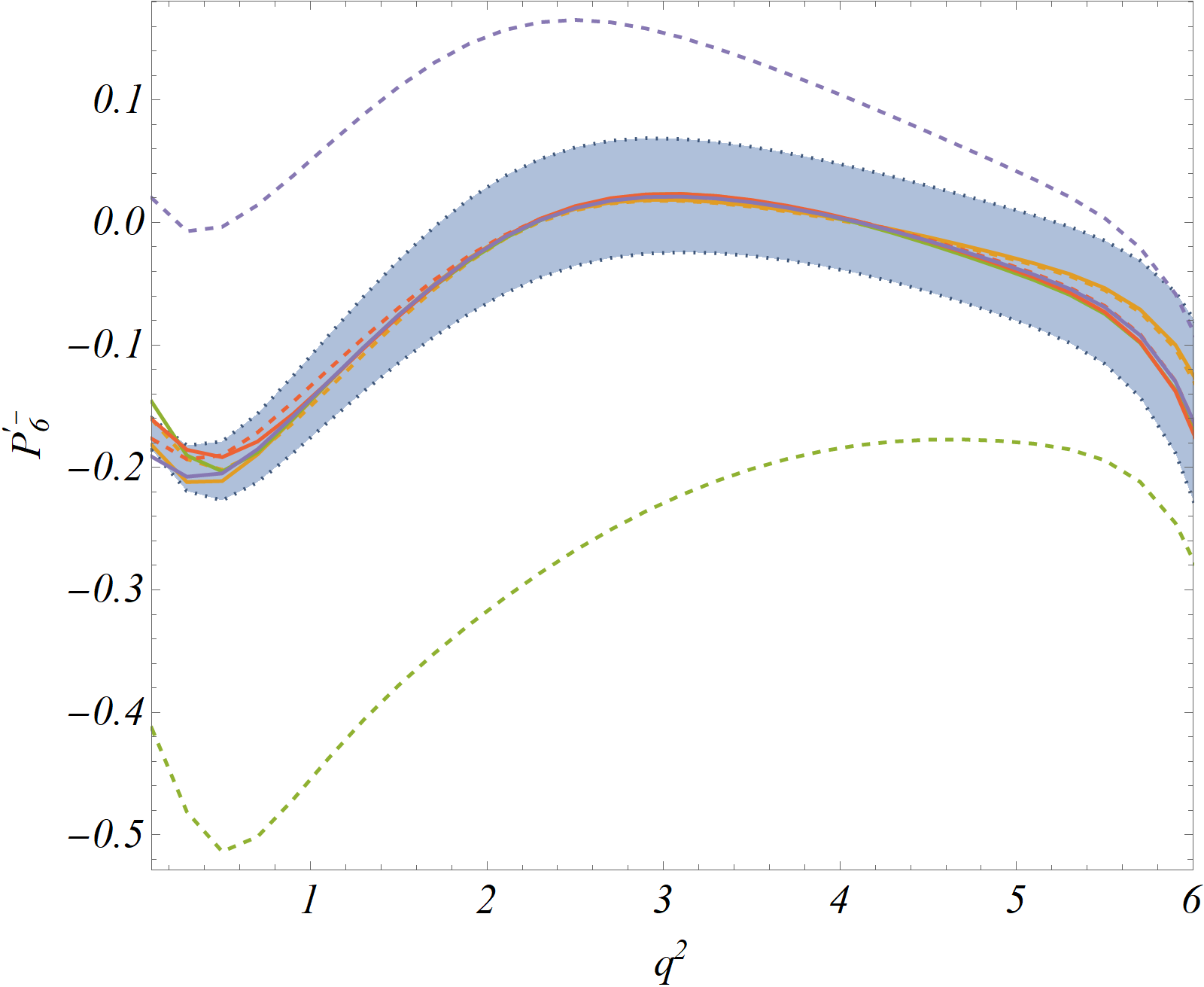}\label{fig:P6mn}}~~~~
		\subfloat[]{\includegraphics[width=0.23\textwidth]{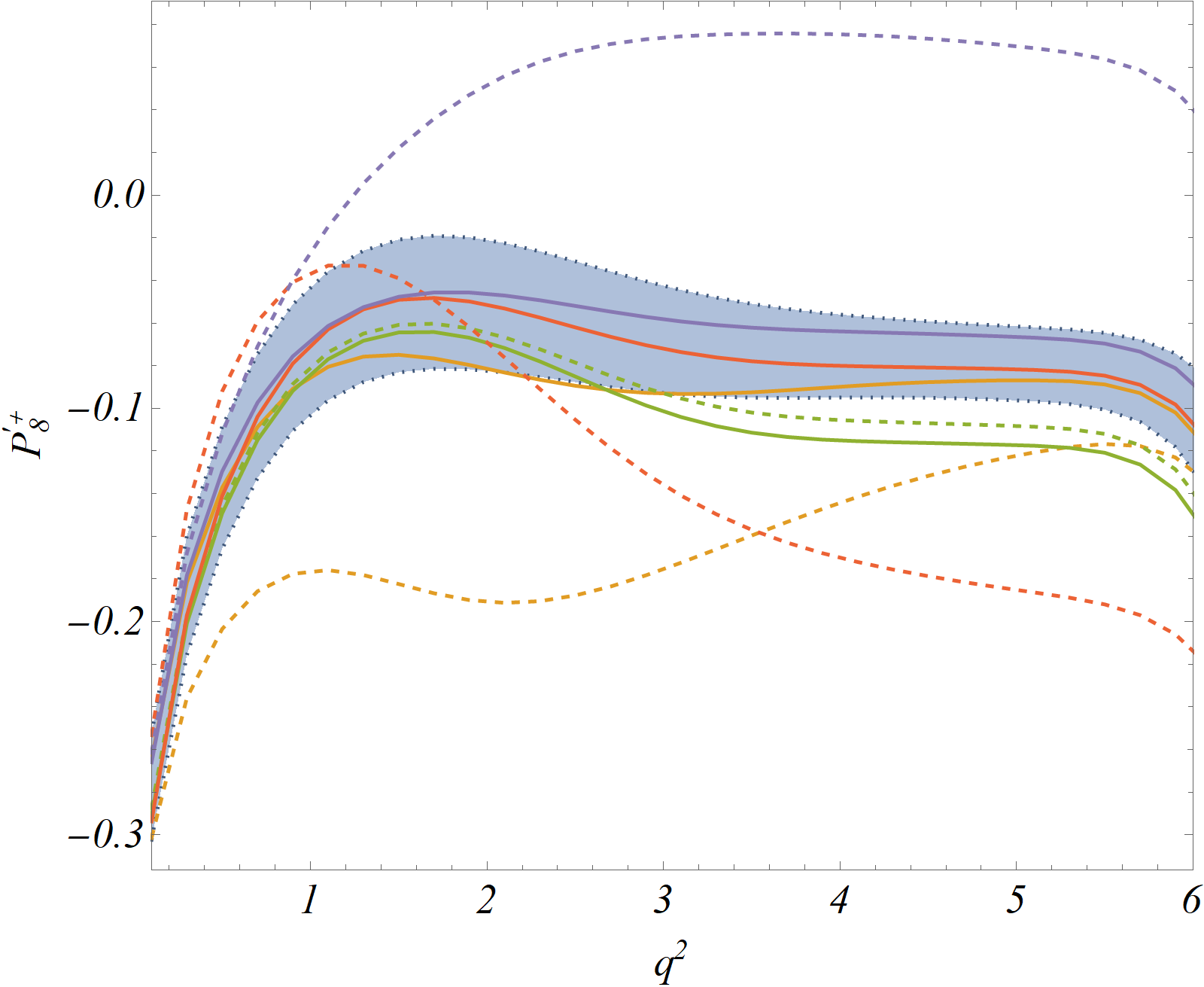}\label{fig:P8pl}}~~~~
		\subfloat[]{\includegraphics[width=0.23\textwidth]{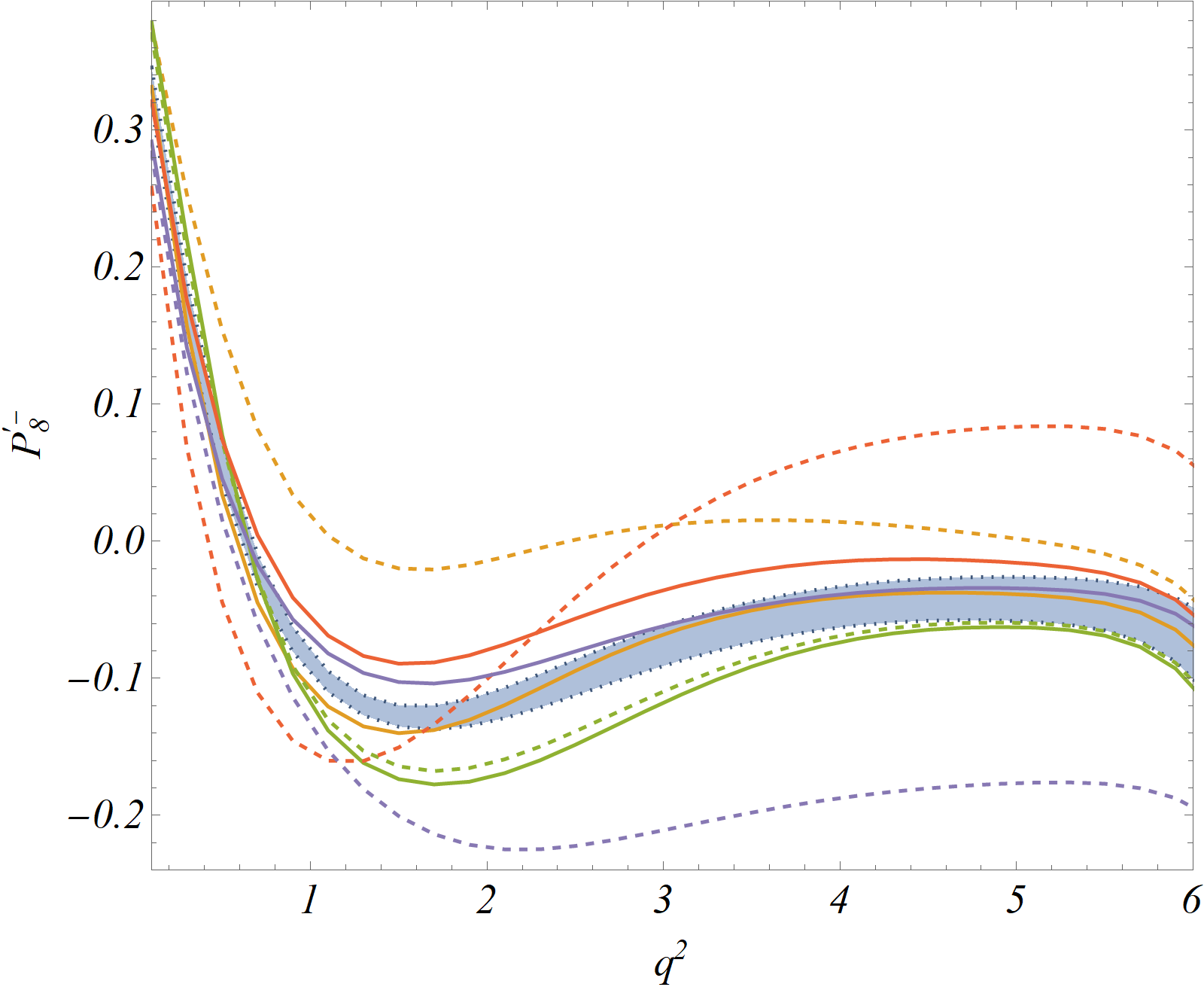}\label{fig:P8mn}}\\
		\caption{The study of the $q^2$ dependencies of a few tagged observables in $B^+\to\rho^+ ll$  and $B^-\to\rho^- ll$ decays, which are measurable at both the LHCb and Belle. The caption will be similar to the one given in fig.~\ref{fig:BRhoobsuntagged}; also, we follow the legends of that figure. }
		\label{fig:BRhoCPtag}
	\end{figure*}

For the charged $B$ decays, we can define the same set of observables for both the LHCb and Belle. Using the fit results given in table \ref{tab:LCSRlatfitres}, we have predicted the angular observables, the longitudinal polarization fraction $F_L$ of the $\rho$ meson, the forward-backward asymmetry $A_{FB}$ associated with the $B^-$ and $B^+$ decays in a few $q^2$-bins which are presented in tables \ref{tab:BMpred} and \ref{tab:BPpred}, respectively. The definitions can be seen from eqs.~\ref{eq:gamBP} and \ref{eq:taggedobsCP}. In addition, we have predicted the corresponding branching fractions and the LFU ratios $R_{\rho}^{\pm} = \Gamma(B^{\pm}\to \rho^{\pm}\mu\mu)/\Gamma(B^{\pm}\to \rho^{\pm} e e )$ in the SM. Also, we have tested the NP sensitivities of all these observables defined above.
	
In the SM, the observables $P_1^{-(+)}$ and $P_3^{-(+)}$ are negligibly small due to smallness of $J_3({\bar J_3})$ and $J_9({\bar J_9})$. The angular coefficients $J_3$ and $J_9$ are defined as follows: 
		\begin{align}
		J_3 &= \frac{1}{2} (1- 4 \frac{m_\mu^2}{q^2}) \left[|A^L_\perp|^2 - |A^L_{||}|^2 + (L\to R )\right] \\
		J_9 &= (1- 4 \frac{m_\mu^2}{q^2}) \left[\text{Im}(A^L_\perp A^{L*}_{||})  + (L\to R )\right].
		\end{align}
Therefore, $J_3$ is small due to the partial cancellation between the modulus of the two transversity amplitudes, whereas $J_9 \propto  Im(A^L_\perp A^{L^*}_{||})$ which is negligibly small. In a couple of other observables, there are differences between the SM predictions in the $B^+ \to\rho^+\mu\mu$ and $B^-\to \rho^-\mu\mu$ decays. These semileptonic decays are sensitive to the weak phases $\beta$ and $\gamma$, respectively, which will change the sign between the amplitudes in $B^+$ and $B^-$ decays. However, for the differences we have noted, the strong phases also play an essential role. As was discussed earlier, the strong phases will contribute to the longitudinal component of the helicity amplitude, $A_{0 L, R}$, via the hard-spectator corrections originating from the leading order annihilation diagrams. The angular coefficients, $J_{1c,2c,6c}$ and $J_{4,7,8}$ are mostly affected by these contributions and we see differences mentioned above only in $A_{FB}$, $F_L$, $P_4^{\prime}$, $P_5^{\prime}$, $P_6^{\prime}$, and $P_8^{\prime}$. All these observables are sensitive to $A_0$ via the respective angular coefficients $J$ or $\bar{J}$.  
	
	
	\begin{table}[t]
		\centering
		\begin{tabular}{*{5}{c}}
			\hline
			$\left.\text{Bin[}\text{GeV}^2\right]$  &  $\text{[0.1-1]}$  &  $\text{[1-2]}$  &  $\text{[2-4]}$  &  $\text{[4-6]}$  \\
			\hline
			$\langle A_5 \rangle$ & $\text{-0.0023(50)}$  &  $\text{0.0016(75)}$  &  $\text{0.0071(71)}$  &  $\text{0.0158(67)}$     \\
			\hline
			$\langle A_{6s} \rangle$ & $\text{0.0028(23)}$  &  $\text{0.0065(64)}$  &  $\text{0.00062(854)}$  &  $\text{-0.019(12)}$     \\
			\hline
			$\langle A_8 \rangle$ & $\text{0.0143(43)}$  &  $\text{0.0062(39)}$  &  $\text{0.0033(17)}$  &  $\text{0.00186(42)}$     \\
			\hline
			$\langle A_9 \rangle$ & $\text{0.000051(45)}$  &  $\text{0.000112(91)}$  &  $\text{0.000117(82)}$  &  $\text{0.000087(53)}$     \\
			\hline
			$\langle P_1 \rangle$ & $\text{0.0028(23)}$  &  $\text{0.0027(42)}$  &  $\text{-0.044(19)}$  &  $\text{-0.084(33)}$     \\
			\hline
			$\langle P_4^{'} \rangle$ & $\text{0.2044(41)}$  &  $\text{0.047(11)}$  &  $\text{-0.287(15)}$  &  $\text{-0.457(11)}$     \\
			\hline
			$\langle P_6^{'} \rangle$  & $\text{-0.114(26)}$  &  $\text{-0.163(36)}$  &  $\text{-0.165(39)}$  &  $\text{-0.157(42)}$     \\
			\hline
			$\langle BR \rangle \times 10^9$ & $\text{1.26(10)}$  &  $\text{0.613(51)}$  &  $\text{1.083(96)}$  &  $\text{1.19(10)}$     \\
			\hline
			$\langle F_L \rangle$ & $\text{0.278(26)}$  &  $\text{0.697(26)}$  &  $\text{0.786(19)}$  &  $\text{0.701(24)}$     \\
			\hline
			$ \langle R_{\rho} \rangle$ & $\text{0.98245(39)}$  &  $\text{0.99660(29)}$  &  $\text{0.99652(34)}$  &  $\text{0.99694(25)}$     \\
			\hline
		\end{tabular}
		\caption{\small Predictions of observables, in the SM, for $\bar{B}(B) \to \rho^0\ell \ell$ decays measurable at LHCb and Belle.} 
		\label{tab:B0predLHCb}
	\end{table}

	\begin{table}[t]
		\centering
		\begin{tabular}{*{5}{c}}
			\hline			
			$\left.\text{Bin[}\text{GeV}^2\right]$  &  $\text{[0.1-1]}$  &  $\text{[1-2]}$  &  $\text{[2-4]}$  &  $\text{[4-6]}$  \\
			\hline
			$\langle A_{CP} \rangle$ & $\text{-0.034(12)}$  &  $\text{-0.0325(53)}$  &  $\text{-0.01670(88)}$  &  $\text{-0.0115(20)}$  \\
			\hline
			$\langle A_3 \rangle$ & $\text{0.0000012(195)}$  &  $\text{-0.000028(62)}$  &  $\text{0.000015(86)}$  &  $\text{0.00028(18)}$  \\
			\hline
			$\langle A_4 \rangle$ & $\text{-0.0075(18)}$  &  $\text{-0.0060(23)}$  &  $\text{-0.00048(206)}$  &  $\text{0.0040(21)}$  \\
			\hline
			$\langle A_7 \rangle$ &  $\text{-0.0025(30)}$  &  $\text{-0.0024(26)}$  &  $\text{-0.00020(107)}$  &  $\text{-0.00015(55)}$  \\
			\hline
			$\langle P_2 \rangle$ &  $\text{-0.0545(17)}$  &  $\text{-0.2238(64)}$  &  $\text{-0.101(15)}$  &  $\text{0.1617(83)}$  \\
			\hline
			$\langle P_3 \rangle$ & $\text{0.00021(13)}$  &  $\text{0.00148(95)}$  &  $\text{0.0031(18)}$  &  $\text{0.0032(18)}$  \\
			\hline
			$ \langle P_5^{'} \rangle$ & $\text{0.3365(43)}$  &  $\text{0.166(13)}$  &  $\text{-0.241(19)}$  &  $\text{-0.479(13)}$  \\
			\hline
			$ \langle P_8^{'} \rangle$ &  $\text{-0.0205(69)}$  &  $\text{-0.036(10)}$  &  $\text{-0.040(10)}$  &  $\text{-0.039(10)}$  \\
			\hline
			$\langle A_{FB} \rangle$ & $\text{-0.0591(27)}$  &  $\text{-0.1017(95)}$  &  $\text{-0.0324(57)}$  &  $\text{0.0725(73)}$  \\
			\hline
		\end{tabular}
		\caption{\small Predictions of observables, in the SM, for $\bar{B}(B) \to \rho^0\ell \ell$ decays measurable only at Belle.}
		\label{tab:B0predbelle}
	\end{table}

	In table \ref{tab:Bchavpred}, we have shown the predicted values of the CP-averaged ($P_i, P_i^{\prime}$) and CP-asymmetric ($A_i$) observables which are obtained from the angular coefficients defined in eq.~\ref{eq:d4Gammacharged}. The respective observables are defined in eqs.~\ref{eq:untaggedobsACP} and \ref{eq:untaggedobsCPA}, respectively. In the SM, the numerical estimates are obtained using the fit results given in table \ref{tab:LCSRlatfitres}. The angular coefficients $J_3$ and $J_9$ are small in the SM, which results in the smallness of $A_3$ and $A_9$. As is evident, apart from $A_3$ and $A_9$, the other CP-asymmetric observables aren't suppressed in the SM for $b \to d \ell \ell$ transitions, as happens in the case of $b \to s \ell \ell$, where the term containing the CP-violating phase is doubly Cabibbo-suppressed. The asymmetric observables are sensitive to the difference between $J$ and $\bar{J}$ whereas the optimized observables are proportional to $J$ (for $B^-$) or $\bar{J}$ (for $B^+$). Therefore, the pattern of the numerical values for the asymmetric observables depicted in table~\ref{tab:Bchavpred} can be gauged from the corresponding values for the optimized observables provided in tables~\ref{tab:BMpred} and~\ref{tab:BPpred}. For eg., among all the CP-asymmetric observables, $A_4$ has the maximum value in terms of magnitude, and is one or two orders enhanced w.r.t the rest of the asymmetric observables for $q^2\leq 2$. This is due to the large difference between the values of $J_4$ and $\bar{J}_4$ for the $B^-$ and $B^+$ modes respectively in this region, and this effect can be anticipated beforehand if one notes the values of $\langle {P_4^{'}}^- \rangle$ (table~\ref{tab:BMpred}) and $ \langle {P_4^{'}}^+ \rangle$ (table~\ref{tab:BPpred}). In fig. \ref{fig:BRhoobsuntagged}, we show the variations of the observables listed in table \ref{tab:Bchavpred} with $q^2$ which are presented by blue thick regions. Among all the CP-asymmetric observables, $A_{CP}$ and $A_4$ have magnitudes of order 10$^{-1}$; the rest are suppressed by one or two orders of magnitudes. This is due to the strong dependencies of the relevant observables on $J_4 (\bar{J_4})$ and $J_{1c,2c} (\bar{J}_{1c,2c})$ which are sensitive to the strong phases in $A_0^{L,R}$. The $q^2$ dependence of the strong phase in $\lambda_{B,+}^{-1}$ will partially contribute to the $q^2$ variations of $A_0^{L,R}$. We have presented the $q^2$ variations of the CP averaged observables given in eq.~\ref{eq:untaggedobsCPA}. Almost similar kinds of $q^2$ variations are obtained for the related observables in $B^{\pm}$ decays defined in eq.~\ref{eq:taggedobsCP} which we have not presented separately. We have presented a few of them in fig.~\ref{fig:BRhoCPtag} for which the $q^2$ variations are slightly different than the corresponding observables obtained from the CP-averaged distributions. At the $q^2$ $\to$ 0 or $B \to V \gamma$ limit, the amplitude is enhanced by the photon pole. The longitudinal contribution to the $B\to\rho\ell \ell$ decay
		rate is suppressed by a power of $q^2$
		relative to the transverse contribution in this limit. 
	
	We have discussed in the sub-section \ref{subsubsec:Btorhointro} that we have followed the theory convention for the decay geometries, which is different (for the tagged decay rates) than what has been followed for $\bar{B} (B)\to \bar{K^*}(K^*)\ell^+\ell^+$ decays at the LHCb. Due to this difference in the convention, there will be differences in the sign of a few extracted angular coefficients. Therefore, while comparing our predicted numbers with the respective measurements, careful inspections are required if there are differences in the conventions. For a detail see the discussion in the sub-section \ref{subsubsec:Btorhointro}, also see the ref.~\cite{Gratrex:2015hna}.

	\begin{table}[t]
		\centering
		\begin{tabular}{*{5}{c}}
			\hline
			$\left.\text{Bin[}\text{GeV}^2\right]$  &  $\text{[0.1-1]}$  &  $\text{[1-2]}$  &  $\text{[2-4]}$  &  $\text{[4-6]}$  \\
			\hline
			$\langle P_1^0 \rangle$  & $\text{0.0039(32)}$  &  $\text{0.0034(51)}$  &  $\text{-0.048(21)}$  &  $\text{-0.087(34)}$     \\
			\hline
			$\langle P_2^0 \rangle$  & $\text{-0.0787(33)}$  &  $\text{-0.267(14)}$  &  $\text{-0.112(30)}$  &  $\text{0.197(19)}$     \\
			\hline
			$\langle P_3^0 \rangle$  & $\text{0.00040(29)}$  &  $\text{0.0021(15)}$  &  $\text{0.0040(23)}$  &  $\text{0.0036(20)}$     \\
			\hline
			$\langle P_4^{\prime 0} \rangle$  & $\text{0.2544(67)}$  &  $\text{0.063(13)}$  &  $\text{-0.296(16)}$  &  $\text{-0.468(10)}$     \\
			\hline
			$\langle P_5^{\prime 0} \rangle$  & $\text{0.393(11)}$  &  $\text{0.171(20)}$  &  $\text{-0.267(25)}$  &  $\text{-0.516(17)}$     \\
			\hline
			$\langle P_6^{\prime 0} \rangle$  & $\text{-0.125(38)}$  &  $\text{-0.166(45)}$  &  $\text{-0.170(42)}$  &  $\text{-0.157(41)}$     \\
			\hline
			$\langle P_8^{\prime 0} \rangle$  & $\text{-0.061(20)}$  &  $\text{-0.052(20)}$  &  $\text{-0.050(15)}$  &  $\text{-0.043(11)}$     \\
			\hline
			$\langle {BR}^0 \rangle \times 10^9$ & $\text{2.60(21)}$  &  $\text{1.27(10)}$  &  $\text{2.20(19)}$  &  $\text{2.41(20)}$     \\
			\hline
			$\langle A^0_{FB} \rangle$ & $\text{-0.0592(32)}$  &  $\text{-0.103(11)}$  &  $\text{-0.0323(92)}$  &  $\text{0.086(12)}$     \\
			\hline
			$\langle F^0_L \rangle$ & $\text{0.283(25)}$  &  $\text{0.698(25)}$  &  $\text{0.788(18)}$  &  $\text{0.699(24)}$     \\
			\hline
			$ \langle R^0_{\rho} \rangle$ & $\text{0.98257(39)}$  &  $\text{0.99670(29)}$  &  $\text{0.99657(33)}$  &  $\text{0.99697(25)}$     \\
			\hline
		\end{tabular}
		\caption{\small Predictions of observables, in the SM, for $B^0\to \rho^0\ell \ell$ decays, which are measurable at Belle.} 
		\label{tab:B0predBelle}
	\end{table}
	
	\begin{table}[t]
		\centering
		\begin{tabular}{*{5}{c}}
			\hline
			$\left.\text{Bin[}\text{GeV}^2\right]$  &  $\text{[0.1-1]}$  &  $\text{[1-2]}$  &  $\text{[2-4]}$  &  $\text{[4-6]}$  \\
			\hline
			$\langle \bar{P}_1^0 \rangle$  & $\text{0.0041(33)}$  &  $\text{0.0031(49)}$  &  $\text{-0.048(21)}$  &  $\text{-0.086(34)}$     \\
			\hline
			$\langle \bar{P}_2^0 \rangle$ &  $\text{-0.0770(22)}$  &  $\text{-0.2559(41)}$  &  $\text{-0.110(22)}$  &  $\text{0.137(21)}$     \\
			\hline
			$\langle \bar{P}_3^0 \rangle$  & $\text{0.00021(12)}$  &  $\text{0.00132(81)}$  &  $\text{0.0028(16)}$  &  $\text{0.0031(17)}$     \\
			\hline
			$\langle \bar{P}_4^{\prime 0} \rangle$  & $\text{0.2334(35)}$  &  $\text{0.037(12)}$  &  $\text{-0.306(17)}$  &  $\text{-0.463(13)}$     \\
			\hline
			$\langle \bar{P}_5^{\prime 0} \rangle$  & $\text{0.412(10)}$  &  $\text{0.189(22)}$  &  $\text{-0.238(29)}$  &  $\text{-0.459(21)}$     \\
			\hline
			$\langle \bar{P}_6^{'0}\rangle$ & $\text{-0.148(23)}$  &  $\text{-0.187(33)}$  &  $\text{-0.176(40)}$  &  $\text{-0.163(45)}$     \\
			\hline
			$\langle \bar{P}_8^{\prime 0}\rangle$ & $\text{0.0143(41)}$  &  $\text{-0.0252(25)}$  &  $\text{-0.0345(69)}$  &  $\text{-0.036(10)}$    \\
			\hline
			$\langle \bar{BR}^0 \rangle \times 10^9$  & $\text{2.43(21)}$  &  $\text{1.19(10)}$  &  $\text{2.13(19)}$  &  $\text{2.35(20)}$     \\
			\hline
			$\langle \bar{A}^0_{FB} \rangle$  &  $\text{-0.0589(25)}$  &  $\text{-0.1001(98)}$  &  $\text{-0.0324(79)}$  &  $\text{0.059(10)}$     \\
			\hline
			$\langle \bar{F}^0_L \rangle$ & $\text{0.273(26)}$  &  $\text{0.696(27)}$  &  $\text{0.784(19)}$  &  $\text{0.703(24)}$     \\
			\hline
			$ \langle \bar{R}^0_{\rho} \rangle$ & $\text{0.98232(39)}$  &  $\text{0.99650(32)}$  &  $\text{0.99646(35)}$  &  $\text{0.99690(25)}$     \\
			\hline
		\end{tabular}
		\caption{\small Predictions of observables, in the SM, for $\bar{B}^0\to \rho^0\ell \ell$ decays measurable at Belle.} 
		\label{tab:Bbar0predBelle}
	\end{table}


	We have also tested the NP sensitivities of all the observables described above, which are shown in the figs.~\ref{fig:BRhoobsuntagged} and \ref{fig:BRhoCPtag}, respectively. Like in $B\to \pi\ell\ell$ decays, we consider the NP effects in four additional operators defined in eq.~ \ref{eq:effopr}. Also, the corresponding effective Hamiltonian is given in eq.~\ref{eq:Heff} with the modified $\mathcal{H}_{eff}^{t}$ defined in eq.~\ref{eq:hefftm}. Apart from a few observables, which are shown in fig.~\ref{fig:BRhoCPtag}, the NP sensitivities of the observables associated with the $B^+$ or $B^-$ decays are similar to the one obtained from the CP-averaged rates; hence we have not shown them separately. The observables like $P_{2,3}^{\pm}, {P_6^{\prime}}^{\pm}, {P_8^{\prime}}^{\pm}$ have different dependencies on a couple of new WCs which are shown in fig.~\ref{fig:BRhoCPtag} which can be compared with the figs.~\ref{fig:P2P}, \ref{fig:P3P}, \ref{fig:P6prP}, and \ref{fig:P8prP}, respectively. The information contained in those figures has been summarised in table \ref{tab:NPsensrho}. Apart from $A_{CP}$, rest of the CP-asymmetric observables are sensitive to the imaginary parts of one or more NP scenarios. However, the $A_{CP}$ is sensitive to all four scenarios, only to the real components, and in a limited region: $0.5 \le q^2 \le 2 $ GeV$^2$. Similarly, the LFU ratio $R_{\rho}$ is sensitive to all the NP scenarios, both to the real and imaginary components. Note that there are observables sensitive to only a particular type of NP scenario. For example, all the three observables, $A_{FB}$, $F_L$, and $P_2$ are sensitive only to $\Delta C_9$, while $P_3$ is sensitive only to $Re(C_9^{\prime})$, on the other hand $A_3$ and $A_6$ are sensitive only to $Im(C_{9}^{\prime})$ and $Im(\Delta C_{10})$, respectively. Hence, any deviations from the SM in one or more of these observables will indicate the presence of the respective new interaction. Also, for the observables sensitive to multiple NP scenarios, a comparative study of the respective deviations in the measured values of the observables will help distinguish a particular type of NP scenario from the rest. The $q^2$ variations of different observables are different in different NP scenarios and are very much indicative.  
	
	Note that it is hard to separate the contributions of NP from respective SM predictions in the observables $P_6^{\prime}$ and $P_8^{\prime}$ obtained from the CP-averaged decay rate. Instead, the observables ${P^{\prime}}_{6,8}^{\pm}$ are sensitive to different new interactions. The respective $q^2$ distributions are shown in figs.~\ref{fig:P6pl}, \ref{fig:P6mn}, \ref{fig:P8pl}, and \ref{fig:P8mn}, respectively. For both these observables, the NP dependencies in the $B^+$ and $B^-$ channels are opposite; hence, in the observables obtained from the CP-averaged rates, those dependencies are lost due to a relative cancellation. We observe a similar kind of pattern in $P_3$ and $P_3^{\pm}$ which can be seen from table \ref{tab:NPsensrho} or from a comparison of the $q^2$ variations in figs.~\ref{fig:P3P}, \ref{fig:P3pl} and \ref{fig:P3mn}. $P_3$ is sensitive only to $Re(C_9^{\prime})$ while $P_3^{\pm}$ are sensitive to $Im(C_{9,10}^{\prime})$ in addition to $Re(C_9^{\prime})$. Due to a relative cancellation between the observables $P_3^+$ and $P_3^-$, the sensitivities towards the respective imaginary components are lost in their corresponding CP-average. Like $P_2$, $P_2^+$ is sensitive to $Re(\Delta C_9)$ and has a little sensitivity to the imaginary part of $\Delta C_{10}$. However, the predicted value is slightly lower than the corresponding SM predictions, which is hard to probe. More precise estimates are required. On the other hand, $P_2^-$ is sensitive to both $Re(\Delta C_9)$ and $Im(\Delta C_{10})$ and in both these scenarios, the magnitude of the predicted values is higher than the respective SM predictions. Here, again due to a relative cancellation the sensitivity towards $Im(\Delta C_{10})$ is lost in $P_2$.  
	
	In accordance to our presentation of the NP dependence of the observables for the $B\to\pi$ sector; we also tabulate the NP dependence of the observables for the $B\to\rho$ sector corresponding to the same four $q^2$ bins and for benchmark combinations of the real and imaginary NP WCs. For the charged modes, the observables corresponding to the untagged case that can be measured at both Belle and LHCb are provided in tables~\ref{tab:appBtorhopmasynotag1},~\ref{tab:appBtorhopmasynotag2},~\ref{tab:appBtorhopmACPnotag1},~\ref{tab:appBtorhopmACPnotag2},~\ref{tab:appBtorhopmoptnotag1},~\ref{tab:appBtorhopmoptnotag2},~\ref{tab:appBtorhopmbrnotag1} \&~\ref{tab:appBtorhopmbrnotag2}. As discussed previously, we refrain from providing all the observables for the tagged case measurable only at LHCb and highlight only those which have NP dependence different than the corresponding observables obtained from the CP-averaged distribution. Such observables corresponding to the charged modes are displayed in tables~\ref{tab:appBtorhopmP2P3tag1},~\ref{tab:appBtorhopmP2P3tag2},~\ref{tab:appBtorhopmP6P8tag1} and~\ref{tab:appBtorhopmP6P8tag2}. From a comparison with the respective SM predictions, one can look for possible deviations which can be statistically significant. Note that we have considered the numerical values of the new WCs $\approx$ 1. We have not considered values larger than one. However, the trend or the pattern of NP effects, which are also dependent on $q^2$ regions, is apparent in the respective predictions.
	
	As we have mentioned earlier, in case of $B^0\to\rho^0 (\to \pi \pi)\ell \ell$ decay, the $\rho$ meson is reconstructed via the decay, $\rho^0 \to \pi^+\pi^-$, to a flavor-non-specific final state. Therefore, the final state can arise from the decay of both $B^0$ and $\bar{B^0}$ mesons. The interference between $B^0-\bar{B}^0$ oscillations and decay processes lead to the time dependencies in the decay amplitudes. These time dependent helicity amplitudes will also impact the corresponsing angular coefficients $J_i$ and $\tilde{J_i}$ which are presented as: 
	\begin{equation}
	J_i(t) = J_i(A_H\to A_H(t)), \ \ \ \  \tilde{J_i}(t) = J_i(A_H\to \tilde{A}_H(t)). 
	\end{equation}   
	The time dependencies of the CP-averaged angular coefficients defined in eqs.~\ref{eq:rateplus} and \ref{eq:rateminus} are given by 
	\begin{align}
	J_i(t) +\tilde{J_i}(t) &= e^{-\Gamma t}\left[(J_i + \tilde{J_i}) \cosh(y\Gamma t) - h_i \sinh(y \Gamma t) \right], \nonumber \\
	J_i(t) - \tilde{J_i}(t) &= e^{-\Gamma t}\left[(J_i - \tilde{J_i}) \cosh(x \Gamma t) - s_i \sinh(x \Gamma t) \right], 
	\end{align}
	where $x = \Delta{m}/\Gamma$ and $y = \Delta{\Gamma}/2\Gamma$. $\Delta{\Gamma} = \Gamma_L - \Gamma_H$ and $\Delta{m} = M_H - M_L$ are the lifetime and mass differences between the mass eigenstates. For $B_d^0$ decays, the lifetime difference is negligible. The detailed mathematical expression for the $s_i$ and $h_i$ are given in ref.~\cite{Descotes-Genon:2015hea}. The time-integrated CP-averaged rates and CP-asymmetries, as measured at hadronic machines and $B$-factories are defined in terms of the modified angular functions \cite{Descotes-Genon:2015hea} 
	
	\begin{subequations}\label{eq:timeaverages}
		\begin{align}
		&\Big{<}J_i + \tilde{J_i} \Big{>} _\text{LHCb} = \frac{1}{\Gamma} \biggl{[} \frac{J_i + \tilde{J_i}}{1-y^2} - \frac{y}{1-y^2} \times h_i \biggl{]} \,, \\
		& \nonumber \\
		&\Big{<}J_i - \tilde{J_i} \Big{>}_\text{LHCb} = \frac{1}{\Gamma} \biggl{[} \frac{J_i - \tilde{J_i}}{1+x^2} - \frac{x}{1+x^2} \times s_i \biggl{]} \,, \\
		& \nonumber \\
		&\Big{<}J_i + \tilde{J_i} \Big{>} _\text{Belle} = \frac{2}{\Gamma} \biggl{[} \frac{J_i + \tilde{J_i}}{1-y^2} \biggl{]} \,, \\
		& \nonumber \\
		&\Big{<}J_i - \tilde{J_i} \Big{>} _\text{Belle} = \frac{2}{\Gamma} \biggl{[} \frac{J_i - \tilde{J_i}}{1+x^2} \biggl{]},
		\end{align}
	\end{subequations}

	\begin{figure*}[t]
	\small
	\centering
	\subfloat[]{\includegraphics[width=0.23\textwidth]{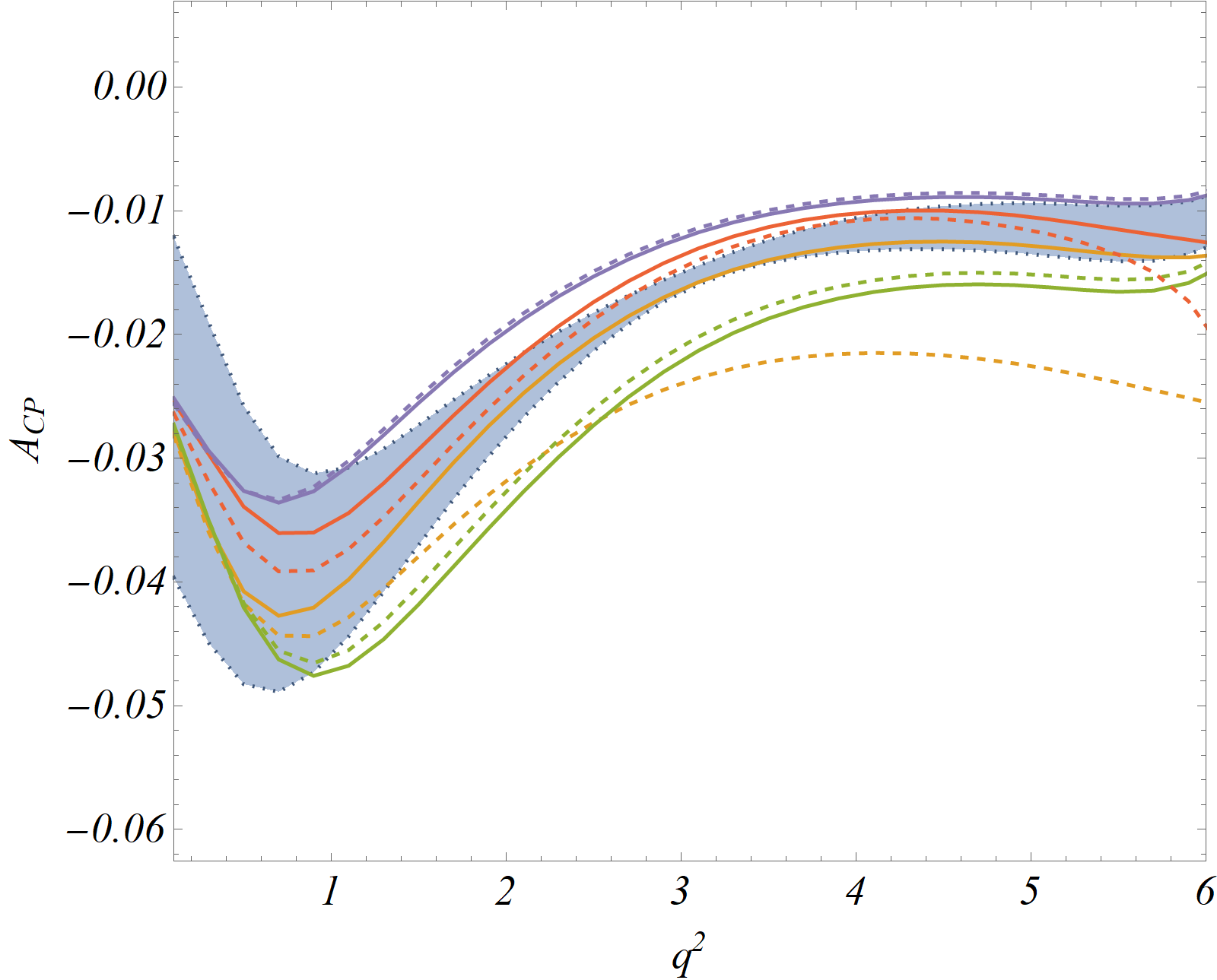}\label{fig:ACPneutral}}~~~
	\subfloat[]{\includegraphics[width=0.23\textwidth]{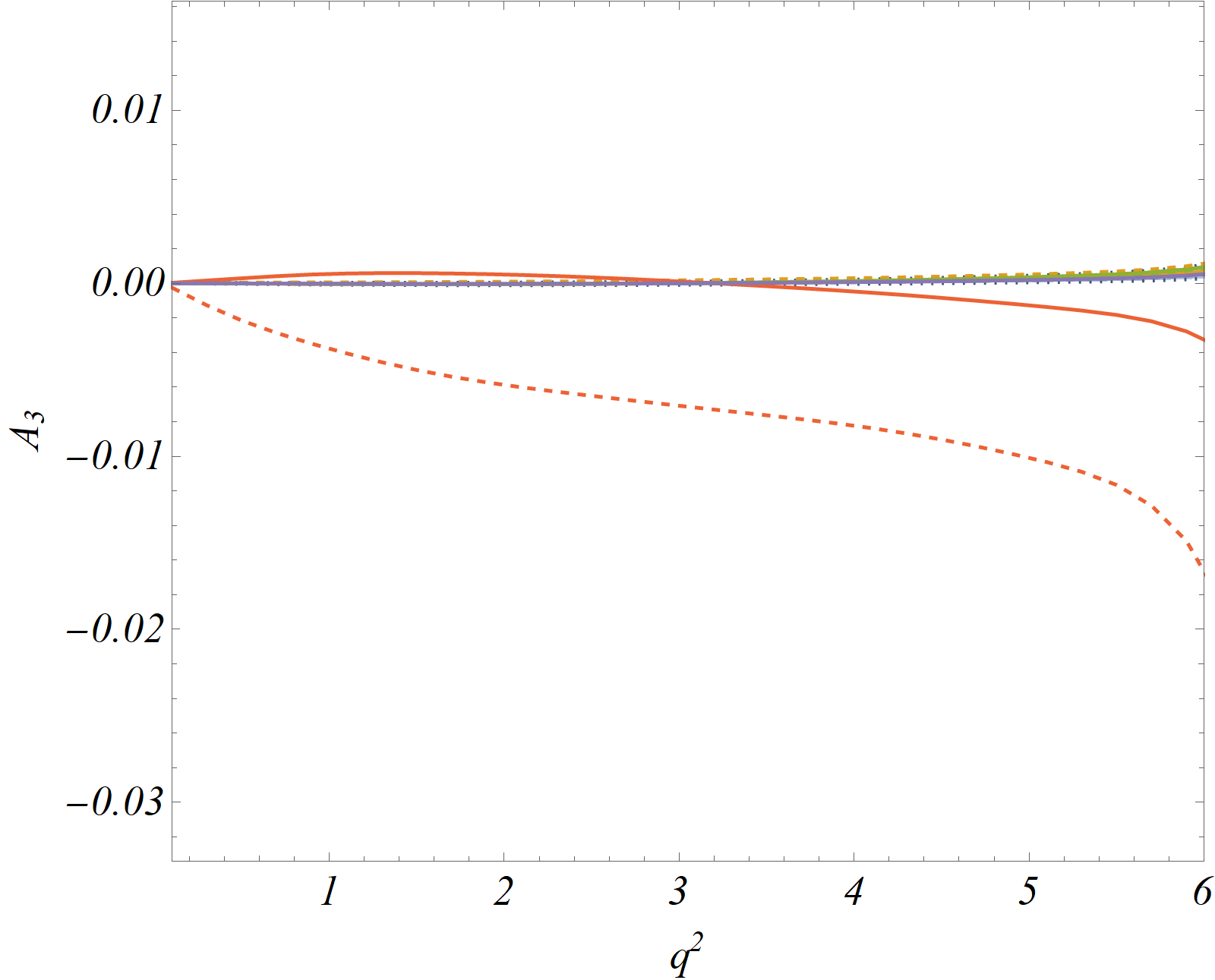}\label{fig:A3neutral}}~~~
	\subfloat[]{\includegraphics[width=0.23\textwidth]{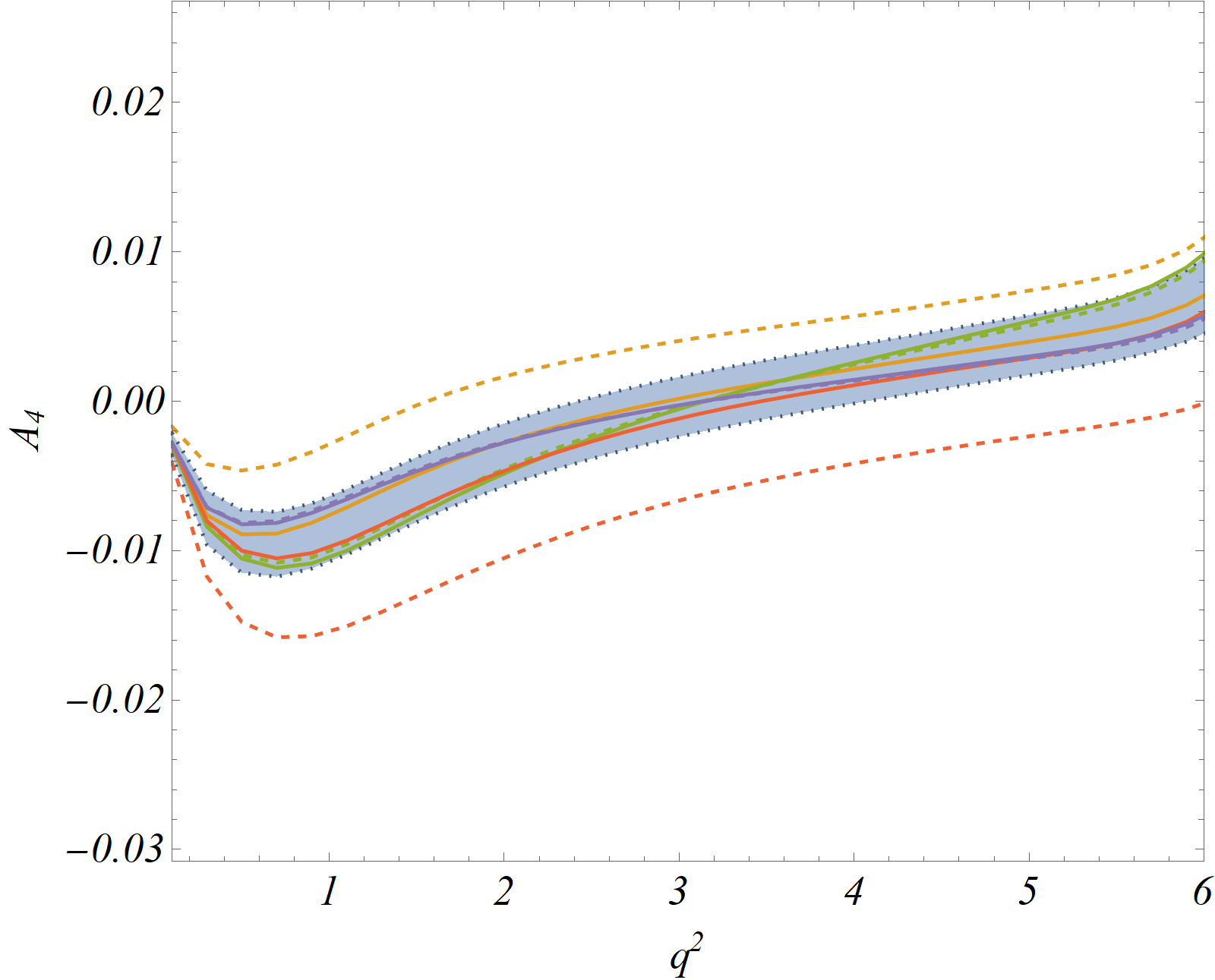}\label{fig:A4neutral}}~~~
	\subfloat[]{\includegraphics[width=0.23\textwidth]{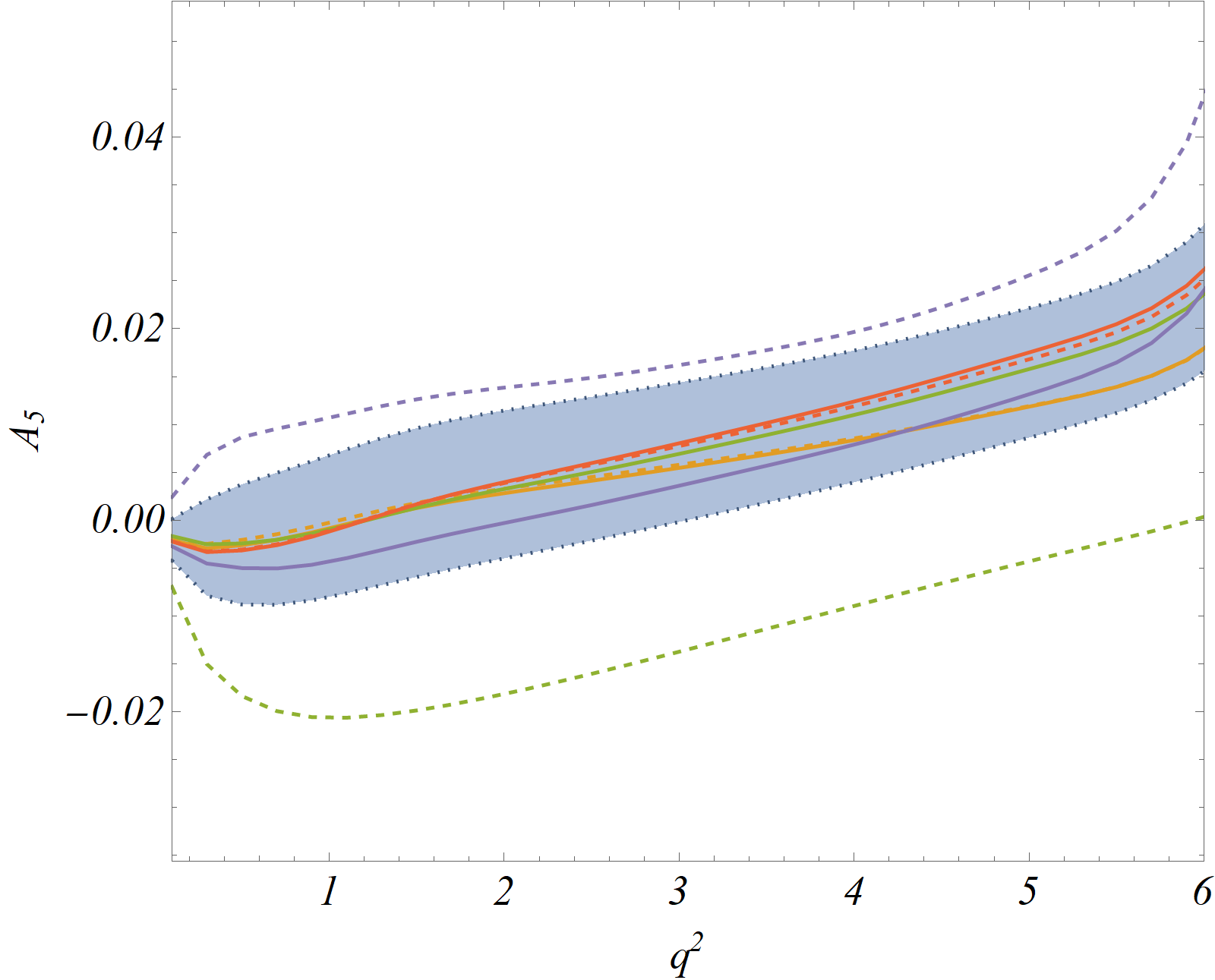}\label{fig:A5LHCb0}}\\
	\subfloat[]{\includegraphics[width=0.23\textwidth]{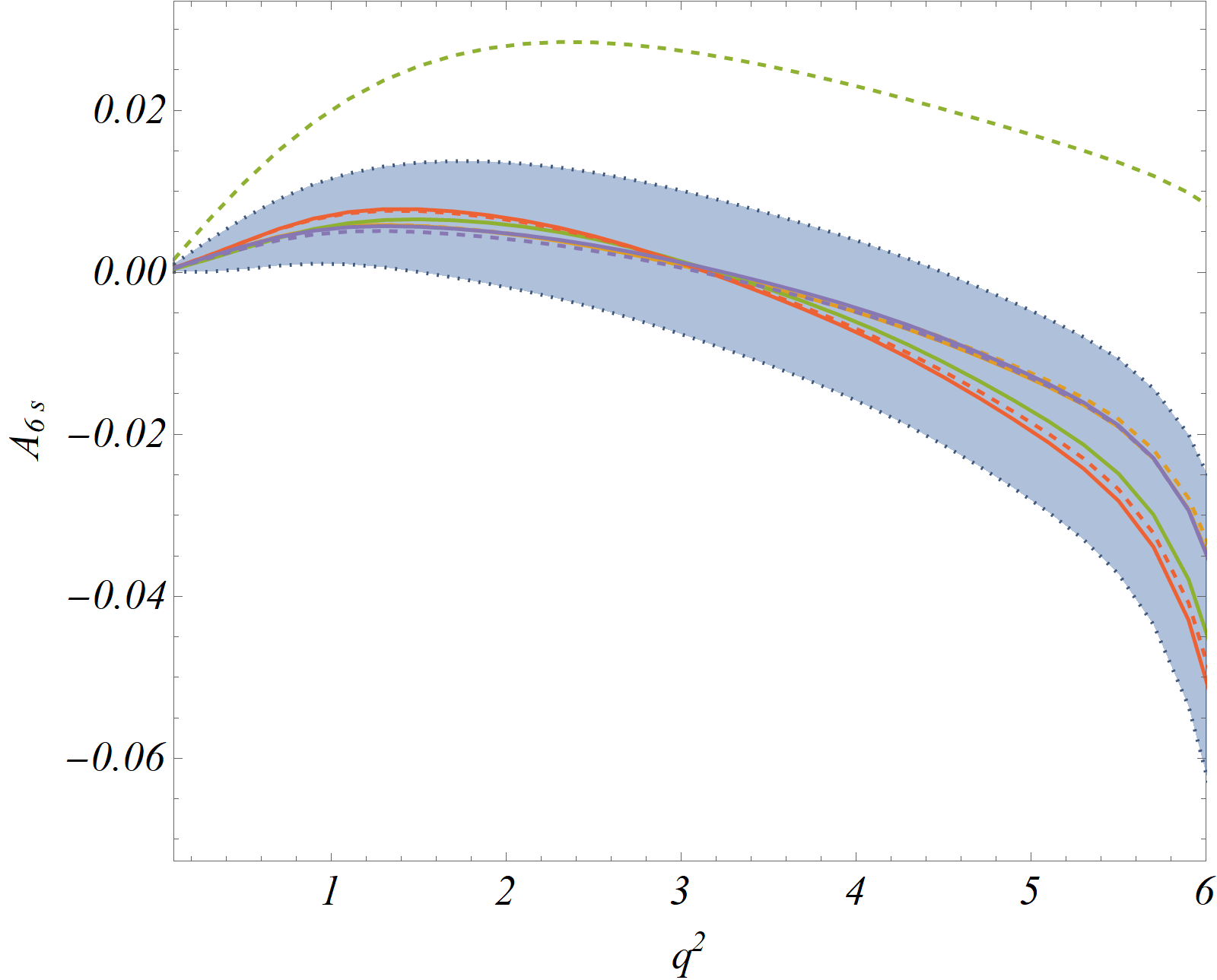}\label{fig:A6sLHCb0}}~~~
	\subfloat[]{\includegraphics[width=0.23\textwidth]{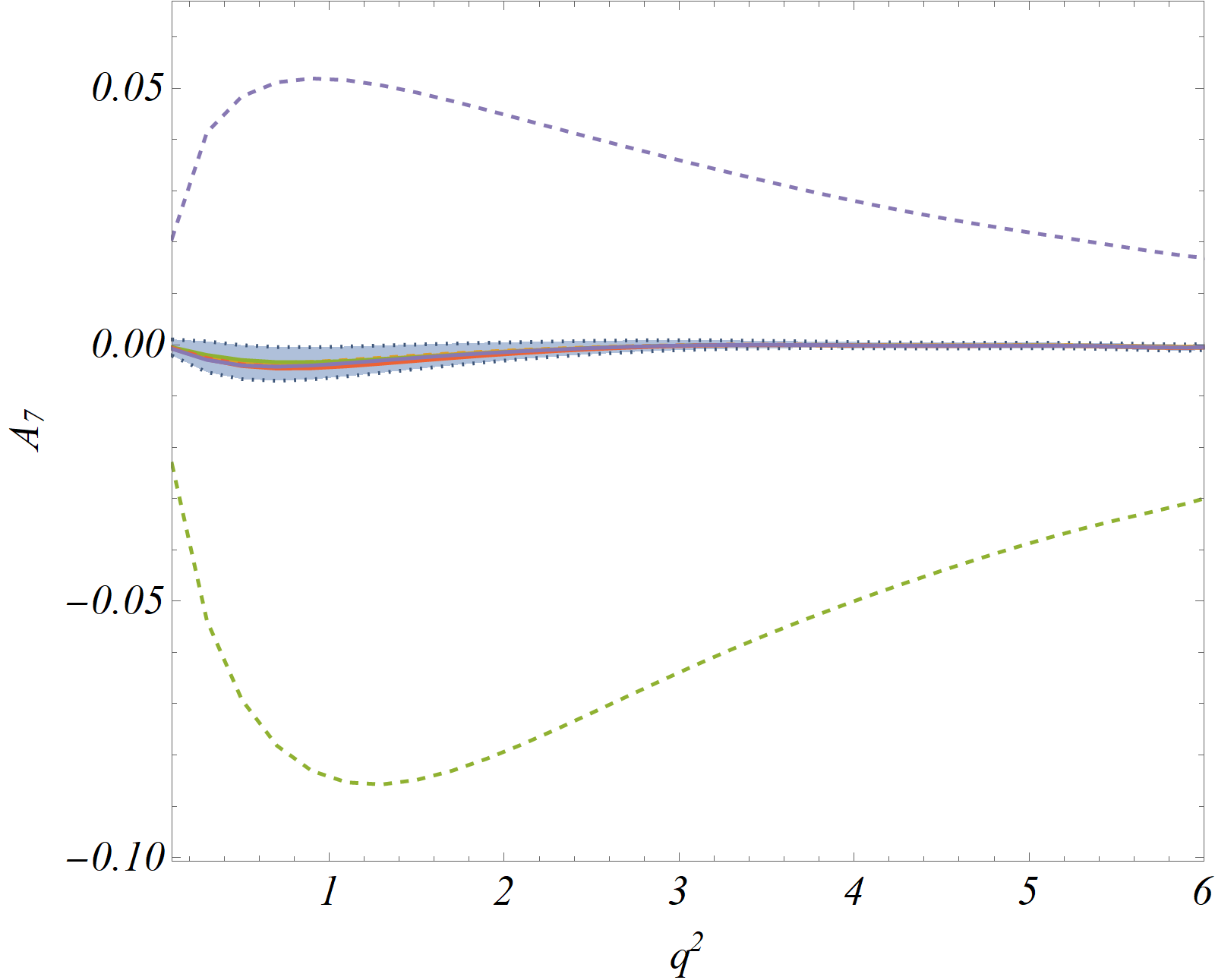}\label{fig:A7neutral}}~~~
	\subfloat[]{\includegraphics[width=0.23\textwidth]{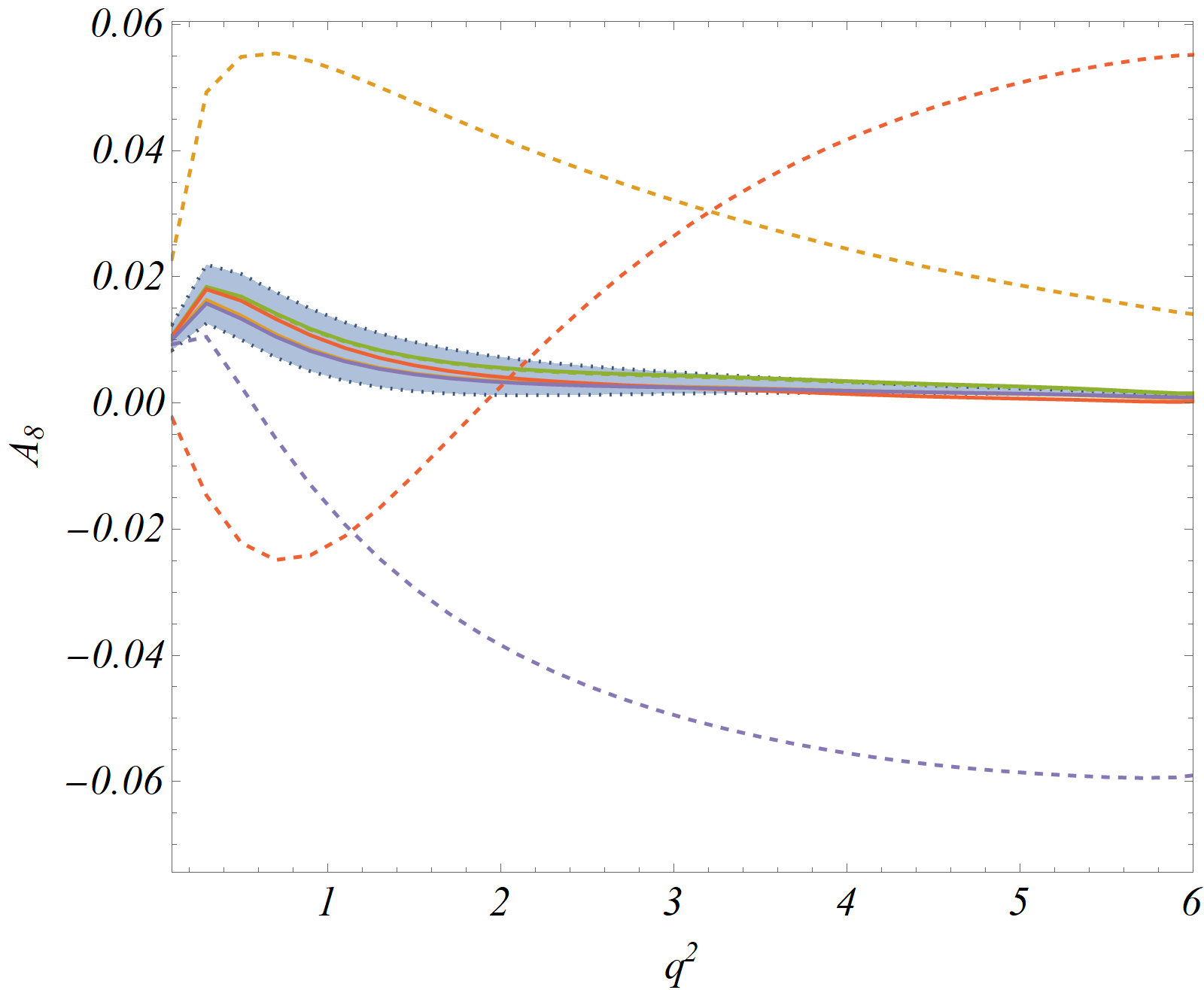}\label{fig:A8LHCb0}}~~~
	\subfloat[]{\includegraphics[width=0.23\textwidth]{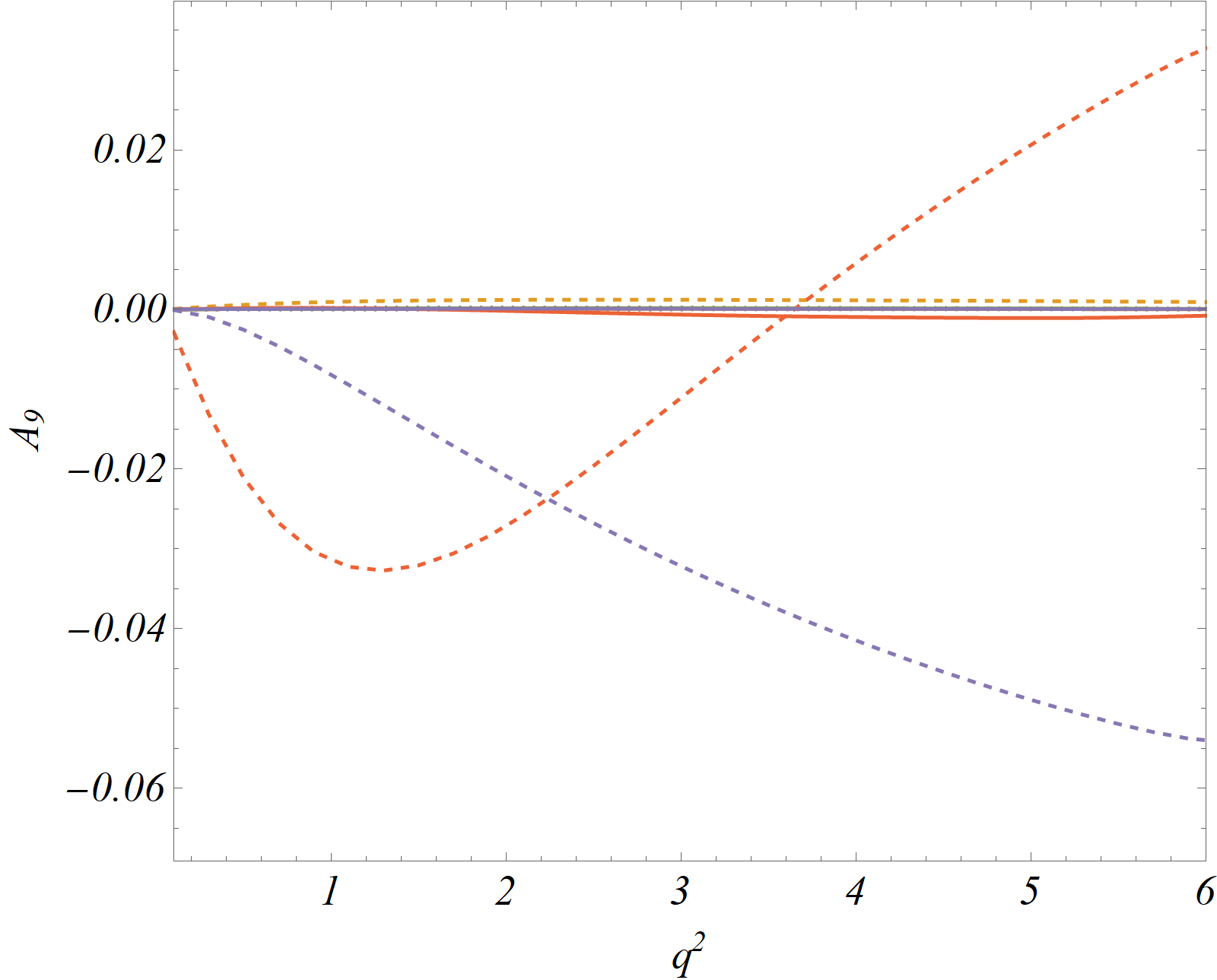}\label{fig:A9LHCb0}}\\
	\subfloat[]{\includegraphics[width=0.23\textwidth]{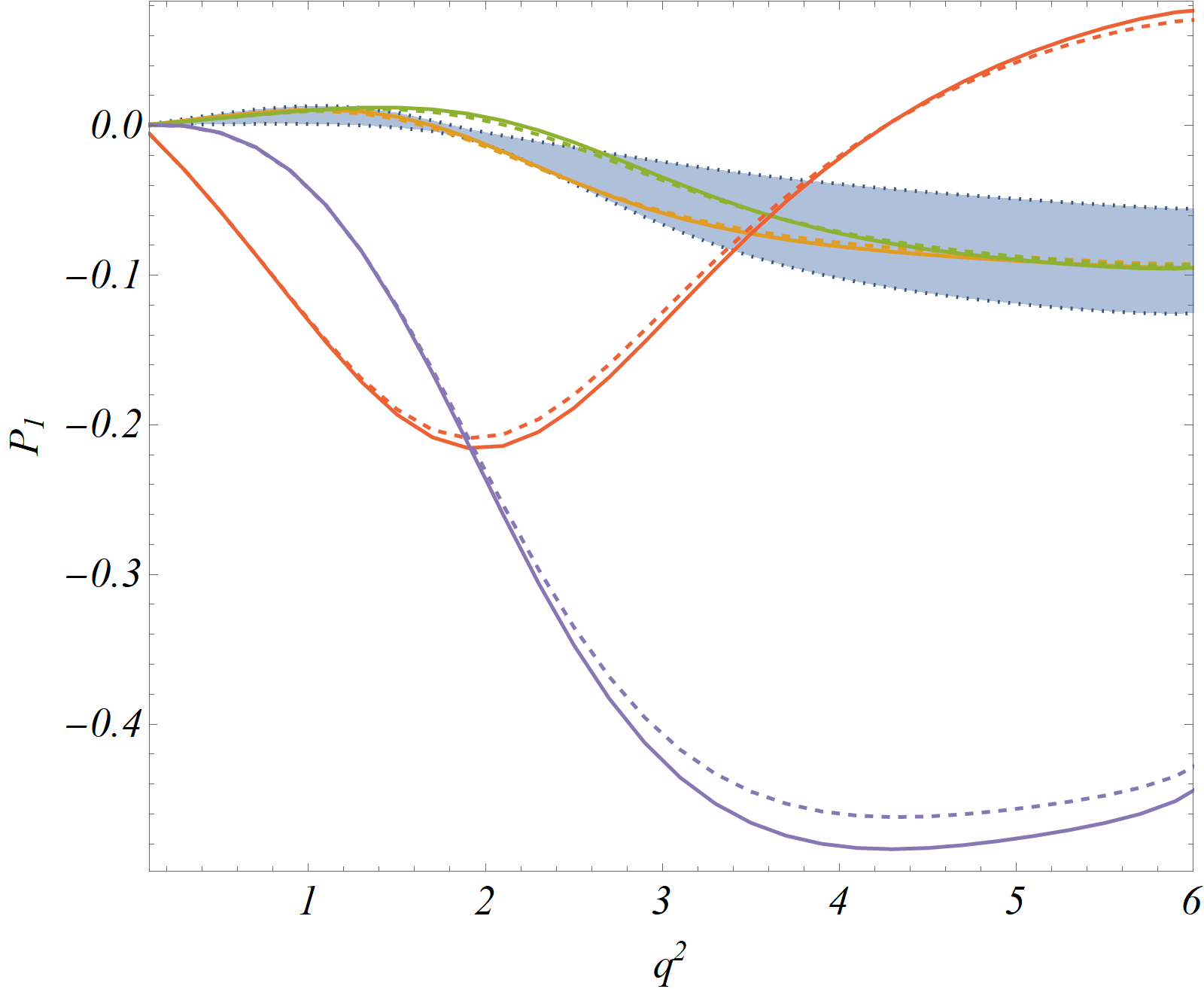}\label{fig:P1LHCb0}}~~~
	\subfloat[]{\includegraphics[width=0.23\textwidth]{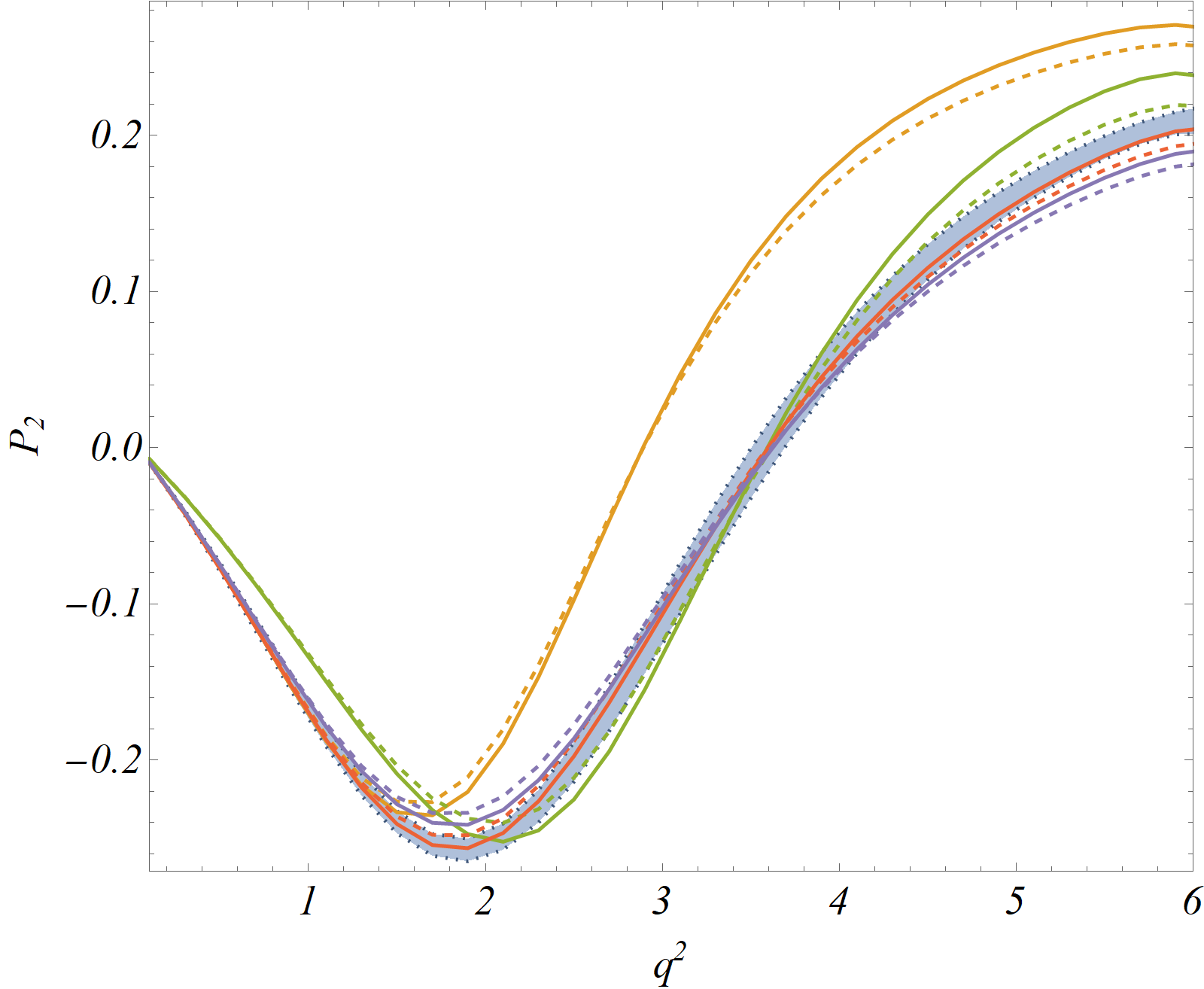}\label{fig:P2neutral}}~~~
	\subfloat[]{\includegraphics[width=0.23\textwidth]{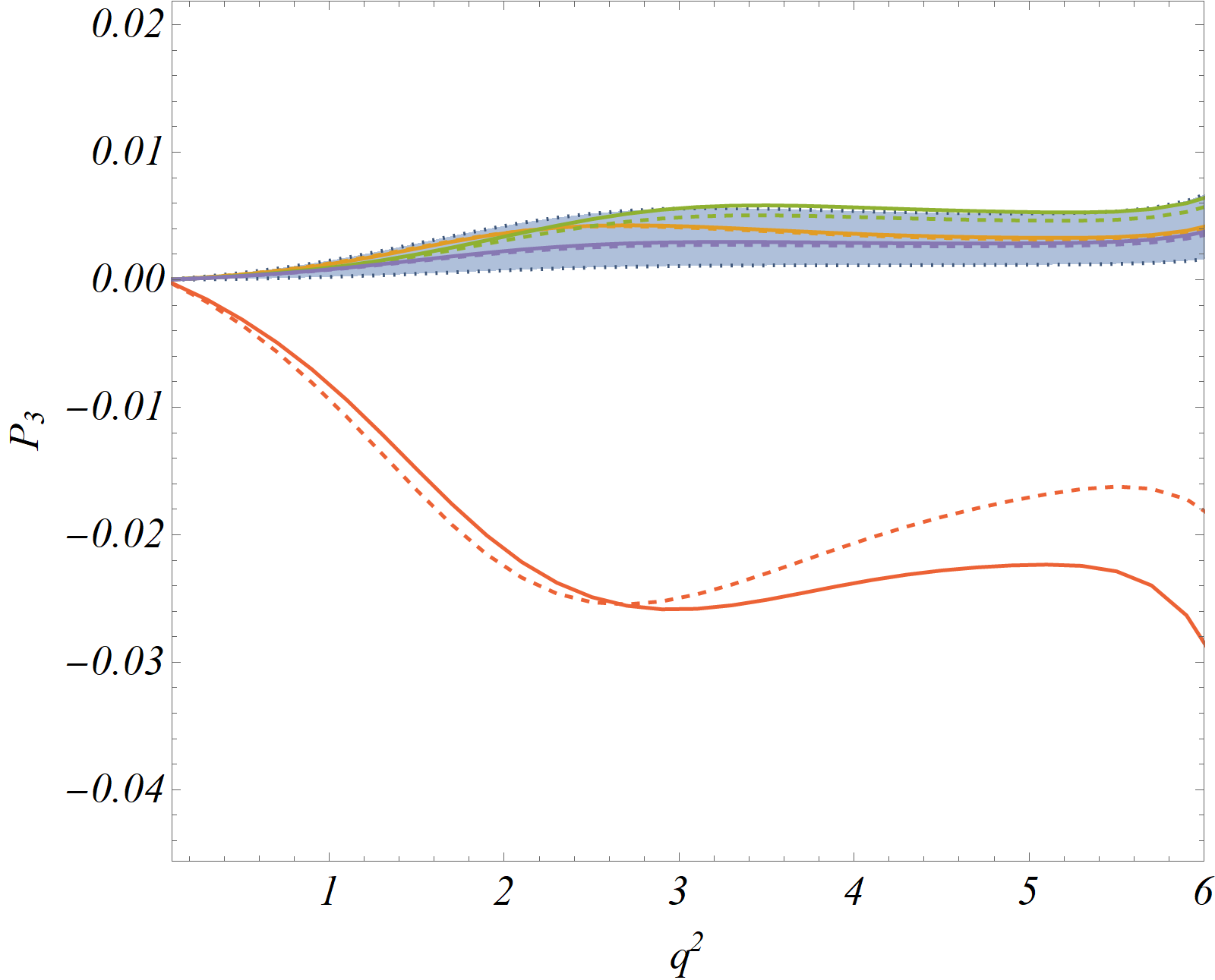}\label{fig:P3neutral}}~~~
	\subfloat[]{\includegraphics[width=0.23\textwidth]{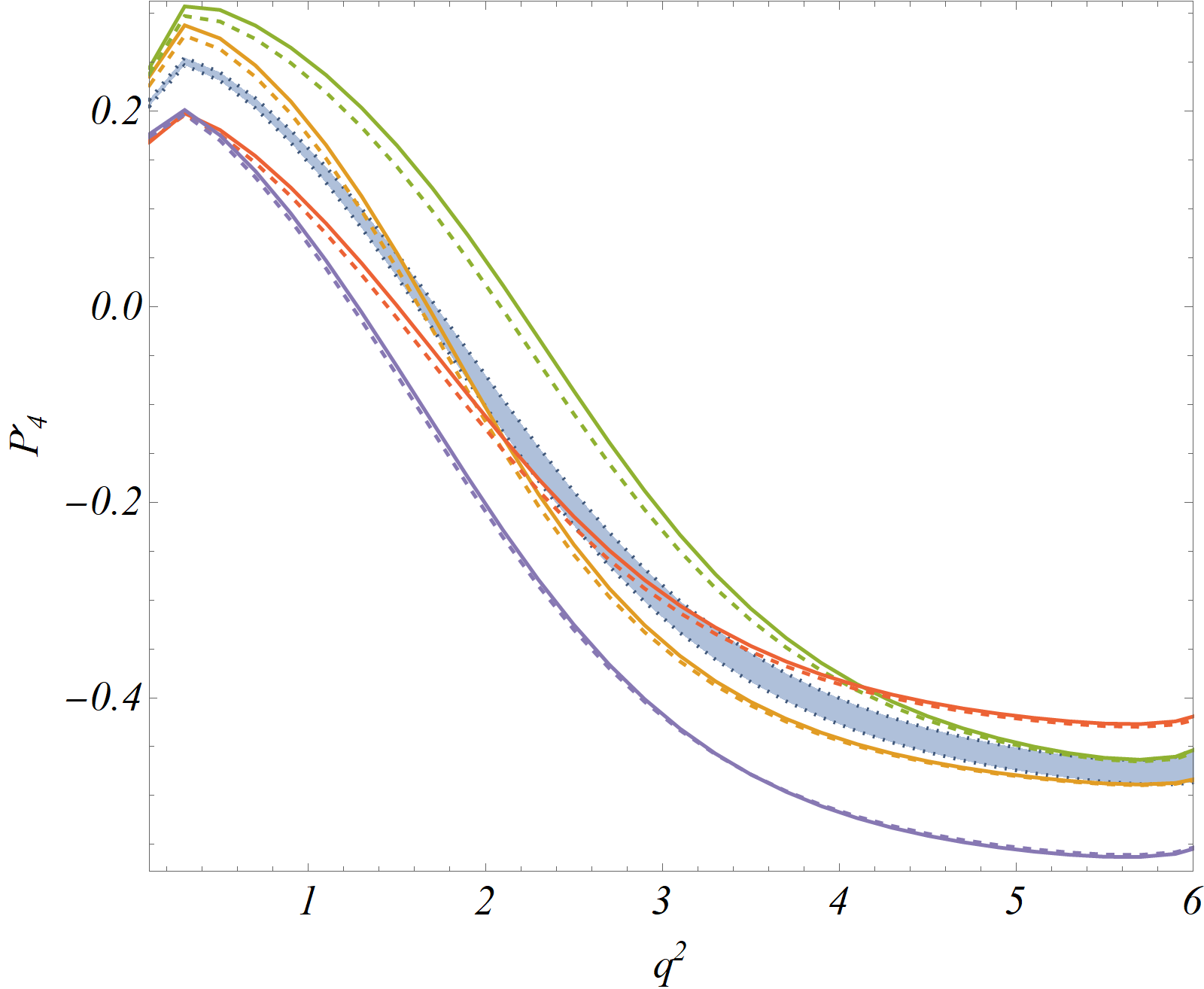}\label{fig:P4prLHCb0}}\\
	\subfloat[]{\includegraphics[width=0.23\textwidth]{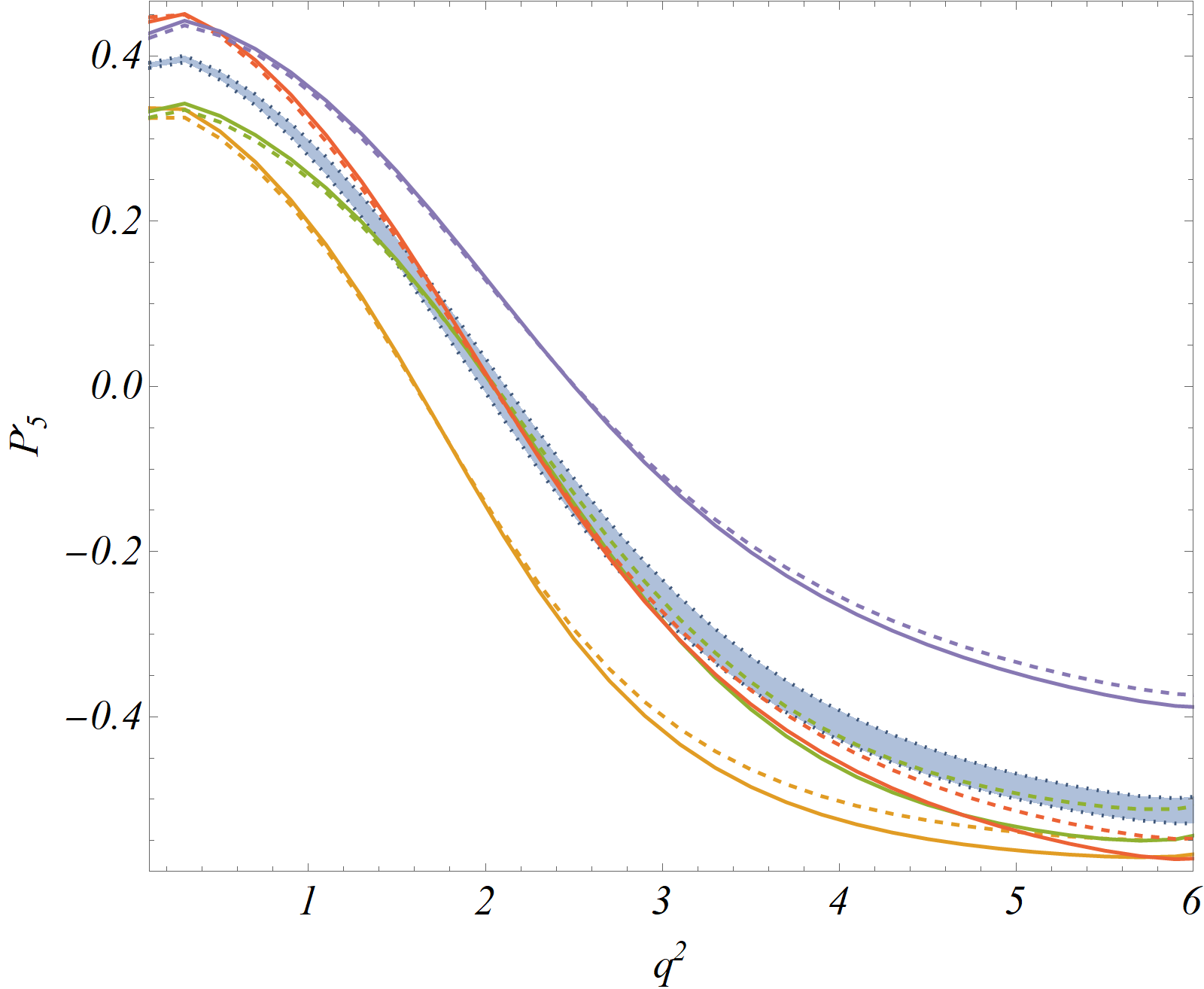}\label{fig:P5neutral}}~~~
	\subfloat[]{\includegraphics[width=0.23\textwidth]{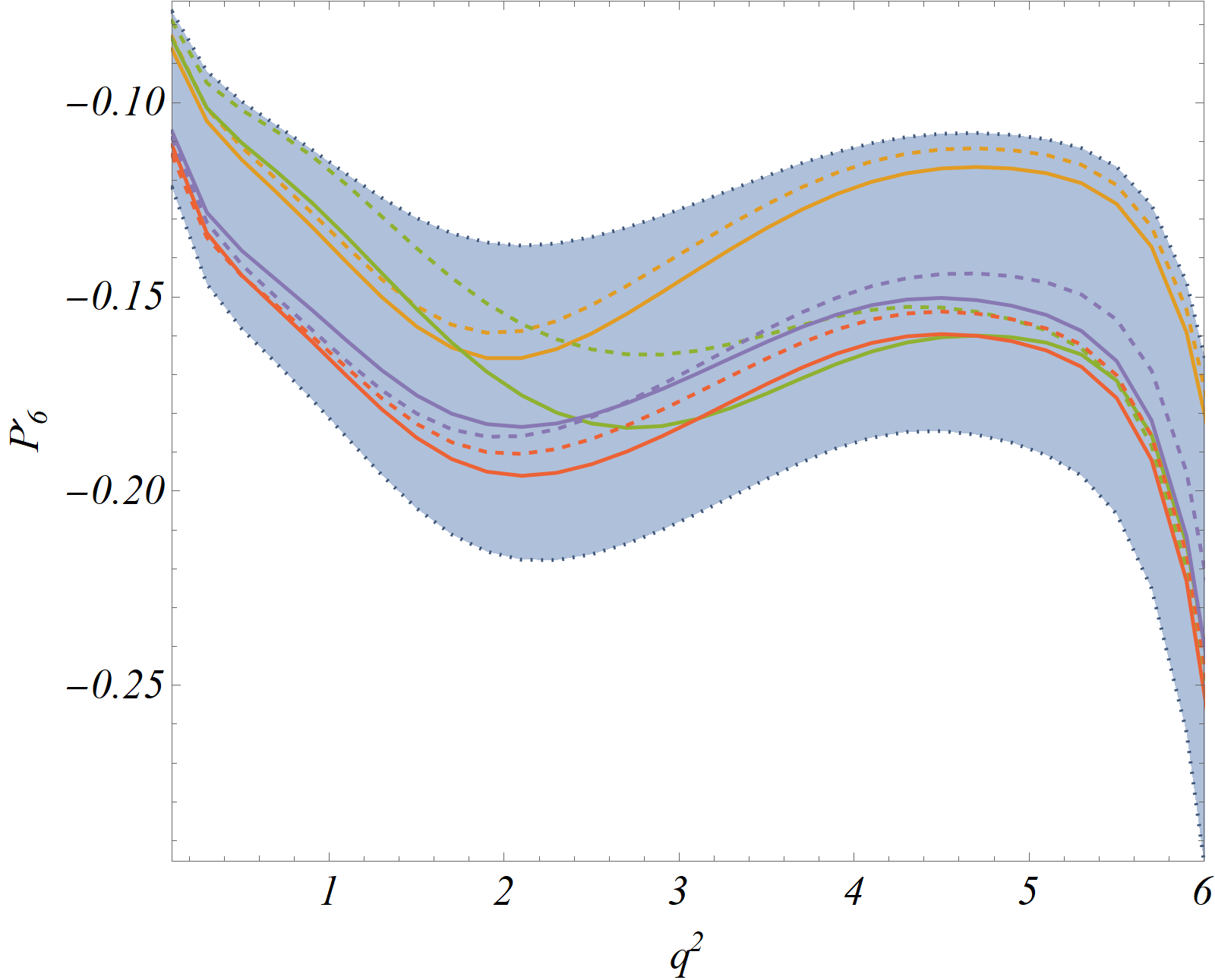}\label{fig:P6prLHCb0}}~~~
	\subfloat[]{\includegraphics[width=0.23\textwidth]{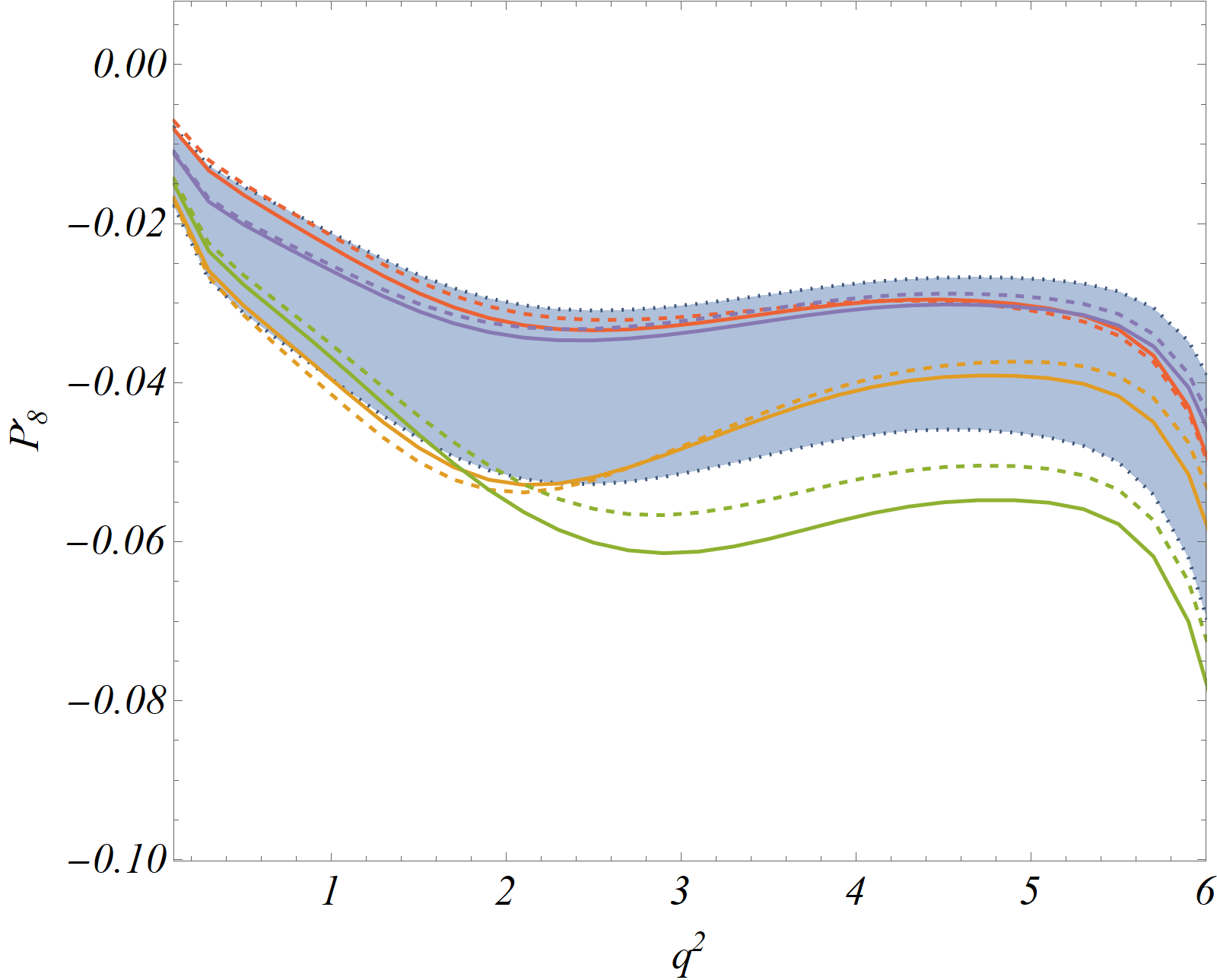}\label{fig:P8neutral}}~~~
	\subfloat[]{\includegraphics[width=0.23\textwidth]{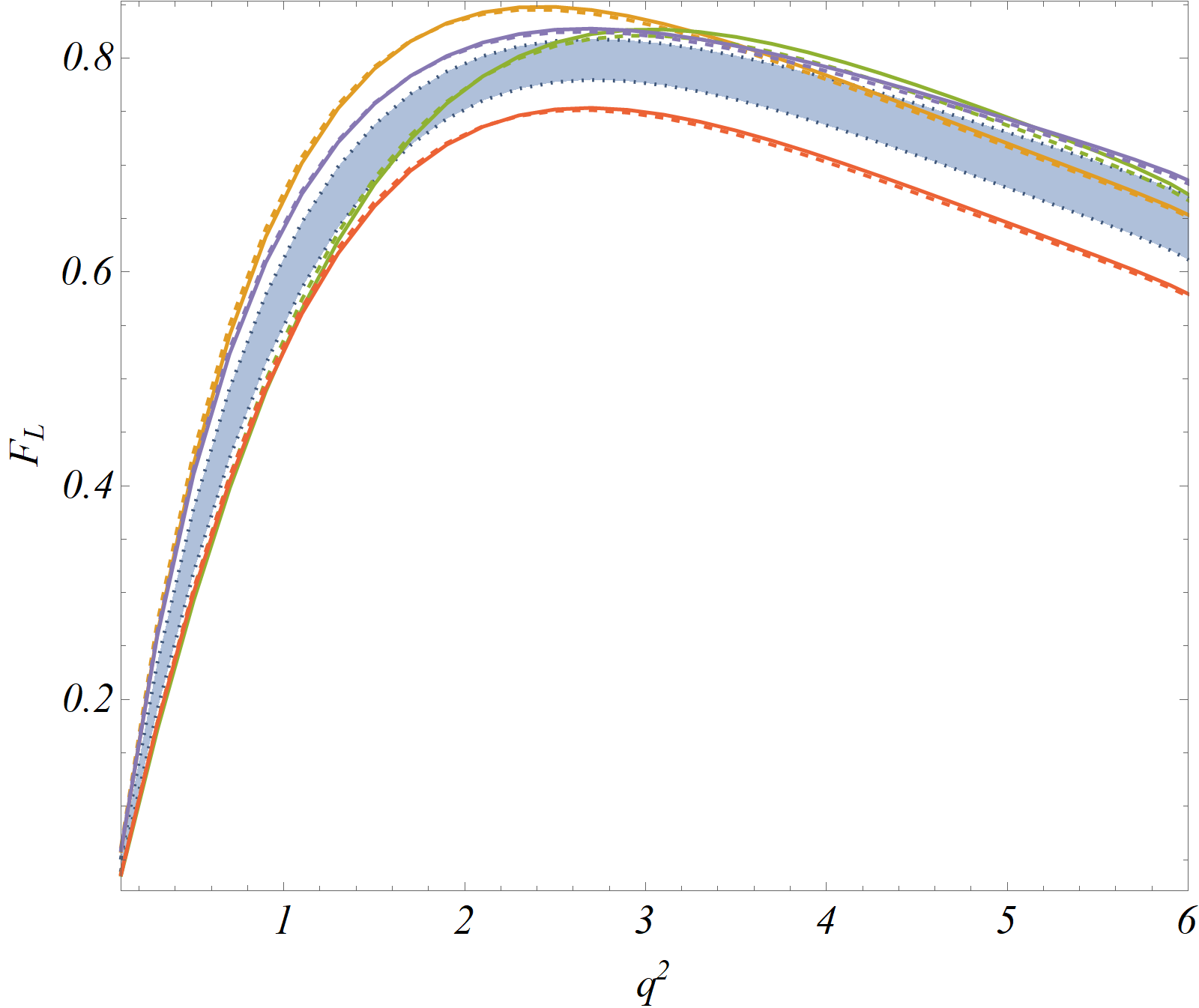}\label{fig:FLLHCb0}}\\
	\subfloat[]{\includegraphics[width=0.23\textwidth]{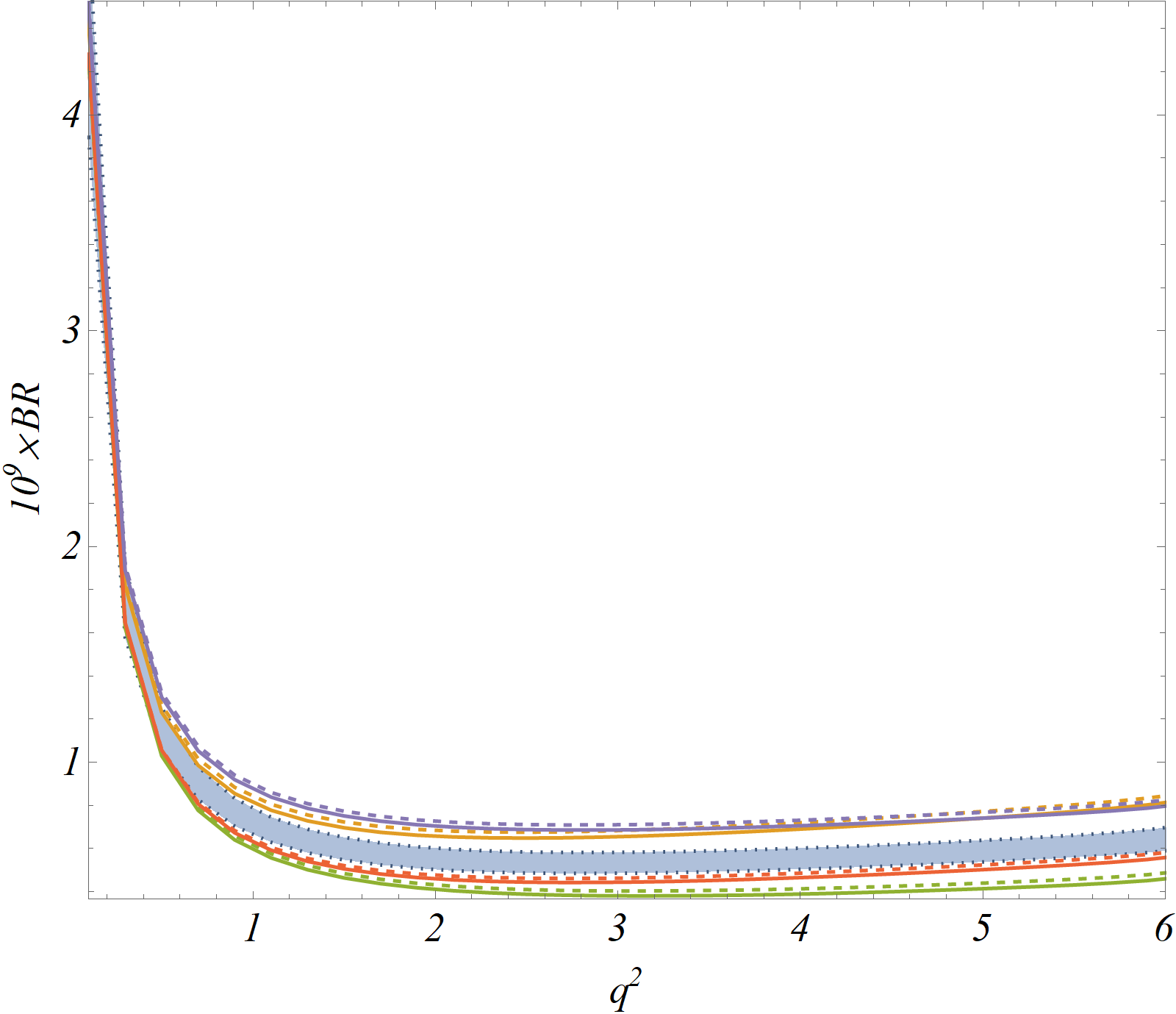}\label{fig:BRfracLHCb0}}~~~
	\subfloat[]{\includegraphics[width=0.23\textwidth]{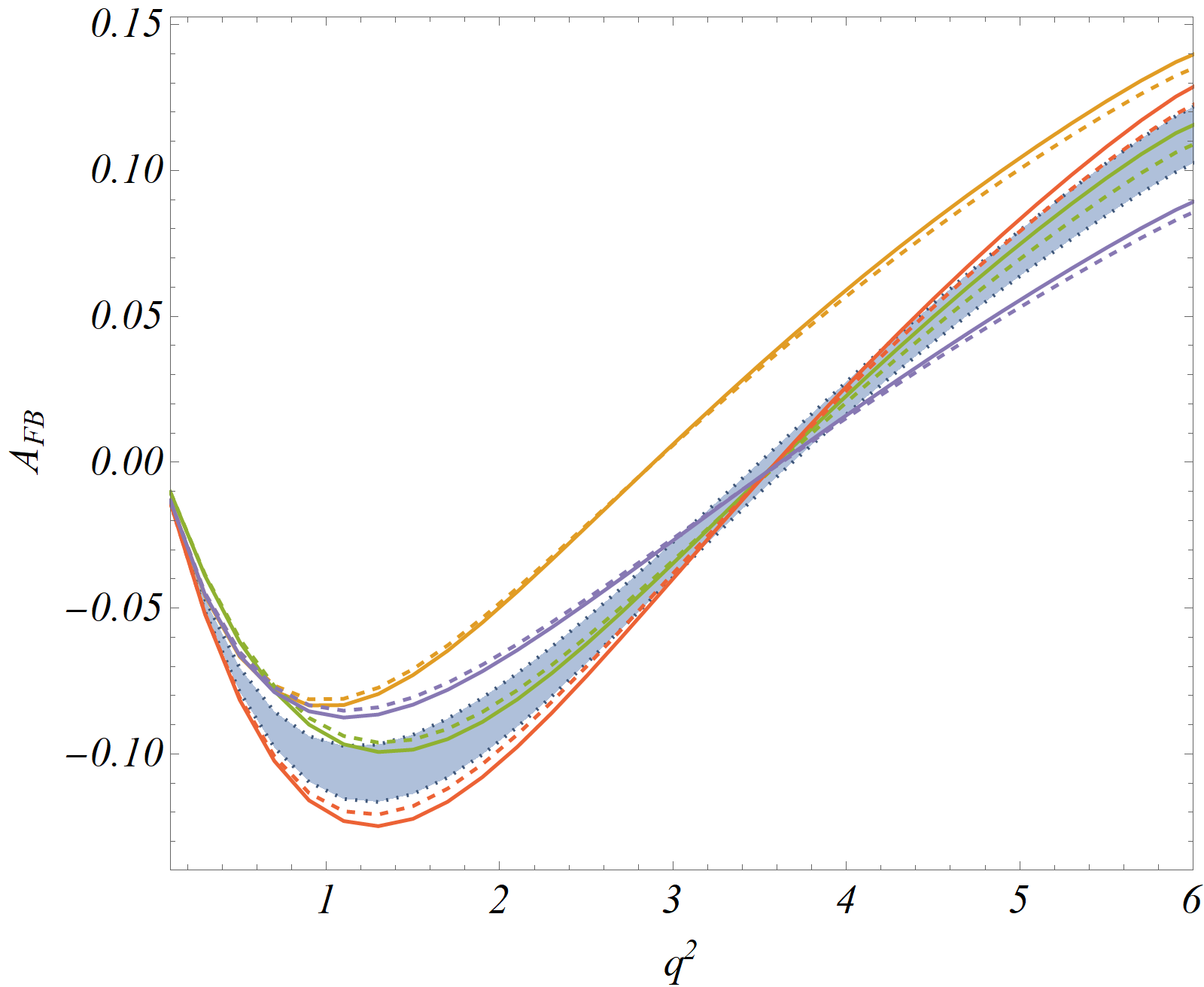}\label{fig:AFBneutral}}~~~
	\subfloat[]{\includegraphics[width=0.23\textwidth]{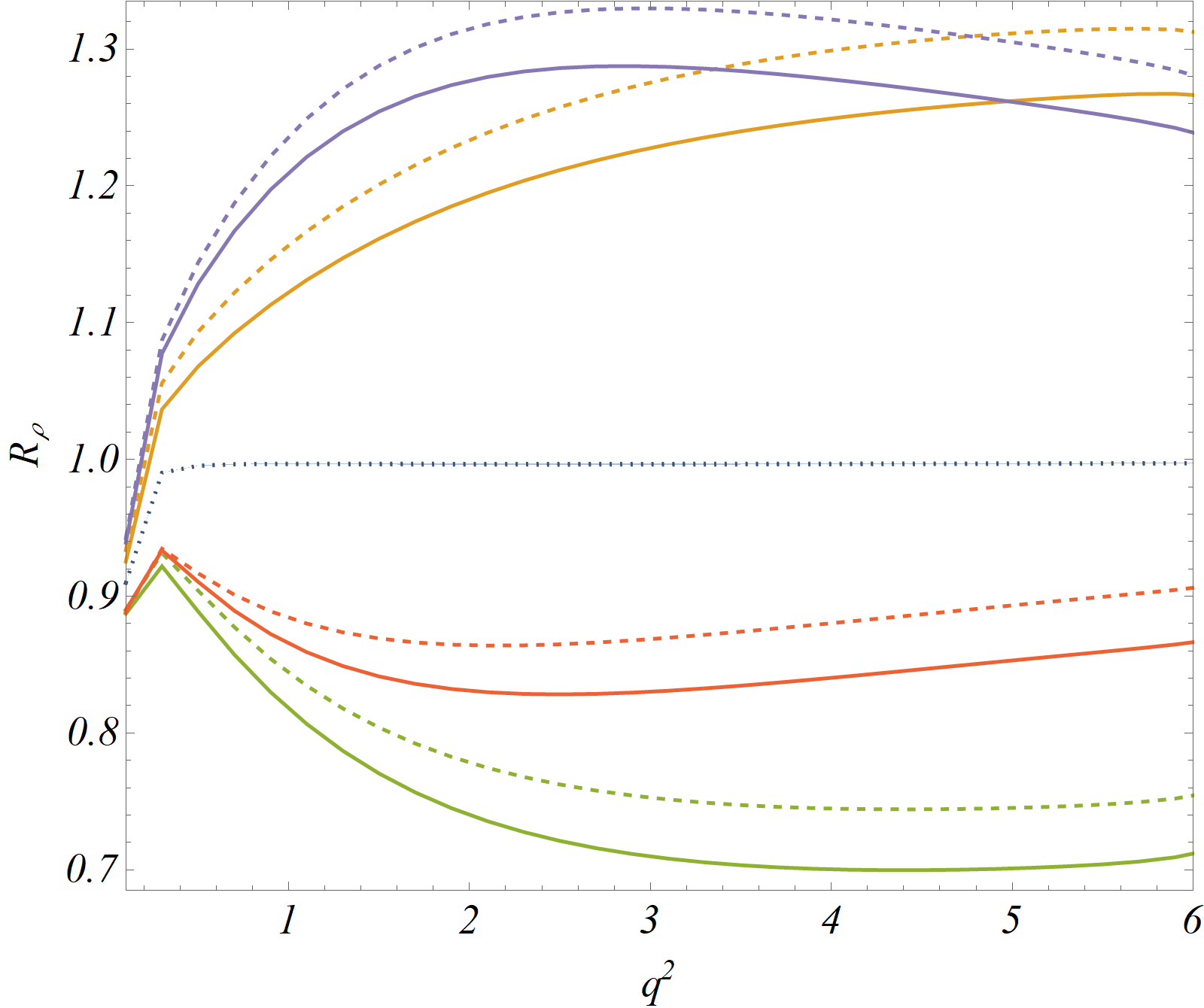}\label{fig:RhoLHCb0}}\\
	\caption{The $q^2$ dependence for the CP-averaged and CP-asymmetric observables in $B^0(\bar{B^0})\to\rho^0 ll$ decays in the SM and in the NP scenarios. Among the listed obsevables $A_{5,6s,8,9}$, $P_{1}$, $P_{4,6}^{\prime}$, $F_L$ and Branching ratio are measurable at both LHCb and Belle while the rest are measurable only at the Belle. The legends are similar to the one used in fig.~\ref{fig:BRhoobsuntagged}.}
	\label{fig:BRhoneutral}
\end{figure*}

\begin{figure*}[htbp]
	\small
	\centering
	\subfloat[]{\includegraphics[width=0.23\textwidth]{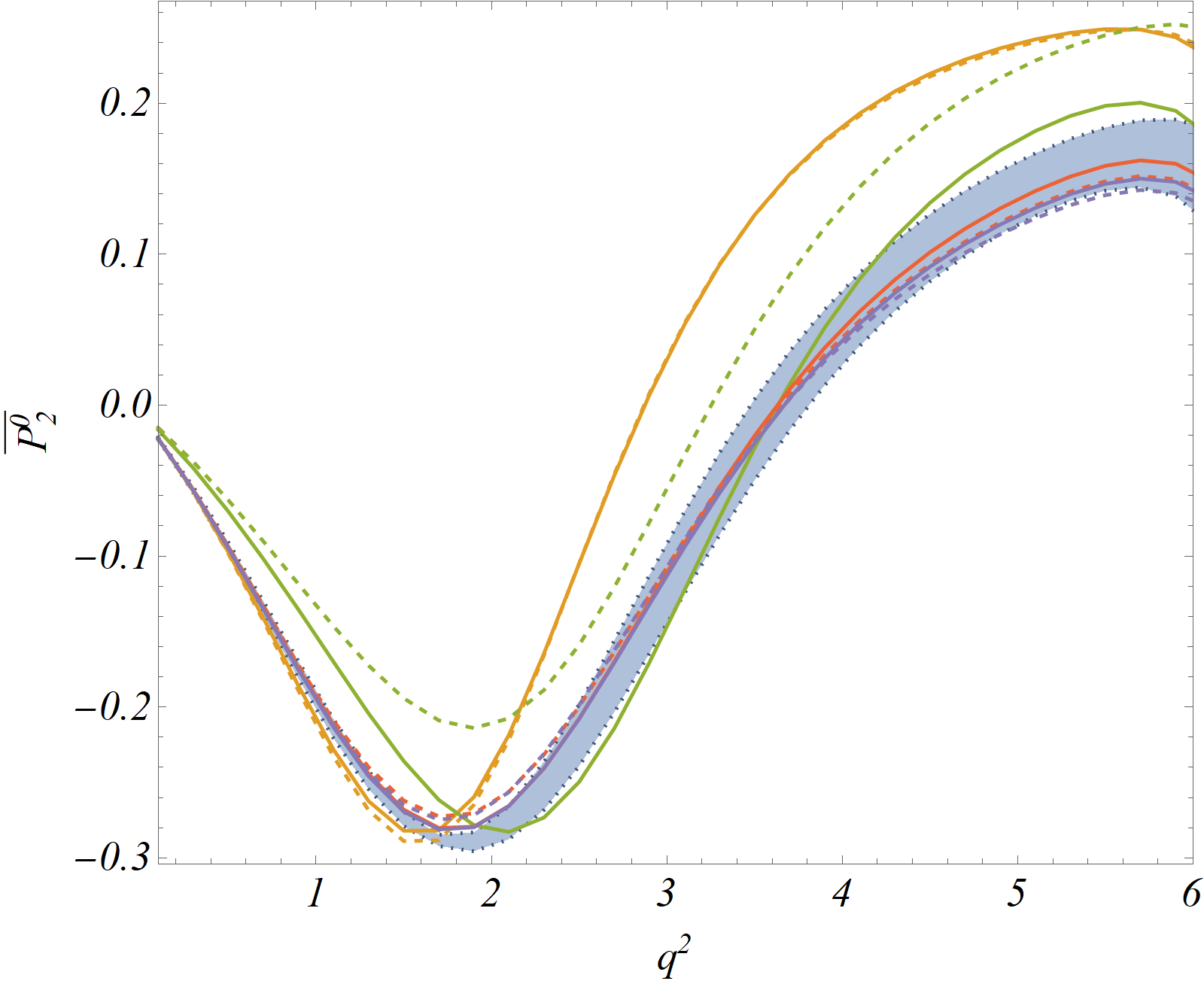}\label{fig:P2neutral0}}~~~
	\subfloat[]{\includegraphics[width=0.23\textwidth]{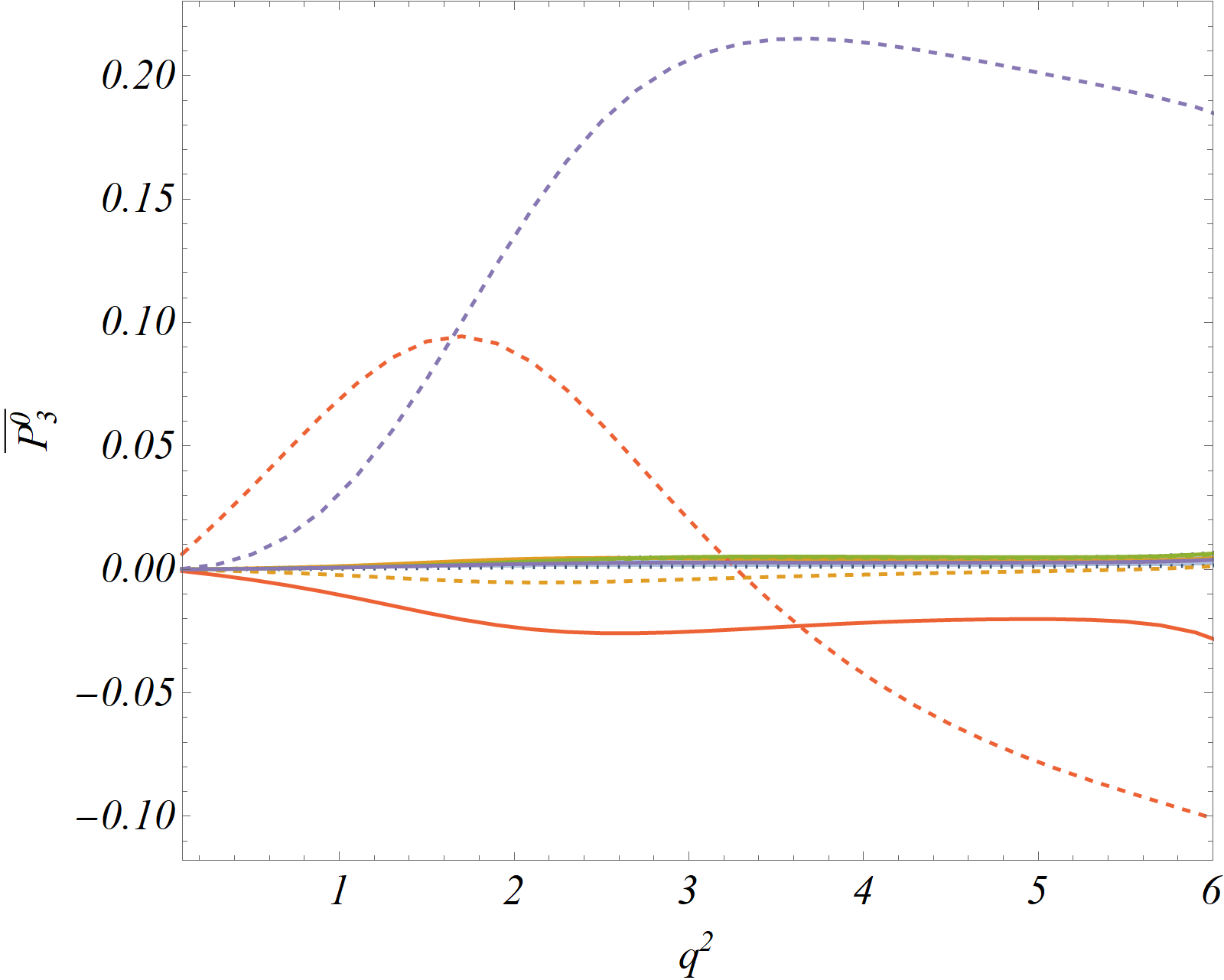}\label{fig:P30Belle}}~~~
	\subfloat[]{\includegraphics[width=0.23\textwidth]{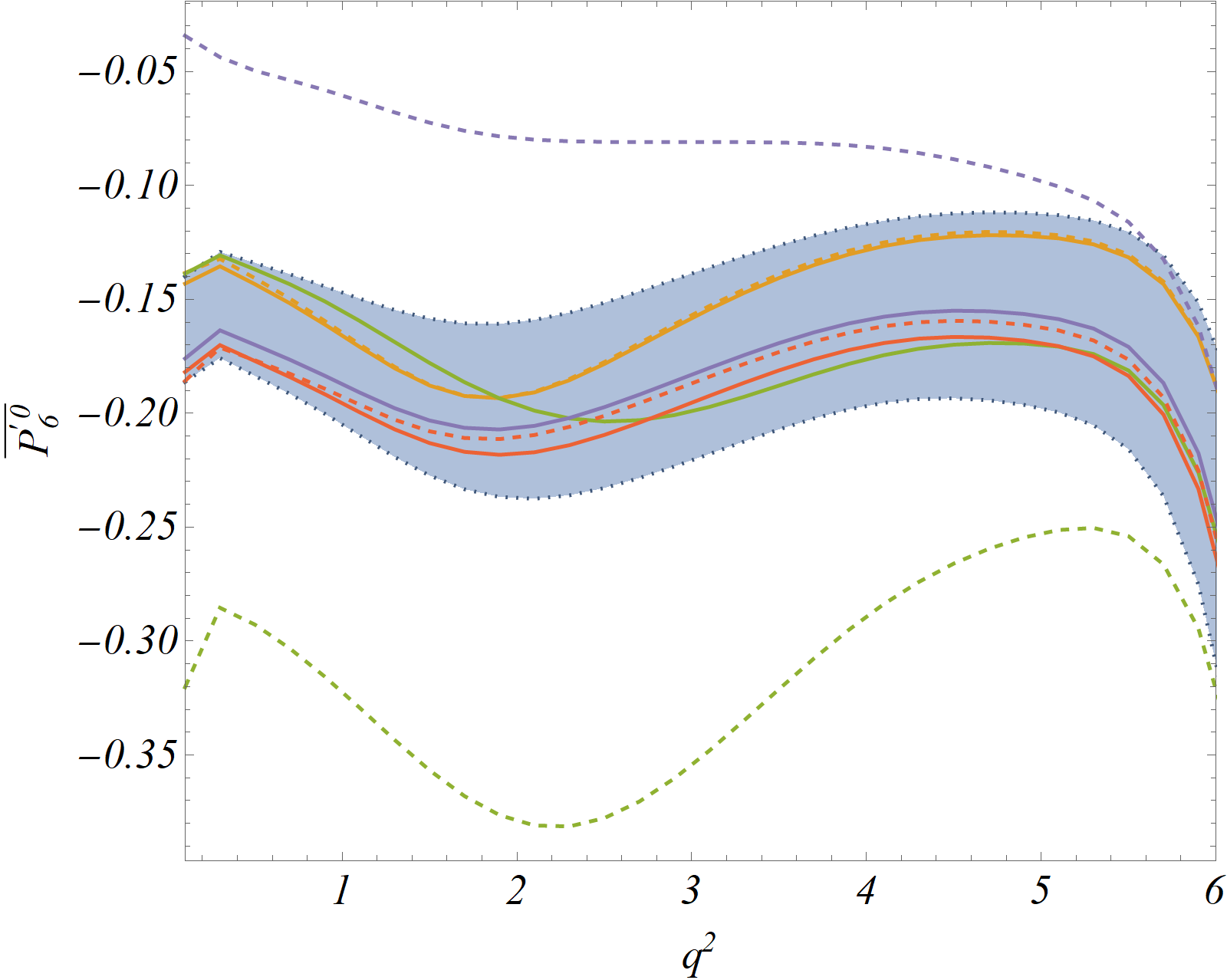}\label{fig:P6pr0Belle}}~~~
	\subfloat[]{\includegraphics[width=0.23\textwidth]{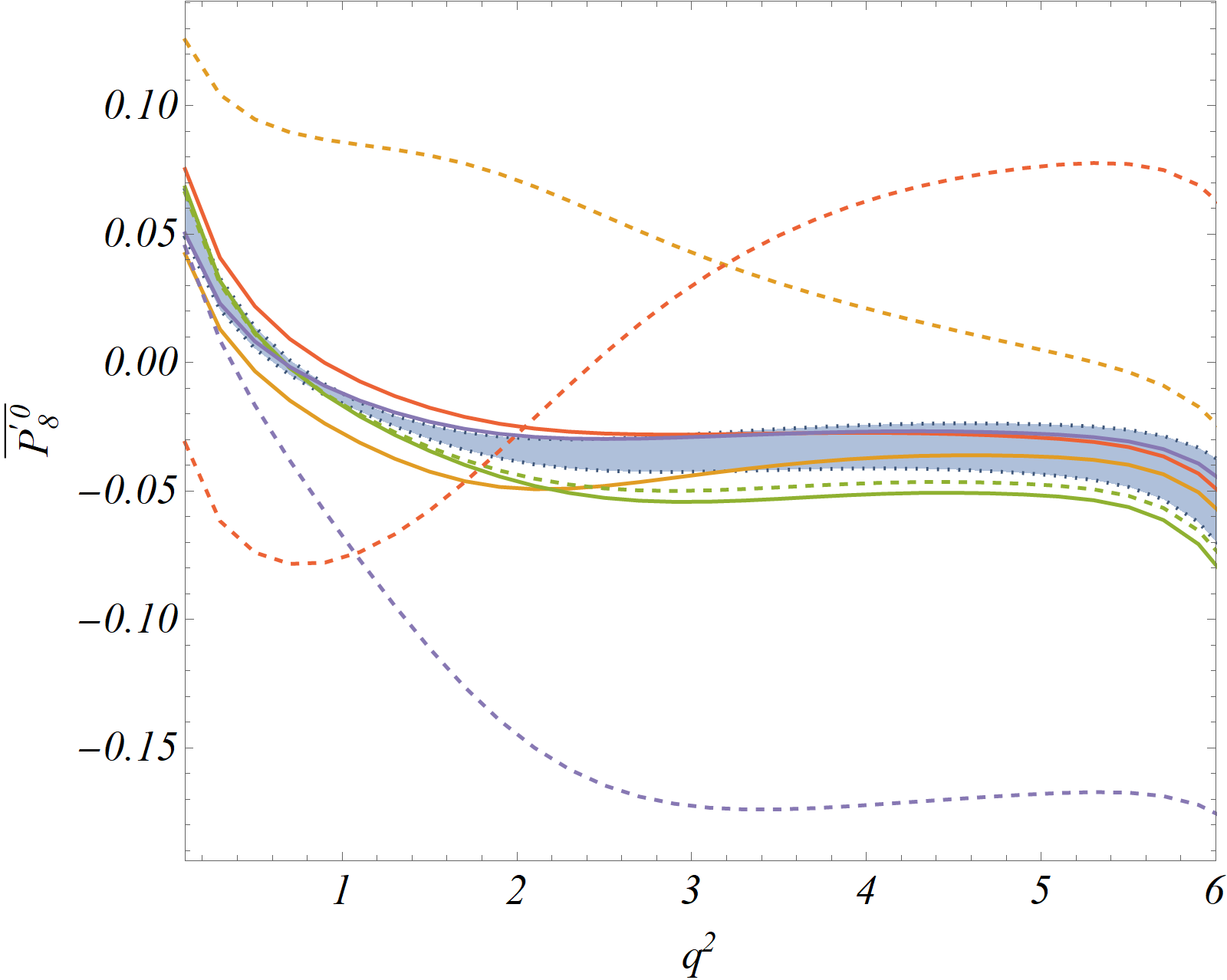}\label{fig:P8pr0Belle}}\\
	\subfloat[]{\includegraphics[width=0.23\textwidth]{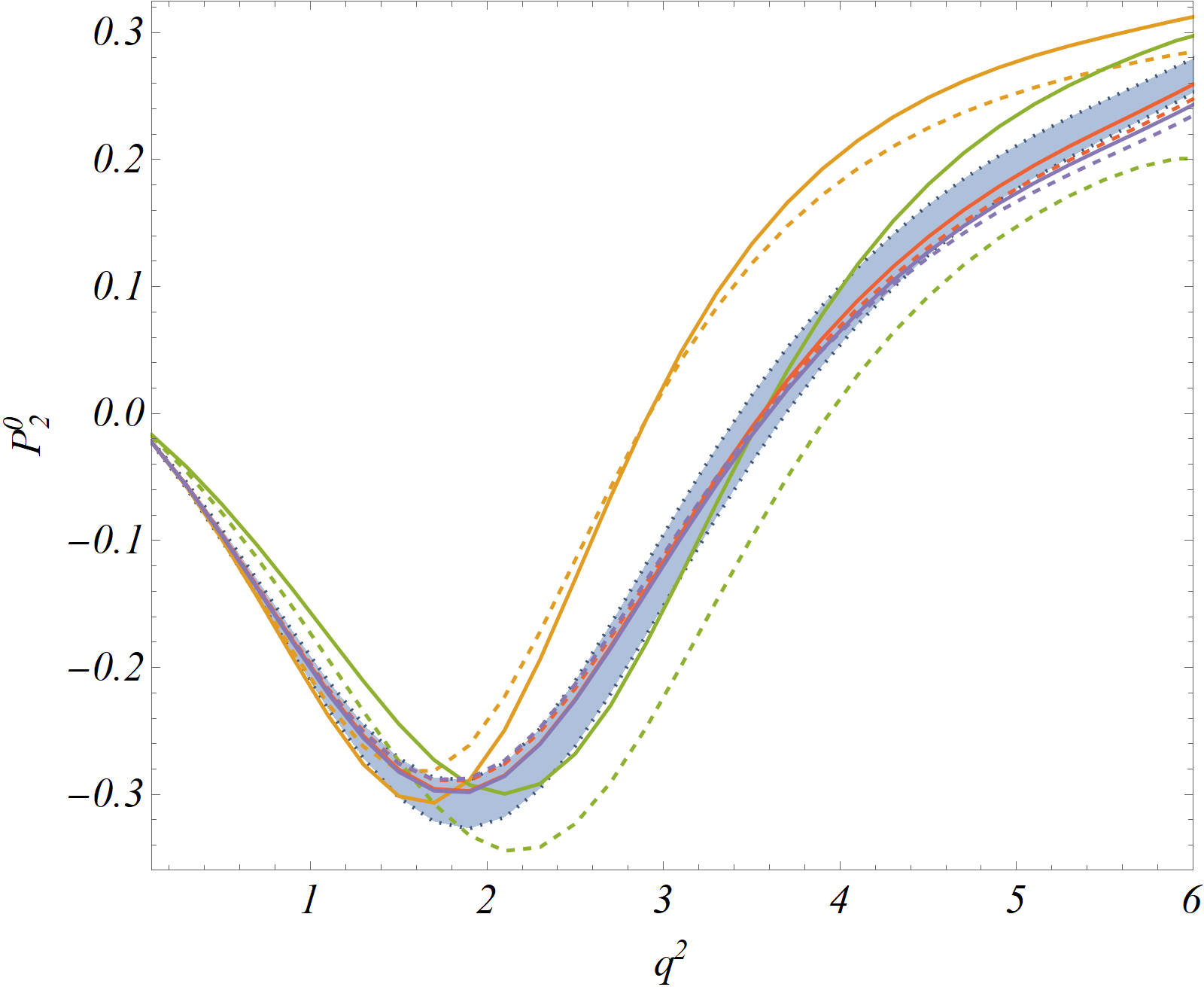}\label{fig:P2neutral_0bar}}~~~
	\subfloat[]{\includegraphics[width=0.23\textwidth]{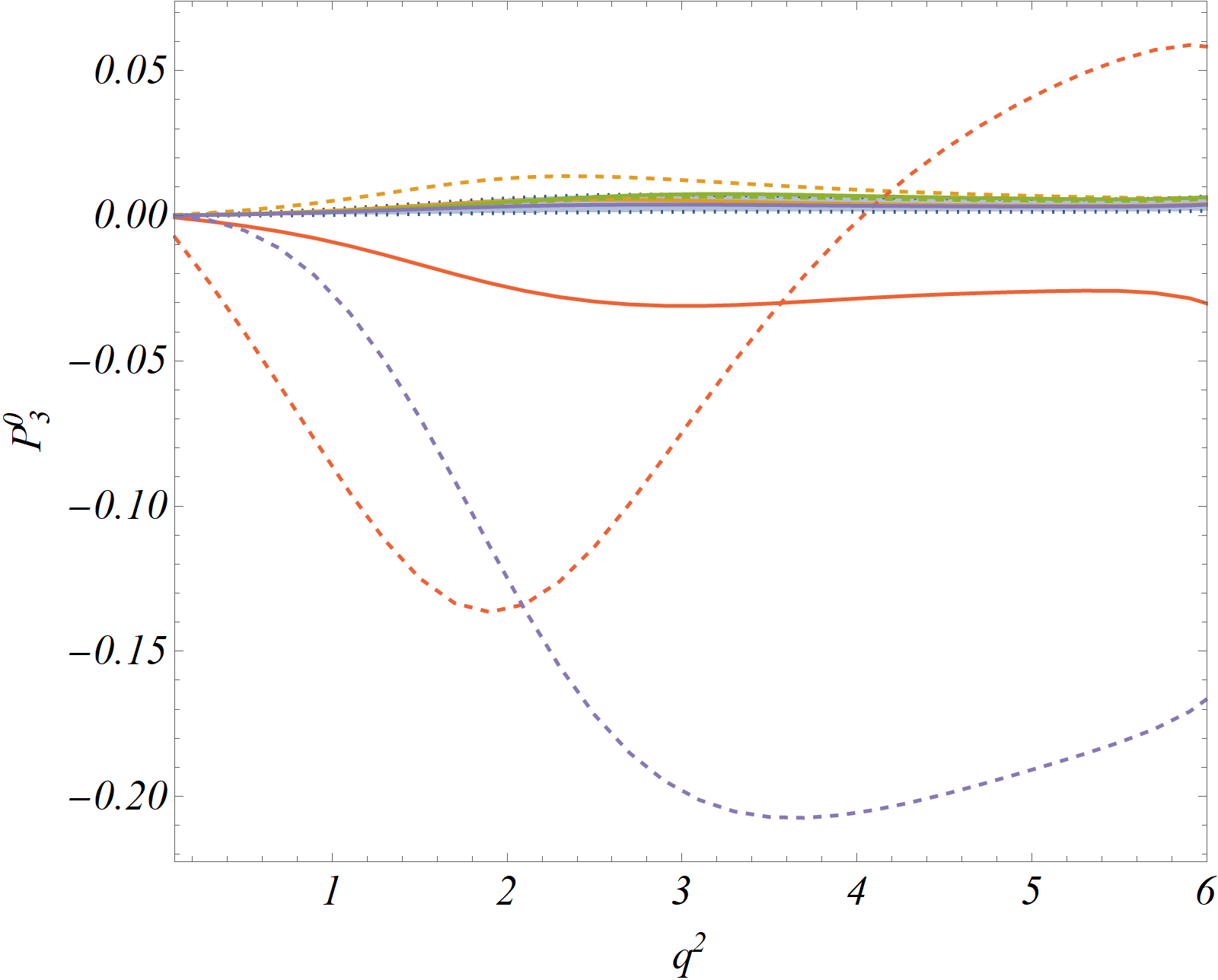}\label{fig:P3bar0Belle}}~~~
	\subfloat[]{\includegraphics[width=0.23\textwidth]{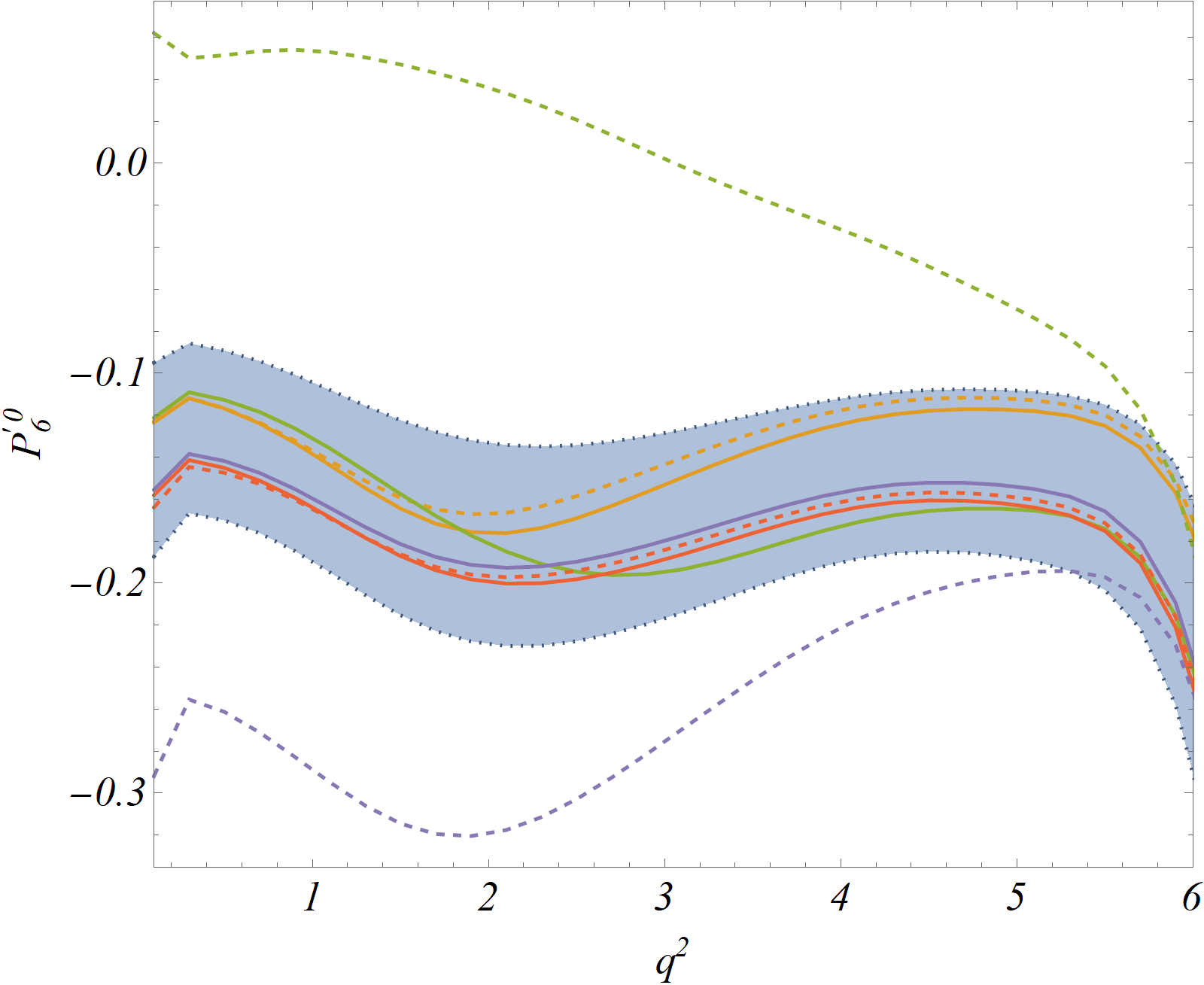}\label{fig:P6prbar0Belle}}~~~
	\subfloat[]{\includegraphics[width=0.23\textwidth]{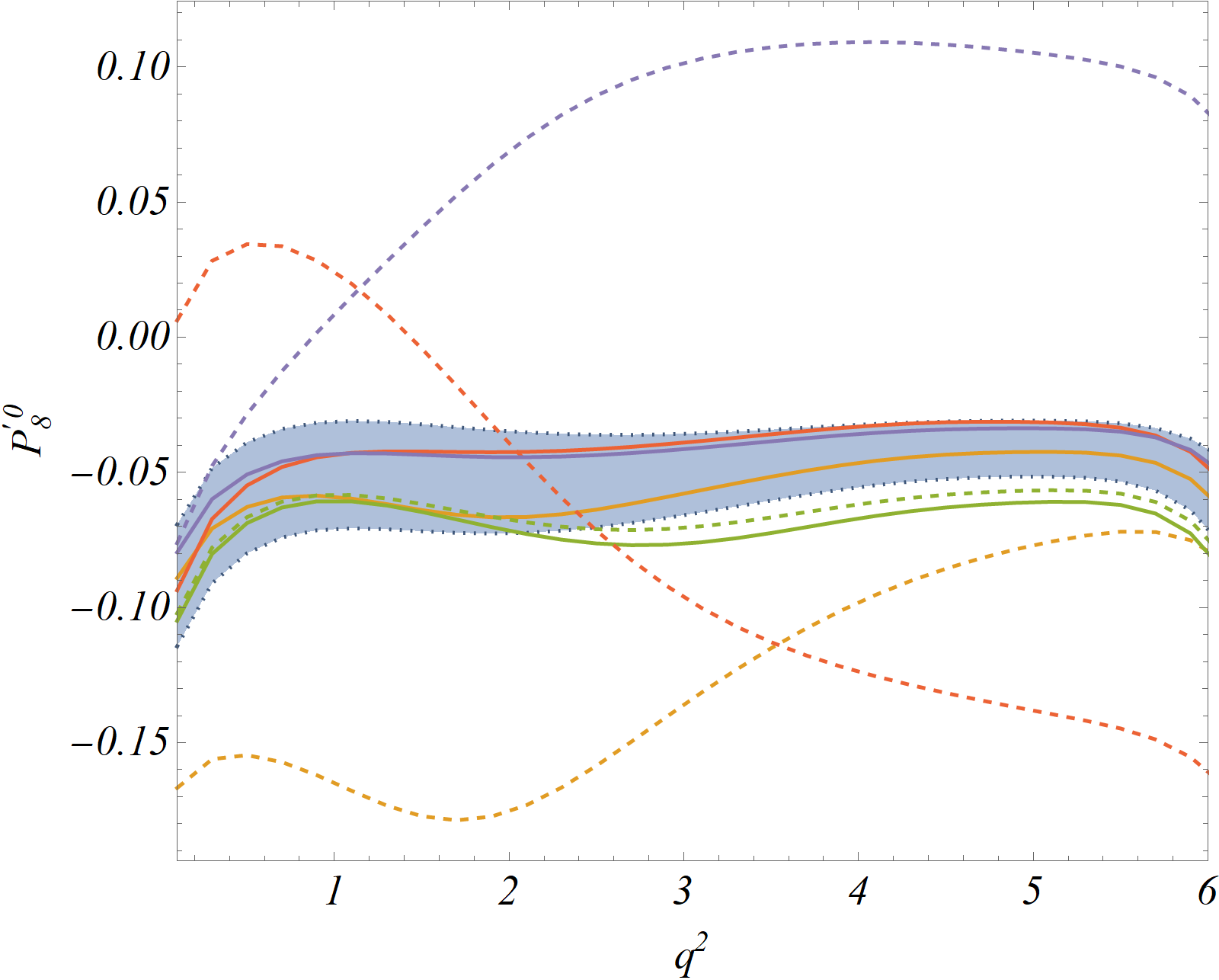}\label{fig:P8prbar0Belle}}\\
	\caption{The $q^2$ dependence for the $P_2^0(\bar{P}_2^0)$, $P_3^0(\bar{P}_3^0)$, $P_6^{\prime 0}(\bar{P}_6^{\prime 0})$, $P_8^{\prime 0}(\bar{P}_8^{\prime 0})$ associated with  $\bar{B}(B)\to\rho^0 ll$ decays measurable only at the Belle. }
	\label{fig:BRhoneutralBelle}
\end{figure*}
	
	At Belle, the flavour of the decaying $B^0$ meson can be tagged and thus, the corresponding angular coefficients $J_i$ or $\tilde{J_i}$ can be extracted. Whereas at LHCb, tagging is not possible for the decay chain as mentioned above. Hence an untagged measurement of the differential decay rate yields the CP averaged angular coefficients $(J_i + \tilde{J}_i)$ which are defined in eq.~\ref{eq:rateplus}. Therefore, from the above time averages it is only possible to estimate $\Big{<}J_i + \tilde{J_i} \Big{>}$. Hence, at the LHCb, apart from the branching fraction and LFUV observable $R_{\rho}$, the following observables can be measured: $A_{5,6s,8,9}$, $P_1$, $P^{\prime}_{4,6}$ and $F_L$. This set of observables can also be measured at Belle. Note that in the limit $y\to 0$, which is true for the $B^0$ system, the measured values will be the same in both experiments except the branching fraction, for which the value measured by Belle will be twice that of the one measured by LHCb\footnote{Note that for the time average, at the LHCb, the integration over time will be over the range $0 \le t\le \infty$ while that for Belle will be  $-\infty \le t \le \infty$.}. However, at the Belle, apart from these observables a few additional observables, resulting from the time average $\Big{<}J_i - \tilde{J_i} \Big{>}$, can be measured which are the following: $A_{CP}$, $A_{3,4,7}$, $P_{2,3}$, $P^{\prime}_{5,8}$, and $A_{FB}$. The definitions of these observables are given in eqs.~\ref{eq:gamBPav}, \ref{eq:untaggedobsACP}, and in \ref{eq:untaggedobsCPA}, respectively. The respective predictions in the SM for the observables measuarable at the LHCb and Belle are given in tables \ref{tab:B0predLHCb} and \ref{tab:B0predbelle}, respectively. We provide those predictions in four $q^2$-bins and the results are obtained using the fit results given in table \ref{tab:LCSRlatfitres}.

	As we mentioned, in the Belle experiment, tagging is possible. Hence it is possible to measure the observables defined in eqs.~\ref{eq:gamBP} and \ref{eq:taggedobsCP} for $B^0\to \rho^0\ell\ell$ and $\bar{B^0} \to \rho^0\ell\ell$ decays. We have presented the respective predictions in the SM in tables \ref{tab:B0predBelle} and \ref{tab:Bbar0predBelle}, respectively. Unlike the $B^{\pm} \to \rho^{\pm}\ell\ell$ decays, the SM predictions for the observables defined for $B^0$ and $\bar{B^0}$ decays are pretty much consistent with each other. This is due to the fact that the hard-spectator contribution from the leading order WA diagrams ($\mathcal T_{||}^{(u), WA}$) are negligible for $B^0(\bar{B^0}) \to \rho^0\ell\ell$ decays which plays a major role for the differences in the predictions for a few observables in $B^+$ and $B^-$ decays. This could be the reason for $A_4$ in neutral $B$ decays being two orders of magnitude suppressed relative to that in the $B^{\pm}$ decays.

	In the SM and in a few NP scenarios, the respective $q^2$ variations of the observables listed in tables \ref{tab:B0predLHCb} and \ref{tab:B0predbelle} are given in fig.~\ref{fig:BRhoneutral}. We note that for the neutral $B$ decays, the variations of CP-asymmetric observables with $q^2$ are less significant than the charged $B$ decays, particularly in the observables $A_{CP}$ and $A_{4,5,7,8}$. For the rest of the observables, the $q^2$ variations are almost similar to that observed for the charged $B$ decays shown in fig.~\ref{fig:BRhoobsuntagged}. Apart from $A_{CP}$ and $A_5$, the NP sensitivities for the rest of the observables in the neutral $B$ decays are almost identical to the respective observables in $B^{\pm}$ decays shown in table \ref{tab:NPsensrho}. For the neutral $B$ decays, $A_{CP}$ is sensitive to $Im(\Delta C_9)$ and $Re(\Delta C_{10})$ whereas $A_5$ is sensitive only to $Im(\Delta C_{10})$, the NP sensitivities of the rest of the observables can be seen from table \ref{tab:NPsensrho}. Like the $B^+$ and $B^-$ decays, the observables $P^0_{2,3} (\bar{P}^0_{2,3})$ and $P^{\prime 0}_{6,8} (\bar{P}_{6,8}^{\prime 0})$ are senstive to the following NP scenarios: 
	\begin{align}
	P^0_{2}, \bar{P}^0_{2} & \to Re(\Delta C_9), Im(\Delta C_{10}), \ \ \ P^0_{3}, \bar{P}^0_{3} \to Re(C^{\prime}_9), Im(C^{\prime}_9), Im(C^{\prime}_{10}), \nn \\
	P^{\prime 0}_{6}, \bar{P}^{\prime 0}_{6} & \to Im(\Delta C_{10}), Im(C^{\prime}_{10}) \ \ \ P^{\prime 0}_{8}, \bar{P}^{\prime 0}_{8} \to Im(\Delta C_9), Im(C^{\prime}_9), Im(C^{\prime}_{10}). 
	\end{align}   
	The CP optimized observables, $P_2$ and $P_3$, obtained from the CP-averaged distribution are sensitive only to $Re(\Delta C_9)$ and $Re(C^{\prime}_9)$, respectively. 
	We can see from fig.~\ref{fig:BRhoneutralBelle} that in the NP scenarios mentioned above, significant deviations in the predictions of $P^{\prime 0}_{6,8}$ and $\bar{P}_{6,8}^{\prime 0}$ are possible, which are lost due to relative cancellations in their CP-averaged observables $P^{\prime}_{6,8}$. The respective plots are shown in figs.~\ref{fig:P6prLHCb0} and \ref{fig:P8neutral}, respectively. Therefore, along with the observables obtained from the CP-averaged distribution, if tagging is possible, the measurement of the tagged observables is equally important in the context of NP searches. 

	Like the charged $B$ decays, in tables~\ref{tab:appBtorhozP1P4P6notag1},~\ref{tab:appBtorhozP1P4P6notag2}, \ref{tab:appBtorhozassymnotag1},~\ref{tab:appBtorhozassymnotag2},~\ref{tab:appBtorhozbrnotag1},~\ref{tab:appBtorhozbrnotag2},~\ref{tab:appBtorhozACPnotag1},~\ref{tab:appBtorhozACPnotag2},~\ref{tab:appBtorhozoptotag1}\&~\ref{tab:appBtorhozoptotag2} we provide the predictions in the NP scenarios for the observables associated with the neutral modes. Also, due to the same reason as discussed in the case of the charged $B$ decays, for the $B^0$ and $\bar{B^0}$ decays we have given the predictions in the NP scenarios for $P_{2,3}^0 (\bar{P}_{2,3}^0)$, $P_{6,8}^{\prime 0} (\bar{P}_{6,8}^{\prime 0})$ in  tables~\ref{tab:appBtorhozP2P3tag1},~\ref{tab:appBtorhozP2P3tag2},~\ref{tab:appBtorhozP6P8tag1} \&~\ref{tab:appBtorhozP6P8tag2}. All these predictions can be tested in future measurements.

\begin{figure*}[t]
	\small
	\centering
	\subfloat[]{\includegraphics[width=0.23\textwidth]{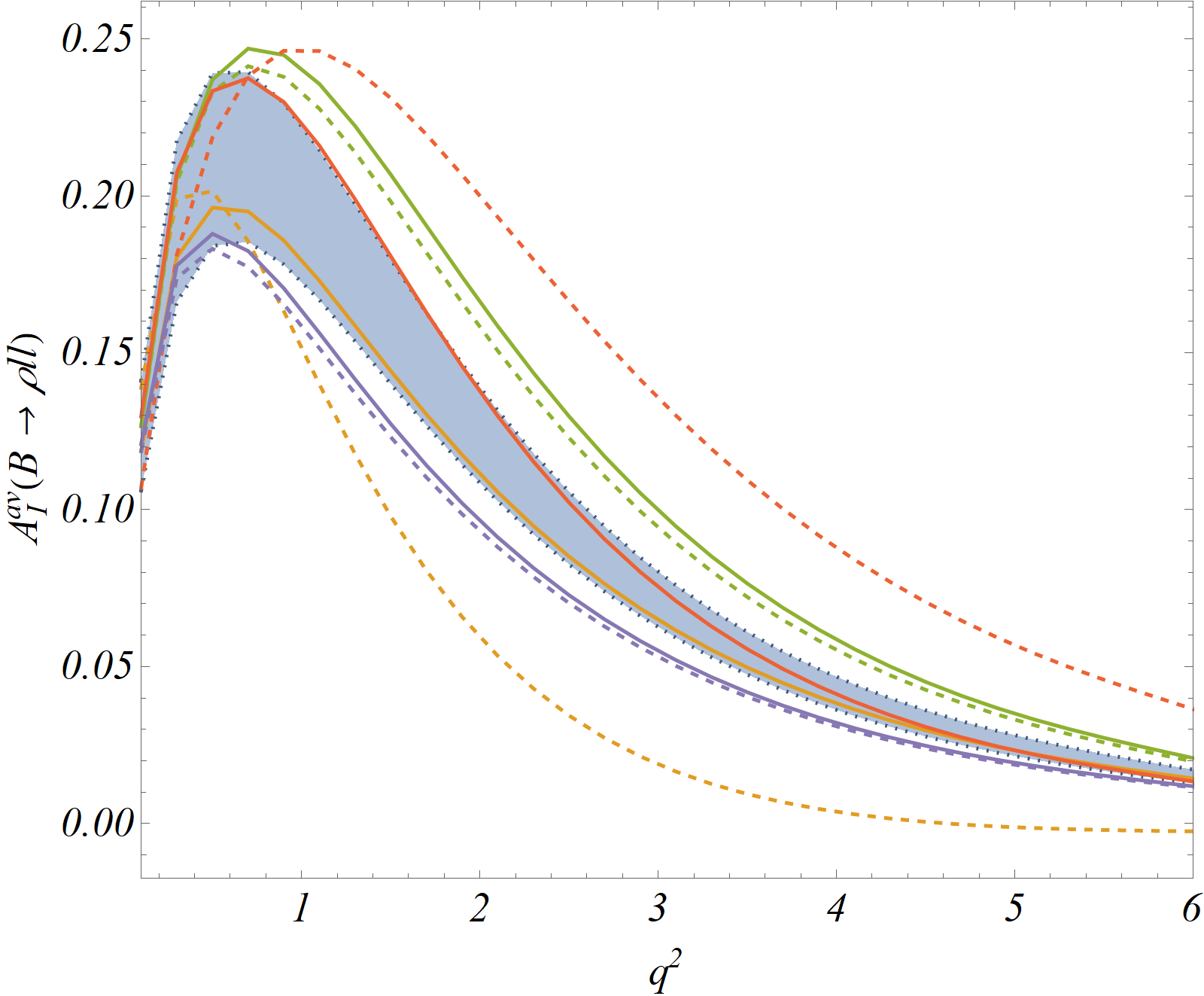}\label{fig:AIBrhoBeneke}}~~~
	\subfloat[]{\includegraphics[width=0.23\textwidth]{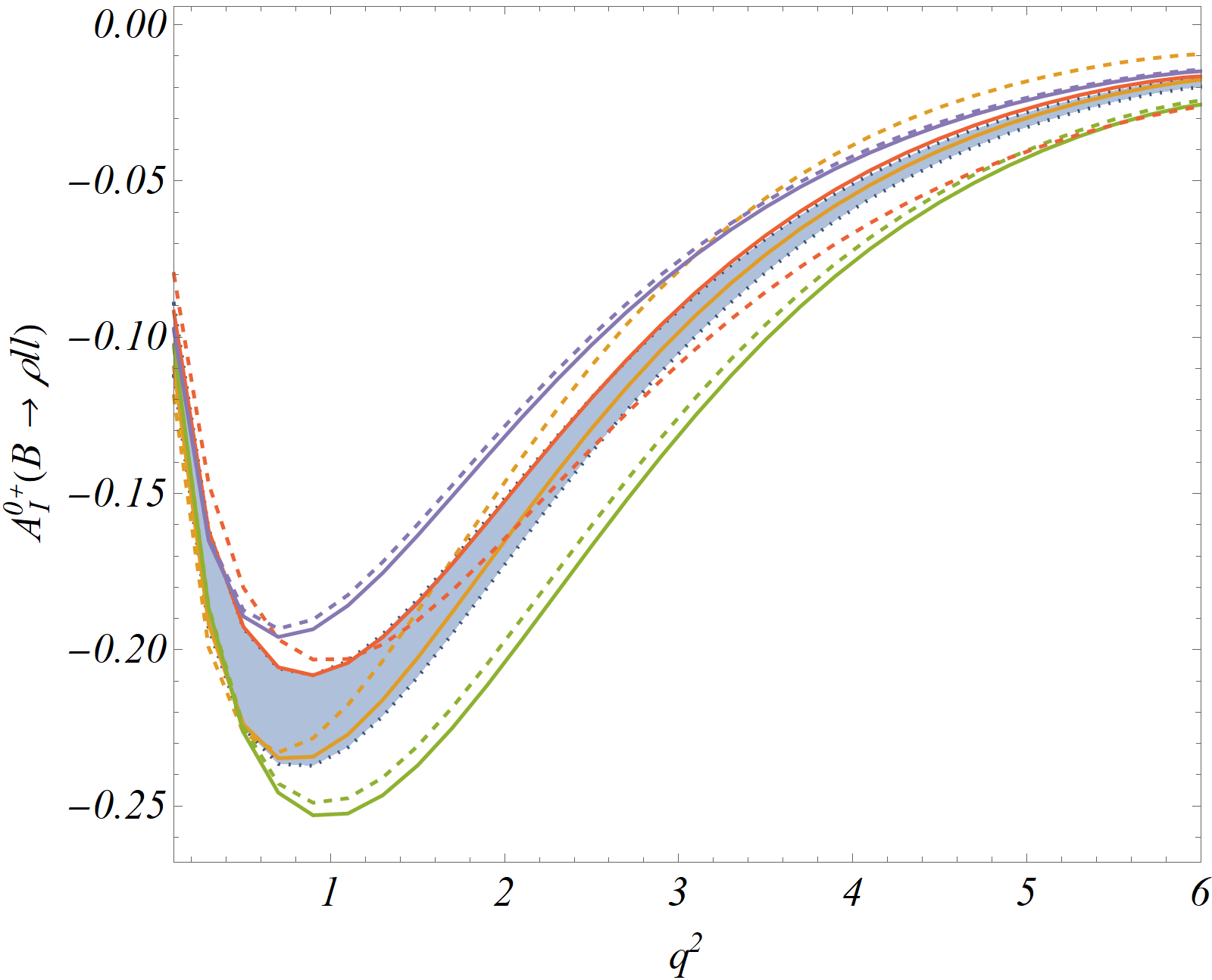}\label{fig:AIBrhoplus}}~~~
	\subfloat[]{\includegraphics[width=0.23\textwidth]{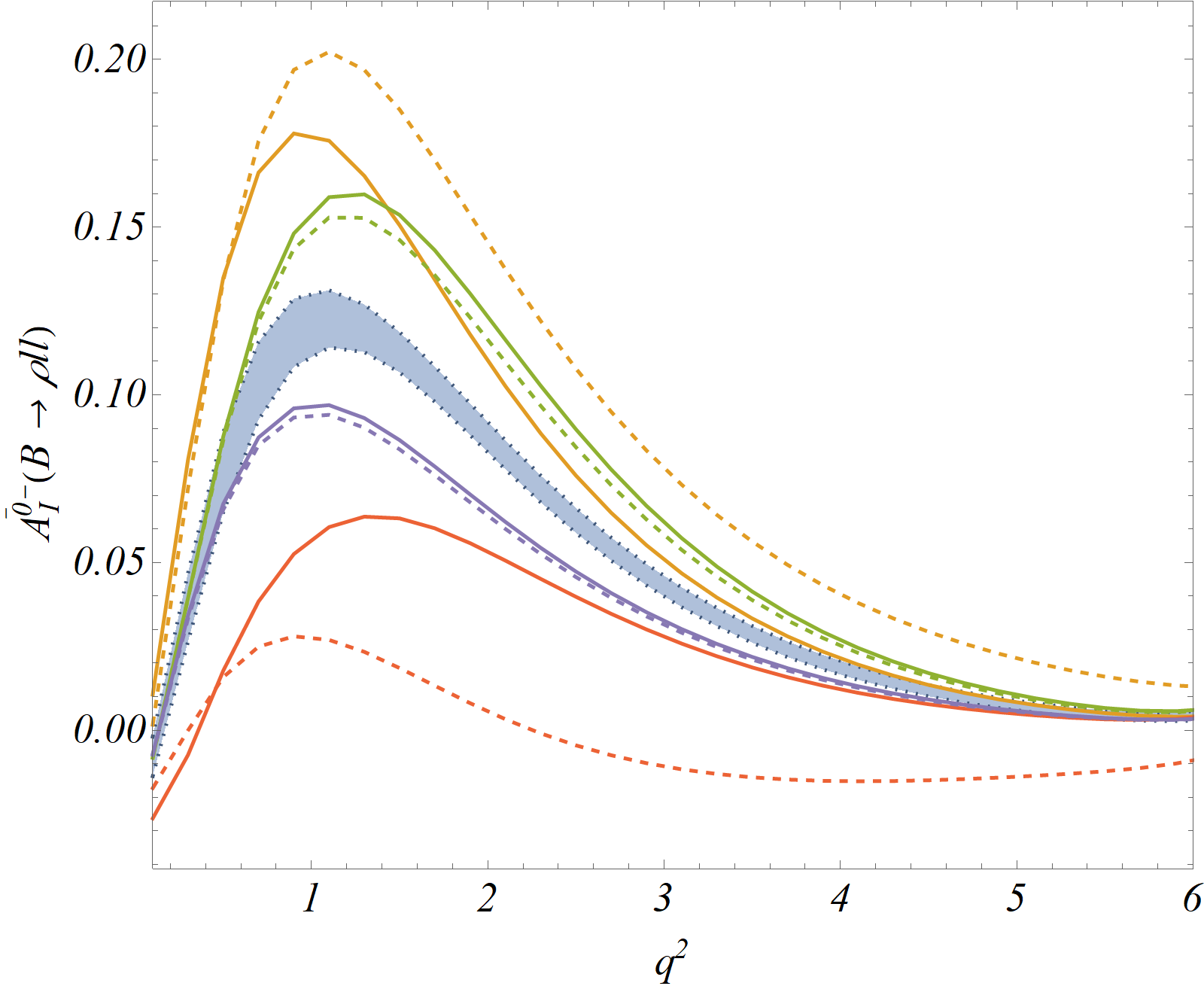}\label{fig:AIBrhozero}}~~~
	\subfloat[]{\includegraphics[width=0.23\textwidth]{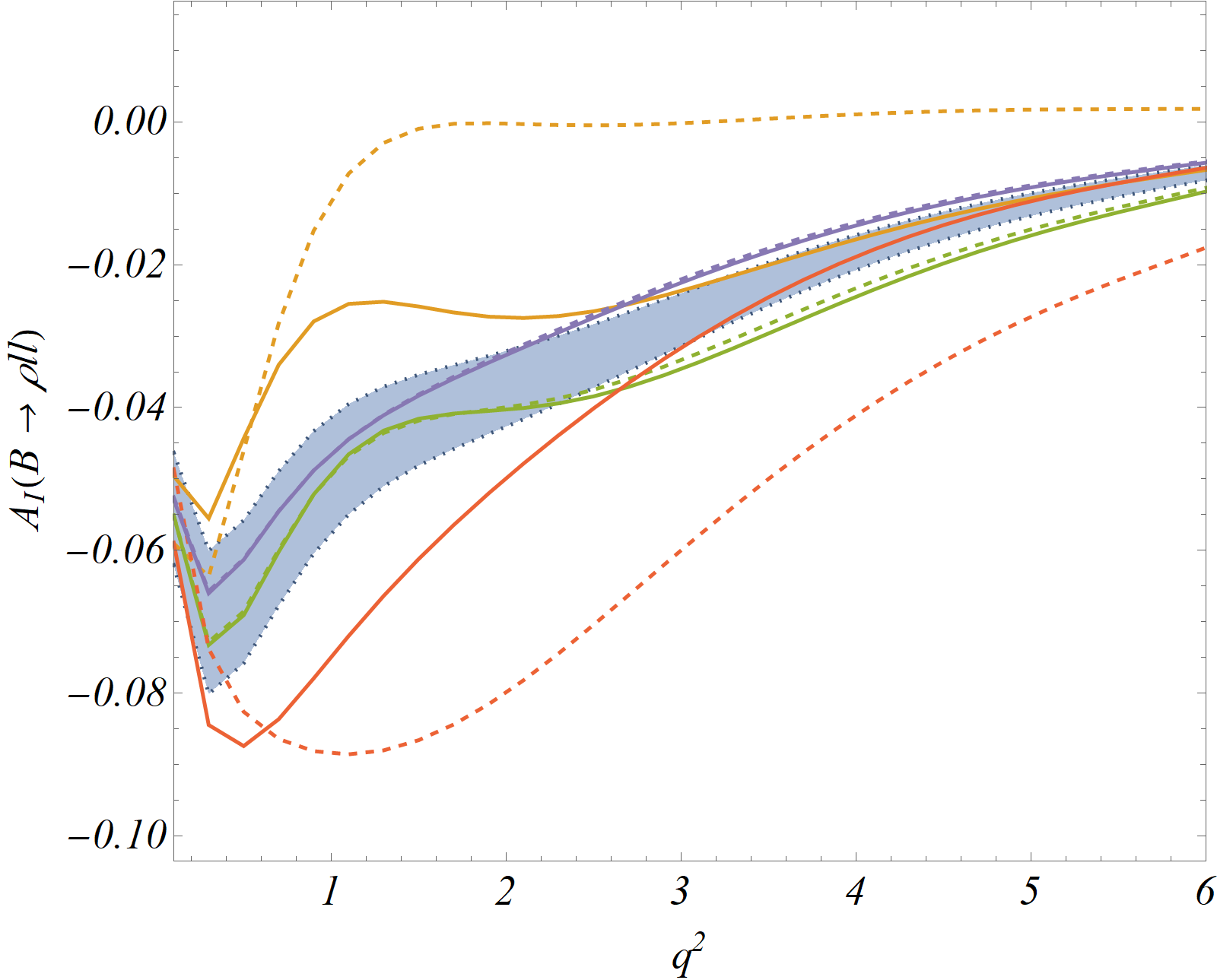}\label{fig:AIBrho1}}\\
	\caption{Isospin asymmetries in $B \to \rho\ell\ell$ decays measurable at the LHCb.}
	\label{fig:isospiasyrho}
\end{figure*}	

\begin{table}[t]
	\small
	\centering
	\renewcommand*{\arraystretch}{1.1}
	\begin{tabular}{*{5}{c}}
		\hline
		$\left.\text{Bin[}\text{GeV}^2\right]$  &  $\text{[0.1-1]}$  &  $\text{[1-2]}$  &  $\text{[2-4]}$  &  $\text{[4-6]}$  \\
		\hline
			$\langle A_{I}^{av} \rangle$ & 0.185(24) & 0.161(20) & 0.0748(90) & 0.0255(34) \\ 
		$\langle A_{I}^{\bar{0}-} \rangle$ & 0.0517(92) & 0.1102(55) & 0.0453(30) & 0.0082(15) \\
		$\langle A_{I}^{0+} \rangle$ & -0.180(15) & -0.195(12) & -0.1023(71) & -0.0319(25) \\
		$\langle A_{I} \rangle$ & -0.0642(87) & -0.0425(57) & -0.0285(35) & -0.0118(17) \\
		\hline
	\end{tabular}
	\caption{\small SM Predictions of isospin observables for $B\to \rho \ell \ell$ defined in eqs.~\ref{eq:IASroman} and \ref{eq:IASbeneke}, respectively.}
	\label{tab:isospinSM}
\end{table}

 Here, we would like to point out that our predictions of the LFU observables, like $R_{\pi/\rho}$, do not include any QED corrections. In the SM, the QED corrections, which include the soft and collinear singularities, could be a dominant source of LFU violation \cite{Isidori:2020acz}. The soft singularities will cancel in the differential rate. However, the cancellation of the hard collinear singularities of the form $\frac{\alpha}{\pi}$ln$(\frac{m_l}{m_B})$ is not obvious. 
	In ref.~\cite{Isidori:2020acz}, it has been shown that with a proper choice of the kinetic variables for fully photon-inclusive observables, there is a possibility of the cancellation of these hard collinear singularities. As for example, in the $B\to K\ell_1\ell_2$ decays the $q_0^2 = (P_B - P_K)^2$ differential distribution is free from ln$(m_l)$ divergences. Also, it has been observed that for the charged modes for the $q_0^2$ distributions, there will be additional collinear logs ln$(m_K)$, which do not cancel. All the observations made on the QED corrections on $B\to K (K^*)\ell^+\ell^-$ modes are relevant for $B\to \pi(\rho) \ell^+\ell^-$ decays. In the experimental analyses, QED corrections are implemented via the photon shower algorithm PHOTOS. However, note that PHOTOS do not include the correction associated with the term dependent on ln$(m_K)$ or ln$(m_\pi)$, which is related to the structure dependence. Therefore special care needs to be taken into account while extracting LFU observable from the charged modes. 
	It has been commented in \cite{Bordone:2021olx} that the $|R_{\pi}^{SM}-1|$ $<$ 0.01 remains true in all the $q^2$ regions even after considering the QED effects as long as $R_{\pi}^{SM}$ is extracted in a  photon-inclusive way.

\paragraph{Isospin asymmetries in $B\to \rho\ell\ell$ decays:}
We have also predicted the isospin asymmetries associated with the $B \to \rho\ell\ell$ transitions in the SM as well as in the different NP scenarios as discussed above. The following are the definitions of the relevant observables given in \cite{Lyon:2013gba}: 

\begin{align}\label{eq:IASroman}
A_I^{0+} (B\to \rho\ell\ell) &= \frac{2 d \Gamma(B^0\to \rho^0\ell\ell)/dq^2 - d \Gamma(B^+\to \rho^+\ell\ell)/dq^2}{2 d \Gamma(B^0\to \rho^0\ell\ell)/dq^2 + d \Gamma(B^+\to \rho^+\ell\ell)/dq^2}, \\ 
A_I^{\bar{0}-} (B\to \rho\ell\ell) &= \frac{2 d \Gamma(\bar{B}^0\to \rho^0\ell\ell)/dq^2 - d \Gamma(B^-\to \rho^-\ell\ell)/dq^2}{2 d \Gamma(\bar{B}^0\to \rho^0\ell\ell)/dq^2 + d \Gamma(B^-\to \rho^-\ell\ell)/dq^2},  \\ 
A_I (B\to \rho\ell\ell) &= \frac{1}{2} (A_I^{0+} + A_I^{\bar{0}-})
\end{align} 

Also, we have predicted the isospin asymmetry defined in \cite{Beneke:2004dp}
	\begin{equation}\label{eq:IASbeneke}
	A_I^{av} (B\to \rho\ell\ell) =\frac{d \Gamma(B^+\to \rho^+\ell\ell)/dq^2 + d \Gamma(B^-\to \rho^-\ell\ell)/dq^2}  {2 (d \Gamma(B^0\to \rho^0\ell\ell)/dq^2 + d \Gamma(\bar{B}^0\to \rho^0\ell\ell)/dq^2)} -1 . 
	\end{equation}
Using the fit results given in table~\ref{tab:LCSRlatfitres} we obtain the $q^2$-distributions of these isospin observables along with the respective errors in the SM (thick blue regions), which are shown in fig.~\ref{fig:isospiasyrho}. The $q^2$ shape, we have obtained for $A_I^{av}$ (fig.~\ref{fig:AIBrhoBeneke}), is in good agreement with that obtained in \cite{Beneke:2004dp}. This is as per the expectations since we are following the treatment of the refs.~ \cite{Beneke:2001at,Beneke:2004dp} for the non-factorizable corrections, in particular, the WA contributions to the respective helicity amplitudes which play an essential role in determining the relative size of the isospin asymmetries defined above. The corresponding predictions of $A_I^{av}$ in the SM in a few $q^2$-bins are given in table \ref{tab:isospinSM}. Note that $A_I^{av}$ can be as large as 20\% in the low $q^2 (\lsim 2 GeV^2)$ regions, which is mainly due to the sizeable strong and weak phases ($\alpha$) in the longitudinal amplitudes $( \propto e^{\pm i \alpha} C_{9,||}^{u})$ at the leading order in $\alpha_s$ and $1/m_b$ for the $B^{\pm}\to \rho^{\pm}\ell\ell$ decays. 

In figs.~\ref{fig:AIBrhoplus} and \ref{fig:AIBrhozero}, we have shown the $q^2$ distributions of the $A_I^{0+}$ and $A_I^{\bar{0}-}$, respectively. Note that in the region of low $q^2 \lsim 3$ GeV$^{2}$, the magnitudes of these asymmetries could be as large as 10\% to 20\%. The reason for such large asymmetries is the sizeable strong and weak phases in longitudinal amplitudes in the $\Gamma(B^{\pm} \to \rho^{\pm}\ell\ell)$, which is mainly due to the large imaginary contribution in the leading order WA contribution to these amplitudes. The corresponding predictions in a few small $q^2$-bins are shown in table \ref{tab:isospinSM}. Note that in the low-$q^2$ regions, we have got relatively larger values for the $A_I^{0+}$ and $A_I^{\bar{0}-}$ as compared to those presented in the ref.~\cite{Lyon:2013gba} though we agree with the same at the high-$q^2$ ($\gsim$ 4 GeV$^2$) presented in this work. This difference could be primarily due to our different treatments of the WA contributions. In ref.~\cite{Lyon:2013gba}, the WA contributions are calculated in the LCSR with an extended basis of dimension six four quark operators for WA. Also, we have shown the $q^2$ variation, and the predictions in the $q^2$-bins for the average of $A_I^{0+}$ and $A_I^{\bar{0}-}$, defined by $A_I$ in eq.~\ref{eq:IASroman}. As expected, the magnitude of $A_I$ is very small due to the relative cancellation of the large contributions in $A_I^{0+}$ and $A_I^{\bar{0}-}$. Our result for $A_I(q^2)$ has an agreement with the one presented in \cite{Lyon:2013gba}. 

\begin{table}[t]
	\small
	\centering
	\renewcommand*{\arraystretch}{1.1}
	\begin{tabular}{*{2}{c}}
		\hline
		Observables  & NP sensitivities   \\
		\hline
		$\langle A_{I}^{av} \rangle$ &  $Im(\Delta C_{9})$, $Im(C_{9}^{\prime})$ for $ 1 \le  q^2 \le 6$ GeV$^2$  \\ 
		\hline 
		$\langle A_{I}^{0+} \rangle$ &  $Re(\Delta C_{10})$, $Re(C^{\prime}_{10})$ for $ 1 \le  q^2 \le 3$ GeV$^2$ \\
		\hline 
		$\langle A_{I}^{\bar{0}-}  \rangle$ & $Im(\Delta C_9)$, $Im(C^{\prime}_9)$ for $ 1 \le  q^2 \le 6$ GeV$^2$  \\
		& $Re(\Delta C_9)$, $Re(C^{\prime}_9)$, $Re(\Delta C_{10})$, $Re(C^{\prime}_{10})$ for $ 1 \le  q^2 \le 3$ GeV$^2$ \\ 
		\hline 
		$\langle A_{I} \rangle$ & $Im(\Delta C_{9})$, $Im(C_{9}^{\prime})$ for $ 1 \le  q^2 \le 6$ GeV$^2$  \\
		\hline
	\end{tabular}
	\caption{\small NP sensitivities of the isospin observables for $B\to \rho \ell \ell$ defined in eqs.~\ref{eq:IASroman} and \ref{eq:IASbeneke}, respectively.}
	\label{tab:isospinNP}
\end{table}

In fig.~\ref{fig:isospiasyrho}, we have also shown NP sensitivities of the isospin asymmetric observables defined above. Like before we have considered the scenarios with $\Delta C_9$, $\Delta C_{10}$, $C_{9,10}^{\prime}$. The respective predictions in a few $q^2$-bins are given in tables \ref{tab:isospinNP1} and \ref{tab:isospinNP2}, respectively. The respective sensitivities have been summarised in table \ref{tab:isospinNP}. It can be seen from the tables mentioned above and from the figs.~\ref{fig:AIBrhoBeneke}, \ref{fig:AIBrho1} and  \ref{fig:AIBrhozero} that the averaged isospin asymmetries $A_I^{av}$, $A_I$ and $A_{I}^{\bar{0}-}$ are sensitive to both $Im(\Delta C_{9})$ and $Im(C_{9}^{\prime})$ for $q^2 \ge 1$ GeV$^2$. However, the resulting impact of both these contributions is similar in $A_I$ and $A_{I}^{\bar{0}-}$ while the impacts are reversed in $A_I^{av}$. Hence a comparative study of the future measurements of these three observables will be helpful to distinguish the contributions from $Im(\Delta C_{9})$ and $Im(C_{9}^{\prime})$. Also, from $A_{I}^{0+}$ and $A_{I}^{\bar{0}-}$, the contributions of $Re(\Delta C_9)$, $Re(C^{\prime}_9)$, $Re(\Delta C_{10})$, and $Re(C^{\prime}_{10})$ can be distinguished from one another.        
	
Finally, we will remark on the impact of the interference of the charm resonances with the operators $\mathcal{O}_9$ and $\mathcal{O}_9^{\prime}$ on the test of new physics sensitivities of certain observables which are in particular sensitive to $Re(\Delta C_9)$, $Re(C_9^{\prime})$ or to both. Due to the charm loop resonances, the modifications in $C_9^{eff}$ and $C_9^{\prime}$ will be given as 
\begin{align}
C_9^{eff} &= C_9 + \eta_c a_{fac} h_c(q^2) + ....., \nonumber \\
C_9^{\prime eff} &= C_9^{\prime} + \eta_c^{\prime} a_{fac} h_c(q^2) + .....,
\end{align}	
where at the scale $\mu = m_b$, $C_9 \approx 4$ and $a_{fac} \approx 0.6$, respectively. In the absence of any new physics contribution in $C_9^{\prime}$, the SM contribution will be simply given by $ C_9^{\prime eff} = \eta_c^{\prime} a_{fac} h_c(q^2)$. In our analysis, the predictions of the observables are obtained in the standard scenario with $\eta_c = 1$ and $\eta_c^{\prime} = 0$, respectively. In the ref.~\cite{Lyon:2014hpa}, the factors $\eta_c$ and $\eta_c^{\prime}$ are fitted from the data. It is shown that to explain the BESII data on $e^+e^- \to hadrons$ \cite{BES:2001ckj,BES:2007zwq} and the resonance structures found by LHCb in $B\to K\mu^+\mu^-$ decays in the low recoil region \cite{LHCb:2013lowresonances}, the $\eta_c + \eta_c^{\prime}$ should have a large negative value $-2.5$. Note that the $B\to K\ell^+\ell^-$ is sensitive to the combination $C_9^{eff} + C_9^{\prime eff}$. To understand the impact of this observation, in ref.~\cite{Lyon:2014hpa} three different scenarios had been considered: $ (\eta_c=-2.5,\eta_c^{\prime}=0)$, $ (\eta_c=0,\eta_c^{\prime}=-2.5)$, or $ (\eta_c=-1.25,\eta_c^{\prime}=-1.25)$, respectively. In such a situation, the predictions given in this analysis for the scenario $Re(\Delta C_9) = -1$, and for $ Re(C_9^{\prime}) = -1$ will be hard to distinguish from the predictions obtained in the SM with $\eta_c = -2.5$ or $\eta_c^{\prime} = -2.5$. For an illustration, we have predicted a couple of observables, which are sensitive to $Re(\Delta C_9)$ or $Re(C_9^{\prime})$, in the SM with $(\eta_c = -2.5,\eta_c^{\prime}=0)$ or $(\eta_c=0, \eta_c^{\prime} = -2.5)$, respectively. The respective predictions are given in the appendix in the table \ref{tab:charmscalingBR} for branching fractions in $B\to \pi\ell\ell$ decays, and in table \ref{tab:charmscalingCPav} for a few CP averaged observables in $B^{\pm}\to \rho^{\pm}\ell\ell$.      

However, it is important to note that in \cite{Lyon:2014hpa}, the analysis is done with the assumption of no NP contributions in $B\to K\mu\mu$. Hence the conclusion may change based on these considerations. It was pointed out in ref.~\cite{Descotes-Genon:2014uoa} that such a large negative value of $\eta_c + \eta_c^{\prime}$, though explains the observation on $P_5^{\prime}$ in the $q^2 = [4.30,8.68]$ {GeV}$^2$ bin, will not be consistent with many other measurements. More concrete pieces of evidence are required to conclude it further. As a more realistic case, one should probably do a simultaneous fit of $\eta_c$ or $\eta_c^{\prime}$ alongside the new contributions in $\mathcal{O}_9$ and $\mathcal{O}_9^{\prime}$, respectively. In the fit with new physics, the data might allow the solution $\eta_c\ne 1$ and $\eta_c^{\prime} \ne 0$.     
	
\section{Summary}\label{summary}
In this paper, we have analysed the decay modes $B\to\pi (\rho)\ell\nu_{\ell}$ and $B\to \pi(\rho)\ell^+\ell^-$ using all the available experimental inputs on the branching fractions, and the lattice and LCSR inputs on the form factors. We fit the coefficients parametrizing the respective form factors and the CKM element $|V_{ub}|$ using all these inputs and the available correlations between them. Using this fit result, for $B\to \pi\ell\ell$ decays we have predicted a few asymmetric observables like CP and isospin-asymmetries along with the respective branching fractions. Also, we have predicted LFU ratios like $R^{-/0}_{\pi}$. We have checked the $q^2$ distributions of all these observables and predicted the values, along with the respective errors, in a few $q^2$-bins. As in the case of the $B\to K^*\ell\ell$ decays, from the angular distribution of $B^{\pm(0)}\to \rho^{\pm (0)}\ell\ell$ decays we have obtained the CP-averaged and CP-asymmetric observables along with the respective branching fractions and $R^{\pm,0}_{\rho}$. Here also we have predicted all these observables in the SM in a few $q^2$-bins using the fit results for the form factors.
For the charged $B$ decays, the observables are defined and predicted for both the $B^+$ and $B^-$ decays following the tagging method, which can be measured both at the LHCb and Belle. Also, we have predicted the associated untagged observables. On the contrary, for the neutral $B$ decays, tagging is possible only at Belle but not at LHCb. Hence, there are observables measurable both at Belle and LHCb. There are also observables measurable only at Belle. We separate and predict them in the SM in a few $q^2$-bins. 

For all the observables mentioned above, we have checked the sensitivities towards some new contributions from a few additional operators beyond the SM. To look for deviations, we have studied the individual $q^2$ variations of all these observables in the different NP scenarios and compared them with the respective distributions in the SM. Also, for a few benchmarks in the different NP scenarios, the predictions are given in the $q^2$-bins, which can be compared to the respective SM predictions. Many observables are sensitive to the NP contributions we have considered, and one can distinguish the effects from the respective SM predictions. A comparative study of the respective NP sensitivities makes it possible to identify the effect of a particular type of scenario from the rest. Therefore, in case we observe deviations in the measured values from the respective SM predictions, looking at the pattern of the results, it will be possible to identify the type of NP scenario. Furthermore, we have noted that a few observables defined from a tagged analysis are sensitive to some NP scenarios. The respective sensitivities are lost due to a relative cancellation while defining the related CP-averaged observables, which could be obtained from the untagged analysis. In the context of NP searches, the measurement of the tagged observables is equally important as the related CP-averaged observables.

	\acknowledgments
	We would like to thank Tim Gershon and Roman Zwicky for some useful remarks and discussion. 
	This work of SN is supported by the Science and Engineering Research Board, Govt. of India, under the grant CRG/2018/001260. 
	
	\appendix
	
	\section{Correlations}
	\subsection{Synthetic data for Form Factors:}
	In this section we provide the correlations corresponding to the synthetic data generated from the fit results reported by refs.~\cite{FermilabLattice:2015cdh,Lattice:2015tia} (corresponding to the inputs provided in table~\ref{tab:MILCsynth}) and~\cite{Straub:2015ica} (corresponding to those in table~\ref{tab:Bharuchasynth}) in eqs.~\ref{eq:MILCcorr} and \ref{eq:Bharuchacorr} respectively. In eq.~\ref{eq:MILCcorr}, the correlated quantities are arranged as: $f_+(19.0)$, $f_+(20.5)$, $f_+(22.6)$, $f_+(25.1)$, $f_0(19.0)$, $f_0(22.6)$, $f_0(25.1)$, $f_T(19.0)$, $f_T(20.5)$, $f_T(22.6)$, $f_T(25.1)$. In eq~\ref{eq:Bharuchacorr} the corresponding arrangement is: $A_0(5)$, $A_0(10)$, $A_1(0)$, $A_1(5)$, $A_1(10)$, $A_2(0)$, $A_2(5)$, $A_2(10)$, $V(0)$, $V(5)$, $V(10)$, $T_1(0)$, $T_1(5)$, $T_1(10)$, $T_2(5)$, $T_2(10)$, $T_3(0)$, $T_3(5)$, $T_3(10)$.
	
	\begin{equation}
	\begin{pmatrix}	
	$1.$  &  $0.93$  &  $0.61$  &  $0.19$  &  $0.48$  &  $0.19$  &  $0.07$  &  $0.63$  &  $0.6$  &  $0.33$  &  $0.02$  \\
	$0.93$  &  $1.$  &  $0.84$  &  $0.36$  &  $0.32$  &  $0.19$  &  $0.09$  &  $0.6$  &  $0.62$  &  $0.41$  &  $0.13$  \\
	$0.61$  &  $0.84$  &  $1.$  &  $0.71$  &  $0.11$  &  $0.19$  &  $0.14$  &  $0.4$  &  $0.49$  &  $0.47$  &  $0.37$  \\
	$0.19$  &  $0.36$  &  $0.71$  &  $1.$  &  $0.17$  &  $0.21$  &  $0.2$  &  $0.08$  &  $0.17$  &  $0.36$  &  $0.56$  \\
	$0.48$  &  $0.32$  &  $0.11$  &  $0.17$  &  $1.$  &  $0.77$  &  $0.55$  &  $0.34$  &  $0.33$  &  $0.21$  &  $0.12$  \\
	$0.19$  &  $0.19$  &  $0.19$  &  $0.21$  &  $0.77$  &  $1.$  &  $0.92$  &  $0.21$  &  $0.28$  &  $0.31$  &  $0.25$  \\
	$0.07$  &  $0.09$  &  $0.14$  &  $0.2$  &  $0.55$  &  $0.92$  &  $1.$  &  $0.14$  &  $0.24$  &  $0.31$  &  $0.27$  \\
	$0.63$  &  $0.6$  &  $0.4$  &  $0.08$  &  $0.34$  &  $0.21$  &  $0.14$  &  $1.$  &  $0.89$  &  $0.49$  &  $0.25$  \\
	$0.6$  &  $0.62$  &  $0.49$  &  $0.17$  &  $0.33$  &  $0.28$  &  $0.24$  &  $0.89$  &  $1.$  &  $0.82$  &  $0.43$  \\
	$0.33$  &  $0.41$  &  $0.47$  &  $0.36$  &  $0.21$  &  $0.31$  &  $0.31$  &  $0.49$  &  $0.82$  &  $1.$  &  $0.73$  \\
	$0.02$  &  $0.13$  &  $0.37$  &  $0.56$  &  $0.12$  &  $0.25$  &  $0.27$  &  $0.25$  &  $0.43$  &  $0.73$  &  $1.$  \\
	\end{pmatrix}
	\label{eq:MILCcorr}
	\end{equation}
	\begin{equation}
	\begin{pmatrix}
	$1.$  &  $0.96$  &  $0.17$  &  $0.19$  &  $0.21$  &  $-0.23$  &  $-0.29$  &  $-0.37$  &  $0.19$  &  $0.24$  &  $0.29$  &  $0.18$  &  $0.23$  &  $0.28$  &  $0.21$  &  $0.22$  &  $-0.11$  &  $-0.14$  &  $-0.18$  \\
	$0.96$  &  $1.$  &  $0.19$  &  $0.23$  &  $0.25$  &  $-0.17$  &  $-0.23$  &  $-0.32$  &  $0.23$  &  $0.3$  &  $0.37$  &  $0.2$  &  $0.26$  &  $0.35$  &  $0.23$  &  $0.25$  &  $-0.07$  &  $-0.1$  &  $-0.15$  \\
	$0.17$  &  $0.19$  &  $1.$  &  $0.99$  &  $0.94$  &  $0.92$  &  $0.88$  &  $0.81$  &  $0.9$  &  $0.87$  &  $0.74$  &  $0.88$  &  $0.86$  &  $0.74$  &  $0.86$  &  $0.82$  &  $0.8$  &  $0.76$  &  $0.69$  \\
	$0.19$  &  $0.23$  &  $0.99$  &  $1.$  &  $0.98$  &  $0.9$  &  $0.88$  &  $0.82$  &  $0.88$  &  $0.89$  &  $0.81$  &  $0.87$  &  $0.88$  &  $0.82$  &  $0.88$  &  $0.87$  &  $0.78$  &  $0.76$  &  $0.71$  \\
	$0.21$  &  $0.25$  &  $0.94$  &  $0.98$  &  $1.$  &  $0.85$  &  $0.86$  &  $0.82$  &  $0.84$  &  $0.88$  &  $0.86$  &  $0.83$  &  $0.87$  &  $0.86$  &  $0.87$  &  $0.89$  &  $0.74$  &  $0.74$  &  $0.71$  \\
	$-0.23$  &  $-0.17$  &  $0.92$  &  $0.9$  &  $0.85$  &  $1.$  &  $0.99$  &  $0.94$  &  $0.81$  &  $0.77$  &  $0.63$  &  $0.8$  &  $0.76$  &  $0.64$  &  $0.77$  &  $0.74$  &  $0.83$  &  $0.8$  &  $0.75$  \\
	$-0.29$  &  $-0.23$  &  $0.88$  &  $0.88$  &  $0.86$  &  $0.99$  &  $1.$  &  $0.98$  &  $0.77$  &  $0.76$  &  $0.65$  &  $0.77$  &  $0.75$  &  $0.66$  &  $0.76$  &  $0.75$  &  $0.82$  &  $0.81$  &  $0.78$  \\
	$-0.37$  &  $-0.32$  &  $0.81$  &  $0.82$  &  $0.82$  &  $0.94$  &  $0.98$  &  $1.$  &  $0.69$  &  $0.69$  &  $0.62$  &  $0.7$  &  $0.7$  &  $0.63$  &  $0.71$  &  $0.71$  &  $0.77$  &  $0.79$  &  $0.79$  \\
	$0.19$  &  $0.23$  &  $0.9$  &  $0.88$  &  $0.84$  &  $0.81$  &  $0.77$  &  $0.69$  &  $1.$  &  $0.97$  &  $0.82$  &  $0.9$  &  $0.89$  &  $0.78$  &  $0.89$  &  $0.85$  &  $0.81$  &  $0.77$  &  $0.7$  \\
	$0.24$  &  $0.3$  &  $0.87$  &  $0.89$  &  $0.88$  &  $0.77$  &  $0.76$  &  $0.69$  &  $0.97$  &  $1.$  &  $0.93$  &  $0.88$  &  $0.91$  &  $0.88$  &  $0.89$  &  $0.89$  &  $0.77$  &  $0.76$  &  $0.7$  \\
	$0.29$  &  $0.37$  &  $0.74$  &  $0.81$  &  $0.86$  &  $0.63$  &  $0.65$  &  $0.62$  &  $0.82$  &  $0.93$  &  $1.$  &  $0.72$  &  $0.82$  &  $0.93$  &  $0.79$  &  $0.85$  &  $0.6$  &  $0.62$  &  $0.61$  \\
	$0.18$  &  $0.2$  &  $0.88$  &  $0.87$  &  $0.83$  &  $0.8$  &  $0.77$  &  $0.7$  &  $0.9$  &  $0.88$  &  $0.72$  &  $1.$  &  $0.98$  &  $0.84$  &  $0.99$  &  $0.94$  &  $0.92$  &  $0.88$  &  $0.8$  \\
	7	$0.23$  &  $0.26$  &  $0.86$  &  $0.88$  &  $0.87$  &  $0.76$  &  $0.75$  &  $0.7$  &  $0.89$  &  $0.91$  &  $0.82$  &  $0.98$  &  $1.$  &  $0.93$  &  $1.$  &  $0.99$  &  $0.88$  &  $0.86$  &  $0.8$  \\
	$0.28$  &  $0.35$  &  $0.74$  &  $0.82$  &  $0.86$  &  $0.64$  &  $0.66$  &  $0.63$  &  $0.78$  &  $0.88$  &  $0.93$  &  $0.84$  &  $0.93$  &  $1.$  &  $0.91$  &  $0.96$  &  $0.72$  &  $0.74$  &  $0.72$  \\
	$0.21$  &  $0.23$  &  $0.86$  &  $0.88$  &  $0.87$  &  $0.77$  &  $0.76$  &  $0.71$  &  $0.89$  &  $0.89$  &  $0.79$  &  $0.99$  &  $1.$  &  $0.91$  &  $1.$  &  $0.98$  &  $0.9$  &  $0.88$  &  $0.82$  \\
	$0.22$  &  $0.25$  &  $0.82$  &  $0.87$  &  $0.89$  &  $0.74$  &  $0.75$  &  $0.71$  &  $0.85$  &  $0.89$  &  $0.85$  &  $0.94$  &  $0.99$  &  $0.96$  &  $0.98$  &  $1.$  &  $0.85$  &  $0.86$  &  $0.82$  \\
	$-0.11$  &  $-0.07$  &  $0.8$  &  $0.78$  &  $0.74$  &  $0.83$  &  $0.82$  &  $0.77$  &  $0.81$  &  $0.77$  &  $0.6$  &  $0.92$  &  $0.88$  &  $0.72$  &  $0.9$  &  $0.85$  &  $1.$  &  $0.99$  &  $0.94$  \\
	$-0.14$  &  $-0.1$  &  $0.76$  &  $0.76$  &  $0.74$  &  $0.8$  &  $0.81$  &  $0.79$  &  $0.77$  &  $0.76$  &  $0.62$  &  $0.88$  &  $0.86$  &  $0.74$  &  $0.88$  &  $0.86$  &  $0.99$  &  $1.$  &  $0.98$  \\
	$-0.18$  &  $-0.15$  &  $0.69$  &  $0.71$  &  $0.71$  &  $0.75$  &  $0.78$  &  $0.79$  &  $0.7$  &  $0.7$  &  $0.61$  &  $0.8$  &  $0.8$  &  $0.72$  &  $0.82$  &  $0.82$  &  $0.94$  &  $0.98$  &  $1.$  \\
	\end{pmatrix}
	\label{eq:Bharuchacorr}
	\end{equation}
	\subsection{Fit correlations:}
	We provide the correlations between the fit parameters provided in table~\ref{tab:LCSRlatfitres} as a .json file named ``fitcorrB2pi.json".
	\newpage

	\section{The predictions in a few benchmark NP scenarios}
	In this section, we present the numerical estimates for the various observables for $B\to\rho,\pi ll$ decays corresponding to a few benchmark value for the real and imaginary parts of the NP Wilson coefficients $\Delta C_{9,10}$ and $C_{9,10}^\prime$ taken one at a time.

	\begin{table}[t]
		\tiny
		\centering
		\begin{tabular}{|c|c|c|c|c|c|c|}
			\hline
			\multirow{2}{*}{\textbf{Bin}} & \multirow{2}{*}{$C_i^{NP}$} &\multirow{2}{*}{$Re(C_i^{NP})$} &\multirow{2}{*}{$Im(C_i^{NP})$} & \multicolumn{3}{c|}{\textbf{Observables}}\\
			\cline{5-7}
			$$  &  $$  &  $$  &  $$  &  $A_ {\text{CP}}^0$  &  $A_ {\text{CP}}^+$  &  $A_I$  \\\hline
			$\text{0.1-1}$  &  $\text{$\Delta $C}_ 9$/$C_ 9^\prime$  &  $1$  &  $1$  &  $\text{-0.258(17)}$  &  $\text{-0.631(16)}$  &  $\text{-0.4533(88)}$  \\
			$$  &  $$  &  $1$  &  $-1$  &  $\text{-0.274(12)}$  &  $\text{-0.660(16)}$  &  $\text{-0.4507(95)}$  \\
			$$  &  $$  &  $-1$  &  $1$  &  $\text{-0.239(20)}$  &  $\text{-0.532(11)}$  &  $\text{-0.500(11)}$  \\
			$$  &  $$  &  $-1$  &  $-1$  &  $\text{-0.263(11)}$  &  $\text{-0.5590(84)}$  &  $\text{-0.504(12)}$  \\
			$$  &  $$  &  $1$  &  $0$  &  $\text{-0.275(15)}$  &  $\text{-0.662(16)}$  &  $\text{-0.4575(99)}$  \\
			$$  &  $$  &  $-1$  &  $0$  &  $\text{-0.262(16)}$  &  $\text{-0.5609(80)}$  &  $\text{-0.511(12)}$  \\
			$$  &  $\text{$\Delta $C}_{10}$/$C_ {10}^\prime$  &  $1$  &  $1$  &  $\text{-0.336(19)}$  &  $\text{-0.720(10)}$  &  $\text{-0.503(12)}$  \\
			$$  &  $$  &  $1$  &  $-1$  &  $\text{-0.336(19)}$  &  $\text{-0.720(10)}$  &  $\text{-0.503(12)}$  \\
			$$  &  $$  &  $-1$  &  $1$  &  $\text{-0.212(12)}$  &  $\text{-0.518(12)}$  &  $\text{-0.4514(91)}$  \\
			$$  &  $$  &  $-1$  &  $-1$  &  $\text{-0.212(12)}$  &  $\text{-0.518(12)}$  &  $\text{-0.4514(91)}$  \\
			$$  &  $$  &  $1$  &  $0$  &  $\text{-0.353(20)}$  &  $\text{-0.742(10)}$  &  $\text{-0.512(13)}$  \\
			$$  &  $$  &  $-1$  &  $0$  &  $\text{-0.219(13)}$  &  $\text{-0.532(13)}$  &  $\text{-0.4568(99)}$  \\\hline
			$\text{1-2}$  &  $\text{$\Delta $C}_ 9$/$C_ 9^\prime$  &  $1$  &  $1$  &  $\text{-0.1895(65)}$  &  $\text{-0.480(18)}$  &  $\text{-0.3755(36)}$  \\
			$$  &  $$  &  $1$  &  $-1$  &  $\text{-0.2115(71)}$  &  $\text{-0.472(17)}$  &  $\text{-0.3999(46)}$  \\
			$$  &  $$  &  $-1$  &  $1$  &  $\text{-0.1694(63)}$  &  $\text{-0.441(15)}$  &  $\text{-0.3887(49)}$  \\
			$$  &  $$  &  $-1$  &  $-1$  &  $\text{-0.2035(68)}$  &  $\text{-0.429(14)}$  &  $\text{-0.4274(65)}$  \\
			$$  &  $$  &  $1$  &  $0$  &  $\text{-0.2070(63)}$  &  $\text{-0.492(17)}$  &  $\text{-0.3904(46)}$  \\
			$$  &  $$  &  $-1$  &  $0$  &  $\text{-0.1948(57)}$  &  $\text{-0.454(13)}$  &  $\text{-0.4126(63)}$  \\
			$$  &  $\text{$\Delta $C}_{10}$/$C_ {10}^\prime$  &  $1$  &  $1$  &  $\text{-0.2558(83)}$  &  $\text{-0.580(18)}$  &  $\text{-0.4127(59)}$  \\
			$$  &  $$  &  $1$  &  $-1$  &  $\text{-0.2558(83)}$  &  $\text{-0.580(18)}$  &  $\text{-0.4127(59)}$  \\
			$$  &  $$  &  $-1$  &  $1$  &  $\text{-0.1571(54)}$  &  $\text{-0.381(14)}$  &  $\text{-0.3844(40)}$  \\
			$$  &  $$  &  $-1$  &  $-1$  &  $\text{-0.1571(54)}$  &  $\text{-0.381(14)}$  &  $\text{-0.3844(40)}$  \\
			$$  &  $$  &  $1$  &  $0$  &  $\text{-0.2687(80)}$  &  $\text{-0.605(17)}$  &  $\text{-0.4179(65)}$  \\
			$$  &  $$  &  $-1$  &  $0$  &  $\text{-0.1623(49)}$  &  $\text{-0.393(14)}$  &  $\text{-0.3870(44)}$  \\\hline
			$\text{2-4}$  &  $\text{$\Delta $C}_ 9$/$C_ 9^\prime$  &  $1$  &  $1$  &  $\text{-0.1749(53)}$  &  $\text{-0.308(11)}$  &  $\text{-0.3451(17)}$  \\
			$$  &  $$  &  $1$  &  $-1$  &  $\text{-0.1845(57)}$  &  $\text{-0.298(11)}$  &  $\text{-0.3718(25)}$  \\
			$$  &  $$  &  $-1$  &  $1$  &  $\text{-0.1622(47)}$  &  $\text{-0.295(11)}$  &  $\text{-0.3431(18)}$  \\
			$$  &  $$  &  $-1$  &  $-1$  &  $\text{-0.1772(57)}$  &  $\text{-0.280(10)}$  &  $\text{-0.3858(33)}$  \\
			$$  &  $$  &  $1$  &  $0$  &  $\text{-0.1855(57)}$  &  $\text{-0.314(11)}$  &  $\text{-0.3598(23)}$  \\
			$$  &  $$  &  $-1$  &  $0$  &  $\text{-0.1778(51)}$  &  $\text{-0.3018(99)}$  &  $\text{-0.3666(28)}$  \\
			$$  &  $\text{$\Delta $C}_{10}$/$C_ {10}^\prime$  &  $1$  &  $1$  &  $\text{-0.2313(69)}$  &  $\text{-0.384(13)}$  &  $\text{-0.3690(28)}$  \\
			$$  &  $$  &  $1$  &  $-1$  &  $\text{-0.2313(69)}$  &  $\text{-0.384(13)}$  &  $\text{-0.3690(28)}$  \\
			$$  &  $$  &  $-1$  &  $1$  &  $\text{-0.1417(43)}$  &  $\text{-0.2425(89)}$  &  $\text{-0.3556(18)}$  \\
			$$  &  $$  &  $-1$  &  $-1$  &  $\text{-0.1417(43)}$  &  $\text{-0.2425(89)}$  &  $\text{-0.3556(18)}$  \\
			$$  &  $$  &  $1$  &  $0$  &  $\text{-0.2429(73)}$  &  $\text{-0.403(13)}$  &  $\text{-0.3715(33)}$  \\
			$$  &  $$  &  $-1$  &  $0$  &  $\text{-0.1463(43)}$  &  $\text{-0.2508(89)}$  &  $\text{-0.3568(21)}$  \\\hline
			$\text{4-6}$  &  $\text{$\Delta $C}_ 9$/$C_ 9^\prime$  &  $1$  &  $1$  &  $\text{-0.1565(54)}$  &  $\text{-0.1974(72)}$  &  $\text{-0.3346(11)}$  \\
			$$  &  $$  &  $1$  &  $-1$  &  $\text{-0.1647(53)}$  &  $\text{-0.1982(68)}$  &  $\text{-0.3525(13)}$  \\
			$$  &  $$  &  $-1$  &  $1$  &  $\text{-0.1470(54)}$  &  $\text{-0.1905(72)}$  &  $\text{-0.3299(11)}$  \\
			$$  &  $$  &  $-1$  &  $-1$  &  $\text{-0.1600(51)}$  &  $\text{-0.1920(65)}$  &  $\text{-0.3588(15)}$  \\
			$$  &  $$  &  $1$  &  $0$  &  $\text{-0.1656(45)}$  &  $\text{-0.2043(63)}$  &  $\text{-0.3441(11)}$  \\
			$$  &  $$  &  $-1$  &  $0$  &  $\text{-0.1607(43)}$  &  $\text{-0.2005(60)}$  &  $\text{-0.3451(13)}$  \\
			$$  &  $\text{$\Delta $C}_{10}$/$C_ {10}^\prime$  &  $1$  &  $1$  &  $\text{-0.2071(67)}$  &  $\text{-0.2544(87)}$  &  $\text{-0.3471(14)}$  \\
			$$  &  $$  &  $1$  &  $-1$  &  $\text{-0.2071(67)}$  &  $\text{-0.2544(87)}$  &  $\text{-0.3471(14)}$  \\
			$$  &  $$  &  $-1$  &  $1$  &  $\text{-0.1275(42)}$  &  $\text{-0.1585(56)}$  &  $\text{-0.34182(89)}$  \\
			$$  &  $$  &  $-1$  &  $-1$  &  $\text{-0.1275(42)}$  &  $\text{-0.1585(56)}$  &  $\text{-0.34182(89)}$  \\
			$$  &  $$  &  $1$  &  $0$  &  $\text{-0.2171(59)}$  &  $\text{-0.2666(80)}$  &  $\text{-0.3481(16)}$  \\
			$$  &  $$  &  $-1$  &  $0$  &  $\text{-0.1315(36)}$  &  $\text{-0.1637(50)}$  &  $\text{-0.34230(100)}$  \\\hline
		\end{tabular}
		\caption{The dependence of the CP and Isospin asymmetry observables corresponding to $B\to\pi ll$ on the real and imaginary parts of the NP WC's $\Delta C_{9,10}$ and $C_{9,10}^\prime$ for the $q^2$ bins $0.1-1$, $1-2$, $2-4$ and $4-6$ $\text{GeV}^2$ for a few benchmark values of the real and imaginary parts of the corresponding WC's. The $+,0$ superscript over $A_{CP}$ denotes the charge of the initial state $B$ and the final state $\pi$ meson.}
		\label{tab:appBtopiasy}
	\end{table}
	
	\begin{table}[t]
		\tiny
		\centering
\begin{tabular}{|c|c|c|c|c|c|c|c|c|c|}
			\hline
			\multirow{2}{*}{\textbf{Bin}} & \multirow{2}{*}{$C_i^{NP}$} &\multirow{2}{*}{$Re(C_i^{NP})$} &\multirow{2}{*}{$Im(C_i^{NP})$} & \multicolumn{6}{c|}{\textbf{Observables}}\\
			\cline{5-10}
			$$  &  $$  &  $$  &  $$  &  $\text{BR(}\bar{B}^0\to \pi ^0\text{ll)}$  &  $\text{BR(}B^0\to \pi ^0\text{ll)}$  &  $\text{BR(}B^-\to \pi ^-\text{ll)}$  &  $\text{BR(}B^+\to \pi ^+\text{ll)}$  &  $\bar{R}_{\pi}^0$  &  $R_{\pi}^-$  \\
			$$  &  $$  &  $$  &  $$  &  $\times 10^8$  &   $\times 10^8$  &   $\times 10^8$  &   $\times 10^8$  &  $$  &  $$  \\\hline
			$\text{0.1-1}$  &  $\text{$\Delta $C}_ 9$/$C_ 9^\prime$  &  $1$  &  $1$  &  $\text{0.0396(28)}$  &  $\text{0.0672(40)}$  &  $\text{0.0565(44)}$  &  $\text{0.2492(95)}$  &  $\text{1.3878(41)}$  &  $\text{1.260(22)}$  \\
			$$  &  $$  &  $1$  &  $-1$  &  $\text{0.0353(27)}$  &  $\text{0.0618(36)}$  &  $\text{0.0470(40)}$  &  $\text{0.2291(85)}$  &  $\text{1.2341(72)}$  &  $\text{1.047(28)}$  \\
			$$  &  $$  &  $-1$  &  $1$  &  $\text{0.0263(18)}$  &  $\text{0.0429(26)}$  &  $\text{0.0523(25)}$  &  $\text{0.1715(63)}$  &  $\text{0.9217(67)}$  &  $\text{1.170(26)}$  \\
			$$  &  $$  &  $-1$  &  $-1$  &  $\text{0.0219(16)}$  &  $\text{0.0375(22)}$  &  $\text{0.0428(21)}$  &  $\text{0.1513(51)}$  &  $\text{0.7680(67)}$  &  $\text{0.957(23)}$  \\
			$$  &  $$  &  $1$  &  $0$  &  $\text{0.0354(26)}$  &  $\text{0.0622(34)}$  &  $\text{0.0479(38)}$  &  $\text{0.2346(83)}$  &  $\text{1.25(14)}$  &  $\text{1.07(11)}$  \\
\hline
		\end{tabular}
		\caption{The dependence of the Branching ratio and $R_{\pi}$ on the NP WC's $\Delta C_{9,10}$ and $C_{9,10}^\prime$ for the $q^2$ bins $0.1-1$, $1-2$, $2-4$ and $4-6$ $\text{GeV}^2$ for a few benchmark values of the real and imaginary parts of the corresponding WC's. The $-,0$ superscript over $R_{\pi}$ denotes the charge of the initial state $B$ and the final state $\pi$ meson.}
		\label{tab:appBtopibrR}
	\end{table}

	\begin{table}[ht]
		\centering
		\tiny
		\renewcommand*{\arraystretch}{1.2}
		\begin{tabular}{|c|c|c|c|c|c|c|c|c|}
			\hline
			\multirow{2}{*}{\textbf{Bin}} & \multirow{2}{*}{$C_i^{NP}$} &\multirow{2}{*}{$Re(C_i^{NP})$} &\multirow{2}{*}{$Im(C_i^{NP})$} & \multicolumn{5}{c|}{\textbf{Observables}}\\
			\cline{5-9}
			$$  &  $$  &  $$  &  $$  &  $A_4$  &  $A_5$  &  $A_{6s}$  &  $A_7$  &  $A_8$  \\\hline
			$\text{0.1-1}$  &  $\text{$\Delta $C}_9$  &  $1$  &  $1$  &  $\text{-0.0862(71)}$  &  $\text{0.0240(40)}$  &  $\text{0.0022(19)}$  &  $\text{-0.0052(37)}$  &  $\text{0.0752(68)}$  \\
			$$  &  $$  &  $1$  &  $-1$  &  $\text{-0.0995(56)}$  &  $\text{0.0249(42)}$  &  $\text{0.0023(20)}$  &  $\text{-0.0054(38)}$  &  $\text{0.0173(79)}$  \\
			$$  &  $$  &  $-1$  &  $1$  &  $\text{-0.0995(68)}$  &  $\text{0.0259(42)}$  &  $\text{0.0024(20)}$  &  $\text{-0.0056(40)}$  &  $\text{0.0821(63)}$  \\
			$$  &  $$  &  $-1$  &  $-1$  &  $\text{-0.1144(52)}$  &  $\text{0.0269(44)}$  &  $\text{0.0025(21)}$  &  $\text{-0.0058(41)}$  &  $\text{0.0199(76)}$  \\
			$$  &  $$  &  $1$  &  $0$  &  $\text{-0.0937(63)}$  &  $\text{0.0247(41)}$  &  $\text{0.0023(19)}$  &  $\text{-0.0053(38)}$  &  $\text{0.0472(74)}$  \\
			$$  &  $$  &  $-1$  &  $0$  &  $\text{-0.1081(60)}$  &  $\text{0.0267(44)}$  &  $\text{0.0025(21)}$  &  $\text{-0.0058(41)}$  &  $\text{0.0522(70)}$  \\
\hline
		\end{tabular}
		\caption{The dependence of the asymmetric observables corresponding to $B^{\pm}\to\rho^{\pm} ll$ on the real and imaginary parts of the NP WC's $\Delta C_{9,10}$ and $C_{9,10}^\prime$ for the $q^2$ bins $0.1-1$ and $1-2$ $\text{GeV}^2$ for a few benchmark values of the real and imaginary parts of the corresponding WC's.}
		\label{tab:appBtorhopmasynotag1}
	\end{table}
	
	\begin{table}[ht]
		\centering
		\tiny
		\renewcommand*{\arraystretch}{1.2}
		\begin{tabular}{|c|c|c|c|c|c|c|c|c|}
			\hline
			\multirow{2}{*}{\textbf{Bin}} & \multirow{2}{*}{$C_i^{NP}$} &\multirow{2}{*}{$Re(C_i^{NP})$} &\multirow{2}{*}{$Im(C_i^{NP})$} & \multicolumn{5}{c|}{\textbf{Observables}}\\
			\cline{5-9}
			$$  &  $$  &  $$  &  $$  &  $A_4$  &  $A_5$  &  $A_{6s}$  &  $A_7$  &  $A_8$  \\\hline
			$\text{2-4}$  &  $\text{$\Delta $C}_9$  &  $1$  &  $1$  &  $\text{0.00090(380)}$  &  $\text{0.0439(67)}$  &  $\text{0.00048(655)}$  &  $\text{0.0484(42)}$  &  $\text{0.0368(28)}$  \\
			$$  &  $$  &  $1$  &  $-1$  &  $\text{-0.0117(15)}$  &  $\text{0.0409(63)}$  &  $\text{0.00043(610)}$  &  $\text{0.0451(38)}$  &  $\text{-0.0218(32)}$  \\
			$$  &  $$  &  $-1$  &  $1$  &  $\text{-0.0159(41)}$  &  $\text{0.0627(92)}$  &  $\text{0.00069(929)}$  &  $\text{0.0691(53)}$  &  $\text{0.0360(32)}$  \\
			$$  &  $$  &  $-1$  &  $-1$  &  $\text{-0.0318(30)}$  &  $\text{0.0567(83)}$  &  $\text{0.00059(841)}$  &  $\text{0.0624(46)}$  &  $\text{-0.0451(41)}$  \\
			$$  &  $$  &  $1$  &  $0$  &  $\text{-0.0058(23)}$  &  $\text{0.0437(67)}$  &  $\text{0.00047(652)}$  &  $\text{0.0482(41)}$  &  $\text{0.0067(28)}$  \\
			$$  &  $$  &  $-1$  &  $0$  &  $\text{-0.0253(28)}$  &  $\text{0.0622(91)}$  &  $\text{0.00066(924)}$  &  $\text{0.0686(51)}$  &  $\text{-0.0069(38)}$  \\
\hline
		\end{tabular}
		\caption{The dependence of the asymmetric observables corresponding to $B^{\pm}\to\rho^{\pm} ll$ on the real and imaginary parts of the NP WC's $\Delta C_{9,10}$ and $C_{9,10}^\prime$ for the $q^2$ bins $2-4$ and $4-6$ $\text{GeV}^2$ for a few benchmark values of the real and imaginary parts of the corresponding WC's.}
		\label{tab:appBtorhopmasynotag2}
	\end{table}
	
	\begin{table}[ht]
		\tiny
		\centering
		\begin{tabular}{|c|c|c|c|c|c|c|}
			\hline
			\multirow{2}{*}{\textbf{Bin}} & \multirow{2}{*}{$C_i^{NP}$} &\multirow{2}{*}{$Re(C_i^{NP})$} &\multirow{2}{*}{$Im(C_i^{NP})$} & \multicolumn{3}{c|}{\textbf{Observables}}\\
			\cline{5-7}
			
			$$  &  $$  &  $$  &  $$  &  $A_3$  &  $A_9$  &  $A_{\text{CP}}$  \\
			\hline
			$\text{0.1-1}$  &  $\text{$\Delta $C}_9$  &  $1$  &  $1$  &  $\text{0.000051(47)}$  &  $\text{0.00040(24)}$  &  $\text{-0.342(25)}$  \\
			$$  &  $$  &  $1$  &  $-1$  &  $\text{-0.000066(48)}$  &  $\text{-0.00034(21)}$  &  $\text{-0.329(27)}$  \\
			$$  &  $$  &  $-1$  &  $1$  &  $\text{0.000072(45)}$  &  $\text{0.00044(25)}$  &  $\text{-0.216(22)}$  \\
			$$  &  $$  &  $-1$  &  $-1$  &  $\text{-0.000053(40)}$  &  $\text{-0.00036(22)}$  &  $\text{-0.197(25)}$  \\
			$$  &  $$  &  $1$  &  $0$  &  $\text{  -0.0000065(326)}$  &  $\text{0.000037(40)}$  &  $\text{-0.339(26)}$  \\
			$$  &  $$  &  $-1$  &  $0$  &  $\text{0.000011(20)}$  &  $\text{0.000048(37)}$  &  $\text{-0.210(24)}$  \\
\hline
		\end{tabular}
		\caption{The dependence of the asymmetric observables corresponding to $B^{\pm}\to\rho^{\pm} ll$ on the real and imaginary parts of the NP WC's $\Delta C_{9,10}$ and $C_{9,10}^\prime$ for the $q^2$ bins $0.1-1$ and $1-2$ $\text{GeV}^2$ for a few benchmark values of the real and imaginary parts of the corresponding WC's.}
		\label{tab:appBtorhopmACPnotag1}
	\end{table}

	\begin{table}[ht]
		\tiny
		\centering
		\begin{tabular}{|c|c|c|c|c|c|c|}
			\hline
			\multirow{2}{*}{\textbf{Bin}} & \multirow{2}{*}{$C_i^{NP}$} &\multirow{2}{*}{$Re(C_i^{NP})$} &\multirow{2}{*}{$Im(C_i^{NP})$} & \multicolumn{3}{c|}{\textbf{Observables}}\\
			\cline{5-7}
			
			$$  &  $$  &  $$  &  $$  &  $A_3$  &  $A_9$  &  $A_{\text{CP}}$  \\
			\hline
			
			$\text{2-4}$  &  $\text{$\Delta $C}_9$  &  $1$  &  $1$  &  $\text{0.00029(23)}$  &  $\text{0.00117(61)}$  &  $\text{-0.202(12)}$  \\
			$$  &  $$  &  $1$  &  $-1$  &  $\text{-0.00022(11)}$  &  $\text{-0.00095(51)}$  &  $\text{-0.1535(94)}$  \\
			$$  &  $$  &  $-1$  &  $1$  &  $\text{0.00038(24)}$  &  $\text{0.00172(90)}$  &  $\text{-0.1704(84)}$  \\
			$$  &  $$  &  $-1$  &  $-1$  &  $\text{-0.00032(12)}$  &  $\text{-0.00129(68)}$  &  $\text{-0.106(12)}$  \\
			$$  &  $$  &  $1$  &  $0$  &  $\text{0.000027(138)}$  &  $\text{0.000074(65)}$  &  $\text{-0.183(10)}$  \\
			$$  &  $$  &  $-1$  &  $0$  &  $\text{0.000012(101)}$  &  $\text{0.000149(91)}$  &  $\text{-0.1428(87)}$  \\
\hline
		\end{tabular}
			\caption{The dependence of the asymmetric observables corresponding to $B^{\pm}\to\rho^{\pm} ll$ on the real and imaginary parts of the NP WC's $\Delta C_{9,10}$ and $C_{9,10}^\prime$ for the $q^2$ bins $2-4$ and $4-6$ $\text{GeV}^2$ for a few benchmark values of the real and imaginary parts of the corresponding WC's.}
		\label{tab:appBtorhopmACPnotag2}
	\end{table}
	
	\begin{table}[ht]
		\centering
		\tiny
		\renewcommand*{\arraystretch}{1.2}
		\begin{tabular}{|c|c|c|c|c|c|c|c|c|c|c|}
			\hline
			\multirow{2}{*}{\textbf{Bin}} & \multirow{2}{*}{$C_i^{NP}$} &\multirow{2}{*}{$Re(C_i^{NP})$} &\multirow{2}{*}{$Im(C_i^{NP})$} & \multicolumn{7}{c|}{\textbf{Observables}}\\
			\cline{5-11}
			$$  &  $$  &  $$  &  $$  &  $P_1$  &  $P_2$  &  $P_3$  &  $P_4^\prime$  &  $P_5^\prime$  &  $P_6^\prime$  &  $P_8^\prime$  \\\hline		
			$\text{0.1-1}$  &  $\text{$\Delta $C}_9$  &  $1$  &  $1$  &  $\text{0.0035(26)}$  &  $\text{-0.0800(26)}$  &  $\text{0.00044(27)}$  &  $\text{0.178(10)}$  &  $\text{0.3473(88)}$  &  $\text{-0.072(18)}$  &  $\text{-0.068(12)}$  \\
			$$  &  $$  &  $1$  &  $-1$  &  $\text{0.0036(27)}$  &  $\text{-0.0813(25)}$  &  $\text{0.00042(29)}$  &  $\text{0.1865(82)}$  &  $\text{0.3609(89)}$  &  $\text{-0.075(19)}$  &  $\text{-0.053(14)}$  \\
			$$  &  $$  &  $-1$  &  $1$  &  $\text{0.0016(15)}$  &  $\text{-0.0825(26)}$  &  $\text{0.00023(14)}$  &  $\text{0.099(11)}$  &  $\text{0.467(14)}$  &  $\text{-0.081(21)}$  &  $\text{-0.0565(90)}$  \\
			$$  &  $$  &  $-1$  &  $-1$  &  $\text{0.0016(16)}$  &  $\text{-0.0838(25)}$  &  $\text{0.00021(14)}$  &  $\text{0.1061(87)}$  &  $\text{0.491(14)}$  &  $\text{-0.085(22)}$  &  $\text{-0.040(12)}$  \\
			$$  &  $$  &  $1$  &  $0$  &  $\text{0.0036(27)}$  &  $\text{-0.0810(26)}$  &  $\text{0.00043(29)}$  &  $\text{0.1864(92)}$  &  $\text{0.3583(91)}$  &  $\text{-0.074(19)}$  &  $\text{-0.061(13)}$  \\
			$$  &  $$  &  $-1$  &  $0$  &  $\text{0.0016(16)}$  &  $\text{-0.0835(26)}$  &  $\text{0.00022(14)}$  &  $\text{0.1065(99)}$  &  $\text{0.486(14)}$  &  $\text{-0.085(22)}$  &  $\text{-0.049(10)}$  \\
\hline
		\end{tabular}
		\caption{The dependence of the optimized observables corresponding to $B^{\pm}\to\rho^{\pm} ll$ on the real and imaginary parts of the NP WC's $\Delta C_{9,10}$ and $C_{9,10}^\prime$ for the $q^2$ bins $0.1-1$ and $1-2$ $\text{GeV}^2$ for a few benchmark values of the real and imaginary parts of the corresponding WC's.}
		\label{tab:appBtorhopmoptnotag1}
	\end{table}
	\begin{table}[ht]
		\centering
		\tiny
		\renewcommand*{\arraystretch}{1.2}
		\begin{tabular}{|c|c|c|c|c|c|c|c|c|c|c|}
			\hline
			\multirow{2}{*}{\textbf{Bin}} & \multirow{2}{*}{$C_i^{NP}$} &\multirow{2}{*}{$Re(C_i^{NP})$} &\multirow{2}{*}{$Im(C_i^{NP})$} & \multicolumn{7}{c|}{\textbf{Observables}}\\
			\cline{5-11}
			$$  &  $$  &  $$  &  $$  &  $P_1$  &  $P_2$  &  $P_3$  &  $P_4^\prime$  &  $P_5^\prime$  &  $P_6^\prime$  &  $P_8^\prime$  \\\hline
			$\text{2-4}$  &  $\text{$\Delta $C}_9$  &  $1$  &  $1$  &  $\text{-0.056(25)}$  &  $\text{0.039(23)}$  &  $\text{0.0059(34)}$  &  $\text{-0.311(15)}$  &  $\text{-0.614(28)}$  &  $\text{-0.119(33)}$  &  $\text{-0.089(19)}$  \\
			$$  &  $$  &  $1$  &  $-1$  &  $\text{-0.055(25)}$  &  $\text{0.039(23)}$  &  $\text{0.0058(34)}$  &  $\text{-0.325(15)}$  &  $\text{-0.588(26)}$  &  $\text{-0.115(32)}$  &  $\text{-0.057(22)}$  \\
			$$  &  $$  &  $-1$  &  $1$  &  $\text{-0.036(15)}$  &  $\text{-0.306(17)}$  &  $\text{0.0030(18)}$  &  $\text{-0.253(13)}$  &  $\text{-0.070(26)}$  &  $\text{-0.144(40)}$  &  $\text{-0.071(13)}$  \\
			$$  &  $$  &  $-1$  &  $-1$  &  $\text{-0.035(14)}$  &  $\text{-0.307(15)}$  &  $\text{0.0029(18)}$  &  $\text{-0.268(12)}$  &  $\text{-0.065(24)}$  &  $\text{-0.135(38)}$  &  $\text{-0.032(16)}$  \\
			$$  &  $$  &  $1$  &  $0$  &  $\text{-0.055(25)}$  &  $\text{0.042(25)}$  &  $\text{0.0062(36)}$  &  $\text{-0.312(15)}$  &  $\text{-0.626(28)}$  &  $\text{-0.122(34)}$  &  $\text{-0.076(21)}$  \\
			$$  &  $$  &  $-1$  &  $0$  &  $\text{-0.034(14)}$  &  $\text{-0.322(17)}$  &  $\text{0.0031(19)}$  &  $\text{-0.249(13)}$  &  $\text{-0.071(26)}$  &  $\text{-0.146(41)}$  &  $\text{-0.053(15)}$  \\
\hline
		\end{tabular}
		\caption{The dependence of the optimized observables corresponding to $B^{\pm}\to\rho^{\pm} ll$ on the real and imaginary parts of the NP WC's $\Delta C_{9,10}$ and $C_{9,10}^\prime$ for the $q^2$ bins $2-4$ and $4-6$ $\text{GeV}^2$ for a few benchmark values of the real and imaginary parts of the corresponding WC's.}
		\label{tab:appBtorhopmoptnotag2}
	\end{table}
	
	\begin{table}[ht]
		\centering
		\tiny
		\renewcommand*{\arraystretch}{1.2}
		\begin{tabular}{|c|c|c|c|c|c|c|c|}
			\hline
			\multirow{2}{*}{\textbf{Bin}} & \multirow{2}{*}{$C_i^{NP}$} &\multirow{2}{*}{$Re(C_i^{NP})$} &\multirow{2}{*}{$Im(C_i^{NP})$} & \multicolumn{4}{c|}{\textbf{Observables}}\\
			\cline{5-8}
			$$  &  $$  &  $$  &  $$  &  $\text{BR$\times $}10^9$  &  $F_L$  &  $A_{\text{FB}}$  &  $R_{\rho}$  \\\hline
			$\text{0.1-1}$  &  $\text{$\Delta $C}_9$  &  $1$  &  $1$  &  $\text{3.47(23)}$  &  $\text{0.416(26)}$  &  $\text{-0.0701(41)}$  &  $\text{1.061(10)}$  \\
			$$  &  $$  &  $1$  &  $-1$  &  $\text{3.35(22)}$  &  $\text{0.405(26)}$  &  $\text{-0.0726(41)}$  &  $\text{1.0248(89)}$  \\
			$$  &  $$  &  $-1$  &  $1$  &  $\text{3.22(22)}$  &  $\text{0.331(24)}$  &  $\text{-0.0828(40)}$  &  $\text{0.9856(49)}$  \\
			$$  &  $$  &  $-1$  &  $-1$  &  $\text{3.10(22)}$  &  $\text{0.317(24)}$  &  $\text{-0.0859(39)}$  &  $\text{0.9497(64)}$  \\
			$$  &  $$  &  $1$  &  $0$  &  $\text{3.37(23)}$  &  $\text{0.407(26)}$  &  $\text{-0.0721(41)}$  &  $\text{1.0317(86)}$  \\
			$$  &  $$  &  $-1$  &  $0$  &  $\text{3.13(22)}$  &  $\text{0.319(24)}$  &  $\text{-0.0853(39)}$  &  $\text{0.9566(63)}$  \\
\hline
		\end{tabular}
		\caption{The dependence of the Branching Ratio and the observables $F_L$, $A_{FB}$ and $R_{\rho}$ corresponding to $B^{\pm}\to\rho^{\pm} ll$ on the real and imaginary parts of the NP WC's $\Delta C_{9,10}$ and $C_{9,10}^\prime$ for the $q^2$ bins $0.1-1$ and $1-2$ $\text{GeV}^2$ for a few benchmark values of the real and imaginary parts of the corresponding WC's.}
		\label{tab:appBtorhopmbrnotag1}
	\end{table}
	
	\begin{table}[ht]
		\centering
		\tiny
		\renewcommand*{\arraystretch}{1.2}
		\begin{tabular}{|c|c|c|c|c|c|c|c|}
			\hline
			\multirow{2}{*}{\textbf{Bin}} & \multirow{2}{*}{$C_i^{NP}$} &\multirow{2}{*}{$Re(C_i^{NP})$} &\multirow{2}{*}{$Im(C_i^{NP})$} & \multicolumn{4}{c|}{\textbf{Observables}}\\
			\cline{5-8}
			$$  &  $$  &  $$  &  $$  &  $\text{BR$\times $}10^9$  &  $F_L$  &  $A_{\text{FB}}$  &  $R_{\rho}$  \\\hline
			$\text{2-4}$  &  $\text{$\Delta $C}_9$  &  $1$  &  $1$  &  $\text{3.05(27)}$  &  $\text{0.826(15)}$  &  $\text{0.0103(62)}$  &  $\text{1.2117(90)}$  \\
			$$  &  $$  &  $1$  &  $-1$  &  $\text{3.28(28)}$  &  $\text{0.838(14)}$  &  $\text{0.0096(57)}$  &  $\text{1.3017(68)}$  \\
			$$  &  $$  &  $-1$  &  $1$  &  $\text{2.14(17)}$  &  $\text{0.724(22)}$  &  $\text{-0.127(14)}$  &  $\text{0.8490(81)}$  \\
			$$  &  $$  &  $-1$  &  $-1$  &  $\text{2.36(18)}$  &  $\text{0.751(20)}$  &  $\text{-0.115(12)}$  &  $\text{0.9390(97)}$  \\
			$$  &  $$  &  $1$  &  $0$  &  $\text{3.07(27)}$  &  $\text{0.836(14)}$  &  $\text{0.0102(61)}$  &  $\text{1.2174(74)}$  \\
			$$  &  $$  &  $-1$  &  $0$  &  $\text{2.15(17)}$  &  $\text{0.739(21)}$  &  $\text{-0.126(13)}$  &  $\text{0.8547(77)}$  \\
\hline
		\end{tabular}
		\caption{The dependence of the Branching Ratio and the observables $F_L$, $A_{FB}$ and $R_{\rho}$ corresponding to $B^{\pm}\to\rho^{\pm} ll$ on the real and imaginary parts of the NP WC's $\Delta C_{9,10}$ and $C_{9,10}^\prime$ for the $q^2$ bins $2-4$ and $4-6$ $\text{GeV}^2$ for a few benchmark values of the real and imaginary parts of the corresponding WC's.}
		\label{tab:appBtorhopmbrnotag2}
	\end{table}
	
	\begin{table}[ht]
		\centering
		\tiny
		\renewcommand*{\arraystretch}{1.2}
		\begin{tabular}{|c|c|c|c|c|c|c|c|}
			\hline
			\multirow{2}{*}{\textbf{Bin}} & \multirow{2}{*}{$C_i^{NP}$} &\multirow{2}{*}{$Re(C_i^{NP})$} &\multirow{2}{*}{$Im(C_i^{NP})$} & \multicolumn{4}{c|}{\textbf{Observables}}\\
			\cline{5-8}
			$$  &  $$  &  $$  &  $$  &  $P_2^-$  &  $P_2^+$  &  $P_3^-$  &  $P_3^+$     \\
			\hline
			$\text{0.1-1}$  &  $\text{$\Delta $C}_9$  &  $1$  &  $1$  &  $\text{-0.1293(52)}$  &  $\text{-0.1220(39)}$  &  $\text{-0.00042(29)}$  &  $\text{0.0017(11)}$    \\
			$$  &  $$  &  $1$  &  $-1$  &  $\text{-0.1261(52)}$  &  $\text{-0.1247(37)}$  &  $\text{0.00157(96)}$  &  $\text{-0.00022(35)}$    \\
			$$  &  $$  &  $-1$  &  $1$  &  $\text{-0.1256(49)}$  &  $\text{-0.1188(36)}$  &  $\text{-0.00067(42)}$  &  $\text{0.00125(78)}$    \\
			$$  &  $$  &  $-1$  &  $-1$  &  $\text{-0.1228(49)}$  &  $\text{-0.1212(34)}$  &  $\text{0.00110(67)}$  &  $\text{-0.00044(33)}$     \\
			$$  &  $$  &  $1$  &  $0$  &  $\text{-0.1279(52)}$  &  $\text{-0.1235(38)}$  &  $\text{0.00059(37)}$  &  $\text{0.00074(54)}$     \\
			$$  &  $$  &  $-1$  &  $0$  &  $\text{-0.1244(49)}$  &  $\text{-0.1201(35)}$  &  $\text{0.00022(14)}$  &  $\text{0.00041(29)}$     \\
\hline
			\hline
		\end{tabular}
		\caption{The dependence of some of the optimized observables corresponding to $B^+\to\rho^+ ll$ and $B^-\to\rho^- ll$ on the real and imaginary parts of the NP WC's $\Delta C_{9,10}$ and $C_{9,10}^\prime$ for the $q^2$ bins $0.1-1$ and $1-2$ $\text{GeV}^2$ for a few benchmark values of the real and imaginary parts of the corresponding WC's.}
		\label{tab:appBtorhopmP2P3tag1}		
	\end{table}
	
	\begin{table}[ht]
		\centering
		\tiny
		\renewcommand*{\arraystretch}{1.2}
		\begin{tabular}{|c|c|c|c|c|c|c|c|}
			\hline
			\multirow{2}{*}{\textbf{Bin}} & \multirow{2}{*}{$C_i^{NP}$} &\multirow{2}{*}{$Re(C_i^{NP})$} &\multirow{2}{*}{$Im(C_i^{NP})$} & \multicolumn{4}{c|}{\textbf{Observables}}\\
			\cline{5-8}
			$$  &  $$  &  $$  &  $$  &  $P_2^-$  &  $P_2^+$  &  $P_3^-$  &  $P_3^+$     \\
			\hline
			$\text{2-4}$  &  $\text{$\Delta $C}_9$  &  $1$  &  $1$  &  $\text{0.051(35)}$  &  $\text{0.038(31)}$  &  $\text{-0.0011(18)}$  &  $\text{0.0131(70)}$     \\
			$$  &  $$  &  $1$  &  $-1$  &  $\text{0.044(30)}$  &  $\text{0.048(40)}$  &  $\text{0.0124(67)}$  &  $\text{-0.00018(276)}$  \\
			$$  &  $$  &  $-1$  &  $1$  &  $\text{-0.363(19)}$  &  $\text{-0.303(30)}$  &  $\text{-0.0039(23)}$  &  $\text{0.0092(50)}$   \\
			$$  &  $$  &  $-1$  &  $-1$  &  $\text{-0.310(13)}$  &  $\text{-0.361(22)}$  &  $\text{0.0082(45)}$  &  $\text{-0.0027(21)}$    \\
			$$  &  $$  &  $1$  &  $0$  &  $\text{0.050(34)}$  &  $\text{0.045(37)}$  &  $\text{0.0067(38)}$  &  $\text{0.0075(44)}$   \\
			$$  &  $$  &  $-1$  &  $0$  &  $\text{-0.351(15)}$  &  $\text{-0.345(28)}$  &  $\text{0.0028(17)}$  &  $\text{0.0040(25)}$ \\
\hline
			\end{tabular}
			\caption{The dependence of some of the optimized observables corresponding to $B^+\to\rho^+ ll$ and $B^-\to\rho^- ll$ on the real and imaginary parts of the NP WC's $\Delta C_{9,10}$ and $C_{9,10}^\prime$ for the $q^2$ bins $2-4$ and $4-6$ $\text{GeV}^2$ for a few benchmark values of the real and imaginary parts of the corresponding WC's.}
			\label{tab:appBtorhopmP2P3tag2}		
			\end{table}

			\begin{table}[ht]
				\centering
				\tiny
				\renewcommand*{\arraystretch}{1.2}
				\begin{tabular}{|c|c|c|c|c|c|c|c|}
					\hline
					\multirow{2}{*}{\textbf{Bin}} & \multirow{2}{*}{$C_i^{NP}$} &\multirow{2}{*}{$Re(C_i^{NP})$} &\multirow{2}{*}{$Im(C_i^{NP})$} & \multicolumn{4}{c|}{\textbf{Observables}}\\
					\cline{5-8}
					$$  &  $$  &  $$  &  $$  &  $P_6^{'-}$  &  $P_6^{'+}$  &  $P_8^{'-}$  &  $P_8^{'+}$  \\
					\hline
					$\text{0.1-1}$  &  $\text{$\Delta $C}_9$  &  $1$  &  $1$  &  $\text{-0.175(21)}$  &  $\text{-0.059(25)}$  &  $\text{0.180(14)}$  &  $\text{-0.211(26)}$  \\
					$$  &  $$  &  $1$  &  $-1$  &  $\text{-0.181(23)}$  &  $\text{-0.062(27)}$  &  $\text{-0.037(16)}$  &  $\text{-0.085(29)}$  \\
					$$  &  $$  &  $-1$  &  $1$  &  $\text{-0.159(19)}$  &  $\text{-0.068(30)}$  &  $\text{0.202(12)}$  &  $\text{-0.227(24)}$  \\
					$$  &  $$  &  $-1$  &  $-1$  &  $\text{-0.164(20)}$  &  $\text{-0.072(31)}$  &  $\text{0.0052(176)}$  &  $\text{-0.082(27)}$  \\
					$$  &  $$  &  $1$  &  $0$  &  $\text{-0.183(22)}$  &  $\text{-0.061(26)}$  &  $\text{0.075(16)}$  &  $\text{-0.150(28)}$  \\
					$$  &  $$  &  $-1$  &  $0$  &  $\text{-0.166(20)}$  &  $\text{-0.071(31)}$  &  $\text{0.108(15)}$  &  $\text{-0.158(26)}$  \\
\hline
				\end{tabular}
				\caption{The dependence of some of the optimized observables corresponding to $B^+\to\rho^+ ll$ and $B^-\to\rho^- ll$ on the real and imaginary parts of the NP WC's $\Delta C_{9,10}$ and $C_{9,10}^\prime$ for the $q^2$ bins $0.1-1$ and $1-2$ $\text{GeV}^2$ for a few benchmark values of the real and imaginary parts of the corresponding WC's.}
				\label{tab:appBtorhopmP6P8tag1}						
			\end{table}
			
			\begin{table}[ht]
				\centering
				\tiny
				\renewcommand*{\arraystretch}{1.2}
				\begin{tabular}{|c|c|c|c|c|c|c|c|}
					\hline
					\multirow{2}{*}{\textbf{Bin}} & \multirow{2}{*}{$C_i^{NP}$} &\multirow{2}{*}{$Re(C_i^{NP})$} &\multirow{2}{*}{$Im(C_i^{NP})$} & \multicolumn{4}{c|}{\textbf{Observables}}\\
					\cline{5-8}
					$$  &  $$  &  $$  &  $$  &  $P_6^{'-}$  &  $P_6^{'+}$  &  $P_8^{'-}$  &  $P_8^{'+}$  \\
					\hline
					$\text{2-4}$  &  $\text{$\Delta $C}_9$  &  $1$  &  $1$  &  $\text{0.010(42)}$  &  $\text{-0.226(28)}$  &  $\text{0.0096(166)}$  &  $\text{-0.171(20)}$  \\
					$$  &  $$  &  $1$  &  $-1$  &  $\text{0.0094(366)}$  &  $\text{-0.244(36)}$  &  $\text{-0.132(16)}$  &  $\text{0.0026(302)}$  \\
					$$  &  $$  &  $-1$  &  $1$  &  $\text{0.012(49)}$  &  $\text{-0.271(37)}$  &  $\text{0.0114(87)}$  &  $\text{-0.138(17)}$  \\
					$$  &  $$  &  $-1$  &  $-1$  &  $\text{0.011(41)}$  &  $\text{-0.282(45)}$  &  $\text{-0.1499(73)}$  &  $\text{0.072(24)}$  \\
					$$  &  $$  &  $1$  &  $0$  &  $\text{0.010(41)}$  &  $\text{-0.243(33)}$  &  $\text{-0.070(17)}$  &  $\text{-0.090(27)}$  \\
					$$  &  $$  &  $-1$  &  $0$  &  $\text{0.012(47)}$  &  $\text{-0.288(42)}$  &  $\text{-0.0800(85)}$  &  $\text{-0.036(22)}$  \\
\hline
				\end{tabular}
				\caption{The dependence of some of the optimized observables corresponding to $B^+\to\rho^+ ll$ and $B^-\to\rho^- ll$ on the real and imaginary parts of the NP WC's $\Delta C_{9,10}$ and $C_{9,10}^\prime$ for the $q^2$ bins $2-4$ and $4-6$ $\text{GeV}^2$ for a few benchmark values of the real and imaginary parts of the corresponding WC's.}
				\label{tab:appBtorhopmP6P8tag2}						
			\end{table}
			
			
			\begin{table}[ht]
				\centering
				\tiny
				\renewcommand*{\arraystretch}{1.2}
				\begin{tabular}{|c|c|c|c|c|c|c|c|}
					\hline
					\multirow{2}{*}{\textbf{Bin}} & \multirow{2}{*}{$C_i^{NP}$} &\multirow{2}{*}{$Re(C_i^{NP})$} &\multirow{2}{*}{$Im(C_i^{NP})$} & \multicolumn{4}{c|}{\textbf{Observables}}\\
					$$  &  $$  &  $$  &  $$  &  $\overline{P_2^0}$  &  $P_2^0$  &  $\overline{P_3^0}$  &  $P_3^0$  \\
					\hline
					$\text{0.1-1}$  &  $\text{$\Delta $C}_9$  &  $1$  &  $1$  &  $\text{-0.0784(22)}$  &  $\text{-0.0791(36)}$  &  $\text{-0.00067(42)}$  &  $\text{0.00146(91)}$  \\
					$$  &  $$  &  $1$  &  $-1$  &  $\text{-0.0772(21)}$  &  $\text{-0.0803(33)}$  &  $\text{0.00131(80)}$  &  $\text{-0.00046(37)}$  \\
					$$  &  $$  &  $-1$  &  $1$  &  $\text{-0.0764(21)}$  &  $\text{-0.0769(32)}$  &  $\text{-0.00079(48)}$  &  $\text{0.00115(71)}$  \\
					$$  &  $$  &  $-1$  &  $-1$  &  $\text{-0.0753(21)}$  &  $\text{-0.0778(30)}$  &  $\text{0.00097(59)}$  &  $\text{-0.00056(37)}$  \\
					$$  &  $$  &  $1$  &  $0$  &  $\text{-0.0779(22)}$  &  $\text{-0.0798(34)}$  &  $\text{0.00033(20)}$  &  $\text{0.00051(37)}$  \\
					$$  &  $$  &  $-1$  &  $0$  &  $\text{-0.0759(21)}$  &  $\text{-0.0775(31)}$  &  $\text{0.000099(64)}$  &  $\text{0.00030(22)}$  \\
					$$  &  $\text{$\Delta $C}_{10}$  &  $1$  &  $1$  &  $\text{-0.0513(19)}$  &  $\text{-0.0645(18)}$  &  $\text{0.00021(12)}$  &  $\text{0.00040(29)}$  \\
					$$  &  $$  &  $1$  &  $-1$  &  $\text{-0.0636(19)}$  &  $\text{-0.0530(39)}$  &  $\text{0.00021(12)}$  &  $\text{0.00040(29)}$  \\
					$$  &  $$  &  $-1$  &  $1$  &  $\text{-0.0896(27)}$  &  $\text{-0.1034(31)}$  &  $\text{0.00020(12)}$  &  $\text{0.00039(29)}$  \\
					$$  &  $$  &  $-1$  &  $-1$  &  $\text{-0.1016(27)}$  &  $\text{-0.0921(53)}$  &  $\text{0.00020(12)}$  &  $\text{0.00039(29)}$  \\
					$$  &  $$  &  $1$  &  $0$  &  $\text{-0.0576(17)}$  &  $\text{-0.0588(25)}$  &  $\text{0.00021(12)}$  &  $\text{0.00040(29)}$  \\
					$$  &  $$  &  $-1$  &  $0$  &  $\text{-0.0957(26)}$  &  $\text{-0.0979(40)}$  &  $\text{0.00020(12)}$  &  $\text{0.00039(29)}$  \\
					$$  &  $C_9'$  &  $1$  &  $1$  &  $\text{-0.0766(21)}$  &  $\text{-0.0784(32)}$  &  $\text{0.0270(11)}$  &  $\text{-0.0328(11)}$  \\
					$$  &  $$  &  $1$  &  $-1$  &  $\text{-0.0767(21)}$  &  $\text{-0.0783(32)}$  &  $\text{-0.0339(11)}$  &  $\text{0.0269(18)}$  \\
					$$  &  $$  &  $-1$  &  $1$  &  $\text{-0.0768(21)}$  &  $\text{-0.0786(33)}$  &  $\text{0.0344(11)}$  &  $\text{-0.0262(20)}$  \\
					$$  &  $$  &  $-1$  &  $-1$  &  $\text{-0.0769(21)}$  &  $\text{-0.0785(33)}$  &  $\text{-0.0267(11)}$  &  $\text{0.0337(15)}$  \\
					$$  &  $$  &  $1$  &  $0$  &  $\text{-0.0768(21)}$  &  $\text{-0.0785(32)}$  &  $\text{-0.00346(27)}$  &  $\text{-0.0029(13)}$  \\
					$$  &  $$  &  $-1$  &  $0$  &  $\text{-0.0770(22)}$  &  $\text{-0.0787(33)}$  &  $\text{0.00388(29)}$  &  $\text{0.0038(15)}$  \\
					$$  &  $C_{10}'$  &  $1$  &  $1$  &  $\text{-0.0775(22)}$  &  $\text{-0.0789(33)}$  &  $\text{0.00587(43)}$  &  $\text{-0.00499(41)}$  \\
					$$  &  $$  &  $1$  &  $-1$  &  $\text{-0.0772(22)}$  &  $\text{-0.0792(33)}$  &  $\text{-0.00545(42)}$  &  $\text{0.00578(46)}$  \\
					$$  &  $$  &  $-1$  &  $1$  &  $\text{-0.0764(21)}$  &  $\text{-0.0777(32)}$  &  $\text{0.00587(44)}$  &  $\text{-0.00499(41)}$  \\
					$$  &  $$  &  $-1$  &  $-1$  &  $\text{-0.0760(21)}$  &  $\text{-0.0780(32)}$  &  $\text{-0.00546(43)}$  &  $\text{0.00579(46)}$  \\
					$$  &  $$  &  $1$  &  $0$  &  $\text{-0.0774(22)}$  &  $\text{-0.0791(33)}$  &  $\text{0.00021(12)}$  &  $\text{0.00040(29)}$  \\
					$$  &  $$  &  $-1$  &  $0$  &  $\text{-0.0763(21)}$  &  $\text{-0.0780(32)}$  &  $\text{0.00021(12)}$  &  $\text{0.00040(29)}$  \\
					\hline
					$\text{1-2}$  &  $\text{$\Delta $C}_9$  &  $1$  &  $1$  &  $\text{-0.2640(47)}$  &  $\text{-0.257(20)}$  &  $\text{-0.0037(23)}$  &  $\text{0.0084(50)}$  \\
					$$  &  $$  &  $1$  &  $-1$  &  $\text{-0.2417(27)}$  &  $\text{-0.281(14)}$  &  $\text{0.0079(46)}$  &  $\text{-0.0027(21)}$  \\
					$$  &  $$  &  $-1$  &  $1$  &  $\text{-0.2497(47)}$  &  $\text{-0.243(13)}$  &  $\text{-0.0034(20)}$  &  $\text{0.0050(28)}$  \\
					$$  &  $$  &  $-1$  &  $-1$  &  $\text{-0.2355(41)}$  &  $\text{-0.2577(91)}$  &  $\text{0.0043(25)}$  &  $\text{-0.0025(16)}$  \\
					$$  &  $$  &  $1$  &  $0$  &  $\text{-0.2584(35)}$  &  $\text{-0.274(18)}$  &  $\text{0.0024(14)}$  &  $\text{0.0032(22)}$  \\
					$$  &  $$  &  $-1$  &  $0$  &  $\text{-0.2462(46)}$  &  $\text{-0.254(11)}$  &  $\text{0.00060(39)}$  &  $\text{0.00138(96)}$  \\
					$$  &  $\text{$\Delta $C}_{10}$  &  $1$  &  $1$  &  $\text{-0.1798(67)}$  &  $\text{-0.2536(70)}$  &  $\text{0.00149(91)}$  &  $\text{0.0024(16)}$  \\
					$$  &  $$  &  $1$  &  $-1$  &  $\text{-0.2470(79)}$  &  $\text{-0.189(21)}$  &  $\text{0.00149(91)}$  &  $\text{0.0024(16)}$  \\
					$$  &  $$  &  $-1$  &  $1$  &  $\text{-0.2459(20)}$  &  $\text{-0.3097(74)}$  &  $\text{0.00111(68)}$  &  $\text{0.0018(11)}$  \\
					$$  &  $$  &  $-1$  &  $-1$  &  $\text{-0.2962(78)}$  &  $\text{-0.260(21)}$  &  $\text{0.00111(68)}$  &  $\text{0.0018(11)}$  \\
					$$  &  $$  &  $1$  &  $0$  &  $\text{-0.2182(53)}$  &  $\text{-0.226(13)}$  &  $\text{0.00152(93)}$  &  $\text{0.0024(17)}$  \\
					$$  &  $$  &  $-1$  &  $0$  &  $\text{-0.2755(36)}$  &  $\text{-0.290(14)}$  &  $\text{0.00113(69)}$  &  $\text{0.0018(13)}$  \\
					$$  &  $C_9'$  &  $1$  &  $1$  &  $\text{-0.2456(36)}$  &  $\text{-0.257(13)}$  &  $\text{0.0865(43)}$  &  $\text{-0.1169(40)}$  \\
					$$  &  $$  &  $1$  &  $-1$  &  $\text{-0.2460(37)}$  &  $\text{-0.256(13)}$  &  $\text{-0.1191(22)}$  &  $\text{0.0856(81)}$  \\
					$$  &  $$  &  $-1$  &  $1$  &  $\text{-0.2467(35)}$  &  $\text{-0.258(13)}$  &  $\text{0.1220(22)}$  &  $\text{-0.0821(92)}$  \\
					$$  &  $$  &  $-1$  &  $-1$  &  $\text{-0.2472(35)}$  &  $\text{-0.258(13)}$  &  $\text{-0.0845(45)}$  &  $\text{0.1214(50)}$  \\
					$$  &  $$  &  $1$  &  $0$  &  $\text{-0.2504(39)}$  &  $\text{-0.261(14)}$  &  $\text{-0.0165(20)}$  &  $\text{-0.0158(57)}$  \\
					$$  &  $$  &  $-1$  &  $0$  &  $\text{-0.2517(37)}$  &  $\text{-0.263(14)}$  &  $\text{0.0192(23)}$  &  $\text{0.0201(70)}$  \\
					$$  &  $C_{10}'$  &  $1$  &  $1$  &  $\text{-0.2481(38)}$  &  $\text{-0.258(13)}$  &  $\text{0.0726(52)}$  &  $\text{-0.0657(42)}$  \\
					$$  &  $$  &  $1$  &  $-1$  &  $\text{-0.2464(35)}$  &  $\text{-0.259(13)}$  &  $\text{-0.0701(52)}$  &  $\text{0.0698(43)}$  \\
					$$  &  $$  &  $-1$  &  $1$  &  $\text{-0.2463(36)}$  &  $\text{-0.256(13)}$  &  $\text{0.0739(55)}$  &  $\text{-0.0667(42)}$  \\
					$$  &  $$  &  $-1$  &  $-1$  &  $\text{-0.2446(36)}$  &  $\text{-0.257(13)}$  &  $\text{-0.0713(54)}$  &  $\text{0.0709(46)}$  \\
					$$  &  $$  &  $1$  &  $0$  &  $\text{-0.2519(39)}$  &  $\text{-0.263(14)}$  &  $\text{0.00129(78)}$  &  $\text{0.0021(14)}$  \\
					$$  &  $$  &  $-1$  &  $0$  &  $\text{-0.2502(38)}$  &  $\text{-0.261(14)}$  &  $\text{0.00131(80)}$  &  $\text{0.0021(15)}$  \\
					\hline
				\end{tabular}
				\caption{The dependence of some of the optimized observables corresponding to tagged $\bar{B^0}\to\rho^0 ll$ and $B^0\to\rho^0 ll$ on the real and imaginary parts of the NP WC's $\Delta C_{9,10}$ and $C_{9,10}^\prime$ for the $q^2$ bins $0.1-1$ and $1-2$ $\text{GeV}^2$ for a few benchmark values of the real and imaginary parts of the corresponding WC's.}
				\label{tab:appBtorhozP2P3tag1}						
			\end{table}
			
			\begin{table}[ht]
				\centering
				\tiny
				\renewcommand*{\arraystretch}{1.2}
				\begin{tabular}{|c|c|c|c|c|c|c|c|}
					\hline
					\multirow{2}{*}{\textbf{Bin}} & \multirow{2}{*}{$C_i^{NP}$} &\multirow{2}{*}{$Re(C_i^{NP})$} &\multirow{2}{*}{$Im(C_i^{NP})$} & \multicolumn{4}{c|}{\textbf{Observables}}\\
					$$  &  $$  &  $$  &  $$  &  $\overline{P_2^0}$  &  $P_2^0$  &  $\overline{P_3^0}$  &  $P_3^0$  \\
					\hline
					$\text{2-4}$  &  $\text{$\Delta $C}_9$  &  $1$  &  $1$  &  $\text{0.032(27)}$  &  $\text{0.024(26)}$  &  $\text{-0.0037(23)}$  &  $\text{0.0112(60)}$  \\
					$$  &  $$  &  $1$  &  $-1$  &  $\text{0.029(25)}$  &  $\text{0.028(30)}$  &  $\text{0.0105(56)}$  &  $\text{-0.0028(25)}$  \\
					$$  &  $$  &  $-1$  &  $1$  &  $\text{-0.219(16)}$  &  $\text{-0.197(25)}$  &  $\text{-0.0050(28)}$  &  $\text{0.0084(46)}$  \\
					$$  &  $$  &  $-1$  &  $-1$  &  $\text{-0.199(12)}$  &  $\text{-0.221(21)}$  &  $\text{0.0073(40)}$  &  $\text{-0.0039(23)}$  \\
					$$  &  $$  &  $1$  &  $0$  &  $\text{0.032(27)}$  &  $\text{0.028(29)}$  &  $\text{0.0040(22)}$  &  $\text{0.0049(30)}$  \\
					$$  &  $$  &  $-1$  &  $0$  &  $\text{-0.219(14)}$  &  $\text{-0.218(25)}$  &  $\text{0.00150(91)}$  &  $\text{0.0027(17)}$  \\
					$$  &  $\text{$\Delta $C}_{10}$  &  $1$  &  $1$  &  $\text{-0.049(16)}$  &  $\text{-0.201(19)}$  &  $\text{0.0042(23)}$  &  $\text{0.0059(35)}$  \\
					$$  &  $$  &  $1$  &  $-1$  &  $\text{-0.195(36)}$  &  $\text{-0.045(51)}$  &  $\text{0.0042(23)}$  &  $\text{0.0059(35)}$  \\
					$$  &  $$  &  $-1$  &  $1$  &  $\text{-0.057(14)}$  &  $\text{-0.124(18)}$  &  $\text{0.0018(10)}$  &  $\text{0.0025(16)}$  \\
					$$  &  $$  &  $-1$  &  $-1$  &  $\text{-0.120(25)}$  &  $\text{-0.057(32)}$  &  $\text{0.0018(10)}$  &  $\text{0.0025(16)}$  \\
					$$  &  $$  &  $1$  &  $0$  &  $\text{-0.133(24)}$  &  $\text{-0.135(37)}$  &  $\text{0.0045(26)}$  &  $\text{0.0064(39)}$  \\
					$$  &  $$  &  $-1$  &  $0$  &  $\text{-0.092(20)}$  &  $\text{-0.094(25)}$  &  $\text{0.0019(11)}$  &  $\text{0.0026(16)}$  \\
					$$  &  $C_9'$  &  $1$  &  $1$  &  $\text{-0.099(20)}$  &  $\text{-0.101(27)}$  &  $\text{0.0178(63)}$  &  $\text{-0.0691(72)}$  \\
					$$  &  $$  &  $1$  &  $-1$  &  $\text{-0.100(20)}$  &  $\text{-0.101(27)}$  &  $\text{-0.0642(88)}$  &  $\text{0.013(14)}$  \\
					$$  &  $$  &  $-1$  &  $1$  &  $\text{-0.098(20)}$  &  $\text{-0.100(27)}$  &  $\text{0.0679(92)}$  &  $\text{-0.0056(154)}$  \\
					$$  &  $$  &  $-1$  &  $-1$  &  $\text{-0.098(20)}$  &  $\text{-0.099(27)}$  &  $\text{-0.0126(70)}$  &  $\text{0.0746(77)}$  \\
					$$  &  $$  &  $1$  &  $0$  &  $\text{-0.105(21)}$  &  $\text{-0.107(28)}$  &  $\text{-0.0244(52)}$  &  $\text{-0.0295(86)}$  \\
					$$  &  $$  &  $-1$  &  $0$  &  $\text{-0.103(21)}$  &  $\text{-0.105(28)}$  &  $\text{0.0292(60)}$  &  $\text{0.036(10)}$  \\
					$$  &  $C_{10}'$  &  $1$  &  $1$  &  $\text{-0.102(20)}$  &  $\text{-0.101(26)}$  &  $\text{0.1973(51)}$  &  $\text{-0.1887(66)}$  \\
					$$  &  $$  &  $1$  &  $-1$  &  $\text{-0.099(20)}$  &  $\text{-0.103(26)}$  &  $\text{-0.1924(67)}$  &  $\text{0.1956(41)}$  \\
					$$  &  $$  &  $-1$  &  $1$  &  $\text{-0.098(21)}$  &  $\text{-0.097(28)}$  &  $\text{0.2080(69)}$  &  $\text{-0.1987(60)}$  \\
					$$  &  $$  &  $-1$  &  $-1$  &  $\text{-0.095(21)}$  &  $\text{-0.100(27)}$  &  $\text{-0.2029(67)}$  &  $\text{0.2061(63)}$  \\
					$$  &  $$  &  $1$  &  $0$  &  $\text{-0.106(21)}$  &  $\text{-0.107(28)}$  &  $\text{0.0025(15)}$  &  $\text{0.0036(22)}$  \\
					$$  &  $$  &  $-1$  &  $0$  &  $\text{-0.102(22)}$  &  $\text{-0.104(29)}$  &  $\text{0.0027(16)}$  &  $\text{0.0039(23)}$  \\
					\hline
					$\text{4-6}$  &  $\text{$\Delta $C}_9$  &  $1$  &  $1$  &  $\text{0.234(14)}$  &  $\text{0.2532(81)}$  &  $\text{-0.00053(103)}$  &  $\text{0.0067(33)}$  \\
					$$  &  $$  &  $1$  &  $-1$  &  $\text{0.216(18)}$  &  $\text{0.280(14)}$  &  $\text{0.0068(34)}$  &  $\text{-0.000057(1290)}$  \\
					$$  &  $$  &  $-1$  &  $1$  &  $\text{-0.0064(201)}$  &  $\text{0.057(24)}$  &  $\text{-0.0028(16)}$  &  $\text{0.0071(35)}$  \\
					$$  &  $$  &  $-1$  &  $-1$  &  $\text{-0.0053(179)}$  &  $\text{0.065(29)}$  &  $\text{0.0066(33)}$  &  $\text{-0.0019(15)}$  \\
					$$  &  $$  &  $1$  &  $0$  &  $\text{0.236(17)}$  &  $\text{0.278(11)}$  &  $\text{0.0035(19)}$  &  $\text{0.0037(20)}$  \\
					$$  &  $$  &  $-1$  &  $0$  &  $\text{-0.0062(201)}$  &  $\text{0.064(27)}$  &  $\text{0.0023(13)}$  &  $\text{0.0031(17)}$  \\
					$$  &  $\text{$\Delta $C}_{10}$  &  $1$  &  $1$  &  $\text{0.2207(87)}$  &  $\text{0.147(10)}$  &  $\text{0.0047(26)}$  &  $\text{0.0053(28)}$  \\
					$$  &  $$  &  $1$  &  $-1$  &  $\text{0.087(41)}$  &  $\text{0.285(32)}$  &  $\text{0.0047(26)}$  &  $\text{0.0053(28)}$  \\
					$$  &  $$  &  $-1$  &  $1$  &  $\text{0.1379(91)}$  &  $\text{0.129(12)}$  &  $\text{0.0020(11)}$  &  $\text{0.0023(13)}$  \\
					$$  &  $$  &  $-1$  &  $-1$  &  $\text{0.081(23)}$  &  $\text{0.190(25)}$  &  $\text{0.0020(11)}$  &  $\text{0.0023(13)}$  \\
					$$  &  $$  &  $1$  &  $0$  &  $\text{0.169(27)}$  &  $\text{0.236(18)}$  &  $\text{0.0051(28)}$  &  $\text{0.0058(31)}$  \\
					$$  &  $$  &  $-1$  &  $0$  &  $\text{0.114(17)}$  &  $\text{0.165(18)}$  &  $\text{0.0021(11)}$  &  $\text{0.0024(13)}$  \\
					$$  &  $C_9'$  &  $1$  &  $1$  &  $\text{0.125(19)}$  &  $\text{0.181(19)}$  &  $\text{-0.0789(47)}$  &  $\text{0.0410(57)}$  \\
					$$  &  $$  &  $1$  &  $-1$  &  $\text{0.125(19)}$  &  $\text{0.179(18)}$  &  $\text{0.0381(95)}$  &  $\text{-0.0912(88)}$  \\
					$$  &  $$  &  $-1$  &  $1$  &  $\text{0.120(18)}$  &  $\text{0.174(18)}$  &  $\text{-0.031(10)}$  &  $\text{0.095(10)}$  \\
					$$  &  $$  &  $-1$  &  $-1$  &  $\text{0.121(18)}$  &  $\text{0.173(18)}$  &  $\text{0.0819(57)}$  &  $\text{-0.0330(61)}$  \\
					$$  &  $$  &  $1$  &  $0$  &  $\text{0.132(20)}$  &  $\text{0.190(19)}$  &  $\text{-0.0216(66)}$  &  $\text{-0.0267(65)}$  \\
					$$  &  $$  &  $-1$  &  $0$  &  $\text{0.127(19)}$  &  $\text{0.183(19)}$  &  $\text{0.0266(77)}$  &  $\text{0.0323(77)}$  \\
					$$  &  $C_{10}'$  &  $1$  &  $1$  &  $\text{0.115(18)}$  &  $\text{0.170(19)}$  &  $\text{0.1994(31)}$  &  $\text{-0.1876(80)}$  \\
					$$  &  $$  &  $1$  &  $-1$  &  $\text{0.118(17)}$  &  $\text{0.168(18)}$  &  $\text{-0.1940(57)}$  &  $\text{0.1938(55)}$  \\
					$$  &  $$  &  $-1$  &  $1$  &  $\text{0.128(20)}$  &  $\text{0.186(19)}$  &  $\text{0.2125(47)}$  &  $\text{-0.1996(81)}$  \\
					$$  &  $$  &  $-1$  &  $-1$  &  $\text{0.131(19)}$  &  $\text{0.184(18)}$  &  $\text{-0.2068(47)}$  &  $\text{0.2063(74)}$  \\
					$$  &  $$  &  $1$  &  $0$  &  $\text{0.123(19)}$  &  $\text{0.178(19)}$  &  $\text{0.0028(15)}$  &  $\text{0.0033(18)}$  \\
					$$  &  $$  &  $-1$  &  $0$  &  $\text{0.137(21)}$  &  $\text{0.195(19)}$  &  $\text{0.0031(17)}$  &  $\text{0.0035(20)}$  \\
					\hline
				\end{tabular}
				\caption{The dependence of some of the optimized observables corresponding to tagged $\bar{B^0}\to\rho^0 ll$ and $B^0\to\rho^0 ll$ on the real and imaginary parts of the NP WC's $\Delta C_{9,10}$ and $C_{9,10}^\prime$ for the $q^2$ bins $2-4$ and $4-6$ $\text{GeV}^2$ for a few benchmark values of the real and imaginary parts of the corresponding WC's.}
				\label{tab:appBtorhozP2P3tag2}										
			\end{table}
			
			\begin{table}[ht]
				\centering
				\tiny
				\renewcommand*{\arraystretch}{1.2}
				\begin{tabular}{|c|c|c|c|c|c|c|c|}
					\hline
					\multirow{2}{*}{\textbf{Bin}} & \multirow{2}{*}{$C_i^{NP}$} &\multirow{2}{*}{$Re(C_i^{NP})$} &\multirow{2}{*}{$Im(C_i^{NP})$} & \multicolumn{4}{c|}{\textbf{Observables}}\\
					$$  &  $$  &  $$  &  $$  &  $\overline{P_6^{'0}}$  &  $P_6^{'0}$  &  $\overline{P_8^{'0}}$  &  $P_8^{'0}$  \\
					\hline
					$\text{0.1-1}$  &  $\text{$\Delta $C}_9$  &  $1$  &  $1$  &  $\text{-0.131(20)}$  &  $\text{-0.110(33)}$  &  $\text{0.0924(28)}$  &  $\text{-0.148(20)}$  \\
					$$  &  $$  &  $1$  &  $-1$  &  $\text{-0.134(21)}$  &  $\text{-0.113(35)}$  &  $\text{-0.0900(33)}$  &  $\text{0.026(22)}$  \\
					$$  &  $$  &  $-1$  &  $1$  &  $\text{-0.155(24)}$  &  $\text{-0.132(40)}$  &  $\text{0.1345(45)}$  &  $\text{-0.159(16)}$  \\
					$$  &  $$  &  $-1$  &  $-1$  &  $\text{-0.161(25)}$  &  $\text{-0.136(42)}$  &  $\text{-0.0821(54)}$  &  $\text{0.050(18)}$  \\
					$$  &  $$  &  $1$  &  $0$  &  $\text{-0.135(21)}$  &  $\text{-0.113(35)}$  &  $\text{0.0021(32)}$  &  $\text{-0.063(21)}$  \\
					$$  &  $$  &  $-1$  &  $0$  &  $\text{-0.162(25)}$  &  $\text{-0.137(42)}$  &  $\text{0.0288(53)}$  &  $\text{-0.057(18)}$  \\
					$$  &  $\text{$\Delta $C}_{10}$  &  $1$  &  $1$  &  $\text{-0.278(22)}$  &  $\text{0.050(35)}$  &  $\text{0.0164(47)}$  &  $\text{-0.068(22)}$  \\
					$$  &  $$  &  $1$  &  $-1$  &  $\text{0.027(18)}$  &  $\text{-0.258(29)}$  &  $\text{0.0164(47)}$  &  $\text{-0.068(22)}$  \\
					$$  &  $$  &  $-1$  &  $1$  &  $\text{-0.270(26)}$  &  $\text{-0.018(44)}$  &  $\text{0.0121(35)}$  &  $\text{-0.052(17)}$  \\
					$$  &  $$  &  $-1$  &  $-1$  &  $\text{-0.045(23)}$  &  $\text{-0.251(38)}$  &  $\text{0.0121(35)}$  &  $\text{-0.052(17)}$  \\
					$$  &  $$  &  $1$  &  $0$  &  $\text{-0.129(20)}$  &  $\text{-0.107(33)}$  &  $\text{0.0168(49)}$  &  $\text{-0.070(23)}$  \\
					$$  &  $$  &  $-1$  &  $0$  &  $\text{-0.160(25)}$  &  $\text{-0.136(42)}$  &  $\text{0.0122(35)}$  &  $\text{-0.053(17)}$  \\
					$$  &  $C_9'$  &  $1$  &  $1$  &  $\text{-0.166(26)}$  &  $\text{-0.140(43)}$  &  $\text{-0.0625(47)}$  &  $\text{0.027(18)}$  \\
					$$  &  $$  &  $1$  &  $-1$  &  $\text{-0.159(24)}$  &  $\text{-0.137(41)}$  &  $\text{0.1105(51)}$  &  $\text{-0.137(16)}$  \\
					$$  &  $$  &  $-1$  &  $1$  &  $\text{-0.130(20)}$  &  $\text{-0.109(33)}$  &  $\text{-0.0652(28)}$  &  $\text{0.0031(211)}$  \\
					$$  &  $$  &  $-1$  &  $-1$  &  $\text{-0.127(20)}$  &  $\text{-0.108(33)}$  &  $\text{0.0722(35)}$  &  $\text{-0.125(20)}$  \\
					$$  &  $$  &  $1$  &  $0$  &  $\text{-0.167(26)}$  &  $\text{-0.141(43)}$  &  $\text{0.0264(52)}$  &  $\text{-0.057(18)}$  \\
					$$  &  $$  &  $-1$  &  $0$  &  $\text{-0.130(20)}$  &  $\text{-0.110(34)}$  &  $\text{0.0045(33)}$  &  $\text{-0.062(21)}$  \\
					$$  &  $C_{10}'$  &  $1$  &  $1$  &  $\text{-0.045(23)}$  &  $\text{-0.253(39)}$  &  $\text{-0.0116(39)}$  &  $\text{-0.030(17)}$  \\
					$$  &  $$  &  $1$  &  $-1$  &  $\text{-0.271(26)}$  &  $\text{-0.018(44)}$  &  $\text{0.0358(32)}$  &  $\text{-0.074(17)}$  \\
					$$  &  $$  &  $-1$  &  $1$  &  $\text{0.027(17)}$  &  $\text{-0.257(29)}$  &  $\text{-0.0156(52)}$  &  $\text{-0.039(22)}$  \\
					$$  &  $$  &  $-1$  &  $-1$  &  $\text{-0.277(22)}$  &  $\text{0.049(35)}$  &  $\text{0.0481(45)}$  &  $\text{-0.097(22)}$  \\
					$$  &  $$  &  $1$  &  $0$  &  $\text{-0.161(25)}$  &  $\text{-0.137(42)}$  &  $\text{0.0123(35)}$  &  $\text{-0.053(17)}$  \\
					$$  &  $$  &  $-1$  &  $0$  &  $\text{-0.128(20)}$  &  $\text{-0.106(33)}$  &  $\text{0.0167(49)}$  &  $\text{-0.070(23)}$  \\
					\hline
					$\text{1-2}$  &  $\text{$\Delta $C}_9$  &  $1$  &  $1$  &  $\text{-0.181(32)}$  &  $\text{-0.154(40)}$  &  $\text{0.0798(52)}$  &  $\text{-0.174(19)}$  \\
					$$  &  $$  &  $1$  &  $-1$  &  $\text{-0.175(31)}$  &  $\text{-0.162(45)}$  &  $\text{-0.1536(28)}$  &  $\text{0.056(25)}$  \\
					$$  &  $$  &  $-1$  &  $1$  &  $\text{-0.185(33)}$  &  $\text{-0.161(43)}$  &  $\text{0.1122(14)}$  &  $\text{-0.157(13)}$  \\
					$$  &  $$  &  $-1$  &  $-1$  &  $\text{-0.182(33)}$  &  $\text{-0.167(47)}$  &  $\text{-0.1294(16)}$  &  $\text{0.083(16)}$  \\
					$$  &  $$  &  $1$  &  $0$  &  $\text{-0.183(32)}$  &  $\text{-0.162(43)}$  &  $\text{-0.0399(42)}$  &  $\text{-0.063(23)}$  \\
					$$  &  $$  &  $-1$  &  $0$  &  $\text{-0.190(34)}$  &  $\text{-0.169(46)}$  &  $\text{-0.0097(13)}$  &  $\text{-0.040(16)}$  \\
					$$  &  $\text{$\Delta $C}_{10}$  &  $1$  &  $1$  &  $\text{-0.349(31)}$  &  $\text{0.047(45)}$  &  $\text{-0.0304(30)}$  &  $\text{-0.062(23)}$  \\
					$$  &  $$  &  $1$  &  $-1$  &  $\text{0.014(28)}$  &  $\text{-0.340(34)}$  &  $\text{-0.0304(30)}$  &  $\text{-0.062(23)}$  \\
					$$  &  $$  &  $-1$  &  $1$  &  $\text{-0.300(34)}$  &  $\text{-0.036(48)}$  &  $\text{-0.0196(20)}$  &  $\text{-0.041(16)}$  \\
					$$  &  $$  &  $-1$  &  $-1$  &  $\text{-0.066(32)}$  &  $\text{-0.292(41)}$  &  $\text{-0.0196(20)}$  &  $\text{-0.041(16)}$  \\
					$$  &  $$  &  $1$  &  $0$  &  $\text{-0.174(31)}$  &  $\text{-0.151(41)}$  &  $\text{-0.0316(31)}$  &  $\text{-0.064(24)}$  \\
					$$  &  $$  &  $-1$  &  $0$  &  $\text{-0.187(34)}$  &  $\text{-0.168(46)}$  &  $\text{-0.0200(20)}$  &  $\text{-0.042(16)}$  \\
					$$  &  $C_9'$  &  $1$  &  $1$  &  $\text{-0.203(36)}$  &  $\text{-0.180(49)}$  &  $\text{-0.0563(35)}$  &  $\text{-0.0037(170)}$  \\
					$$  &  $$  &  $1$  &  $-1$  &  $\text{-0.200(35)}$  &  $\text{-0.179(48)}$  &  $\text{0.0255(38)}$  &  $\text{-0.079(15)}$  \\
					$$  &  $$  &  $-1$  &  $1$  &  $\text{-0.160(29)}$  &  $\text{-0.141(38)}$  &  $\text{-0.0634(49)}$  &  $\text{-0.026(22)}$  \\
					$$  &  $$  &  $-1$  &  $-1$  &  $\text{-0.158(28)}$  &  $\text{-0.141(38)}$  &  $\text{0.0013(35)}$  &  $\text{-0.086(20)}$  \\
					$$  &  $$  &  $1$  &  $0$  &  $\text{-0.208(37)}$  &  $\text{-0.185(50)}$  &  $\text{-0.0156(15)}$  &  $\text{-0.043(16)}$  \\
					$$  &  $$  &  $-1$  &  $0$  &  $\text{-0.163(29)}$  &  $\text{-0.144(39)}$  &  $\text{-0.0317(34)}$  &  $\text{-0.058(22)}$  \\
					$$  &  $C_{10}'$  &  $1$  &  $1$  &  $\text{-0.070(34)}$  &  $\text{-0.310(43)}$  &  $\text{-0.1070(34)}$  &  $\text{0.038(17)}$  \\
					$$  &  $$  &  $1$  &  $-1$  &  $\text{-0.320(36)}$  &  $\text{-0.039(51)}$  &  $\text{0.0653(43)}$  &  $\text{-0.126(17)}$  \\
					$$  &  $$  &  $-1$  &  $1$  &  $\text{0.013(26)}$  &  $\text{-0.317(32)}$  &  $\text{-0.1448(44)}$  &  $\text{0.051(22)}$  \\
					$$  &  $$  &  $-1$  &  $-1$  &  $\text{-0.324(30)}$  &  $\text{0.044(42)}$  &  $\text{0.0883(56)}$  &  $\text{-0.167(22)}$  \\
					$$  &  $$  &  $1$  &  $0$  &  $\text{-0.200(35)}$  &  $\text{-0.179(48)}$  &  $\text{-0.0214(21)}$  &  $\text{-0.045(17)}$  \\
					$$  &  $$  &  $-1$  &  $0$  &  $\text{-0.161(29)}$  &  $\text{-0.141(38)}$  &  $\text{-0.0293(30)}$  &  $\text{-0.060(23)}$  \\
					\hline
				\end{tabular}
				\caption{The dependence of some of the optimized observables corresponding to tagged $\bar{B^0}\to\rho^0 ll$ and $B^0\to\rho^0 ll$ on the real and imaginary parts of the NP WC's $\Delta C_{9,10}$ and $C_{9,10}^\prime$ for the $q^2$ bins $0.1-1$ and $1-2$ $\text{GeV}^2$ for a few benchmark values of the real and imaginary parts of the corresponding WC's.}
				\label{tab:appBtorhozP6P8tag1}										
			\end{table}

			\begin{table}[ht]
				\centering
				\tiny
				\renewcommand*{\arraystretch}{1.2}
				\begin{tabular}{|c|c|c|c|c|c|c|c|}
					\hline
					\multirow{2}{*}{\textbf{Bin}} & \multirow{2}{*}{$C_i^{NP}$} &\multirow{2}{*}{$Re(C_i^{NP})$} &\multirow{2}{*}{$Im(C_i^{NP})$} & \multicolumn{4}{c|}{\textbf{Observables}}\\
					$$  &  $$  &  $$  &  $$  &  $\overline{P_6^{'0}}$  &  $P_6^{'0}$  &  $\overline{P_8^{'0}}$  &  $P_8^{'0}$  \\
					\hline
					$\text{2-4}$  &  $\text{$\Delta $C}_9$  &  $1$  &  $1$  &  $\text{-0.155(36)}$  &  $\text{-0.140(33)}$  &  $\text{0.0426(88)}$  &  $\text{-0.134(12)}$  \\
					$$  &  $$  &  $1$  &  $-1$  &  $\text{-0.147(33)}$  &  $\text{-0.151(39)}$  &  $\text{-0.1219(82)}$  &  $\text{0.031(19)}$  \\
					$$  &  $$  &  $-1$  &  $1$  &  $\text{-0.180(41)}$  &  $\text{-0.166(41)}$  &  $\text{0.0787(45)}$  &  $\text{-0.1321(79)}$  \\
					$$  &  $$  &  $-1$  &  $-1$  &  $\text{-0.173(38)}$  &  $\text{-0.178(47)}$  &  $\text{-0.1158(35)}$  &  $\text{0.064(14)}$  \\
					$$  &  $$  &  $1$  &  $0$  &  $\text{-0.157(36)}$  &  $\text{-0.152(37)}$  &  $\text{-0.0435(89)}$  &  $\text{-0.057(16)}$  \\
					$$  &  $$  &  $-1$  &  $0$  &  $\text{-0.185(41)}$  &  $\text{-0.180(46)}$  &  $\text{-0.0214(42)}$  &  $\text{-0.039(12)}$  \\
					$$  &  $\text{$\Delta $C}_{10}$  &  $1$  &  $1$  &  $\text{-0.347(42)}$  &  $\text{0.0032(523)}$  &  $\text{-0.0479(92)}$  &  $\text{-0.069(20)}$  \\
					$$  &  $$  &  $1$  &  $-1$  &  $\text{-0.015(38)}$  &  $\text{-0.351(33)}$  &  $\text{-0.0479(92)}$  &  $\text{-0.069(20)}$  \\
					$$  &  $$  &  $-1$  &  $1$  &  $\text{-0.232(36)}$  &  $\text{-0.059(41)}$  &  $\text{-0.0235(47)}$  &  $\text{-0.035(10)}$  \\
					$$  &  $$  &  $-1$  &  $-1$  &  $\text{-0.069(33)}$  &  $\text{-0.235(33)}$  &  $\text{-0.0235(47)}$  &  $\text{-0.035(10)}$  \\
					$$  &  $$  &  $1$  &  $0$  &  $\text{-0.194(42)}$  &  $\text{-0.186(45)}$  &  $\text{-0.0515(99)}$  &  $\text{-0.074(22)}$  \\
					$$  &  $$  &  $-1$  &  $0$  &  $\text{-0.156(36)}$  &  $\text{-0.152(38)}$  &  $\text{-0.0243(50)}$  &  $\text{-0.036(11)}$  \\
					$$  &  $C_9'$  &  $1$  &  $1$  &  $\text{-0.185(42)}$  &  $\text{-0.180(45)}$  &  $\text{0.0273(83)}$  &  $\text{-0.092(12)}$  \\
					$$  &  $$  &  $1$  &  $-1$  &  $\text{-0.183(41)}$  &  $\text{-0.178(44)}$  &  $\text{-0.0788(47)}$  &  $\text{0.018(11)}$  \\
					$$  &  $$  &  $-1$  &  $1$  &  $\text{-0.144(33)}$  &  $\text{-0.140(35)}$  &  $\text{0.0053(99)}$  &  $\text{-0.097(17)}$  \\
					$$  &  $$  &  $-1$  &  $-1$  &  $\text{-0.143(32)}$  &  $\text{-0.139(35)}$  &  $\text{-0.0774(64)}$  &  $\text{-0.011(16)}$  \\
					$$  &  $$  &  $1$  &  $0$  &  $\text{-0.194(44)}$  &  $\text{-0.188(47)}$  &  $\text{-0.0273(52)}$  &  $\text{-0.039(11)}$  \\
					$$  &  $$  &  $-1$  &  $0$  &  $\text{-0.150(34)}$  &  $\text{-0.145(36)}$  &  $\text{-0.0377(78)}$  &  $\text{-0.056(17)}$  \\
					$$  &  $C_{10}'$  &  $1$  &  $1$  &  $\text{-0.081(38)}$  &  $\text{-0.274(38)}$  &  $\text{-0.1684(57)}$  &  $\text{0.098(13)}$  \\
					$$  &  $$  &  $1$  &  $-1$  &  $\text{-0.271(42)}$  &  $\text{-0.069(48)}$  &  $\text{0.1135(73)}$  &  $\text{-0.178(12)}$  \\
					$$  &  $$  &  $-1$  &  $1$  &  $\text{-0.012(30)}$  &  $\text{-0.276(28)}$  &  $\text{-0.2311(67)}$  &  $\text{0.133(18)}$  \\
					$$  &  $$  &  $-1$  &  $-1$  &  $\text{-0.273(35)}$  &  $\text{0.0023(414)}$  &  $\text{0.1558(94)}$  &  $\text{-0.242(16)}$  \\
					$$  &  $$  &  $1$  &  $0$  &  $\text{-0.183(41)}$  &  $\text{-0.178(44)}$  &  $\text{-0.0286(57)}$  &  $\text{-0.042(12)}$  \\
					$$  &  $$  &  $-1$  &  $0$  &  $\text{-0.150(34)}$  &  $\text{-0.144(36)}$  &  $\text{-0.0397(80)}$  &  $\text{-0.058(17)}$  \\
					\hline
					$\text{4-6}$  &  $\text{$\Delta $C}_9$  &  $1$  &  $1$  &  $\text{-0.129(36)}$  &  $\text{-0.118(30)}$  &  $\text{0.0027(110)}$  &  $\text{-0.0787(79)}$  \\
					$$  &  $$  &  $1$  &  $-1$  &  $\text{-0.124(34)}$  &  $\text{-0.125(33)}$  &  $\text{-0.078(11)}$  &  $\text{-0.0055(138)}$  \\
					$$  &  $$  &  $-1$  &  $1$  &  $\text{-0.185(51)}$  &  $\text{-0.171(44)}$  &  $\text{0.0333(76)}$  &  $\text{-0.0868(50)}$  \\
					$$  &  $$  &  $-1$  &  $-1$  &  $\text{-0.177(47)}$  &  $\text{-0.183(50)}$  &  $\text{-0.0825(68)}$  &  $\text{0.021(12)}$  \\
					$$  &  $$  &  $1$  &  $0$  &  $\text{-0.131(36)}$  &  $\text{-0.126(33)}$  &  $\text{-0.040(11)}$  &  $\text{-0.045(11)}$  \\
					$$  &  $$  &  $-1$  &  $0$  &  $\text{-0.190(52)}$  &  $\text{-0.187(49)}$  &  $\text{-0.0272(77)}$  &  $\text{-0.0368(92)}$  \\
					$$  &  $\text{$\Delta $C}_{10}$  &  $1$  &  $1$  &  $\text{-0.265(49)}$  &  $\text{-0.079(51)}$  &  $\text{-0.050(14)}$  &  $\text{-0.060(14)}$  \\
					$$  &  $$  &  $1$  &  $-1$  &  $\text{-0.071(42)}$  &  $\text{-0.242(31)}$  &  $\text{-0.050(14)}$  &  $\text{-0.060(14)}$  \\
					$$  &  $$  &  $-1$  &  $1$  &  $\text{-0.186(41)}$  &  $\text{-0.095(41)}$  &  $\text{-0.0246(70)}$  &  $\text{-0.0298(76)}$  \\
					$$  &  $$  &  $-1$  &  $-1$  &  $\text{-0.092(37)}$  &  $\text{-0.177(32)}$  &  $\text{-0.0246(70)}$  &  $\text{-0.0298(76)}$  \\
					$$  &  $$  &  $1$  &  $0$  &  $\text{-0.180(49)}$  &  $\text{-0.172(43)}$  &  $\text{-0.054(15)}$  &  $\text{-0.064(15)}$  \\
					$$  &  $$  &  $-1$  &  $0$  &  $\text{-0.144(40)}$  &  $\text{-0.141(37)}$  &  $\text{-0.0255(73)}$  &  $\text{-0.0308(79)}$  \\
					$$  &  $C_9'$  &  $1$  &  $1$  &  $\text{-0.172(47)}$  &  $\text{-0.167(44)}$  &  $\text{0.0735(99)}$  &  $\text{-0.1392(87)}$  \\
					$$  &  $$  &  $1$  &  $-1$  &  $\text{-0.170(47)}$  &  $\text{-0.166(43)}$  &  $\text{-0.1325(70)}$  &  $\text{0.0743(82)}$  \\
					$$  &  $$  &  $-1$  &  $1$  &  $\text{-0.133(37)}$  &  $\text{-0.128(34)}$  &  $\text{0.044(12)}$  &  $\text{-0.128(12)}$  \\
					$$  &  $$  &  $-1$  &  $-1$  &  $\text{-0.132(36)}$  &  $\text{-0.128(34)}$  &  $\text{-0.1151(97)}$  &  $\text{0.036(12)}$  \\
					$$  &  $$  &  $1$  &  $0$  &  $\text{-0.180(49)}$  &  $\text{-0.175(46)}$  &  $\text{-0.0316(85)}$  &  $\text{-0.0338(85)}$  \\
					$$  &  $$  &  $-1$  &  $0$  &  $\text{-0.138(38)}$  &  $\text{-0.133(35)}$  &  $\text{-0.037(11)}$  &  $\text{-0.048(12)}$  \\
					$$  &  $C_{10}'$  &  $1$  &  $1$  &  $\text{-0.107(43)}$  &  $\text{-0.205(36)}$  &  $\text{-0.1689(85)}$  &  $\text{0.102(11)}$  \\
					$$  &  $$  &  $1$  &  $-1$  &  $\text{-0.218(47)}$  &  $\text{-0.111(47)}$  &  $\text{0.1113(92)}$  &  $\text{-0.1717(81)}$  \\
					$$  &  $$  &  $-1$  &  $1$  &  $\text{-0.056(33)}$  &  $\text{-0.192(26)}$  &  $\text{-0.232(11)}$  &  $\text{0.140(14)}$  \\
					$$  &  $$  &  $-1$  &  $-1$  &  $\text{-0.207(39)}$  &  $\text{-0.063(41)}$  &  $\text{0.153(12)}$  &  $\text{-0.234(10)}$  \\
					$$  &  $$  &  $1$  &  $0$  &  $\text{-0.169(47)}$  &  $\text{-0.164(43)}$  &  $\text{-0.0299(85)}$  &  $\text{-0.0360(91)}$  \\
					$$  &  $$  &  $-1$  &  $0$  &  $\text{-0.139(38)}$  &  $\text{-0.134(35)}$  &  $\text{-0.042(12)}$  &  $\text{-0.050(12)}$  \\
					\hline
				\end{tabular}
				\caption{The dependence of some of the optimized observables corresponding to tagged $\bar{B^0}\to\rho^0 ll$ and $B^0\to\rho^0 ll$ on the real and imaginary parts of the NP WC's $\Delta C_{9,10}$ and $C_{9,10}^\prime$ for the $q^2$ bins $2-4$ and $4-6$ $\text{GeV}^2$ for a few benchmark values of the real and imaginary parts of the corresponding WC's.}
				\label{tab:appBtorhozP6P8tag2}										
			\end{table}
			
			
			\begin{table}[ht]
				\centering
				\tiny
				\renewcommand*{\arraystretch}{1.2}
\begin{tabular}{|c|c|c|c|c|c|c|}
					\hline
					\multirow{2}{*}{\textbf{Bin}} & \multirow{2}{*}{$C_i^{NP}$} &\multirow{2}{*}{$Re(C_i^{NP})$} &\multirow{2}{*}{$Im(C_i^{NP})$} & \multicolumn{3}{c|}{\textbf{Observables}}\\
					\cline{5-7}
					$$  &  $$  &  $$  &  $$  &  $P_1$  &  $P_4^\prime$  &  $P_6^\prime$  \\\hline		
					$\text{0.1-1}$  &  $\text{$\Delta $C}_ 9$  &  $1$  &  $1$  &  $\text{0.0037(28)}$  &  $\text{0.2265(54)}$  &  $\text{-0.099(22)}$  \\
					$$  &  $$  &  $1$  &  $-1$  &  $\text{0.0037(28)}$  &  $\text{0.2326(34)}$  &  $\text{-0.101(23)}$  \\
					$$  &  $$  &  $-1$  &  $1$  &  $\text{0.0017(16)}$  &  $\text{0.1430(59)}$  &  $\text{-0.121(27)}$  \\
					$$  &  $$  &  $-1$  &  $-1$  &  $\text{0.0017(16)}$  &  $\text{0.1494(31)}$  &  $\text{-0.125(29)}$  \\
					$$  &  $$  &  $1$  &  $0$  &  $\text{0.0038(28)}$  &  $\text{0.2361(42)}$  &  $\text{-0.102(23)}$  \\
					$$  &  $$  &  $-1$  &  $0$  &  $\text{0.0017(17)}$  &  $\text{0.1533(44)}$  &  $\text{-0.126(29)}$  \\
\hline
				\end{tabular}
				\caption{The dependence of the optimized observables corresponding to $\bar{B}(B)\to\rho^0 ll$ on the real and imaginary parts of the NP WC's $\Delta C_{9,10}$ and $C_{9,10}^\prime$ for the $q^2$ bins $0.1-1$ and $1-2$ $\text{GeV}^2$ for a few benchmark values of the real and imaginary parts of the corresponding WC's.}
				\label{tab:appBtorhozP1P4P6notag1}						
			\end{table}
			
			\begin{table}[ht]
				\centering
				\tiny
				\renewcommand*{\arraystretch}{1.2}
				\begin{tabular}{|c|c|c|c|c|c|c|}
					\hline
					\multirow{2}{*}{\textbf{Bin}} & \multirow{2}{*}{$C_i^{NP}$} &\multirow{2}{*}{$Re(C_i^{NP})$} &\multirow{2}{*}{$Im(C_i^{NP})$} & \multicolumn{3}{c|}{\textbf{Observables}}\\
					\cline{5-7}
					$$  &  $$  &  $$  &  $$  &  $P_1$  &  $P_4^\prime$  &  $P_6^\prime$  \\\hline		
					$\text{2-4}$  &  $\text{$\Delta $C}_ 9$  &  $1$  &  $1$  &  $\text{-0.056(25)}$  &  $\text{-0.331(15)}$  &  $\text{-0.138(33)}$  \\
					$$  &  $$  &  $1$  &  $-1$  &  $\text{-0.056(25)}$  &  $\text{-0.330(15)}$  &  $\text{-0.140(34)}$  \\
					$$  &  $$  &  $-1$  &  $1$  &  $\text{-0.036(15)}$  &  $\text{-0.280(13)}$  &  $\text{-0.167(39)}$  \\
					$$  &  $$  &  $-1$  &  $-1$  &  $\text{-0.035(14)}$  &  $\text{-0.278(13)}$  &  $\text{-0.169(41)}$  \\
					$$  &  $$  &  $1$  &  $0$  &  $\text{-0.056(25)}$  &  $\text{-0.325(16)}$  &  $\text{-0.145(34)}$  \\
					$$  &  $$  &  $-1$  &  $0$  &  $\text{-0.034(14)}$  &  $\text{-0.268(13)}$  &  $\text{-0.176(42)}$  \\
\hline
				\end{tabular}
				\caption{The dependence of the optimized observables corresponding to $\bar{B}(B)\to\rho^0 ll$ on the real and imaginary parts of the NP WC's $\Delta C_{9,10}$ and $C_{9,10}^\prime$ for the $q^2$ bins $2-4$ and $4-6$ $\text{GeV}^2$ for a few benchmark values of the real and imaginary parts of the corresponding WC's.}
				\label{tab:appBtorhozP1P4P6notag2}										
			\end{table}
			
			\begin{table}[ht]
				\centering
				\tiny
				\renewcommand*{\arraystretch}{1.2}
				\begin{tabular}{|c|c|c|c|c|c|c|c|}
					\hline
					\multirow{2}{*}{\textbf{Bin}} & \multirow{2}{*}{$C_i^{NP}$} &\multirow{2}{*}{$Re(C_i^{NP})$} &\multirow{2}{*}{$Im(C_i^{NP})$} & \multicolumn{4}{c|}{\textbf{Observables}}\\
					\cline{5-8}
					$$  &  $$  &  $$  &  $$  &  $A_5$  &  $A_{6s}$  &  $A_ 8$  &  $A_ 9$  \\\hline
					$\text{0.1-1}$  &  $\text{$\Delta $C}_ 9$  &  $1$  &  $1$  &  $\text{-0.0022(46)}$  &  $\text{0.0026(22)}$  &  $\text{0.0474(42)}$  &  $\text{0.00047(27)}$  \\
					$$  &  $$  &  $1$  &  $-1$  &  $\text{-0.0022(47)}$  &  $\text{0.0027(23)}$  &  $\text{-0.0221(47)}$  &  $\text{-0.00039(24)}$  \\
					$$  &  $$  &  $-1$  &  $1$  &  $\text{-0.0024(51)}$  &  $\text{0.0029(23)}$  &  $\text{0.0529(36)}$  &  $\text{0.00052(31)}$  \\
					$$  &  $$  &  $-1$  &  $-1$  &  $\text{-0.0024(52)}$  &  $\text{0.0029(25)}$  &  $\text{-0.0233(43)}$  &  $\text{-0.00042(25)}$  \\
					$$  &  $$  &  $1$  &  $0$  &  $\text{-0.0022(47)}$  &  $\text{0.0027(23)}$  &  $\text{0.0130(46)}$  &  $\text{0.000044(47)}$  \\
					$$  &  $$  &  $-1$  &  $0$  &  $\text{-0.0024(52)}$  &  $\text{0.0029(25)}$  &  $\text{0.0153(39)}$  &  $\text{0.000057(44)}$  \\
\hline
				\end{tabular}
				\caption{The dependence of the asymmetric observables corresponding to $\bar{B}(B)\to\rho^0 ll$ on the real and imaginary parts of the NP WC's $\Delta C_{9,10}$ and $C_{9,10}^\prime$ for the $q^2$ bins $0.1-1$ and $1-2$ $\text{GeV}^2$ for a few benchmark values of the real and imaginary parts of the corresponding WC's.}
				\label{tab:appBtorhozassymnotag1}						
			\end{table}
			
			\begin{table}[ht]
				\centering
				\tiny
				\renewcommand*{\arraystretch}{1.2}
				\begin{tabular}{|c|c|c|c|c|c|c|c|}
					\hline
					\multirow{2}{*}{\textbf{Bin}} & \multirow{2}{*}{$C_i^{NP}$} &\multirow{2}{*}{$Re(C_i^{NP})$} &\multirow{2}{*}{$Im(C_i^{NP})$} & \multicolumn{4}{c|}{\textbf{Observables}}\\
					\cline{5-8}
					$$  &  $$  &  $$  &  $$  &  $A_5$  &  $A_{6s}$  &  $A_ 8$  &  $A_ 9$  \\\hline
					$\text{2-4}$  &  $\text{$\Delta $C}_ 9$  &  $1$  &  $1$  &  $\text{0.0056(56)}$  &  $\text{0.00050(670)}$  &  $\text{0.0324(18)}$  &  $\text{0.00120(63)}$  \\
					$$  &  $$  &  $1$  &  $-1$  &  $\text{0.0056(57)}$  &  $\text{0.00048(676)}$  &  $\text{-0.0276(23)}$  &  $\text{-0.00106(57)}$  \\
					$$  &  $$  &  $-1$  &  $1$  &  $\text{0.0079(79)}$  &  $\text{0.00071(942)}$  &  $\text{0.0460(17)}$  &  $\text{0.00174(92)}$  \\
					$$  &  $$  &  $-1$  &  $-1$  &  $\text{0.0080(80)}$  &  $\text{0.00068(954)}$  &  $\text{-0.0390(25)}$  &  $\text{-0.00146(78)}$  \\
					$$  &  $$  &  $1$  &  $0$  &  $\text{0.0058(57)}$  &  $\text{0.00051(696)}$  &  $\text{0.0026(16)}$  &  $\text{0.000079(69)}$  \\
					$$  &  $$  &  $-1$  &  $0$  &  $\text{0.0083(83)}$  &  $\text{0.00073(995)}$  &  $\text{0.0039(18)}$  &  $\text{0.000160(98)}$  \\
\hline
				\end{tabular}
				\caption{The dependence of the asymmetric observables corresponding to $\bar{B}(B)\to\rho^0 ll$ on the real and imaginary parts of the NP WC's $\Delta C_{9,10}$ and $C_{9,10}^\prime$ for the $q^2$ bins $2-4$ and $4-6$ $\text{GeV}^2$ for a few benchmark values of the real and imaginary parts of the corresponding WC's.}
				\label{tab:appBtorhozassymnotag2}										
			\end{table}
			
			\begin{table}[ht]
				\centering
				\tiny
				\renewcommand*{\arraystretch}{1.2}
				\begin{tabular}{|c|c|c|c|c|c|c|}
					\hline
					\multirow{2}{*}{\textbf{Bin}} & \multirow{2}{*}{$C_i^{NP}$} &\multirow{2}{*}{$Re(C_i^{NP})$} &\multirow{2}{*}{$Im(C_i^{NP})$} & \multicolumn{3}{c|}{\textbf{Observables}}\\
					\cline{5-7}
					$$  &  $$  &  $$  &  $$  &  $\text{BR$\times $}10^9$  &  $F_L$  &  $R_{\rho}$\\\hline
					$\text{0.1-1}$  &  $\text{$\Delta $C}_ 9$  &  $1$  &  $1$  &  $\text{1.36(11)}$  &  $\text{0.348(28)}$  &  $\text{1.063(12)}$  \\
					$$  &  $$  &  $1$  &  $-1$  &  $\text{1.34(11)}$  &  $\text{0.340(28)}$  &  $\text{1.047(11)}$  \\
					$$  &  $$  &  $-1$  &  $1$  &  $\text{1.24(11)}$  &  $\text{0.238(23)}$  &  $\text{0.9705(65)}$  \\
					$$  &  $$  &  $-1$  &  $-1$  &  $\text{1.22(10)}$  &  $\text{0.228(23)}$  &  $\text{0.9538(72)}$  \\
					$$  &  $$  &  $1$  &  $0$  &  $\text{1.33(11)}$  &  $\text{0.338(28)}$  &  $\text{1.042(10)}$  \\
					$$  &  $$  &  $-1$  &  $0$  &  $\text{1.22(10)}$  &  $\text{0.226(22)}$  &  $\text{0.9490(76)}$  \\
\hline
				\end{tabular}
				\caption{The dependence of the Branching Ratio and the observables $F_L$ and $R_{\rho}$ corresponding to $\bar{B}(B)\to\rho^0 ll$  on the real and imaginary parts of the NP WC's $\Delta C_{9,10}$ and $C_{9,10}^\prime$ for the $q^2$ bins $0.1-1$ and $1-2$ $\text{GeV}^2$ for a few benchmark values of the real and imaginary parts of the corresponding WC's.}
				\label{tab:appBtorhozbrnotag1}										
			\end{table}
			\begin{table}[ht]
				\centering
				\tiny
				\renewcommand*{\arraystretch}{1.2}
				\begin{tabular}{|c|c|c|c|c|c|c|}
					\hline
					\multirow{2}{*}{\textbf{Bin}} & \multirow{2}{*}{$C_i^{NP}$} &\multirow{2}{*}{$Re(C_i^{NP})$} &\multirow{2}{*}{$Im(C_i^{NP})$} & \multicolumn{3}{c|}{\textbf{Observables}}\\
					\cline{5-7}
					$$  &  $$  &  $$  &  $$  &  $\text{BR$\times $}10^9$  &  $F_L$  &  $R_{\rho}$\\\hline
					$\text{2-4}$  &  $\text{$\Delta $C}_ 9$  &  $1$  &  $1$  &  $\text{1.38(13)}$  &  $\text{0.823(15)}$  &  $\text{1.2730(72)}$  \\
					$$  &  $$  &  $1$  &  $-1$  &  $\text{1.37(12)}$  &  $\text{0.823(15)}$  &  $\text{1.2619(75)}$  \\
					$$  &  $$  &  $-1$  &  $1$  &  $\text{0.978(82)}$  &  $\text{0.720(23)}$  &  $\text{0.9001(88)}$  \\
					$$  &  $$  &  $-1$  &  $-1$  &  $\text{0.966(80)}$  &  $\text{0.720(23)}$  &  $\text{0.8890(85)}$  \\
					$$  &  $$  &  $1$  &  $0$  &  $\text{1.33(12)}$  &  $\text{0.827(15)}$  &  $\text{1.2252(74)}$  \\
					$$  &  $$  &  $-1$  &  $0$  &  $\text{0.926(77)}$  &  $\text{0.721(23)}$  &  $\text{0.8523(79)}$  \\
\hline
				\end{tabular}
				\caption{The dependence of the Branching Ratio and the observables $F_L$  and $R_{\rho}$ corresponding to $\bar{B}(B)\to\rho^0 ll$ on the real and imaginary parts of the NP WC's $\Delta C_{9,10}$ and $C_{9,10}^\prime$ for the $q^2$ bins $2-4$ and $4-6$ $\text{GeV}^2$ for a few benchmark values of the real and imaginary parts of the corresponding WC's.}
				\label{tab:appBtorhozbrnotag2}										
			\end{table}
			
			\begin{table}[htbp]
				\centering
				\tiny
				\renewcommand*{\arraystretch}{1.2}
				\begin{tabular}{|c|c|c|c|c|c|c|c|}
					\hline
					\multirow{2}{*}{\textbf{Bin}} & \multirow{2}{*}{$C_i^{NP}$} &\multirow{2}{*}{$Re(C_i^{NP})$} &\multirow{2}{*}{$Im(C_i^{NP})$} & \multicolumn{4}{c|}{\textbf{Observables}}\\
					\cline{5-8}
					$$  &  $$  &  $$  &  $$  &  $A_{\text{CP}}$  &  $A_3$  &  $A_4$  &  $A_7$  \\
					\hline
					$\text{0.1-1}$  &  $\text{$\Delta $C}_9$  &  $1$  &  $1$  &  $\text{-0.0373(98)}$  &  $\text{0.000038(35)}$  &  $\text{-0.0037(28)}$  &  $ \text{-0.0023(27)}$  \\
					$$  &  $$  &  $1$  &  $-1$  &  $\text{-0.035(12)}$  &  $\text{-0.000048(35)}$  &  $\text{-0.0105(11)}$  &  $\text{-0.0023(28)}$  \\
					$$  &  $$  &  $-1$  &  $1$  &  $\text{-0.031(11)}$  &  $\text{0.000054(33)}$  &  $\text{-0.0037(23)}$  &  $\text{-0.0025(30)}$  \\
					$$  &  $$  &  $-1$  &  $-1$  &  $\text{-0.028(14)}$  &  $\text{-0.000040(30)}$  &  $\text{-0.01123(72)}$  &  $\text{-0.0026(30)}$  \\
					$$  &  $$  &  $1$  &  $0$  &  $\text{-0.036(11)}$  &  $\text{ -0.0000049(240)}$  &  $\text{-0.0072(21)}$  &  $\text{-0.0024(28)}$  \\
					$$  &  $$  &  $-1$  &  $0$  &  $\text{-0.030(13)}$  &  $\text{      0.0000079(149)}$  &  $\text{-0.0076(15)}$  &  $\text{-0.0026(31)}$  \\
\hline
					\end{tabular}
					\caption{The dependence of the asymmetric observables corresponding to $\bar{B}^0(B^0)\to\rho^0 ll$ which can be measured only at Belle on the real and imaginary parts of the NP WC's $\Delta C_{9,10}$ and $C_{9,10}^\prime$ for the $q^2$ bins $0.1-1$ and $1-2$ $\text{GeV}^2$ for a few benchmark values of the real and imaginary parts of the corresponding WC's.}
					\label{tab:appBtorhozACPnotag1}						
					\end{table}
					
					\begin{table}[htbp]
						\centering
						\tiny
						\renewcommand*{\arraystretch}{1.2}
						\begin{tabular}{|c|c|c|c|c|c|c|c|}
							\hline
							\multirow{2}{*}{\textbf{Bin}} & \multirow{2}{*}{$C_i^{NP}$} &\multirow{2}{*}{$Re(C_i^{NP})$} &\multirow{2}{*}{$Im(C_i^{NP})$} & \multicolumn{4}{c|}{\textbf{Observables}}\\
							\cline{5-8}	
							$$  &  $$  &  $$  &  $$  &  $A_{\text{CP}}$  &  $A_3$  &  $A_4$  &  $A_7$  \\
							\hline
							$\text{2-4}$  &  $\text{$\Delta $C}_9$  &  $1$  &  $1$  &  $\text{-0.0250(39)}$  &  $\text{0.00018(14)}$  &  $\text{0.0039(32)}$  &  $\text{-0.00016(84)}$  \\
							$$  &  $$  &  $1$  &  $-1$  &  $\text{-0.0083(23)}$  &  $\text{-0.000150(78)}$  &  $\text{-0.0037(10)}$  &  $\text{-0.00016(85)}$  \\
							$$  &  $$  &  $-1$  &  $1$  &  $\text{-0.0253(23)}$  &  $\text{0.00024(14)}$  &  $\text{0.0041(32)}$  &  $\text{-0.00022(119)}$  \\
							$$  &  $$  &  $-1$  &  $-1$  &  $\text{-0.0017(62)}$  &  $\text{-0.000230(94)}$  &  $\text{-0.00669(40)}$  &  $\text{-0.00022(119)}$  \\
							$$  &  $$  &  $1$  &  $0$  &  $\text{-0.0173(13)}$  &  $\text{0.000018(92)}$  &  $\text{0.00013(215)}$  &  $\text{-0.00016(87)}$  \\
							$$  &  $$  &  $-1$  &  $0$  &  $\text{-0.0143(23)}$  &  $\text{      0.0000082(685)}$  &  $\text{-0.0013(17)}$  &  $\text{-0.00023(125)}$  \\
\hline
						\end{tabular}
						\caption{The dependence of the asymmetric observables corresponding to $\bar{B^0}(B^0)\to\rho^0 ll$ which can be measured only at Belle on the real and imaginary parts of the NP WC's $\Delta C_{9,10}$ and $C_{9,10}^\prime$ for the $q^2$ bins $2-4$ and $4-6$ $\text{GeV}^2$ for a few benchmark values of the real and imaginary parts of the corresponding WC's.}
					\label{tab:appBtorhozACPnotag2}												
					\end{table}
					
					\begin{table}[htbp]
						\centering
						\tiny
						\renewcommand*{\arraystretch}{1.2}
						\begin{tabular}{|c|c|c|c|c|c|c|c|c|}
							\hline
							\multirow{2}{*}{\textbf{Bin}} & \multirow{2}{*}{$C_i^{NP}$} &\multirow{2}{*}{$Re(C_i^{NP})$} &\multirow{2}{*}{$Im(C_i^{NP})$} & \multicolumn{5}{c|}{\textbf{Observables}}\\
							\cline{5-9}	
							$$  &  $$  &  $$  &  $$  &  $A_{\text{FB}}$  &  $P_2$  &  $P_3$  &  $P_5'$  &  $P_8'$  \\
							\hline
							$\text{0.1-1}$  &  $\text{$\Delta $C}_9$  &  $1$  &  $1$  &  $\text{-0.0520(28)}$  &  $\text{-0.0532(17)}$  &  $\text{0.00029(19)}$  &  $\text{0.2665(44)}$  &  $\text{-0.0267(69)}$  \\
							$$  &  $$  &  $1$  &  $-1$  &  $\text{-0.0529(28)}$  &  $\text{-0.0534(16)}$  &  $\text{0.00028(19)}$  &  $\text{0.2722(46)}$  &  $\text{-0.0245(87)}$  \\
							$$  &  $$  &  $-1$  &  $1$  &  $\text{-0.0625(26)}$  &  $\text{-0.0547(17)}$  &  $\text{0.00015(10)}$  &  $\text{0.3898(41)}$  &  $\text{-0.0144(43)}$  \\
							$$  &  $$  &  $-1$  &  $-1$  &  $\text{-0.0636(25)}$  &  $\text{-0.0550(17)}$  &  $\text{0.00014(10)}$  &  $\text{0.4027(41)}$  &  $\text{-0.0113(64)}$  \\
							$$  &  $$  &  $1$  &  $0$  &  $\text{-0.0531(28)}$  &  $\text{-0.0535(17)}$  &  $\text{0.00028(19)}$  &  $\text{0.2738(45)}$  &  $\text{-0.0260(79)}$  \\
							$$  &  $$  &  $-1$  &  $0$  &  $\text{-0.0640(25)}$  &  $\text{-0.0551(17)}$  &  $\text{0.00014(10)}$  &  $\text{0.4064(40)}$  &  $\text{-0.0132(55)}$  \\
\hline
						\end{tabular}
						\caption{The dependence of the optimized observables corresponding to $\bar{B^0}(B^0)\to\rho^0 ll$ which can be measured only at Belle on the real and imaginary parts of the NP WC's $\Delta C_{9,10}$ and $C_{9,10}^\prime$ for the $q^2$ bins $0.1-1$ and $1-2$ $\text{GeV}^2$ for a few benchmark values of the real and imaginary parts of the corresponding WC's.}
						\label{tab:appBtorhozoptotag1}						
					\end{table}
					
					\begin{table}[htbp]
						\centering
						\tiny
						\renewcommand*{\arraystretch}{1.2}
						\begin{tabular}{|c|c|c|c|c|c|c|c|c|}
							\hline
							\multirow{2}{*}{\textbf{Bin}} & \multirow{2}{*}{$C_i^{NP}$} &\multirow{2}{*}{$Re(C_i^{NP})$} &\multirow{2}{*}{$Im(C_i^{NP})$} & \multicolumn{5}{c|}{\textbf{Observables}}\\
							\cline{5-9}	
							
							$$  &  $$  &  $$  &  $$  &  $A_{\text{FB}}$  &  $P_2$  &  $P_3$  &  $P_5'$  &  $P_8'$  \\
							\hline		
							$\text{2-4}$  &  $\text{$\Delta $C}_9$  &  $1$  &  $1$  &  $\text{0.0066(40)}$  &  $\text{0.025(15)}$  &  $\text{0.0037(21)}$  &  $\text{-0.380(18)}$  &  $\text{-0.047(10)}$  \\
							$$  &  $$  &  $1$  &  $-1$  &  $\text{0.0066(40)}$  &  $\text{0.025(15)}$  &  $\text{0.0037(22)}$  &  $\text{-0.384(17)}$  &  $\text{-0.044(13)}$  \\
							$$  &  $$  &  $-1$  &  $1$  &  $\text{-0.0809(88)}$  &  $\text{-0.193(10)}$  &  $\text{0.0019(11)}$  &  $\text{-0.030(16)}$  &  $\text{-0.0294(60)}$  \\
							$$  &  $$  &  $-1$  &  $-1$  &  $\text{-0.0820(88)}$  &  $\text{-0.1950(98)}$  &  $\text{0.0018(11)}$  &  $\text{-0.030(16)}$  &  $\text{-0.0262(89)}$  \\
							$$  &  $$  &  $1$  &  $0$  &  $\text{0.0068(41)}$  &  $\text{0.026(16)}$  &  $\text{0.0039(23)}$  &  $\text{-0.398(18)}$  &  $\text{-0.047(12)}$  \\
							$$  &  $$  &  $-1$  &  $0$  &  $\text{-0.0855(92)}$  &  $\text{-0.204(10)}$  &  $\text{0.0020(11)}$  &  $\text{-0.032(17)}$  &  $\text{-0.0292(78)}$  \\
\hline
							\end{tabular}
							\caption{The dependence of the optimized observables corresponding to $\bar{B^0}(B^0)\to\rho^0 ll$ which can be measured only at Belle on the real and imaginary parts of the NP WC's $\Delta C_{9,10}$ and $C_{9,10}^\prime$ for the $q^2$ bins $2-4$ and $4-6$ $\text{GeV}^2$ for a few benchmark values of the real and imaginary parts of the corresponding WC's.}
							\label{tab:appBtorhozoptotag2}						
							\end{table}

	\begin{table}[t]
		\centering
		\tiny
		\renewcommand*{\arraystretch}{1.2}
		\begin{tabular}{|c|c|c|c|c|c|c|c|}
			\hline
			\multirow{2}{*}{\textbf{Bin}} & \multirow{2}{*}{$C_i^{NP}$} &\multirow{2}{*}{$Re(C_i^{NP})$} &\multirow{2}{*}{$Im(C_i^{NP})$} & \multicolumn{4}{c|}{\textbf{Observables}}\\
			$$  &  $$  &  $$  &  $$  &  $A_I^{\bar{0}-}$  &  $A_I^{0+}$  &  $A_I$  &  $A_I^{\text{av}}$  \\
			\hline
			$\text{0.1-1}$  &  $\text{$\Delta $C}_9$  &  $1$  &  $1$  &  $\text{0.095(10)}$  &  $\text{-0.199(14)}$  &  $\text{-0.0517(75)}$  &  $\text{0.181(23)}$  \\
			$$  &  $$  &  $1$  &  $-1$  &  $\text{0.097(10)}$  &  $\text{-0.187(14)}$  &  $\text{-0.0451(76)}$  &  $\text{0.160(22)}$  \\
			$$  &  $$  &  $-1$  &  $1$  &  $\text{0.0045(85)}$  &  $\text{-0.165(14)}$  &  $\text{-0.0801(97)}$  &  $\text{0.203(26)}$  \\
			$$  &  $$  &  $-1$  &  $-1$  &  $\text{0.0048(82)}$  &  $\text{-0.150(14)}$  &  $\text{-0.0724(95)}$  &  $\text{0.180(25)}$  \\
			$$  &  $$  &  $1$  &  $0$  &  $\text{0.097(11)}$  &  $\text{-0.195(14)}$  &  $\text{-0.0487(76)}$  &  $\text{0.173(23)}$  \\
			$$  &  $$  &  $-1$  &  $0$  &  $\text{0.0047(84)}$  &  $\text{-0.159(14)}$  &  $\text{-0.0772(97)}$  &  $\text{0.194(26)}$  \\
\hline
		\end{tabular}
		\caption{The dependence of the isospin observables corresponding to $B\to\rho ll$ on the real and imaginary parts of the NP WC's $\Delta C_{9,10}$ and $C_{9,10}^\prime$ for the $q^2$ bins $0.1-1$ and $1-2$ $\text{GeV}^2$ for a few benchmark values of the real and imaginary parts of the corresponding WC's.}
		\label{tab:isospinNP1}			
	\end{table}
	
	\begin{table}[t]
		\centering
		\tiny
		\renewcommand*{\arraystretch}{1.2}
		\begin{tabular}{|c|c|c|c|c|c|c|c|}
			\hline
			\multirow{2}{*}{\textbf{Bin}} & \multirow{2}{*}{$C_i^{NP}$} &\multirow{2}{*}{$Re(C_i^{NP})$} &\multirow{2}{*}{$Im(C_i^{NP})$} & \multicolumn{4}{c|}{\textbf{Observables}}\\
			$$  &  $$  &  $$  &  $$  &  $A_I^{\bar{0}-}$  &  $A_I^{0+}$  &  $A_I$  &  $A_I^{\text{av}}$  \\
			\hline
			$\text{2-4}$  &  $\text{$\Delta $C}_9$  &  $1$  &  $1$  &  $\text{0.0813(50)}$  &  $\text{-0.0837(62)}$  &  $\text{-0.0012(26)}$  &  $\text{0.0228(60)}$  \\
			$$  &  $$  &  $1$  &  $-1$  &  $\text{0.0252(30)}$  &  $\text{-0.1158(75)}$  &  $\text{-0.0453(41)}$  &  $\text{0.108(10)}$  \\
			$$  &  $$  &  $-1$  &  $1$  &  $\text{0.0660(30)}$  &  $\text{-0.0656(57)}$  &  $\text{0.00023(236)}$  &  $\text{0.0136(57)}$  \\
			$$  &  $$  &  $-1$  &  $-1$  &  $\text{-0.0086(35)}$  &  $\text{-0.1118(75)}$  &  $\text{-0.0602(51)}$  &  $\text{0.135(12)}$  \\
			$$  &  $$  &  $1$  &  $0$  &  $\text{0.0542(37)}$  &  $\text{-0.1027(70)}$  &  $\text{-0.0243(32)}$  &  $\text{0.0677(83)}$  \\
			$$  &  $$  &  $-1$  &  $0$  &  $\text{0.0282(23)}$  &  $\text{-0.0926(68)}$  &  $\text{-0.0322(38)}$  &  $\text{0.0776(94)}$  \\
\hline
		\end{tabular}
		\caption{The dependence of the isospin observables corresponding to $B\to\rho ll$ on the real and imaginary parts of the NP WC's $\Delta C_{9,10}$ and $C_{9,10}^\prime$ for the $q^2$ bins $2-4$ and $4-6$ $\text{GeV}^2$ for a few benchmark values of the real and imaginary parts of the corresponding WC's.}
		\label{tab:isospinNP2}				
	\end{table}

	\begin{table}
		\centering
		\begin{tabular}{*{4}{|c|}}
			\hline
			$\left.\text{Bin[}\text{GeV}^2\right]$  & $(\eta_c, \eta_c^{\prime})$  & $\text{[2-4]}$  &  $\text{[4-6]}$  \\
			\hline
			$B(B^0\to\pi^0\ell \ell) \times 10^8$ & (-1.25, -1.25)   & 0.0663(33) & 0.0609(28) \\
			$B(\bar{B}^0\to\pi^0\ell \ell) \times 10^8$ & (-1.25, -1.25)   &  0.0496(27) & 0.0486(25) \\
			$B(B^-\to\pi^-\ell \ell) \times 10^8$ &  (-1.25, -1.25)   & 0.0941(57) & 0.0982(55) \\
			$B(B^+\to\pi^+\ell \ell) \times 10^8$ & (-1.25, -1.25)   & 0.1610(71) & 0.1333(64) \\
			\hline
		\end{tabular}
		\caption{\small Predictions of branching fractions for $B^{\pm} \to \pi^{\pm} \ell \ell$ decays on rescaling the charm vacuum polarisation.}
		\label{tab:charmscalingBR}
	\end{table}
	\begin{table}
		\centering
		\begin{tabular}{*{4}{|c|}}
			\hline
			$\left.\text{Bin[}\text{GeV}^2\right]$  &  $(\eta_c, \eta_c^{\prime})$  &  $\text{[2-4]}$  &  $\text{[4-6]}$  \\
			\hline
			$\langle P_1 \rangle$ & (0, -2.5)   &   0.101(22) & -0.150(31)   \\
			$\langle P_2 \rangle$ & (-2.5, 0)   & -0.387(13) & -0.181(20) \\
			$\langle P_3 \rangle$ & (0, -2.5)   & 0.050(14) & 0.066(19) \\
			$\langle P_5^{'} \rangle$ &  (-2.5, 0)   & 0.158(27) & -0.106(27) \\
			\hline
		\end{tabular}
		\caption{\small Predictions of observables for $B^{\pm} \to \rho^{\pm} \ell \ell$ decays on rescaling the charm vacuum polarisation by a factor of -2.5.}
		\label{tab:charmscalingCPav}
	\end{table}

							\pagebreak

\bibliographystyle{JHEP}
\bibliography{ref_ASSI}

\end{document}